%% file: phd.tex
\documentclass[layoutnew]{tktla}

\usepackage{hyperref}
\usepackage{mine}

\newcommand{\NP}{\textsc{NP}}
\newcommand{\ZPP}{\textsc{ZPP}}
\newcommand{\APX}{\textsc{APX}}

\newcommand{\degr}[1]{\mathit{deg}(#1)}
\newcommand{\prv}[1]{\mathit{prev}\left[#1\right]}
\newcommand{\nxt}[1]{\mathit{next}\left[#1\right]}
\newcommand{\rk}[1]{\mathit{rank}(#1)}
\newcommand{\lbl}[1]{\mathit{label}\ifthenelse{\equal{#1}{}}{}{(#1)}}
\newcommand{\Nbr}[2]{\mathit{N}_{#2}(#1)}

\newcommand{\Apriori}{\textsc{Apriori}}

\newcommand{\DB}{\mathcal{D}}
\newcommand{\Pat}{\mathcal{P}}
\newcommand{\SubPat}{\mathcal{S}}
\newcommand{\Chain}{\mathcal{C}}
\newcommand{\AntiChain}{\mathcal{A}}
\newcommand{\Items}{\mathcal{I}}
\newcommand{\Rules}{\mathcal{R}}
\newcommand{\Max}[1]{\mathit{Max}(#1)}
\newcommand{\Min}[1]{\mathit{Min}(#1)}
\newcommand{\Cl}[1]{\mathit{Cl}(#1)}
\newcommand{\Gen}[1]{\mathit{Gen}(#1)}

\newcommand{\Fr}{\mathcal{F}}
\newcommand{\Mx}{\mathcal{M}}
\newcommand{\Maximal}[1]{\mathcal{FM}(#1)}
\newcommand{\Minimal}[1]{\mathcal{IM}(#1)}
\newcommand{\Free}[1]{\mathcal{FG}(#1)}
\newcommand{\Closed}[1]{\mathcal{FC}(#1)}
\newcommand{\Frequent}[1]{\mathcal{F}(#1)}
\newcommand{\NDI}[1]{\mathcal{N}(#1)}
\newcommand{\Ch}{\mathcal{C}h}
\newcommand{\Candidates}{\mathcal{K}}
\newcommand{\Maximum}[1]{\mathit{M}(#1)}

\newcommand{\ipredicate}{\mathit{q}}
\newcommand{\imeasure}{\phi}
\newcommand{\ipred}[1]{\ipredicate(#1)}
\newcommand{\imeas}[1]{\imeasure(#1)}

\newcommand{\simplification}{\psi}
\newcommand{\simp}[1]{\simplification(#1)}
\newcommand{\estimation}{\psi}
\newcommand{\est}[2]{\estimation_{#2}\ifthenelse{\equal{#1}{}}{}{(#1)}}

\newcommand{\tid}[1]{\mathit{tid}(#1)}
\newcommand{\eqmap}[1]{\pi\ifthenelse{\equal{#1}{}}{}{(#1)}}
\newcommand{\tile}[1]{\tau(#1)}
\newcommand{\area}[1]{\mathit{area}(#1)}

\newcommand{\Tiling}{\mathcal{T}}

\newcommand{\discretization}{\gamma}
\newcommand{\disc}[3]{\discretization^{#2}_{#3}\ifthenelse{\equal{#1}{}}{}{(#1)}}
\newcommand{\error}{\ell}
\newcommand{\err}[2]{\error_{#2}(#1)}
\newcommand{\weight}{\mathit{w}}
\newcommand{\wg}[1]{\weight\ifthenelse{\equal{#1}{}}{}{(#1)}}
\newcommand{\funct}{\mathit{f}}
\newcommand{\fun}[1]{\funct(#1)}
\newcommand{\distance}{\mathit{d}}
\newcommand{\dist}[1]{\distance(#1)}

\newcommand{\freq}{\mathit{fr}}
\newcommand{\fr}[1]{\freq(#1)}
\newcommand{\fru}[1]{\overline{\freq}\ifthenelse{\equal{#1}{}}{}{(#1)}}
\newcommand{\frl}[1]{\underline{\freq}\ifthenelse{\equal{#1}{}}{}{(#1)}}
\newcommand{\cover}[1]{\mathit{cover}\ifthenelse{\equal{#1}{}}{}{(#1)}}
\newcommand{\support}{\mathit{supp}}
\newcommand{\supp}[1]{\support(#1)}
\newcommand{\acc}[1]{\mathit{acc}\ifthenelse{\equal{#1}{}}{}{(#1)}}
\newcommand{\cl}[1]{\mathit{cl}\ifthenelse{\equal{#1}{}}{}{(#1)}}
\newcommand{\assoc}[2]{#1 \Rightarrow #2}
\newcommand{\pr}[1]{\mathit{pr}(#1)}
\newcommand{\sz}[1]{\mathit{s}\ifthenelse{\equal{#1}{}}{}{(#1)}}
\newcommand{\bin}[1]{\mathit{bin}(#1)}
\newcommand{\occ}[1]{\mathit{occ}(#1)}
\newcommand{\cnt}[1]{\mathit{count}(#1)}

\newcommand{\ch}[3]{\mathit{ch}^{#2}_{#3}\ifthenelse{\equal{#1}{}}{}{(#1)}}
\newcommand{\cch}[3]{\mathit{cch}^{#2}_{#3}\ifthenelse{\equal{#1}{}}{}{(#1)}}
\newcommand{\sch}[3]{\mathit{sch}^{#2}_{#3}\ifthenelse{\equal{#1}{}}{}{(#1)}}

\newcommand{\exend}{\hfill$\Box$}

\title{Summarization Techniques for Pattern Collections in Data Mining}
\author{Taneli Mielik{\"a}inen}
\authorcontact{Taneli.Mielikainen@iki.fi\par
  http://www.iki.fi/Taneli.Mielikainen/}
\pubtime{April}{2005}
\reportno{1}
\isbnpaperback{952-10-2436-4}
\isbnpdf{952-10-2437-2}
\issn{1238-8645}
\printhouse{Helsinki University Printing House}
\pubpages{\pageref{lastpage}}
\generalterms{Algorithms, Theory, Experimentation}
\additionalkeywords{Pattern Discovery, Condensed Representations of Pattern Collections, Post-Processing of Data Mining Results}
\crcshort{E.4, H.2.8, I.2, I.2.4}
\crclong{
 \item[E.4] Coding and Information Theory: Data Compaction and Compression
 \item[H.2.8] Database Applications: Data Mining
 \item[I.2] Artificial Intelligence
 \item[I.2.4] Knowledge Representation Formalisms and Methods 
}

\permissionnotice{ To be presented, with the permission of the Faculty
 of Science of the University of Helsinki, for public criticism in
 Auditorium B123, Exactum, on May 27th, 2005, at noon.  }

\begin{document}

\frontmatter

\maketitle

\include{abstract}

\include{acknowledgements}

\tableofcontents

\mainmatter

%
%

\include{introduction}

\include{discovery}
\include{views}
\include{tradeoffs}
\include{chains}
\include{profiles}
\include{ipd}
\include{conclusions}

%
%

\bibliographystyle{alpha}
\bibliography{mine}

\label{lastpage}

\backmatter\appendix

\include{asarja}

\end{document}

%% file: abstract.tex
\begin{abstract}
Discovering patterns from data is an important task in data mining.
There exist techniques to find large collections of many kinds of
patterns from data very efficiently. A collection of patterns can be
regarded as a summary of the data. A major difficulty with patterns is
that pattern collections summarizing the data well are often very
large.

In this dissertation we describe methods for summarizing pattern
collections in order to make them also more understandable.  More
specifically, we focus on the following themes:

\begin{description}
\item[Quality value simplifications.]  We study simplifications of
pattern collections based on simplifying the quality values of the
patterns. Especially, we study simplification by discretization.

\item[Pattern orderings.]  It is difficult to find a suitable
trade-off between the accuracy of the representation and its size. As
a solution to this problem, we suggest that patterns could be ordered
in such a way that each prefix of the \emph{pattern ordering} gives a
good summary of the whole collection.

\item[Pattern chains and antichains.]  Virtually all pattern
collections have natural underlying partial orders.  We exploit the
partial orders over pattern collections by clustering the patterns
into chains and antichains.

\item[Change profiles.]  We describe how patterns can be related to
each other by comparing how their quality values change with respect
to their common neighborhoods, i.e., by comparing their \emph{change
profiles}.

\item[Inverse pattern discovery.]  As the patterns are often used to
summarize data, it is natural to ask whether the original data set can
be deduced from the pattern collection.  We study the computational
complexity of such problems.
\end{description}

\end{abstract}

%% file: acknowledgements.tex
\begin{acknowledgements}
I am most grateful to my supervisors Heikki Mannila and Esko Ukkonen
for their tireless guidance, patience and encouragement throughout my
studies.  Their insightful comments were most useful to help me
improve the thesis manuscript considerably.  I wish to thank also
Jean-Fran\c{c}ois Boulicaut and Dimitrios Gunopulos for serving as the
reviewers of the manuscript of the thesis and for their helpful
comments.

The studies have been carried out at the Department of Computer Science
of the University of Helsinki that has provided me an excellent
working environment. I wish to thank especially the computing
facilities staff of the department for ensuring the fluent operation
of the computing facilities and for their endurance to help me with my
numerous problems.  Financial support by Helsinki Graduate School for
Computer Science and Engineering, From Data to Knowledge research unit,
and HIIT Basic Research Unit are gratefully acknowledged.

Many friends and colleagues have provided me invaluable support.
Especially Floris Geerts, Bart Goethals, Matti K{\"a}{\"a}ri{\"a}inen,
Ari Rantanen and Janne Ravantti have hopefully affected the
dissertation considerably by their insightful comments and refreshing
discussions about the topic.  Also the intellectual support by Tapio
Elomaa, Patrik Flor{\'e}en and Hannu Toivonen were most valuable.

I am most indebted to my parents Marketta and Kari for their support
and encouragement.  Their frequent queries about my studies and the
state of the dissertation were most helpful.  The cultural support
provided by Antti Nissinen is greatly appreciated.  My deepest love
and gratitude belong to Satu, Iida and Lauri. Their unconditional
support and love were vital also for this work.
\end{acknowledgements}

%% file: introduction.tex

\chapter{Introduction \label{c:i}}


\begin{quotation}
``But what kind of authentic and valuable information do you
require?'' asked Klapaucius.

``All kinds, as long as it's true'', replied the pirate. ``You never
can tell what facts may come in handy. I already have a few hundred
wells and cellars full of them, but there's room for twice again as
much. So out with it; tell me everything you know, and I'll jot it
down. But make it snappy!''  
\begin{flushright}
{\em Stanislaw Lem: The Cyberiad (1974)}
\end{flushright}
\end{quotation}

Mankind has achieved an impressive ability to store
data~\cite{a:riedel03}. The capacity of digital data storage has
doubled every nine months for at least a decade~\cite{a:fayyad02}.
Furthermore, our skills and interest to collect data are also
remarkable~\cite{m:lyman03}.

Our ability to process the collected data is not so impressive. In
fact, there is a real danger that we construct \emph{write-only data
stores} that cannot be exploited using current
technologies~\cite{a:fayyad02}. Besides constructing \emph{data tombs}
that contain snapshots of our world for the tomb raiders of the
forthcoming generations, this is not very useful.  It can be said that
we are in a \emph{data rich but information poor}
situation~\cite{b:han01}.

In addition to the immense amount of data being collected, the data is
becoming increasingly complex and diverse~\cite{a:fayyad01,a:smyth02}:
companies collect data about their customers to maximize their expected
profit~\cite{a:kohavi02}, scientists gather large repositories of
observations to better understand nature~\cite{a:han02} and
governments of many countries are collecting vast amounts of data to
ensure the homeland security which has been recognized to be a very
important issue due to the globalization of conflicts and
terrorism~\cite{a:yen04}. When several different data repositories are
combined, the data concerning even only a single person can be
tremendously large and complex.

Due to the weakness of the current techniques to exploit large data
repositories and the complexity of the data being collected, a new
discipline known as \emph{data mining} is emerging in the intersection
of artificial intelligence, databases and statistics. The current
working definition of this new field is the following~\cite{b:hand01}:

\begin{quote}
Data mining is the analysis of (often large) observational data sets
to find unsuspected relationships and to summarize the data in novel
ways that are both understandable and useful to the data owner.
\end{quote}

On one hand this definition is acceptable for a large variety of data
mining scholars. On the other hand its interpretation depends on
several imprecise concepts: The meanings of the words 'unsuspected',
'understandable' and 'useful' depend on the context. Also the words
'relationships' and 'summarize' have vast number of different
interpretations. This indeterminacy in general seems to be inherent to
data mining since the actual goal is in practice determined by the
task at hand.

Albeit the inherent vagueness of the definition, the field of data
mining can be elucidated by arranging the techniques to groups of
similar approaches. The techniques can be divided roughly to two
parts, namely to \emph{global} and \emph{local} methods.

Global methods concern constructing and manipulating \emph{global
models} that describe the entire data.  Global models comprise most of
the classical statistical methods. For example, the Gaussian
distribution function is a particularly well-known global model for
real-valued data. The focus in the data mining research of global
methods has been on developing and scaling up global modeling
techniques to very large data sets.

Local methods focus on \emph{discovering patterns} from data. Patterns
are parsimonious summaries of subsets of data~\cite{a:fayyad02}. The
rule ``People who buy diapers tend to buy beer'' is a classical
example of such pattern. In contrast to global modeling approach,
pattern discovery as a discipline in its own right is relatively
new~\cite{i:hand02}.  (The term 'discovery' has recently been
criticized in the context of data mining to be misleading since data
mining is based on scientific principles and it can be argued that
science does not discover facts by induction but rather invents
theories that are then checked against experience~\cite{i:piscopo02}.
The term is used, however, in this dissertation because of its
established use in data mining literature.)

The global and local methods can be summarized in the following way.
The global modeling approach views data mining as the task of
\emph{approximating the joint probability distribution} whereas the
pattern discovery can be summarized in the slogan: \emph{data mining
is the technology of fast counting}~\cite{i:mannila02}.

The distinction to global models and local patterns is not strict.
Although a Gaussian distribution is usually considered as a global model,
it can be also a pattern: each Gaussian distribution in a mixture of
Gaussians is assumed to describe only a part of the data.

This work focuses on pattern discovery. There exist effective
techniques to discover many kinds of
patterns~\cite{p:fimi03,a:mannila97:levelwise}. Due to that fact the
question of how the discovered patterns could actually be exploited is
becoming increasingly important. Often the answer to that question is
tightly coupled with the particular application. Many problems,
obstacles and characteristics, however, are shared with different
applications.

A very important application of patterns is to summarize given data as
a collection of patterns, possibly augmented with some auxiliary
information such as the quality values of the patterns.
Unfortunately, often the size of the pattern collection that
faithfully represents the aspects of the data considered to be
relevant is very large.  Thus, in addition to data tombs, there is a
risk of constructing also \emph{pattern tombs}.

\section{The Contributions and the Organization}

The main purpose of this dissertation is to study how to summarize
pattern collections by exploiting the structure of the collections and
the quality values of the patterns. The rest of the dissertation is
organized as follows.

\begin{description}
\item[Chapter~\ref{c:discovery}] provides an introduction to pattern
discovery that is sufficient to follow the rest of the dissertation.
It contains a systematic derivation of a general framework for pattern
discovery, a brief overview of the current state of pattern discovery
and descriptions of the most important (condensed) representations of
pattern collections. Furthermore, some technical challenges of pattern
exploitation are briefly discussed.

\item[Chapter~\ref{c:views}] concerns simplifying pattern collections
by simplifying the quality values of the patterns.  The only
assumption needed about the pattern collection is that there is a
quality value associated to each pattern.

We illustrate the idea of constraining the quality values of the
patterns by discretizing the frequencies of frequent itemsets. We
examine the effect of discretizing frequencies to the accuracies of
association rules and propose algorithms for computing optimal
discretizations with respect to several loss functions.  We show
empirically that discretizations with quite small errors can reduce
the representation of the pattern collection considerably.

\item[Chapter~\ref{c:tradeoffs}] focuses on trade-offs between the
size of the pattern collection and its accuracy to describe the data.
The chapter suggests to order the patterns by their abilities to
describe the whole pattern collection with respect to a given loss
function and an estimation method. The obtained ordering is a refining
description of the pattern collection and it requires only a loss
function and an estimation method.

We show that for several pairs of loss functions and estimation
methods, the most informative $k$-subcollection of the patterns can be
approximated within a constant factor by the $k$-prefix of the pattern
ordering for all values of $k$ \emph{simultaneously}.  We illustrate
the pattern orderings by refining approximations closed itemsets and
tilings of transaction databases. We evaluate the condensation
abilities of the pattern orderings empirically by computing refining
approximations of closed frequent itemsets. The results show that
already short prefixes of the orderings of the frequent itemsets are
sufficient to provide reasonably accurate approximations.

\item[Chapter~\ref{c:chains}] is motivated by the fact that a pattern
collection has usually some structure apart from the quality values of
the patterns. Virtually all pattern collections have non-trivial
partial orders over the patterns.  In this chapter we suggest the use
of minimum chain and antichain partitions of partially ordered pattern
collections to figure out the essence of a given pattern collection.

For an arbitrary pattern collection, its chain and antichain
partitions provide clusterings of the collection. The benefit from the
chain partition can be even greater: for many known pattern
collections, each chain in the partition can be described as a single
pattern.  The chain partitions give a partially negative answer to the
question whether a random sample of the data is essentially the best
one can hope.  We evaluate empirically the ability of pattern chains
to condense pattern collections in the case of closed frequent itemset
collections.

\item[Chapter~\ref{c:profiles}] introduces a novel approach to relate
patterns in a pattern collection to each other: patterns are
considered similar if their change profiles are similar, i.e., if
their quality values change similarly with respect to their common
neighbors in a given neighborhood relation. This can be seen as an
attempt to bridge the gap between local and global descriptions of the
data.

A natural way of using similarities is the clustering of
patterns. Unfortunately, clustering based on change profiles turns out
to be computationally very difficult. Because of that, we discuss
advantages and disadvantages of different heuristic approaches to
cluster patterns using change profiles. Furthermore, we demonstrate
that change profiles can determine meaningful (hierarchical)
clusterings.  In addition to examining the suitability of change
profiles for comparing patterns, we propose two algorithms for
estimating the quality values of the patterns from their approximate
change profiles.  To see how the approximate change profiles affect
the estimation of the quality values of the patterns, the stability of
the frequency estimates of the frequent itemsets is empirically
evaluated with respect to different kinds of noise.

\item[Chapter~\ref{c:ipd}] studies the problems of inverse pattern
discovery, i.e., finding data that could have generated the
patterns. In particular, the main task considered in the chapter is to
decide whether there exists a database that has the correct
frequencies for a given itemset collection. This question is relevant
in, e.g., privacy-preserving data mining, in quality evaluation of
pattern collections, and in inductive data\-bases.  We show that many
variants of the problem are \NP-hard but some non-trivial special
cases have polynomial-time algorithms.

\item[Chapter~\ref{c:conclusions}] concludes this dissertation. 
\end{description}

%% file: discovery.tex

\chapter{Pattern Discovery \label{c:discovery}}


This chapter provides an introduction to pattern discovery, one of the
two main sub-disciplines of data mining, and its central concepts that
are used through and through this dissertation. A general framework is
derived for pattern discovery, the most important condensed
representations of pattern collections are introduced and the purpose
of patterns in shortly discussed.

\section{The Pattern Discovery Problem}

The goal in pattern discovery is to find interesting patterns from
given data~\cite{i:hand02,i:mannila02}. The task can be defined more
formally as follows:

\begin{problem}[pattern discovery]
Given a class $\Pat$ of patterns and an interestingness predicate
$\ipredicate : \Pat \to \Set{0,1}$ for the pattern class, find the
collection
\begin{displaymath}
\Pat_{\ipredicate}=\Set{p \in \Pat : \ipred{p}=1}
\end{displaymath}
of \emph{interesting} patterns.  Its complement
$\Pat_{\bar{\ipredicate}}=\Pat \setminus
\Pat_{\ipredicate}$ is called the collection of
\emph{uninteresting patterns} in $\Pat$ with respect to $\ipredicate$.
\end{problem}

The pattern discovery problem as defined above consists of only two
parts: the collection $\Pat$ of possibly interesting patterns and the
interestingness predicate $\ipredicate$.

The pattern collection $\Pat$ constitutes \emph{a priori} assumptions
of which patterns could be of interest. The collection $\Pat$ is
usually not represented explicitly since its cardinality can be very
large, sometimes even infinite. For example, the collection of
patterns could consist of all regular expressions over a given
alphabet $\Sigma$. (For an introduction to regular expressions, see
e.g.~\cite{b:hopcroft01}.) This collection is infinite even for the
unary alphabet.

The absence of data from the definition might be a bit confusing at
first. It is omitted on purpose: Often the interestingness predicate
depends on data and the data is usually given as a parameter for the
predicate. This is not true in every case, however, since the
\emph{interestingness} (or, alternatively, the \emph{quality}) of a
pattern can be determined by an expert who has specialized to some
particular data set and the interestingness predicate might be useless
for any other data set regardless of its form. For example, a company
offering industrial espionage that is specialized to investigate power
plants can be rather poor detecting interesting patterns from
gardening data.

Defining a reasonable interestingness predicate is usually a highly
non-trivial task: the interestingness predicate should capture most
truly interesting patterns and only few uninteresting ones. 

Due to these difficulties, a relaxation of an interestingness predicate,
an \emph{interestingness measure}
\begin{displaymath}
\imeasure : \Pat \to [0,1]
\end{displaymath}
expressing the quantitative value $\imeas{p}$ of the interestingness
(or the quality) for each pattern $p \in \Pat$ is used instead of an
interestingness predicate. In this dissertation the value $\imeas{p}$
of $p \in \Pat$ is called the \emph{quality} value of $p$ with respect
to the interestingness measure $\imeasure$, or in short: the quality
of $p$.  Many kinds of interestingness measures have been studied in
the literature, see e.g.~\cite{i:tan02}.

\begin{example}[an interestingness measure]
Let the data set consist of names of recently born children and their
ages (that are assumed to be strictly positive), i.e., let $\DB$ be a
set of pairs $\Tuple{\mathit{name},\mathit{age}} \in \Sigma^* \times \RN_+$.

An interestingness measure $\imeasure$ for the pattern class
$\Pat_{\mathrm{regexp}}$ consisting of all regular expressions could
be defined as follows. Let $\Restrict{\DB}{p}$ be the group of
children whose names satisfy the regular expression $p \in
\Pat_{\mathrm{regexp}}$.  The quality of a pattern $p \in
\Pat_{\mathrm{regexp}}$ is the smallest age of any child in $\DB$
divided by the average ages the children whose names belong to the
regular language $p$, i.e.,
\begin{displaymath} 
\imeas{p,\DB}=\frac{\min \Set{ \mathit{age} :
\Tuple{\mathit{name},\mathit{age}} \in
\DB}}{\Paren{\sum_{\Tuple{\mathit{name},\mathit{age}} \in
\Restrict{\DB}{p}} \mathit{age}}/\Abs{\Set{ \mathit{age} :
\Tuple{\mathit{name},\mathit{age}} \in \Restrict{\DB}{p}}}}.
\end{displaymath}
\exend \end{example}

There are many reasons why interestingness measures are favored over
interestingness predicates. An important reason is that it is often
easier to suggest some degrees of interestingness for the patterns in
the given collection than to partition the patterns into the groups of
strictly interesting and uninteresting ones. In fact, using an
interestingness measure, instead of an interestingness predicate,
partially postpones the difficulty of fixing a suitable
interestingness predicate, since an interestingness measure implicitly
determines an infinite number of interestingness predicates:
\begin{displaymath}
\ipred{p}=
\left\{
\begin{array}{l l}
1 \quad \quad & \mbox{if } \imeas{p} \geq \sigma \\
0 & \mbox{otherwise}.
\end{array}
\right.
\end{displaymath}

In addition to these practical reasons, there are also some more
foundational arguments that support the use of interestingness
measures instead of predicates.  Namely, it can be argued that the
actual goal in pattern discovery is not merely to find a collection of
interesting patterns but to rank the patterns with respect to their
quality values~\cite{i:mielikainen04:idb}. Also, due to the
exploratory nature of data mining, it might not be wise to completely
discard the patterns that seem to be uninteresting, since you never can
tell what patterns may come in handy.  Instead, it could be more useful
just to list the pattern in decreasing order with respect to their
quality values.

Also the interestingness predicates can be defended against the
interestingness measures.  The interestingness predicates determine
collections of patterns whereas the interestingness measures determine
rankings (or gradings). On one hand, the interestingness predicates
can be manipulated and combined by boolean connectives. Furthermore,
the manipulations have direct correspondents in the pattern
collections. Combining rankings corresponding to interestingness
measures, on the other hand, is not so straightforward.

Thus, the interestingness predicates and the interestingness measures
have both strong and weak points. Due to this, the majority of pattern
discovery research has been focused on the combination of
interestingness measures and predicates: they consider discovering
collections of interesting patterns augmented by their quality values.

\section{Frequent Itemsets and Association Rules \label{s:frequent}}

The most prominent example of pattern discovery is discovering (or
mining) \emph{frequent itemsets} from \emph{transaction
databases}~\cite{i:agrawal93,i:mannila02}. 
\begin{definition}[items and itemsets]
A set of possible \emph{items} is denoted by $\Items$.  An
\emph{itemset} $X$ is a subset of $\Items$.  For brevity, an itemset
$X$ consisting items $A_1,A_2,\ldots,A_{\Abs{X}}$ can be written
$A_1A_2\ldots A_{\Abs{X}}$ instead of
$\Set{A_1,A_2,\ldots,A_{\Abs{X}}}$.
\end{definition}

\begin{definition}[transactions and transaction databases] \label{d:tdb}
A \emph{transaction} $t$ is a pair $\Tuple{i,X}$ where $i$ is
a \emph{transaction identifier (tid)} and $X$ is an itemset. The
number of items in the itemset $X$ of a transaction $t=\Tuple{i,X}$ is
denoted by $\Abs{t}$.  

A \emph{transaction database} $\DB$ is a set of transactions.  Each
transaction in $\DB$ has a unique transaction identifier.  The number of
transactions in the transaction database $\DB$ is denoted by $\Abs{\DB}$
and the set of transaction identifiers in $\DB$ by $\tid{\DB}=\Set{i :
\Tuple{i,X} \in \DB}$.  In the context of this dissertation it is
assumed, without loss of generality, that
$\tid{\DB}=\Set{1,\ldots,\Abs{\DB}}$.

The set of occurrences of an itemset $X$ in $\DB$ is the set
\begin{displaymath}
\occ{X,\DB}=\Set{i : \Tuple{i,X} \in \DB}
\end{displaymath}
of transaction identifiers of the transactions $\Tuple{i,X} \in \DB$.
The number of occurrences of $X$ in $\DB$ is denoted by
$\cnt{X,\DB}=\Abs{\occ{X,\DB}}$.
\end{definition}

Another important aspect for frequent itemsets is the definition of
what it means that an itemset is frequent with respect to a
transaction database.

\begin{definition}[covers, supports and frequencies] \label{d:csf}
A transaction $t=\Tuple{i,Y}$ in a transaction database $\DB$ is said
to \emph{cover} or \emph{support} an itemset $X$ if $X \subseteq Y $.
The cover of an itemset $X$ in $\DB$ is the set
\begin{displaymath}
\cover{X,\DB}=\Set{i : \Tuple{i,Y} \in \DB, X \subseteq Y}
\end{displaymath}
of transaction identifiers of the transactions in $\DB$ that
\emph{cover} $X$.  The \emph{support} of $X$ in $\DB$ is denoted by
$\supp{X,\DB}$ and it is equal to the cardinality of the \emph{cover}
of $X$ in $\DB$, i.e., 
\begin{displaymath}
\supp{X,\DB}=\Abs{\cover{X,\DB}}.
\end{displaymath}
The frequency of $X$ in $\DB$ is its support divided by the number of
transactions in $\DB$, i.e.,
\begin{displaymath}
\fr{X,\DB}=\frac{\supp{X,\DB}}{\Abs{\DB}}.
\end{displaymath}

The database $\DB$ can be omitted from the parameters of these
functions when $\DB$ is not known or needed. If there are several
itemset collections $\Fr_1,\ldots,\Fr_m$ with different covers,
supports or frequencies, we denote the cover, the support and the
frequency of an itemset $X$ in the collection $\Fr_i$ ($1 \leq i \leq
m$) by $\cover{X,\Fr_i}$, $\supp{X,\Fr_i}$ and $\fr{X,\Fr_i}$,
respectively.
\end{definition}

Based on these definitions, the frequent itemset mining problem can be
formulated as follows:

\begin{problem}[frequent itemset mining~\cite{i:agrawal93}]
Given a transaction database $\DB$ and a minimum frequency threshold
$\sigma \in \IntLO{0,1}$, find all $\sigma$-frequent itemsets in
$\DB$, i.e., all itemsets such that $\fr{X,\DB}\geq \sigma$.  The
collection of $\sigma$-frequent itemsets is denoted by
$\Frequent{\sigma,\DB}$.
\end{problem}

\begin{example}[frequent itemsets] \label{ex:frequent}
Let the transaction database $\DB$ consist of transactions
$\Tuple{1,ABC}$, $\Tuple{2,AB}$, $\Tuple{3,ABCD}$ and
$\Tuple{4,BC}$. Then the frequencies of itemsets in $\DB$ are as shown
in Table~\ref{t:ex:frequent}.  For example, the collection
$\Frequent{2/4,\DB}$ of $2/4$-frequent itemsets in $\DB$ is
$\Set{\emptyset,A,B,C,AB,AC,BC,ABC}$.

\begin{table}[h!] \centering
\caption{Itemsets and their frequencies in $\DB$. \label{t:ex:frequent}}
\begin{tabular}{|c|c|}
\hline
$X$ & $\fr{X,\DB}$\\
\hline
\hline
$\emptyset$ & $1$ \\
$A$ & $3/4$ \\
$B$ & $1$ \\
$C$ & $3/4$ \\
$AB$ & $3/4$ \\
$AC$ & $2/4$ \\
$BC$ & $3/4$ \\
$ABC$ & $2/4$ \\
$ABCD$ & $1/4$ \\
\hline
\end{tabular}
\end{table}
\exend \end{example}

Probably the most well-known example of frequent itemset mining tasks
is the \emph{market basket analysis}. In that case the items are
products available for sale. Each transaction consists of a
transaction identifier and a subset of the products that typically
corresponds to items bought in a single purchase, i.e., the
transactions are market baskets. (Alternatively each transaction can
correspond to all items bought by a single customer, possibly as
several shopping events.)  Thus, frequent itemsets are the sets of
products that people tend to buy together as a single purchase event.

The frequent itemsets are useful also in text mining. An important
representation of text documents is the so-called bag-of-words model
where a document is represented as a set of stemmed words occurring in
the document. Thus, items correspond to the stemmed words and each
document is a transaction. The frequent itemsets are the sets of
stemmed words that occur frequently together in the documents of the
document collection.

Web mining is yet another application of frequent itemsets.  There
each item could be, for example, a link pointing at (from) a certain
web page and each transaction could the correspond to the links
pointing from (at) a web page. Then the frequent itemsets correspond
to groups of web pages that are referred concurrently by (that refer
concurrently) the same web pages.

\subsection{Real Transaction Databases\label{ss:ds}}
The purpose of data mining is to analyze data. Without data there is
not much data mining.  Also the methods described in this dissertation
are demonstrated using real data and the patterns discovered from the
data. More specifically, in this dissertation, we use the (frequent)
itemsets mined from three transaction databases as running examples of
pattern collections (of interesting patterns). The two main reasons
for this are that many data analysis tasks can be modeled as frequent
itemset mining and frequent itemset mining has been studied very
actively for more than a decade.

We use a course completion database of the computer science students
at the University of Helsinki to illustrate the methods described in
this dissertation.  Each transaction in that database corresponds to a
student and items in a transaction correspond to the courses the
student has passed.  As data cleaning, we removed from the database
the transactions corresponding to students without any passed courses
in computer science. The cleaned database consists of 2405
transactions corresponding to students and 5021 different items
corresponding to courses.

\begin{figure}
\includegraphics[width=\textwidth]{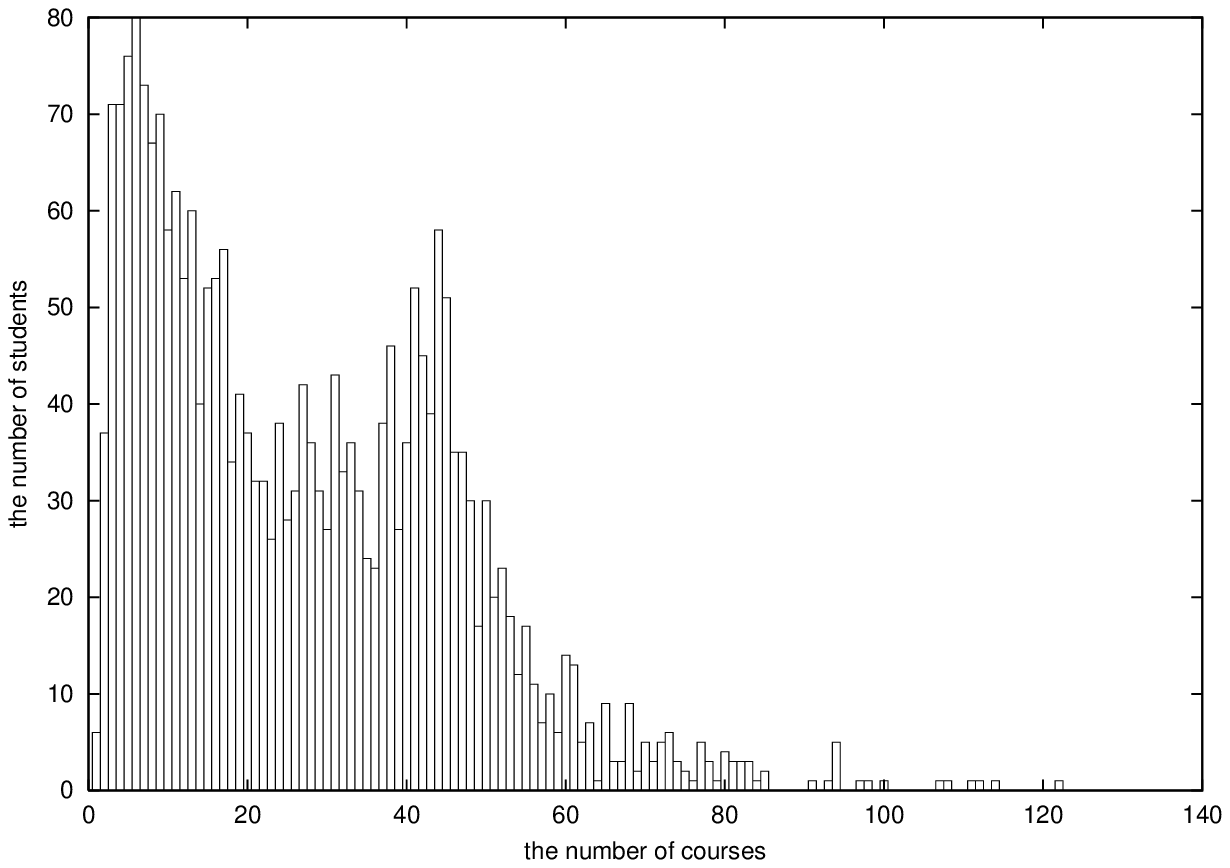}

\includegraphics[width=\textwidth]{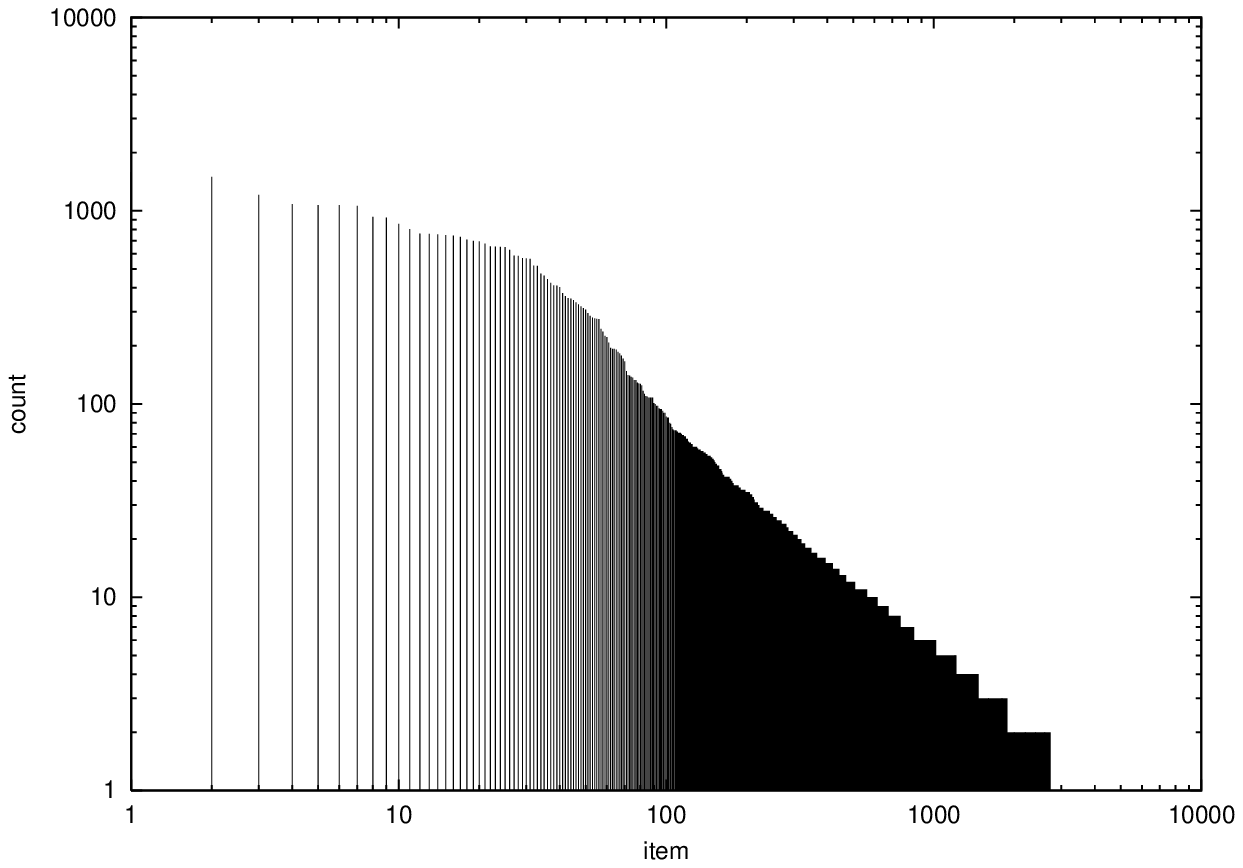}

\caption{The number of transactions of different cardinalities (top) and the
item counts in the course completion database (bottom).\label{f:coursedata}}
\end{figure}

\begin{table} \centering
\caption{The courses in the course completion database that at least a
$0.20$-fraction of the students in the database has passed.  The
columns are the rank of the course with respect to the support, the
number of students that have passed the course, the official course
code and the name of the course, respectively. \label{t:freqitems}}
\begin{tabular}{|c|c|c|p{0.61\textwidth}|}
\hline
rank & count & code & name \\
\hline
\hline
$0$ & $2076$ & $50001$ & Orientation Studies \\
$1$ & $1587$ & $99270$ & Reading Comprehension in English \\
$2$ & $1498$ & $58160$ & Programming Project \\
$3$ & $1210$ & $58123$ & Computer Organization \\
$4$ & $1081$ & $58128$ & Introduction to UNIX \\
$5$ & $1071$ & $58125$ & Information Systems \\
$6$ & $1069$ & $58131$ & Data Structures \\
$7$ & $1060$ & $58161$ & Data Structures Project \\
$8$ & $931$ & $99280$ & English Oral Test \\
$9$ & $920$ & $58127$ & Programming in C \\
$10$ & $856$ & $58122$ & Programming (Pascal) \\
$11$ & $803$ & $99291$ & Oral and Written Skills in the Second Official Language, Swedish \\
$12$ & $763$ & $58162$ & Information Systems Project \\
$13$ & $760$ & $58132$ & Concurrent Systems \\
$14$ & $755$ & $58110$ & Scientific Writing \\
$15$ & $748$ & $58038$ & Database Systems I \\
$16$ & $744$ & $57031$ & Approbatur in Mathematics I \\
$17$ & $733$ & $581259$ & Software Engineering \\
$18$ & $709$ & $57019$ & Discrete Mathematics I \\
$19$ & $697$ & $581330$ & Models for Programming and Computing \\
$20$ & $695$ & $50028$ & Maturity Test in Finnish \\
$21$ & $677$ & $581327$ & Introduction to Application Design \\
$22$ & $655$ & $581326$ & Programming in Java \\
$23$ & $651$ & $581328$ & Introduction to Databases \\
$24$ & $650$ & $581325$ & Introduction to Programming \\
$25$ & $649$ & $581256$ & Teacher Tutoring \\
$26$ & $628$ & $57013$ & Linear Algebra I \\
$27$ & $586$ & $57274$ & Logic I \\
$28$ & $585$ & $580212$ & Introduction to Computing \\
$29$ & $568$ & $581324$ & Introduction to the Use of Computers \\
$30$ & $567$ & $58069$ & Data Communications \\
$31$ & $564$ & $581260$ & Software Engineering Project \\
$32$ & $520$ & $57032$ & Approbatur in Mathematics II \\
$33$ & $519$ & $581329$ & Database Application Project \\
\hline
\end{tabular}
\end{table}

The number of students that have passed certain number of courses and
the the number of students passed each course are shown in
Figure~\ref{f:coursedata}. The courses that at least a $0.20$-fraction
of the student in the course completion database have passed (i.e.,
the $34$ most popular courses) are shown as Table~\ref{t:freqitems}.
The $0.20$-frequent itemsets in the course completion database are
illustrated by Example~\ref{rex:frequent}.

\begin{realexample}[$0.20$-frequent itemsets in the course completion database] \label{rex:frequent}
Let us denote by $\Frequent{\sigma,\DB}[i]$ the $\sigma$-frequent
itemsets in $\DB$ with cardinality $i$. Then the cardinality
distributions of the $0.20$-frequent itemsets in the course completion
database and the most frequent itemsets of
each cardinality in that collection are as shown in
Table~\ref{t:rex:frequent}. 

\begin{table}[h] \centering
\caption{The number of the $0.20$-frequent itemsets of each
cardinality in the course completion database, the most frequent
itemsets of each cardinality and their
supports. \label{t:rex:frequent}}
\begin{tabular}{|c|c|c|c|}
\hline
$i$ & $\Abs{\Frequent{\sigma,\DB}[i]}$ & the largest $X \in \Frequent{\sigma,\DB}[i]$ & $\supp{X,\DB}$ \\
\hline
\hline
$0$ & $1$ & $\emptyset$ & $2405$ \\
$1$ & $34$ & $\{0\}$ & $2076$ \\
$2$ & $188$ & $\{0,1\}$ & $1345$ \\
$3$ & $474$ & $\{2,3,5\}$ & $960$ \\
$4$ & $717$ & $\{0,2,3,5\}$ & $849$ \\
$5$ & $626$ & $\{0,2,3,4,5\}$ & $681$ \\
$6$ & $299$ & $\{0,2,3,5,7,12\}$ & $588$ \\
$7$ & $72$ & $\{2,3,5,7,12,13,15\}$ & $547$ \\
$8$ & $8$ & $\{0,2,3,5,7,12,13,15\}$ & $512$ \\
\hline
\end{tabular}
\end{table}
\exend \end{realexample}

The condensation approaches described in
Chapters~\ref{c:views}--\ref{c:profiles} are quantitatively evaluated
using two data sets from UCI KDD Repository
(\url{http://kdd.ics.uci.edu/}): Internet Usage data consisting of
10104 transactions and 10674 items, and IPUMS Census data consisting
of 88443 transactions and 39954 items.

The transaction database Internet Usage is an example of dense
transaction databases and the transaction database IPUMS Census is a
sparse one: in the Internet Usage database only few $\sigma$-frequent
itemsets are contained exactly in the same transactions whereas in the
IPUMS Census databases many $\sigma$-frequent itemsets are contained
in exactly the same transactions. (This holds for many different
values of $\sigma$). This means also that most of the frequent
itemsets in Internet Usage are closed whereas most of the frequent
itemsets in IPUMS Census are not. (See Definition~\ref{d:closed} for
more details on itemsets being closed.)

\subsection{Computing Frequent Itemsets}

The frequent itemset mining problem has been studied extensively for
more than a decade and several efficient search strategies have been
developed, see
e.g.~\cite{i:agrawal93,i:agrawal96,p:fimi03,a:han04,a:zaki00}.  Most
of the techniques follow the generate-and-test approach: the
collection of frequent itemsets is initialized to consist of the empty
itemset with support equal to the number of transactions in the
database. (This is due to the fact that the empty itemset is contained
in each transaction which means also that its frequency is one.) Then
the collections of itemsets that might be frequent are generated and
tested repeatedly until it is decided that there are no more itemsets
that are not tested but could still be frequent.  The most important
property of frequent itemsets for search space pruning and candidate
generation is the anti-monotonicity of the supports with respect to
the set inclusion relation.
\begin{observation}
If $X \subseteq Y$, then $\supp{X,\DB}\geq \supp{Y,\DB}$.  Thus, all
subitemsets of frequent itemsets are frequent and all superitemsets of
infrequent itemsets are infrequent.
\end{observation}

This observation is largely responsible for the computational
feasibility of the famous frequent itemset mining algorithm
\Apriori~\cite{i:agrawal96} in practice. It or some of its variant
is extensively used in virtually all frequent itemset mining methods.

\subsection{Association Rules}
The itemsets that are frequent in the database are itself summaries of
the database but they can be considered also as side-products of
finding association rules.

\begin{definition}[Association rules]
Let $\DB$ be a transaction database. An \emph{association rule} is an
implication of form $\assoc{X}{Y}$ such that $X,Y \subseteq
\Items$. The itemset $X$ is called the \emph{body} (or the
\emph{antecedent}) of the rule and the itemset $Y$ is known as the
\emph{head} (or the \emph{consequent}) of the rule. 

The \emph{accuracy} of the association rule $\assoc{X}{Y}$ is denoted by
\begin{displaymath}
\acc{\assoc{X}{Y},\DB}=\frac{\fr{X \cup Y,\DB}}{\fr{X,\DB}},
\end{displaymath}
its support $\supp{\assoc{X}{Y},\DB}$ is equal to $\supp{X \cup
Y,\DB}$ and the frequency of the association rule $\assoc{X}{Y}$ is
\begin{displaymath}
\fr{\assoc{X}{Y},\DB}=\frac{\supp{\assoc{X}{Y},\DB}}{\Abs{\DB}}.
\end{displaymath}

An association rule is called \emph{simple} if the head is a
singleton.
\end{definition}

To avoid generating redundant association rules, it is usually assumed
that the body $X$ and the head $Y$ of the rule $\assoc{X}{Y}$ are
disjoint.  Instead of all association rules, typically only the
$\sigma$-frequent association rules, i.e., the association rules with
frequency at least $\sigma$ are computed. The intuition behind this
restriction is that the support of the association rule immediately
tells how many transactions in the database the association rule
concerns. Another reason for concentrating only to $\sigma$-frequent
association rules is that they can be computed from the
$\sigma$-frequent itemsets by a straightforward algorithm
(Algorithm~\ref{a:Association-Rules})~\cite{i:agrawal93}.

\begin{algorithm}
\caption{Association rule mining. \label{a:Association-Rules}}
\begin{algorithmic}[1]
\Input{A collection $\Frequent{\sigma,\DB}$ of $\sigma$-frequent itemsets in a transaction database $\DB$.}
\Output{The collection $\Rules$ of association rules over the collection $\Frequent{\sigma,\DB}$.}
\Function{Association-Rules}{$\Frequent{\sigma,\DB}$}
\State $\Rules \leftarrow \emptyset$
\ForAll{$Y \in \Frequent{\sigma,\DB}$}
\ForAll{$X \subset Y$}
\State $\Rules \leftarrow \Rules \cup \Set{\assoc{X}{Y \setminus X}}$
\EndFor
\EndFor
\State \Return $\Rules$
\EndFunction
\end{algorithmic}
\end{algorithm}

\section{From Frequent Itemsets to Interesting Patterns}

The definition of frequent itemsets readily generalizes to arbitrary
pattern collections $\Pat$ and databases $\DB$ such that the frequency
of a pattern $p \in \Pat$ in the database $\DB$ can be determined. 
Also association rules can be defined for a pattern collection $\Pat$ if
there is a suitable partial order $\preceq$ over the collection.

\begin{definition}[partial order] \label{d:partial}
A \emph{partial order} $\preceq$ is a \emph{transitive},
\emph{antisymmetric} and \emph{reflexive} binary relation, i.e., a
relation $\preceq \subseteq \Pat \times \Pat$ such that $p \preceq p'
\land p' \preceq p'' \Rightarrow p \preceq p''$, $p \preceq p' \land
p' \preceq p \Rightarrow p=p'$ and $p \preceq p$ for all $p,p',p'' \in
\Pat$.  (Note that $p \preceq p'$ is equivalent to $\Tuple{p,p'}\in
\preceq$.) We use the shorthand $p \prec p'$ when $p \preceq p'$ but
$p' \not \preceq p$.

Elements $p,p' \in \Pat$ are called \emph{comparable} with respect to
the partial order $\preceq$ if and only if $p \preceq p'$ or $p'
\preceq p$. If the elements are not comparable, then they are
\emph{incomparable}.  A partial order is a \emph{total order} in
$\Pat$ if and only if all $p,p' \in \Pat$ are comparable.
\end{definition}

Association rules can be defined over the pattern collection $\Pat$
and the partial order $\preceq$ over $\Pat$ if $p \preceq p'$ implies
$\fr{p,\DB}\geq \fr{p',\DB}$ for all $p,p' \in \Pat$.  For example,
itemsets are a special case of this: one such partial order $\preceq$
over the collection of all itemsets $X \subseteq \Items$ is naturally
defined by the set inclusion relation
\begin{displaymath}
X \preceq Y \iff X \subseteq Y
\end{displaymath}
holding for all $X,Y \subseteq \Items$. Then, by the definition of the
frequency of itemsets (Definition~\ref{d:csf}), $X \subseteq Y$
implies $\fr{X,\DB}\geq \fr{Y,\DB}$ for all $X,Y \subseteq \Items$.

\begin{definition}[frequent patterns and their association rules]
Let $\Pat$ be a pattern collection, $\DB$ a database, $\sigma$ a
positive value in the interval $\IntC{0,1}$, and for each pattern $p
\in \Pat$, let $\fr{p,\DB}$ denote the frequency of $p$ in $\DB$. The
collection $\Frequent{\sigma,\DB}$ of $\sigma$-frequent patterns
consists of the patterns $p \in \Pat$ such that $\fr{p,\DB} \geq
\sigma$.

Let $\preceq$ be a partial order over the pattern collection $\Pat$
and let $p \preceq p'$ imply $\fr{p,\DB}\geq \fr{p',\DB}$ for all $p,p'
\in \Pat$.  Then an association rule is a rule $\assoc{p}{p'}$ where
$p,p' \in \Pat$ and $p \preceq p'$.  The accuracy of an association
rule $\assoc{p}{p'}$ is
\begin{displaymath}
\acc{\assoc{p}{p'},\DB}=\frac{\fr{p',\DB}}{\fr{p,\DB}}.
\end{displaymath}

The association rules can be generalized also for incomparable
patterns $p,p' \in \Pat$ by defining
\begin{displaymath}
\acc{\assoc{p}{p'},\DB}=\frac{\fr{p'',\DB}}{\fr{p,\DB}}.
\end{displaymath}
where $p''$ is such a pattern in $\Pat$ that $p,p' \preceq p''$ and
$\fr{p'',\DB} \geq \fr{p''',\DB}$ for all $p''' \in \Pat$ with such
that $p,p' \preceq p'''$.
\end{definition}

\begin{example}[frequent substrings and association rules]
Let $s$ be a string over an alphabet $\Sigma$ and let the frequency of
a string $p=p_{1}\ldots p_{\Abs{p}} \in \Sigma^*$ in $s$ be the number
of its occurrences in $s$ divided by the length of $s$, i.e.,
\begin{displaymath}
\fr{p,s}=\frac{\Abs{\Set{i : p=s_{i+1}\ldots s_{i+\Abs{p}}}}}{\Abs{s}-\Abs{p}+1}.
\end{displaymath}
Furthermore, let the partial order $\preceq$ over the strings in
$\Sigma^*$ be the substring relation, i.e.,
\begin{displaymath}
s \preceq t \iff \exists i \in \Set{0,\ldots,\Abs{t}-\Abs{s}} : s=t_{i+1}\ldots t_{i+\Abs{s}} 
\end{displaymath}
for all $s,t \in \Sigma^*$. As $p \preceq p'$ then implies $\fr{p,s}
\geq \fr{p',s}$, the association rules can be defined for
substrings.

The frequencies of all strings in $s$ can be computed in time
$\Oh{\Abs{s}}$ by constructing a suffix tree or a suffix array of
$s$. (For details on linear-time suffix tree and array constructions,
see e.g.~\cite{a:farach-colton00,a:giegerich97,i:karkkainen03}.)
\exend \end{example}

The previously outlined search strategies to find $\sigma$-frequent
itemsets and association rules have been adapted to many kinds of
patterns such as sequences~\cite{i:wang04,a:zaki01},
episodes~\cite{i:casas-garriga03,i:gwadera03,a:mannila97:episodes},
trees~\cite{i:xiao03,i:zaki02:tree},
graphs~\cite{a:inokuchi03,i:kurakochi01,a:wang02,i:yan02} and
queries~\cite{i:dehaspe01,i:goethals02,i:maloberti03}.

The interestingness predicate obtained by a minimum frequency
threshold determines a \emph{downward closed} pattern collection for
many kinds of patterns.

\begin{definition}[downward closed pattern collections]
A pattern collection $\Pat$ is \emph{downward closed} with respect to
a partial order $\preceq$ and an interestingness predicate
$\ipredicate$ if and only if $p \in \Pat_{\ipredicate}$ implies that
$p' \in \Pat_{\ipredicate}$ for all $p' \preceq p$.
\end{definition}

Many of the pattern discovery techniques are adaptations of the
general levelwise search strategy for downward closed collections of
interesting patterns~\cite{a:mannila97:levelwise}. The search
procedure repeatedly evaluates all patterns whose all subpatterns are
recognized to be interesting.  The procedure is described by
Algorithm~\ref{a:Levelwise} (which is an adaptation
from~\cite{a:mannila97:levelwise}).

\begin{algorithm}
\caption{The levelwise algorithm for discovering interesting patterns. \label{a:Levelwise}}
\begin{algorithmic}[1]
\Input{A pattern collection $\Pat$, a partial order $\preceq$ over
$\Pat$ and an interestingness predicate $\ipredicate : \Pat \to \Set{0,1}$
such that $p \preceq p'$ implies $\ipred{p} \geq \ipred{p'}$ for all
$p,p' \in \Pat$.}
\Output{The collection $\Pat_{\ipredicate}$ of interesting
patterns in $\Pat$.}
\Function{Levelwise}{$\Pat,\preceq,\ipredicate$} 
\State $\Pat_{\ipredicate} \leftarrow \emptyset$ \Comment{No pattern is known to be interesting.}
\State $\Pat' \leftarrow \Pat$ \Comment{All patterns are potentially interesting.}
\Repeat \Comment{Find the minimal still potentially interesting patterns and check whether they are interesting.}
\State $\Candidates \leftarrow \Set{ p \in \Pat' : p' \in \Pat, p' \prec p \Rightarrow p' \in \Pat_{\ipredicate}}$
\State $\Pat_{\ipredicate} \leftarrow \Pat_{\ipredicate} \cup \Set{p \in \Candidates : \ipred{p}=1}$
\State $\Pat' \leftarrow \Pat' \setminus \Candidates$
\Until{$\Candidates=\emptyset$}
\State \Return $\Pat_{\ipredicate}$
\EndFunction
\end{algorithmic}
\end{algorithm}

Algorithm~\ref{a:Levelwise} can be modified in such a way that the
requirement of having downward closed pattern collection can be
relaxed. Specifically, it is sufficient to require that the collection
of potentially interesting patterns that has to be evaluated in the
levelwise search is downward closed in the sense that there is a way
to neglect other patterns in the collection. (For an example, see
subsection~\ref{ss:closedfree}.)

\section{Condensed Representations of Pattern Collections \label{s:condensed}}

A major difficulty in pattern discovery is that the pattern
collections tend to be too large to understand. Fortunately, the
pattern collections contain often redundant information and many
patterns can be inferred from the other patterns. That is, the pattern
collection can be described by its subcollection of irredundant
patterns. The irredundancy of a pattern does not always depend only on
the pattern collection and the interestingness predicate but also on
the other irredundant patterns and the method for inferring all
patterns in the collection from the interesting ones.

In pattern discovery literature such collections of irredundant
patterns are known as \emph{condensed (or concise) representations of
pattern collections}~\cite{i:calders03}, although the condensed
representations in the context of data mining were introduced in a
slightly more general sense as small representations of data that are
accurate enough with respect to a given class of
queries~\cite{i:mannila96}.

\subsection{Maximal and Minimal Patterns \label{ss:maximalminimal}}

Sometimes it is sufficient, for representing the pattern collection,
to store only the \emph{maximal patterns} in the
collection~\cite{a:gunopulos03}.

\begin{definition}[maximal patterns]
A pattern $p \in \Pat$ is \emph{maximal} in the collection $\Pat$ with
respect to the partial order $\prec$ if and only if $p \not \prec p'$
for all $p' \in \Pat$. The collection of maximal patterns in $\Pat$ is
denoted by $\Max{\Pat, \preceq}$.
\end{definition}

It can be shown that the maximal interesting patterns in the
collection determine the whole collection of interesting patterns if
the interesting patterns form a downward closed pattern collection.

\begin{proposition}
The collection $\Max{\Pat_{\ipredicate},\preceq}$ of the maximal
interesting patterns determines the collection
$\Pat_{\ipredicate}$ of interesting patterns if and only if
$\Pat_{\ipredicate}$ is downward closed.
\end{proposition}
\begin{proof}
If the collection $\Pat_{\ipredicate}$ is downward closed, then by the
definition of maximality, for each pattern $p \in \Pat_{\ipredicate}$
there is the maximal pattern in $p' \in \Pat_{\ipredicate}$ such that
$p \preceq p'$.  Furthermore, for each maximal pattern $p' \in
\Pat_{\ipredicate}$ it holds $p \preceq p' \Rightarrow p \in
\Pat_{\ipredicate}$ if $\Pat_{\ipredicate}$ is downward closed.

If the collection $\Pat_\ipredicate$ is not downward closed, then
there is a non-maximal pattern $p$ such that $p \notin
\Pat_{\ipredicate}$ but $p \prec p'$ for some $p' \in
\Pat_{\ipredicate}$.  The maximal patterns in $\Pat_{\ipredicate}$ are
not sufficient to point out that pattern.
\end{proof}

The maximal patterns in the collection of $\sigma$-frequent itemsets,
i.e., the maximal $\sigma$-frequent itemsets in $\DB$, are denoted by
$\Maximal{\sigma,\DB}$. Representing a downward closed collection of
patterns by the maximal patterns in the collection can reduce the
space consumption drastically. For example, the number
$\Abs{\Maximal{\sigma,\DB}}$ of maximal frequent itemsets can be
exponentially smaller than the number $\Abs{\Frequent{\sigma,\DB}}$ of
all frequent itemsets.

\begin{example}[the number of $\sigma$-frequent itemsets versus the number of maximal $\sigma$-frequent itemsets]
Let us consider a transaction database $\DB$ consisting only of one
tuple $\Tuple{1,\Items}$. For this database and all possible minimum
frequency thresholds $\sigma \in [0,1]$ we have:
$\Abs{\Maximal{\sigma,\DB}}=1$ and
$\Abs{\Frequent{\sigma,\DB}}=2^{\Abs{\Items}}$.
\exend \end{example}

\begin{realexample}[maximal $0.20$-frequent itemsets in the course completion database] \label{rex:maximal}
Let us denote the collection of the maximal $\sigma$-frequent itemsets
in $\DB$ with cardinality $i$ by $\Maximal{\sigma,\DB}[i]$. Then the
cardinality distributions of the maximal $0.20$-frequent itemsets in
the course completion database (see Subsection~\ref{ss:ds}) and the
most frequent itemsets of each cardinality are as shown in
Table~\ref{t:rex:maximal}

\begin{table}[h] \centering
\caption{The number of the maximal $0.20$-frequent itemsets of each cardinality in
the course completion database, the most frequent itemsets of each
cardinality and their supports. \label{t:rex:maximal}}
\begin{tabular}{|c|c|c|c|}
\hline
$i$ & $\Abs{\Maximal{\sigma,\DB}[i]}$ & the largest $X \in \Maximal{\sigma,\DB}[i]$ & $\supp{X,\DB}$ \\
\hline
\hline
$0$ & $0$ & - & - \\
$1$ & $1$ & $\{33\}$ & $519$ \\
$2$ & $21$ & $\{2,26\}$ & $547$ \\
$3$ & $41$ & $\{0,2,19\}$ & $529$ \\
$4$ & $58$ & $\{0,2,7,17\}$ & $553$ \\
$5$ & $38$ & $\{2,3,5,9,10\}$ & $511$ \\
$6$ & $66$ & $\{0,1,2,3,4,5\}$ & $550$ \\
$7$ & $20$ & $\{0,2,3,5,7,14,20\}$ & $508$ \\
$8$ & $8$ & $\{0,2,3,5,7,12,13,15\}$ & $512$ \\
\hline
\end{tabular}
\end{table}
\exend \end{realexample}

Due to the potential reduction in the number of itemsets needed to
find, several search strategies for finding only the maximal frequent
itemsets have been
developed~\cite{i:burdick01,i:boros02,i:bayardo98,i:gouda01,p:fimi03,a:gunopulos03,i:satoh03}.

It is not clear, however, whether the maximal interesting patterns are
the most concise subcollection of patterns to represent the interesting
patterns.  The collection could be represented also by the \emph{minimal}
uninteresting patterns.

\begin{definition}[minimal patterns]
A pattern $p \in \Pat$ is \emph{minimal} in the collection $\Pat$ with
respect to the partial order $\prec$ if and only if $p' \not \prec p$
for all $p' \in \Pat$. The collection of minimal patterns in $\Pat$ is
denoted by $\Min{\Pat,\preceq}$.
\end{definition}

As in the case of the maximal interesting patterns, it is easy to see
that the minimal uninteresting patterns uniquely determine the
collection of the interesting patterns if the pattern collection is
downward closed.

The collection of minimal $\sigma$-infrequent itemsets in $\DB$ is
denoted by $\Minimal{\sigma,\DB}$. It is much more difficult to relate
the number of minimal uninteresting patterns to the number of
interesting patterns, even when the collection of interesting patterns
is downward closed. In fact, for a downward collection
$\Pat_{\ipredicate}$ of interesting patterns patterns the number
$\Abs{\Min{\Pat_{\bar{\ipredicate}},\preceq}}$ of uninteresting
patterns cannot be bounded very well in general from above nor from
below by the number $\Abs{\Pat_{\ipredicate}}$ of interesting patterns
and the number $\Abs{\Max{\Pat_{\ipredicate},\preceq}}$ of maximal
interesting patterns.

Bounding the number of the minimal infrequent itemsets by the number
of frequent itemsets is also slightly more complex than bounding the
number of maximal frequent itemsets.

\begin{example}[the number of $\sigma$-frequent itemsets versus the number of minimal $\sigma$-infrequent itemsets]
The number $\Abs{\Minimal{\sigma,\DB}}$ of minimal infrequent itemsets
can be $\Abs{\Items}$ times larger than the number of
$\Abs{\Frequent{\sigma,\DB}}$ frequent itemsets.  

Namely, let the transaction database consists of transaction
$\Tuple{1,\emptyset}$. Then $\Frequent{\sigma,\DB}=\Set{\emptyset}$
but $\Minimal{\sigma,\DB}=\Set{ \Set{A} : A \in \Items}$.  This is
also the worst case since each frequent itemset $X \in
\Frequent{\sigma,\DB}$ can have at most $\Abs{\Items}$ superitemsets
in $\Minimal{\sigma,\DB}$.

If the collection $\Minimal{\sigma,\DB}$ is empty, then
$\Abs{\Frequent{\sigma,\DB}}>c\Abs{\Minimal{\sigma,\DB}}$ for all
values $c \in \mathbb{R}$. Otherwise, let the transaction database
$\DB$ consist of one transaction with itemset $\Items \setminus
\Set{A}$ for each $A \in \Items$ and let $\sigma=1/\Abs{\DB}$.  Then
$\Abs{\Minimal{\sigma,\DB}}$ is exponentially smaller than
$\Abs{\Frequent{\sigma,\DB}}$.  \exend \end{example}

It is known that the number $\Abs{\Maximal{\sigma,\DB}}$ of maximal
itemset can be bounded from above by $\Paren{\Abs{\Items}-\sigma
\Abs{\DB}+1}\Abs{\Minimal{\sigma,\DB}}$ if $\Minimal{\sigma,\DB}$ is
not empty~\cite{i:boros02}. Furthermore, it is clear that
$\Abs{\Minimal{\sigma,\DB}} \leq
\Abs{\Items}\Abs{\Maximal{\sigma,\DB}}$ for all minimum frequency
thresholds $\sigma \in [0,1]$.

The collection $\Minimal{\sigma,\DB}$ can be obtained from
$\Maximal{\sigma,\DB}$ by generating all minimal hypergraph
transversals in the hypergraph
\begin{displaymath}
\Set{\Items \setminus X : X \in \Maximal{\sigma,\DB}},
\end{displaymath}
i.e., in the hypergraph consisting of the complements of the maximal
$\sigma$-frequent itemsets in $\DB$~\cite{a:mannila97:levelwise}.

The slack in the bounds between the number of the maximal frequent and
the number of the minimal infrequent itemsets implies that it cannot
be decided in advance without seeing the data which of the
representations --- $\Maximal{\sigma,\DB}$ or $\Minimal{\sigma,\DB}$
--- is better. In practice, the smaller of the collections
$\Maximal{\sigma,\DB}$ and $\Minimal{\sigma,\DB}$ can be chosen. Each
maximal frequent and each minimal infrequent itemset determines its
subitemsets to be frequent and superitemsets to be
infrequent. Sometimes one can obtain a representation for
$\Frequent{\sigma,\DB}$ that is smaller than $\Maximal{\sigma,\DB}$ or
$\Minimal{\sigma,\DB}$ by choosing some itemsets from
$\Maximal{\sigma,\DB}$ and some from $\Minimal{\sigma,\DB}$ in such a
way that the chosen itemsets determine the collection
$\Frequent{\sigma,\DB}$ uniquely~\cite{i:mielikainen04:ssi}.

Sometimes it is not sufficient to represent only the collection of
interesting patterns but also the quality values for the patterns are
needed as well. For example, the accuracy of an association rule
$\assoc{X}{Y}$ depends on the frequencies of the frequent itemsets $X$
and $X \cup Y$. One solution is to determine the pattern collection as
described above and describe the quality values in the collection of
interesting patterns separately. The quality values can be
represented, e.g., by a simplified
database~\cite{i:mielikainen03:faosi} or by a random sample of
transactions from the database~\cite{i:mielikainen04:ssi}. In these
approaches, however, the condensed representation is not a
subcollection of the patterns anymore. Thus, a different approach is
required if the condensed representation of the pattern collection is
required to consist of patterns.

\subsection{Closed and Free Patterns \label{ss:closedfree}}

For the rest of the chapter we shall focus on interestingness measures
$\imeasure$ such that $p \preceq p'$ implies $\imeas{p}\geq \imeas{p'}$
for all $p,p' \in \Pat$, i.e., to anti-monotone interestingness
measures.  Then maximal interesting patterns and their quality values
determine lower bounds for all other interesting patterns as well.
The highest lower bound obtainable for the quality value of a pattern $p$
from the quality values of the maximal patterns is
\begin{displaymath}
\max \Set{ \imeas{p'} : p \preceq p' \in \Max{\Pat_{\ipredicate},\preceq}}.
\end{displaymath}

The patterns $p$ with the quality value matching with the maximum
quality value of the maximal interesting patterns that are
superpatterns of $p$ can be removed from the collection of potentially
irredundant patterns if the maximal interesting patterns are decided
to be irredundant. An exact representation for the collection of
interesting patterns can be obtained by repeating these operations.
The collection of the irredundant patterns obtained by the previous
procedure is called the collection of \emph{closed} interesting
patterns~\cite{i:zaki98}.

\begin{definition}[closed patterns] \label{d:closed}
A pattern $p \in \Pat$ is \emph{closed} in the collection $\Pat$ with
respect to the partial order $\prec$ and the interestingness measure
$\imeasure$ if and only if $p \prec p'$ implies $\imeas{p}>\imeas{p'}$
for all $p' \in \Pat$.  The collection of closed patterns in $\Pat$ is
denoted by $\Cl{\Pat,\preceq,\imeasure}$. For brevity, $\preceq$ and
$\imeasure$ can be omitted when they are clear from the context.
\end{definition}

The collection of closed $\sigma$-frequent itemsets in $\DB$ is
denoted by $\Closed{\sigma,\DB}$. One procedure for detecting the
closed patterns (for a given pattern collection $\Pat$, a partial
order $\preceq$ and an interestingness measure $\imeasure$) is
described as Algorithm~\ref{a:Closed-Patterns}.
\begin{algorithm}
\caption{Detection of closed patterns \label{a:Closed-Patterns}}
\begin{algorithmic}[1]
\Input{A collection $\Pat$ of patterns, a partial order $\preceq$ over
$\Pat$ and an interestingness measure $\imeasure$.}
\Output{The collection $\Cl{\Pat,\preceq,\imeasure}$ of patterns in $\Pat$
that are closed with respect to $\imeasure$.}
\Function{Closed-Patterns}{$\Pat,\prec,\imeasure$}
\State $\Candidates \leftarrow \Pat$
\While{$\Candidates\neq \emptyset$}
\State $\Candidates' \leftarrow \Max{\Candidates,\preceq}$
\State $\Cl{\Pat,\preceq,\imeasure} \leftarrow \Cl{\Pat,\preceq,\imeasure} \cup \Candidates'$
\State $\Candidates \leftarrow \Candidates \setminus \Candidates'$
\State $\Candidates \leftarrow \Set{p \in \Candidates : p' \in \Candidates', p \prec p' \Rightarrow \imeas{p}>\imeas{p'}}$
\EndWhile
\State \Return $\Cl{\Pat,\preceq,\imeasure}$
\EndFunction
\end{algorithmic}
\end{algorithm}

\begin{example}[closed frequent itemsets] \label{ex:closed}
Let the transaction database $\DB$ be the same as in
Example~\ref{ex:frequent},
i.e.,
\begin{displaymath}
\DB=\Set{\Tuple{1,ABC},\Tuple{2,AB},\Tuple{3,ABCD},\Tuple{4,BC}}.
\end{displaymath}
Then 
\begin{displaymath}
\Closed{2/4,\DB}=\Set{B,AB,BC,ABC}=\Frequent{2/4,\DB}\setminus \Set{\emptyset,A,C,AC}.
\end{displaymath}
\exend \end{example}

\begin{realexample}[closed $0.20$-frequent itemsets in the course completion database] \label{rex:closed}
Let us denote the collection of the closed $\sigma$-frequent itemsets
in $\DB$ with cardinality $i$ by $\Closed{\sigma,\DB}[i]$. Then the
cardinality distributions of the closed $0.20$-frequent itemsets in
the course completion database (see Subsection~\ref{ss:ds}) and the
most frequent closed itemsets of each cardinality are as shown in
Table~\ref{t:rex:closed}.

\begin{table}[h!] \centering
\caption{The number of the closed $0.20$-frequent itemsets of each
cardinality in the course completion database, the most frequent
closed itemsets of each cardinality and their
supports. \label{t:rex:closed}}
\begin{tabular}{|c|c|c|c|}
\hline
$i$ & $\Abs{\Closed{\sigma,\DB}[i]}$ & the largest $X \in \Closed{\sigma,\DB}[i]$ & $\supp{X,\DB}$ \\
\hline
\hline
$0$ & $1$ & $\emptyset$ & $2405$ \\
$1$ & $34$ & $\{0\}$ & $2076$ \\
$2$ & $186$ & $\{0,1\}$ & $1345$ \\
$3$ & $454$ & $\{2,3,5\}$ & $960$ \\
$4$ & $638$ & $\{0,2,3,5\}$ & $849$ \\
$5$ & $519$ & $\{0,2,3,4,5\}$ & $681$ \\
$6$ & $238$ & $\{0,2,3,5,7,12\}$ & $588$ \\
$7$ & $58$ & $\{2,3,5,7,12,13,15\}$ & $547$ \\
$8$ & $8$ & $\{0,2,3,5,7,12,13,15\}$ & $512$ \\
\hline
\end{tabular}
\end{table}
\exend \end{realexample}

It is a natural question whether the closed interesting patterns could
be discovered immediately without generating all interesting
patterns. For many kinds of frequent closed patterns this question has
been answered positively; there exist methods for mining directly,
e.g., closed frequent
itemsets~\cite{i:pasquier99,i:pan03,i:wang03,i:zaki02:charm}, closed
frequent sequences~\cite{i:wang04,i:yan03}, and closed frequent
graphs~\cite{i:yan04} from data. Recently it has been shown that
frequent closed itemsets can be found in time polynomial in the size
of the output~\cite{i:uno04}.

The number $\Abs{\Cl{\Pat_{\ipredicate}}}$ of closed interesting
patterns is at most the number $\Abs{\Pat_{\ipredicate}}$ of all
interesting patterns and at least the number
$\Abs{\Max{\Pat_{\ipredicate}}}$ of maximal interesting
patterns, since $\Pat_{\ipredicate} \supseteq \Cl{\Pat_{\ipredicate}}
\supseteq \Max{\Pat_{\ipredicate}}$. Tighter bounds for the number of
closed interesting patterns depend on the properties of the pattern
collection $\Pat$.

\begin{example}[the number of $\sigma$-frequent itemsets versus the number of closed $\sigma$-frequent itemsets]
Similarly to the maximal frequent itemsets, the number of closed
frequent itemsets in the transaction database
$\DB=\Set{\Tuple{1,\Items}}$ is exponentially smaller than the number
of all frequent itemsets for all minimum frequency thresholds $\sigma
\in \IntLO{0,1}$.  \exend \end{example}

However, the number of closed frequent sets can be exponentially
larger than the number of maximal itemsets.

\begin{example}[the number of maximal $\sigma$-frequent itemsets versus the number of closed $\sigma$-frequent itemsets]
Let $\DB$ consist of one transaction for each subset of size
$\Abs{\Items}-1$ of $\Items$ and
$\Ceil{\sigma/(1-\sigma)}\Abs{\Items}$ transactions consisting of
the itemset $\Items$.  Then $\Maximal{\sigma,\DB}=\Set{\Items}$ but
$\Closed{\sigma,\DB}=\Set{X \subseteq \Items}=2^{\Items}$.  \exend
\end{example}

\begin{realexample}[comparing all, closed and maximal $\sigma$-fre\-quent itemsets in the course completion database] \label{rex:acm}
Let us consider the course completion database (see
Subsection~\ref{ss:ds}). In that transaction database, the number of all,
closed and maximal $\sigma$-frequent itemsets for several different
minimum frequency thresholds $\sigma$ are as shown in
Table~\ref{rex:t:acm}.

\begin{table}[h!] \centering
\caption{The number of all, closed and maximal $\sigma$-frequent
itemsets in the course completion database for several different
minimum frequency thresholds $\sigma$.\label{rex:t:acm}}
\begin{tabular}{|c||c|c|c|}
\hline
$\sigma$ & all & closed & maximal \\
\hline
\hline
0.50 & 7 & 7 & 3 \\
0.40 & 18 & 18 & 10 \\
0.30 &103 & 103 & 28 \\
0.25 & 363 & 360 & 80 \\
0.20 & 2419 & 2136 & 253 \\
0.15 & 19585 & 12399 & 857 \\
0.10 & 208047 & 82752 & 4456 \\
0.05 & 5214764 & 918604 & 43386 \\
0.04 & 12785998 & 1700946 & 80266 \\
0.03 & 38415247 & 3544444 & 172170 \\
0.02 & 167578070 & 8486933 & 414730 \\
0.01 & 1715382996 & 23850242 & 1157338 \\
\hline
\end{tabular}
\end{table}

The number of maximal $\sigma$-frequent
itemsets is quite low compared even to the number of closed
$\sigma$-frequent itemsets. The number of closed $\sigma$-frequent
itemsets is also often considerably smaller than the number of all
$\sigma$-frequent itemsets, especially for low values of $\sigma$.
\exend \end{realexample}

A desirable property of closed frequent itemsets is that they can be
defined by closures of the itemsets.  A \emph{closure} of an itemset
$X$ in a transaction database $\DB$ is the intersection of the
transactions in $\DB$ containing $X$, i.e.,
\begin{displaymath}
\cl{X,\DB}=\bigcap_{\Tuple{i,Y} \in \DB, Y \supseteq X} Y.
\end{displaymath}

Clearly, there is unique closure $\cl{X,\DB}$ in the transaction
database $\DB$ for each itemset $X$. It can be shown that each closed
itemset is its own
closure~\cite{b:ganter99,i:kryszkiewicz01,i:pasquier99}.  Thus, the
collection of closed $\sigma$-frequent itemsets can be expressed
alternatively as
\begin{displaymath}
\Closed{\sigma,\DB}=\Set{X \subseteq \Items: \cl{X,\DB}=X, \fr{X,\DB}\geq \sigma}.
\end{displaymath}

In fact, this is often used as a definition of a closed itemset.  In
this dissertation, however, the closed patterns are not defined using
closures; the reason is that it is not clear in the case of other
pattern collections than frequent itemsets whether the closure can be
defined in a natural way and when it is unique.

The levelwise algorithm (Algorithm~\ref{a:Levelwise}) can be adapted
to mine also closed itemsets: Let $\mathcal{FI}(\sigma,\DB)$ denote
the collection of all $\sigma$-frequent items in $\DB$ and let
$\mathcal{FC}_k$ be the collection of the closed frequent itemsets at
level $k$. The level for closed itemsets is the length of the shortest
path from the itemset to the closure of the empty itemset in the
partial order defined by the set inclusion relation. Thus, the zeroth
level consists of the closure of the empty itemset. The collection of
potentially frequent closed itemsets at level $k$ ($k\geq 1$) consists
of closures of $X \cup \Set{A}$ for each frequent closed itemset $X$
in level $k-1$ and each frequent item $A \notin X$.  The adaptation of
Algorithm~\ref{a:Levelwise} for frequent closed itemset mining is
described as Algorithm~\ref{a:Closures}.

\begin{algorithm}[h]
\caption{The levelwise algorithm for discovering frequent closed itemsets in a transaction database. \label{a:Closures}}
\begin{algorithmic}[1]
\Input{A transaction database $\DB$ and a minimum frequency threshold $\sigma \in \IntLO{0,1}$.}
\Output{The collection $\Closed{\sigma,\DB}$ of $\sigma$-frequent closed itemsets in $\DB$.}
\Function{Closures}{$\sigma,\DB$}
\State $\Items \leftarrow \bigcup_{X \in \DB} X$
\State $\mathcal{FI} \leftarrow \Set{ A \in \Items : \fr{A,\DB} \geq \sigma}$
\State $i \leftarrow 0$
\State $\mathcal{FC}_i \leftarrow \Set{\cl{\emptyset,\DB}}$
\State $\Closed{\sigma,\DB} \leftarrow \mathcal{FC}_0$
\Repeat 
\State $i \leftarrow i+1$
\State $\Candidates \leftarrow \Set{\cl{X \cup \Set{A},\DB} : X \in \mathcal{FC}_{i-1}, A \in \mathcal{FI} \setminus X}$
\State $\mathcal{FC}_{i} \leftarrow \Set{X \in \Candidates : \fr{X,\DB} \geq \sigma} \setminus \Closed{\sigma,\DB}$
\State $\Closed{\sigma,\DB} \leftarrow \Closed{\sigma,\DB} \cup \mathcal{FC}_i$
\Until{$\Candidates=\emptyset$}
\State \Return $\Closed{\sigma,\DB}$
\EndFunction
\end{algorithmic}
\end{algorithm}

The collection of closed interesting patterns can be seen a refinement
of the collection of maximal interesting patterns: a closed
interesting pattern $p$ is a maximal interesting pattern for the
minimum quality value thresholds in the interval 
\begin{displaymath}
\IntLO{\max \Set{\imeas{p'} : p \prec p'}, \imeas{p}}.
\end{displaymath}

A natural relaxation of the closed interesting patterns is to store
maximal interesting patterns for several minimum quality value
thresholds. For example, the collections
\begin{displaymath}
\Maximal{\sigma,\DB},\Maximal{\sigma+\epsilon,\DB},\ldots,\Maximal{\sigma+\Paren{\Ceil{\Paren{1-\sigma}/\epsilon}-1}\epsilon,\DB}
\end{displaymath}
of the maximal frequent itemsets are sufficient for estimating the
frequency of any $\sigma$-frequent itemset in $\DB$ by the maximum
absolute error at most $\epsilon$. Furthermore, the frequencies of the
maximal frequent itemsets are not needed: it is sufficient to know in
which of the collections
$\Maximal{\sigma,\DB},\Maximal{\sigma+\epsilon,\DB},\ldots$ the
maximal pattern belongs to and what is the minimum frequency threshold
for that collection. Then the frequency of an itemset $X$ can be
estimated to be the maximum of the minimum frequency thresholds of the
maximal itemset collections that contain an itemset containing the
itemset $X$.

Algorithm~\ref{a:Closed-Patterns} can be modified to solve this task
of approximating the collection of $\sigma$-frequent closed
itemsets. To approximate especially the collections of the frequent
itemsets, many maximal frequent itemset mining techniques can be
adapted for mining the maximal frequent itemset collections for
several minimum frequency thresholds, see e.g.~\cite{i:pei02}.

An alternative notion of approximating closed frequent itemsets is
proposed in~\cite{i:boulicaut00:closure}. The approach readily
generalizes to any collection of interesting patterns with an
anti-monotone interestingness measure: a pattern is considered to be
$\epsilon$-closed if the absolute difference between its quality value 
and the largest quality value of its superpatterns is more than $\epsilon$.

Finally, an approach based on simplifying interestingness values to
approximate closed interesting patterns is described in
Chapter~\ref{c:views} of this dissertation and another approximation
based on pattern ordering with respect to the informativeness of the
prefixes of the ordering is proposed in Chapter~\ref{c:tradeoffs}.

Instead of defining irredundant patterns to be those that have
strictly higher quality values than any of their superpatterns, the
irredundant patterns could be defined to be those that have strictly
lower quality values than any of their subpatterns. The latter
patterns are called \emph{free patterns}~\cite{a:boulicaut03},
\emph{generators}~\cite{i:pasquier99} or \emph{key
patterns}~\cite{a:bastide00}.

\begin{definition}[free patterns]
A pattern $p \in \Pat$ is \emph{free} in the collection $\Pat$ with
respect to the partial order $\prec$ and the interestingness measure
$\imeasure$ if and only if $p' \prec p$ implies $\imeas{p}<\imeas{p'}$
for all $p' \in \Pat$. The collection of free patterns in $\Pat$ is
denoted by $\Gen{\Pat}$.
\end{definition}

The collection of free $\sigma$-frequent itemsets in $\DB$ is denoted
by $\Free{\sigma,\DB}$.  Unfortunately, the free interesting patterns
$\Gen{\Pat_{\ipredicate}}$ are not always a sufficient representation
for all interesting patterns but also minimal free uninteresting
patterns, i.e., the patterns in the collection
$\Min{\Gen{\Pat_{\bar{\ipredicate}}}}$ are needed.

\begin{example}[free frequent itemsets] \label{ex:free}
Let the transaction database $\DB$ be the same as in
Example~\ref{ex:frequent},
i.e.,
\begin{displaymath}
\DB=\Set{\Tuple{1,ABC},\Tuple{2,AB},\Tuple{3,ABCD},\Tuple{4,BC}}.
\end{displaymath}
Then 
\begin{displaymath}
\Free{1/4,\DB}=\Set{\emptyset,A,C,AC}=\Frequent{1/4,\DB} \setminus \Set{B,AB,BC,ABC}
\end{displaymath}

This is not, however, sufficient to determine the collection of
$1/4$-frequent itemsets in $\DB$ since there is no information about
$B$ nor $D$. The item $B$ is frequent but not free, whereas the item
$D$ is free but not frequent.  \exend \end{example}

As in the case of closed interesting patterns, the number of free
interesting patterns is at most the number of all interesting
patterns.  The number of free interesting itemsets can be smaller than
even the number of maximal interesting or minimal uninteresting
patterns.

In the case of frequent itemsets, the number of free frequent itemsets
is always at least as large as the number of closed frequent itemsets
since each free itemset has a only one closure but several free
itemsets can share the same one. Although the free frequent itemsets
seem to have many disadvantages, they have one major advantage
compared to closed frequent itemsets: collections of free frequent
itemsets are downward closed~\cite{a:boulicaut03}. Thus, closed
frequent itemsets can be discovered from free frequent itemsets by
computing the closures for all free frequent itemsets. Notice that if
free frequent itemsets are used only to compute the closed frequent
itemsets, the minimal free infrequent itemsets are not needed for the
representation since for each closed frequent itemset $X$ there is at
least one free frequent itemset $Y$ such that $X=\cl{Y,\DB}$.

Similarly to closed frequent itemsets, also mining the approximate
free itemset collections based on a few different notions of
approximation has been studied~\cite{a:boulicaut03,i:pei02}.

\subsection{Non-Derivable Itemsets \label{ss:ndi}}

Taking the maximum or the minimum of the quality values of the super-
or subpatterns are rather simple methods of inferring the unknown
quality values but not much more complex inference techniques are
useful with arbitrary anti-monotone interestingness measures. (Note
that this is the case even with arbitrary frequent pattern collections
since the only requirement for frequency is the anti-monotonicity.)
For some pattern collections with suitable interestingness measures it
is possible to find more concise representations.

For example, several more sophisticated condensed representations have
been developed for frequent
itemsets~\cite{i:bykowski01,i:calders03,i:kryszkiewicz01}.  This line
of work can be seen to be culminated on \emph{non-derivable
itemsets}~\cite{i:calders02}. The idea of non-derivable itemsets is to
deduce lower and upper bounds for the frequency of the itemset from
the frequencies of its subitemsets.

\begin{definition}[non-derivable itemsets] \label{d:ndi}
Let $\overline{\freq}$ and $\underline{\freq}$ denote mappings that
give upper and lower bounds for the frequency of any itemset over
$\Items$.  An itemset $X \subseteq \Items$ is \emph{non-derivable}
with respect to the transaction database $\DB$ (and functions
$\overline{\freq}$ and $\underline{\freq}$) if and only if the lower
bound $\frl{X,\DB}$ is strictly smaller than the upper bound
$\fru{X,\DB}$.  The collection of non-derivable itemsets is denoted by
$\NDI{\DB}$.
\end{definition}

One bound for the frequencies can be computed using
inclusion-exclusion~\cite{i:calders02}. (An alternative to
inclusion-exclusion would be to use (integer) linear
programming~\cite{i:bykowski02,i:calders04}. However, if the bounds
for the frequencies are computed from the frequencies of all subitemsets,
then inclusion-exclusion leads to the best possible
solution~\cite{i:calders04:bounds}.)  From the inequality
\begin{displaymath}
\sum_{Y \subseteq Z \subseteq X} (-1)^{|Z \setminus Y|} \fr{Z,d} \geq 0
\end{displaymath}
holding for all $X$ and $Y$, it is possible to derive upper and lower
bounds for the frequency of the itemsets $X$ in
$\DB$~\cite{i:calders03}:
\begin{eqnarray*}
\fru{X,\DB} & = &\min_{Y \subset X} \Set{\sum_{Y \subseteq Z \subset X} (-1)^{{|X \setminus Z|}+1}\fr{Z,\DB} : |X \setminus Y| \mbox{ is odd}} \\
\frl{X,\DB} & = & \max_{Y \subset X} \Set{\sum_{Y \subseteq Z \subset X} (-1)^{{|X \setminus Z|}+1}\fr{Z,\DB} : |X \setminus Y| \mbox{ is even}}
\end{eqnarray*}

The collection of non-derivable itemsets is downward closed. The
largest non-derivable itemset is at most of size $\Floor{ \log_2
\Abs{\DB}}$~\cite{i:calders02}.  To represent frequent itemsets it is
sufficient to store the frequent non-derivable itemsets and the
minimal infrequent non-derivable itemsets with upper bounds to the
frequency at least the minimum frequency threshold.

\begin{example}[non-derivable itemsets] \label{ex:ndi}
Let the transaction database $\DB$ be the same as in
Example~\ref{ex:frequent}, i.e.,
\begin{displaymath}
\DB=\Set{\Tuple{1,ABC},\Tuple{2,AB},\Tuple{3,ABCD},\Tuple{4,BC}}.
\end{displaymath}
Then $\NDI{\DB}=\Set{\emptyset,A,B,C,AC}$.
\exend \end{example}

The approach of non-derivable itemsets is essentially different from
the other condensed representations described, as no additional
assumptions are made about the itemsets with unknown frequencies:
their frequencies can be determined uniquely using, e.g.,
inclusion-exclusion. In contrast, using closed and free itemsets, each
unknown frequency is assumed to be determined exactly as the maximum
frequency of its superitemsets and the minimum frequency of its subitemsets,
respectively.

The problem of finding non-derivable representations for essentially
other pattern classes than itemsets is a very important and still
largely open problem. 

\section{Exploiting Patterns \label{c:i:s:ep}}

The real goal in pattern discovery is rarely just to obtain the
patterns themselves but to use the discovered patterns.

One indisputable use of patterns is to disclose interesting aspects of
the data. The suitability of different ways to represent pattern
collections for the disclosure depends crucially on the actual
application and the goals of data mining in the task at hand.
However, at least the number of patterns and their complexity affect
the understandability of the collection.

In practice, the number of patterns in the representation is strongly
affected by the application and the database.  For example, when
represented explicitly, the itemset collection consisting only of the
itemset $\Items$ is probably easier to understand than the collection
$2^\Items$ of all subsets of $\Items$.  The explicit representation,
however, is not always to most suitable.

\begin{example}[represeting a collection implicitly]
Sometimes the
database $\DB$ can be expected to be so dense that all frequent
itemsets are also closed, i.e.,
$\Frequent{\sigma,\DB}=\Closed{\sigma,\DB}$ under normal circumstances
(with respect to the assumptions). If only the closed frequent
itemsets are being represented, then it is most convenient to describe
the collection by its maximal itemsets and those non-maximal itemsets
that are not closed. Thus, it would be very surprising if the database
$\DB$ happens to be such that the only closed itemset would be
$\Items$, and recognizing exactly that fact from the representation of
the collection, i.e., the collection $2^{\Items} \setminus
\Set{\Items}$, would be quite arduous.
\exend \end{example}

Also the complexity of the representation can have a significant
influence to the understandability. For example, the smallest Turing
machine generating the pattern collection is probably quite an
unintuitive representation. (The length of the encoding of such a
Turing machine is called the Kolmogorov complexity or algorithmic
information of the pattern collection~\cite{b:calude02,b:li97}.)
Similar situations occur also with the condensed representations. For
example, although the number of non-derivable itemsets is usually less
than the number of free frequent itemsets, the collection of the free
frequent itemsets might still be more understandable since for most of
us choosing the minimum value is more natural operation than computing
all possible inclusion-exclusion truncations.

Data mining is an exploratory process to exploit the data. The data or
the patterns derived from the data might not be understandable as
whole and the right questions to be asked about the data are not
always known in advance. Thus, it would be useful to be able to answer
(approximately) to several queries to patterns and data. (A database
capable to support data mining by means of that kind of queries is
often called an \emph{inductive
database}~\cite{i:boulicaut04,a:deraedt03:pid,a:imielinski96,i:mannila97}.)
Three most important aspects of approximate query answering are the
following:
\begin{description}
\item[Representation size.]  The size of the summary structure needed
for answering the queries is very important. In addition to the actual
space required for the storage, the size can affect also the
efficiency of query answering: it is much more expensive to retrieve
patterns from, e.g., tertiary memory than doing small computations
based on patterns in main memory. For example, if all
$\sigma$-frequent itemsets and their frequencies fit into main memory,
then the frequency queries can be answered very efficiently for
$\sigma$-frequent itemsets compared to computing the frequency by
scanning through the complete transaction database that might reside
on an external server with heavy load. There are many ways how pattern
collections can be stored concisely.  For example, representing the
pattern collection and their quality values by listing just the
quality values leads to quite concise
representations~\cite{i:mielikainen04:implicit}.
\item[The efficiency of query answering.] It is not always known in
advance what should be asked about the data. Also, the pattern
collections can be too large to digest completely in one go. Thus,
different viewpoints to data and patterns might be helpful. The
efficient query answering can be provided by efficient index
structures. For example, although the number of closed frequent
itemsets is often considerably smaller than the number of all frequent
itemsets, retrieving the frequency of a given frequent itemset can be
more difficult. If all frequent itemsets are stored, then answering
the frequency query $\fr{X,\DB}$ can be implemented as a membership
query: the frequencies of the frequent itemsets can be stored in a
trie and thus the frequency of a frequent itemset $X$ can be found in
time linear in $\Abs{X}$.  Answering the same query when storing only
the closed frequent itemsets in a trie is much more difficult: in the
worst case the whole trie has to be transversed. This problem can be
relieved by inserting some additional links to the trie. The trie
representations can be generalized to deterministic automata
representations~\cite{i:mielikainen04:automata}.
\item[The accuracy of the answers.] Sometimes approximate answers to
queries are sufficient if they can be provided substantially faster
than the exact answers. Furthermore, it might be too expensive to
store all data (or patterns) and thus exact answers might be
impossible~\cite{i:babcock02}.  A simple approach to answer quite
accurately to many queries is to store a random sample of the data.
For example, storing a random subset $\DB'$ of a transactions in the
transaction database $\DB$ gives good approximations to frequency
queries~\cite{i:toivonen96,i:mielikainen04:ssi}.  Another alternative
is to store some subset of itemsets and estimate the unknown
frequencies from them~\cite{a:kessler02,i:mannila96,a:pavlov03}.  A
natural fusion of these two approaches is use both patterns and data
to represent the structure facilitating the possible
queries~\cite{i:geerts03}.  When the query answers are inaccurate, it
is often valuable to obtain some bounds to the errors. The frequencies
of the frequent itemsets, for example, can be bounded below and above
by, e.g., linear programming and (truncated)
inclusion-exclusion~\cite{i:bykowski02,i:calders02}.
\end{description}

%% file: views.tex

\chapter{Frequency-Based Views to Pattern Collections \label{c:views}}


It is a highly non-trivial task to define an (anti-monotone)
interestingness measure $\imeasure$ such that there is a minimum
quality value threshold $\sigma$ capturing almost all truly
interesting and only few uninteresting patterns in the collection. One
way to augment the interestingness measure is to define additional
constraints for the patterns. The use of constraints is a very
important research topic in pattern discovery but the research has
been concentrated mostly on structural constraints on patterns and
pattern
collections~\cite{i:bonchi03:ExAnte,a:bayardo00,i:deraedt02,i:goethals00:interactive,i:kifer03,a:lakshmanan03,i:mielikainen03:faosi,i:srikant97}.
Typical examples of structural constraints for patterns are
constraints for items and itemsets: an interesting itemset can be
required or forbidden to contain certain items or itemsets. Other
typical constraints for pattern collections are monotone and
anti-monotone constraints such as minimum and maximum frequency
thresholds, or minimum and maximum cardinality constraints for the
itemsets.

\begin{example}[constraints in itemset mining]
Let the set $\Items$ of items be products sold in a grocery store. The
transaction database $\DB$ could then consist of transactions
corresponding to purchases of customers that have bought something
from the shop at least three times. As a constrained itemset mining
task, we could be interested to find itemsets that
\begin{enumerate}
\item
do not contain garlic,
\item
consist of at least seven products,
\item
contain at least two vegetables or bread and sour milk, and
\item
cost at most ten euros.
\end{enumerate}
These constraints attempt to characterize global travelers that are
likely to become low-profit regular customers.

The first and the third constraint are examples of constraints for
items or itemsets. The second and the fourth constraints are examples
of anti-monotone and monotone constraints, respectively.

Clearly, all constraints could be expressed as boolean combinations of
item constraints, since that is sufficient for defining any
subcollection of $2^\Items$ and all constraints define a subcollection
of $2^\Items$.  However, that would not be very intuitive and also it
could be computationally very demanding to find all satisfying truth
assignments (corresponding to itemsets) for an arbitrary boolean
formula.  \exend \end{example}

In this chapter we propose a complementary approach to further
restrict and sharpen the collection of interesting patterns. The
approach is based on simplifying the quality values of the patterns
and it can be seen as a natural generalization of characterizing the
interesting patterns by a minimum quality value threshold $\sigma$ for
the quality values of the patterns. The quality value simplifications
can be adapted easily to pattern classes of various kind since they
depend only on the quality values of the interesting patterns and not
on the structural properties of the patterns. Simplifying the quality
values is suitable for interactive pattern discovery as
post-processing of a pattern collection containing the potentially
interesting patterns.  For example, in the case of itemsets, the
collection of potentially interesting patterns usually consists of the
$\sigma$-frequent itemsets for the smallest possible minimum frequency
threshold $\sigma$ such that the frequent itemset mining is still
feasible in practice.

In addition to making the collection more understandable in general,
the simplifications of the quality values can be used to reduce the
number of interesting patterns by discretizing the quality values and
removing the patterns whose discretized quality values can be inferred
(approximately) from the quality values of the patterns that are not
removed. Although there might be more powerful ways to condense the
collection of interesting patterns, the great virtue of discretization
is its conceptual simplicity: it is relatively understandable how the
discretization simplifies the structure of the quality values in the
collection of interesting patterns.

This chapter is based on the article ``Frequency-Based Views to
Pattern Collections''~\cite{i:mielikainen03:views}. For brevity, we
consider for the rest of the chapter frequencies instead of arbitrary
quality values.

\section{Frequency-Based Views}

A \emph{simplification of frequencies} is a mapping $\simplification :
\IntC{0,1} \to I$, where $I$ is a collection of non-overlapping intervals
covering the interval $\IntC{0,1}$, i.e.,
\begin{displaymath}
I \subset
\Set{\IntC{a,b},\IntRO{a,b},\IntLO{a,b},\IntO{a,b} \subseteq \IntC{0,1}}
\end{displaymath}
such that $\bigcup I = \IntC{0,1}$ and $i \cap j=\emptyset$ for all
$i,j \in I$. 

\begin{example}[frequent patterns] \label{ex:fp}
The collection $\Frequent{\sigma,\DB}$ of $\sigma$-frequent
patterns can be defined using frequency simplifications as follows:
\begin{displaymath}
\simp{\fr{p,\DB}}= \left\{ 
\begin{array}{ll}
\fr{p,\DB} \quad & \mbox{if } \fr{p,\DB} \geq \sigma \mbox{ and} \\
\IntRO{0,\sigma} & \mbox{otherwise.}
\end{array} 
\right.
\end{displaymath}
\exend \end{example}

There are several immediate applications of frequency
simplifications. They can be used, for example, to focus on some
particular frequency-based property of the pattern class.
\begin{example}[focusing on some frequencies] \label{ex:focus}
First, example~\ref{ex:fp} is an example of focusing on some
frequencies.

As a second example, the data analyst might be interested only in very
frequent (e.g., the frequency is at least $1-\epsilon$) and very
infrequent (e.g., the frequency is at most $\epsilon$) patterns. Then
the patterns in the interval $\IntO{\epsilon,1-\epsilon}$ could be
neglected or their frequencies could be mapped all to the interval
$\IntO{\epsilon,1-\epsilon}$. Thus, the corresponding frequency
simplification is the mapping
\begin{displaymath}
\simp{\fr{p,\DB}}= \left\{ 
\begin{array}{ll}
\IntO{\epsilon,1-\epsilon} & \mbox{if } \fr{p,\DB} \in
\IntO{\epsilon,1-\epsilon} \mbox{ and} \\ 
\fr{p,\DB} & \mbox{otherwise.}
\end{array} \right.
\end{displaymath}

As a third example, let us consider association rules.  The data
analyst might be interested in the rules with accuracy close to $1/2$
(e.g., within some positive constant $\epsilon$), i.e., the
association rules $\assoc{p}{p'}$ ($p,p' \in \Pat, p \preceq p'$) with
no predictive power. Thus, in that case the frequency simplification
$\simp{\acc{p'',\DB}}$ of the association rule $\assoc{p}{p'}$
(denoted by a shorthand $p''$) is
\begin{displaymath}
\simp{\acc{p'',\DB}}= \left\{ 
\begin{array}{ll}
\IntRO{0,1/2-\epsilon} & \mbox{if } \acc{p'',\DB} < 1/2-\epsilon, \\
\IntLO{1/2+\epsilon,1} & \mbox{if } \acc{p'',\DB} > 1/2+\epsilon \mbox{ and} \\
\acc{p'',\DB} & \mbox{otherwise.}
\end{array} \right.
\end{displaymath}
\exend \end{example}

Frequency simplifications are useful also in condensing collections
of frequent patterns. For an example of this, see
Section~\ref{s:conddisc}.  Other potential applications are speeding
up the pattern discovery algorithms, hiding confidential information
about the data from the pattern users, correcting or indicating errors
in data and in frequent patterns, and examining the stability of the
collection of frequent patterns.

Although the frequency simplifications in general may require a
considerable amount of interaction, defining simple mappings from the
unit interval $\IntC{0,1}$ to a collection of its subintervals and
applying the simplification in pattern discovery is often more
tractable than defining complex structural constraints with respect to
definability and computational complexity. Here are some examples of
simple mappings:
\begin{itemize}
\item
Points in a subinterval of $\IntC{0,1}$ can be replaced by the subinterval
itself.
\item
The points can be discretized by a given discretization function.
\item
Affine transformations, logarithms and other mappings can be applied
to the points.
\end{itemize}
Note that the simplification does not have to be applicable to all
points in $\IntC{0,1}$ but only to the finite number of different
frequencies $\fr{p,\DB}$ of the patterns at hand.

The frequency simplifications have clearly certain limitations, as
they focus just on frequencies, neglecting the structural aspects of
the patterns and the pattern collection (although the structure of the
pattern collection can be taken into account indirectly when defining
the simplification). For example, sometimes interesting and
uninteresting patterns can have the same frequency. Nevertheless, the
frequency simplifications can be useful in constrained pattern
discovery as a complementary approach to structural
constraints. Furthermore, the simplifications could be used to aid in
the search for advantageous constraints by revealing properties that
cannot be expressed by the frequencies.

\section{Discretizing Frequencies \label{s:discretization}}

Discretization is an important special case of simplifying
frequencies. In general, discretizations are used especially for two
purposes: reducing noise and decreasing the size of the
representation. As an example of these, let us look at $k$-means
clusterings.

\begin{example}[$k$-means clustering]
The $k$-means clustering of a (finite) point set $P \subseteq \RN^d$
tries to find a set $O$ of $k$ points in $\RN^d$ that minimize the
cost
\begin{displaymath}
\sum_{p \in P} \min_{o \in O} \sum_{i=1}^d \Paren{p_i-o_i}^2.
\end{displaymath}

This objective can be interpreted as trying to find the centers of $k$
Gaussian distributions that would be the most likely to generate the
point set $P$. Thus, each point in $P$ can be considered as a cluster
center plus some Gaussian noise.

The representation of the set $P$ by the cluster centers is clearly
smaller than the original point set $P$. Furthermore, if the centers
of the Gaussian distributions are far enough from each other, then the
points in $P$ can be encoded in smaller space by expressing for each
point $p \in P$ the cluster $o \in O$ where it belongs and the vector
$p-o$.

Note that in practice, the $k$-means clusterings are not always
correct ones, even if the assumption of $k$ Gaussian distributions
generating the set $P$ is true, because the standard algorithm used
for $k$-means clustering (known as \emph{the $k$-means algorithm}) is
a greedy heuristic. Furthermore, even if it were known to which
cluster each of the points in $P$ belongs to, the points in each
cluster rarely provide the correct estimate for the cluster
center. (For more details on $k$-means clustering, see
e.g.~\cite{b:hand01,b:hastie01}.)
\exend \end{example}

A discretization of frequencies can be defined as follows:
\begin{definition}[discretization of frequencies] \label{d:disc}
A \emph{discretization of frequencies} is a mapping $\disc{}{}{}$ from
$\IntC{0,1}$ to a (finite) subset of $\IntC{0,1}$ that preserves the
order of the points. That is, if $x,y \in \IntC{0,1}$ and $x\leq y$
then $\disc{x}{}{}\leq\disc{y}{}{}$. Points in the range of the
discretization function $\disc{}{}{}$ are called the
\emph{discretization points} of $\disc{}{}{}$.
\end{definition}

\begin{example}[discretization of frequencies]
Probably the simplest example of discretization functions is the
mapping $\disc{}{}{}$ that maps all frequencies in $\IntC{0,1}$ to
some constant $c \in \IntC{0,1}$.  Clearly, such $\disc{}{}{}$ is a
mapping from $\IntC{0,1}$ to a finite subset of $\IntC{0,1}$ and $x
\leq y \Rightarrow \disc{x}{}{} \leq \disc{y}{}{}$ for all $x,y \in
\IntC{0,1}$.
\exend \end{example}

One often very important requirement for a good discretization
function is that it should not introduce much error, i.e., the
discretized values should not differ too much from the original
values. In the next subsections we prove data-independent bounds for
the errors in accuracies of association rules with respect to certain
discretization functions of frequencies and give algorithms to
minimize the empirical loss of several loss functions.

To simplify the considerations, the frequencies of the patterns are
assumed to be strictly positive for the rest of the chapter.

\subsection{Loss Functions for Discretization}
The loss functions considered in this section are absolute error and
approximation ratio.

The absolute error for a point $x \in \IntLO{0,1}$ with respect to
a discretization function $\disc{}{}{}$ is
\begin{displaymath}
\err{x,\disc{}{}{}}{a}=\Abs{x-\disc{x}{}{}}
\end{displaymath}
and the maximum absolute error with respect to a discretization
function $\disc{}{}{}$ for a finite set $P \subset (0,1]$ of points is
\begin{equation} \label{eq:erra}
\err{P,\disc{}{}{}}{a}=\max_{x \in P} \err{x,\disc{}{}{}}{a}.
\end{equation}

In addition to the absolute error, also the relative error, i.e., the
approximation ratio is often used to evaluate goodness of the
approximation. The approximation ratio for a point $x \in \IntLO{0,1}$
is
\begin{displaymath}
\err{x,\disc{}{}{}}{r}=\frac{\disc{x}{}{}}{x}
\end{displaymath} 
and the maximum approximation ratio interval with respect to a
discretization function for a finite set $P \subset \IntLO{0,1}$ is
\begin{equation} \label{eq:errr}
\err{P,\disc{}{}{}}{r}=\IntC{
\min_{x \in P} \err{x,\disc{}{}{}}{r} , 
\max_{x \in P} \err{x,\disc{}{}{}}{r}}.
\end{equation}

Let $\err{x,\discretization}{}$ denote the loss for a point $x \in P$
with respect to a given discretization $\discretization$.  Sometimes
the most appropriate error for a point set is not the maximum error
$\max_{x \in P}\err{x,\disc{}{}{}}{}$ but a weighted sum of the errors
of the points in $P$. If the weight function is $\weight : P \to \RN$
then the weighted sum of errors is
\begin{equation} \label{eq:errs}
\err{P,\disc{}{}{}}{\weight}=\sum_{x \in P} \wg{x}\err{x,\disc{}{}{}}{}.
\end{equation}

In the next few subsections we derive efficient algorithms for
minimizing these loss functions defined by Equation~\ref{eq:erra},
Equation~\ref{eq:errr} and Equation~\ref{eq:errs}.

\subsection{Data-Independent Discretization}

In this subsection we show that the discretization functions
\begin{equation}
\disc{x}{\epsilon}{a}=\epsilon+2\epsilon\Floor{ \frac{x}{2\epsilon} } \label{eq:disc:a}
\end{equation}
and
\begin{equation}
\disc{x}{\epsilon}{r}=\Paren{1-\epsilon}^{1+2\Floor{ \Paren{\ln x}/\Paren{2\ln\Paren{1-\epsilon}}}} \label{eq:disc:e}
\end{equation}
are the worst case optimal discretization functions with respect to
the maximum absolute error and the maximum approximation ratio
interval, respectively.  Furthermore, we bound the maximum absolute
error and the intervals for approximation ratios for the accuracies of
association rules computed using the discretized frequencies.

Let us first study the optimality of the discretization functions.
The discretization function $\disc{}{\epsilon}{a}$ is optimal in the
following sense:
\begin{theorem} \label{t:abs-disc-opt}
Let $P \subset \IntLO{0,1}$ be a finite set.  Then
\begin{displaymath}
\err{P,\disc{}{\epsilon}{a}}{a} \leq \epsilon. 
\end{displaymath}
Furthermore, for any other data-independent discretization function $\disc{}{}{}$ with
less discretization points, $\err{P',\disc{}{}{}}{}>\epsilon$ for some
point set $P' \subset \IntLO{0,1}$ such that $\Abs{P}=\Abs{P'}$.
\end{theorem}
\begin{proof}
For any point $x \in \IntLO{0,1}$, the absolute error
$\err{x,\disc{}{\epsilon}{a}}{a}$ with respect to the discretization
function $\disc{}{\epsilon}{a}$ is at most $\epsilon$ since
\begin{displaymath}
2\epsilon\Floor{ \frac{x}{2\epsilon} } \leq x < 2\epsilon + 2\epsilon \Floor{ \frac{x}{2\epsilon}}
\end{displaymath}
and
\begin{displaymath}
\disc{x}{\epsilon}{a}=\epsilon + 2\epsilon\Floor{ \frac{x}{2\epsilon}}.
\end{displaymath}

Any discretization function $\discretization$ can be considered as a
collection $\discretization^{-1}$ of intervals covering the interval
$\IntLO{0,1}$. Each discretization point can cover an interval of
length at most $2\epsilon$ when the maximum absolute error is allowed
to be at most $\epsilon$. Thus, at least $\Ceil{1/(2\epsilon)}$
discretization points are needed to cover the whole interval
$\IntLO{0,1}$. The discretization function $\disc{}{\epsilon}{a}$ uses
exactly that number of discretization points.
\end{proof}

It can be observed from the proof of Theorem~\ref{t:abs-disc-opt} that
some maximum error bounds $\epsilon$ are unnecessary high.  Thus, the
maximum absolute error bound $\epsilon$ can be decreased without
increasing the number of discretization points.

\begin{corollary}
The bound $\epsilon$ for the maximum absolute error can be decreased
to 
\begin{displaymath}
\frac{1}{2\Ceil{ \frac{1}{2\epsilon}}}
\end{displaymath}
without increasing the number of discretization points when
discretizing by the function $\disc{}{\epsilon}{a}$.
\end{corollary}

The worst case optimality of the discretization function
$\disc{}{\epsilon}{r}$ can be shown as follows:
\begin{theorem}
Let $P \subset \IntLO{0,1}$ be a finite set. Then
\begin{displaymath}
\err{P,\disc{}{\epsilon}{r}}{r} \subseteq \IntC{1-\epsilon,\frac{1}{1-\epsilon}}.
\end{displaymath}
Furthermore, for any other data-independent discretization function $\disc{}{}{}$ with
less discretization points we have
\begin{displaymath}
\err{P',\disc{}{}{}}{} \not \subseteq \IntC{1-\epsilon,\frac{1}{1-\epsilon}}
\end{displaymath}
for some point set $P' \subset \IntLO{0,1}$ such that
$\Abs{P}=\Abs{P'}$.
\end{theorem}
\begin{proof}
Clearly,
\begin{displaymath}
\Floor{\frac{\ln x}{2\ln\Paren{1-\epsilon}}} \leq \frac{\ln
x}{2\ln\Paren{1-\epsilon}} \leq 1+\Floor{\frac{\ln
x}{2\ln\Paren{1-\epsilon}}}
\end{displaymath}
holds for all $x > 0$ and we can write
\begin{displaymath}
x=\Paren{1-\epsilon}^{\Paren{\ln x}/\Paren{\ln
\Paren{1-\epsilon}}}=\Paren{1-\epsilon}^{2\Paren{\ln x}/\Paren{2\ln
\Paren{1-\epsilon}}}.
\end{displaymath}
Thus,
\begin{eqnarray*}
1-\epsilon 
&=& \frac{\Paren{1-\epsilon}^{1+2\Paren{\ln x}/\Paren{2\ln\Paren{1-\epsilon}}}}{x} \\
&\leq& \frac{\Paren{1-\epsilon}^{1+2\Floor{ \Paren{\ln x}/\Paren{2\ln\Paren{1-\epsilon}}}}}{x} \\
&=& \frac{\Paren{1-\epsilon}^{-1+2+2\Floor{ \Paren{\ln x}/\Paren{2\ln\Paren{1-\epsilon}}}}}{x} \\
&\leq& \frac{\Paren{1-\epsilon}^{-1+2\Paren{\ln x}/\Paren{2\ln\Paren{1-\epsilon}}}}{x} = \frac{1}{1-\epsilon}.
\end{eqnarray*}

The discretization function $\disc{}{\epsilon}{r}$ is the worst case
optimal for any interval $\IntC{x,1} \subset \IntLO{0,1}$, since it
defines a partition of $\IntC{x,1}$ with maximally long intervals.
\end{proof}

Furthermore, the discretization function with the maximum absolute and
the maximum relative approximation error guarantees gives guarantees
for the maximum relative and the maximum absolute errors,
respectively, as follows.

\begin{theorem}
A discretization function with the maximum absolute error $\epsilon$
guarantees that a discretization of a point $x \in \IntLO{0,1}$ has
the relative error in the interval $\IntC{1-\epsilon/x,1+\epsilon/x}$.
\end{theorem}
\begin{proof}
By definition, the minimum and the maximum discretization errors of a
discretization function with the maximum absolute error at most
$\epsilon$ are $\Paren{x-\epsilon}/x=1-\epsilon/x$ and
$\Paren{x+\epsilon}/x=1+\epsilon/x$.
\end{proof}

\begin{theorem}
A discretization function with the maximum relative error in the
interval $\IntC{1-\epsilon,1+\epsilon}$ guarantees that a point $x \in
\IntLO{0,1}$ has the maximum absolute error at most
$\epsilon x$.
\end{theorem}
\begin{proof}
The discretization $\disc{x}{}{}$ of $x$ with the maximum relative
error in the interval $\IntC{1-\epsilon,1+\epsilon}$ is in the
interval $\IntC{\Paren{1-\epsilon}x,\Paren{1+\epsilon}x}$. Thus, the
maximum absolute error is
\begin{displaymath}
\max \Set{x-\Paren{1-\epsilon}x,\Paren{1+\epsilon}x-x}=\epsilon x
\end{displaymath}
as claimed.
\end{proof}

An important use of frequent patterns is to discover accurate
association rules. Thus, it would be very useful to be able to bound
the errors for the accuracies of the association rules. For simplicity,
we consider association rules over itemsets although all following
results hold for any pattern collections and quality values.

Let us first study how well the maximum absolute error guarantees for
frequency discretizations transfer to the maximum absolute error
guarantees for the accuracies of association rules.

\begin{theorem}
Let $\disc{}{\epsilon}{}$ be a discretization function with the
maximum absolute error $\epsilon$. The maximum absolute error for the
accuracy of the association rule $\assoc{X}{Y}$ when the frequencies
$\fr{X\cup Y,\DB}$ and $\fr{X,\DB}$ are discretized by
$\disc{}{\epsilon}{}$ is at most
\begin{displaymath}
\min \Set{1,\frac{2\epsilon}{\fr{X,\DB}}}.
\end{displaymath}
\end{theorem}
\begin{proof}
By definition, a discretization function preserves the order of points
in the discretizations. Because $\fr{X \cup Y,\DB}\leq \fr{X,\DB}$, we
have $\disc{\fr{X \cup Y,\DB}}{\epsilon}{} \leq
\disc{\fr{X,\DB}}{\epsilon}{}$.

Since the correct accuracies are always in the interval $\IntC{0,1}$,
the maximum absolute error is at most $1$.

The two extreme cases are
\begin{enumerate}
\item
when $\fr{X \cup Y,\DB}=\fr{X,d}-\delta > 0$ for arbitrary small
$\delta>0$, but $\disc{\fr{X \cup Y,\DB}}{\epsilon}{}=\fr{X \cup
Y,\DB}-\epsilon$ and $\disc{\fr{X ,\DB}}{\epsilon}{}=\fr{X \cup
Y,\DB}+\epsilon$, and
\item
when $\fr{X \cup Y,\DB}=\fr{X,d}-2\epsilon+\delta>0$ for arbitrary small
$\delta>0$, but $\disc{\fr{X \cup
Y,\DB}}{\epsilon}{}=\disc{\fr{X,\DB}}{\epsilon}{}$.
\end{enumerate}

In the first case, the worst case absolute error is at most
\begin{eqnarray*}
&&\Abs{\frac{\fr{X \cup Y,\DB}-\epsilon}{\fr{X \cup Y,\DB}+\epsilon}-\frac{\fr{X \cup Y,\DB}}{\fr{X,\DB}}} \\
& \leq & \Abs{\frac{\fr{X \cup Y,\DB}-\epsilon}{\fr{X,\DB}+\epsilon}-\frac{\fr{X \cup Y,\DB}}{\fr{X,\DB}}} \\ 
& = & \frac{\epsilon\fr{X,\DB}+\epsilon \fr{X \cup Y,\DB}}{\fr{X,\DB}^2+\epsilon \fr{X,\DB}} \\
& \leq & \frac{2\epsilon\fr{X,\DB}}{\fr{X,\DB}^2+\epsilon \fr{X,\DB}} \\
& \leq & \frac{2\epsilon}{\fr{X,\DB}+\epsilon}.
\end{eqnarray*}

In the second case, the absolute error in the worst case is at most
\begin{displaymath}
1-\frac{\fr{X,\DB}-2\epsilon}{\fr{X,\DB}}=\frac{2\epsilon}{\fr{X,\DB}}
\end{displaymath}
when $\fr{X,\DB} \geq 2\epsilon$.

Thus, the second case is larger and gives the upper bound. 
\end{proof}

Note that in the worst case the maximum absolute error can indeed be
$1$ as shown by Example~\ref{ex:tightness-of-the-bounds}.
 
\begin{example}[the tightness of the bound for $\disc{}{\epsilon}{a}$ and any $\epsilon>0$] \label{ex:tightness-of-the-bounds}
Let $\fr{X \cup Y,\DB}=\delta$ and
$\fr{X,\DB}=2\epsilon-\delta$. Then $\disc{\fr{X \cup
Y,\DB}}{\epsilon}{a}=\disc{\fr{X,\DB}}{\epsilon}{a}=\epsilon$. Thus,
\begin{displaymath}
\Abs{1-\frac{\fr{X \cup Y,\DB}}{\fr{X,\DB}}}=
\Abs{1-\frac{\delta}{2\epsilon-\delta}} \to 1
\end{displaymath}
when $\delta \to 0$.
\exend \end{example}

When the maximum absolute error for the frequency discretization
function is bounded, also the maximum relative error for the accuracies
of the association rules computed from discretized and original
frequencies can bounded as follows:
\begin{theorem}
Let $\disc{}{\epsilon}{}$ be a discretization function with the maximum
absolute error $\epsilon$. The approximation ratio for the accuracy of
the association rule $\assoc{X}{Y}$, when the frequencies $\fr{X\cup
Y,\DB}$ and $\fr{X,\DB}$ are discretized using the function
$\disc{}{\epsilon}{}$, is in the interval
\begin{displaymath}
\IntC{\max \Set{0, \frac{\fr{X \cup Y,\DB}-\epsilon}{\fr{X \cup Y,\DB}+\epsilon}},\frac{\fr{X,\DB}}{\fr{X \cup Y,\DB}}}.
\end{displaymath}
\end{theorem}
\begin{proof}
The smallest approximation ratio is obtained when $\fr{X,\DB}=\fr{X
\cup Y,\DB}+\delta$ where $\delta$ is an arbitrary small positive
value, but $\disc{\fr{X \cup Y,\DB}}{\epsilon}{}=\fr{X \cup
Y,\DB}-\epsilon$ and
$\disc{\fr{X,\DB}}{\epsilon}{}=\fr{X,\DB}+\epsilon$.  Then the
approximation ratio is
\begin{eqnarray*}
&&\frac{\fr{X \cup Y,\DB}-\epsilon}{\fr{X,\DB}+\epsilon}\Paren{\frac{\fr{X \cup Y,\DB}}{\fr{X,\DB}}}^{-1} \\
&=&
\frac{\fr{X \cup Y,\DB}\fr{X,\DB}-\epsilon\fr{X,\DB}}{\fr{X \cup Y,\DB}\fr{X,\DB}+\epsilon\fr{X \cup Y,\DB}} \\
&=&
\frac{\fr{X \cup Y,\DB}^2+\delta\fr{X \cup Y,\DB}-\epsilon\fr{X\cup Y,\DB} -\delta\epsilon}{\fr{X \cup Y,\DB}^2+\delta\fr{X \cup Y,\DB}+\epsilon\fr{X \cup Y,\DB}}.
\end{eqnarray*}

If $\delta \to 0$, then
\begin{displaymath}
\frac{\fr{X \cup Y,\DB}-\epsilon}{\fr{X,\DB}+\epsilon}\Paren{\frac{\fr{X \cup Y,\DB}}{\fr{X,\DB}}}^{-1}
\to \frac{\fr{X \cup Y,\DB}-\epsilon}{\fr{X \cup Y,\DB}+\epsilon}.
\end{displaymath}
Note that in that inequality, we assume that $\fr{X,\DB} \geq \fr{X
\cup Y,\DB} \geq \epsilon$ because, by Definition~\ref{d:disc}, all
discretized values are non-negative. Hence, we get the claimed lower
bound.

By the definition of the approximation ratio, the upper bound is
obtained when $\fr{X,\DB}\neq \fr{X \cup Y,\DB}$ but $\disc{\fr{X \cup
Y,\DB}}{\epsilon}{}=\disc{\fr{X,\DB}}{\epsilon}{}$.  The greatest
approximation ratio is obtained when $\fr{X \cup
Y,\DB}=\fr{X,\DB}-2\epsilon+\delta$ for arbitrary small $\delta>0$ but
$\disc{\fr{X \cup
Y,\DB}}{\epsilon}{}=\disc{\fr{X,\DB}}{\epsilon}{}$. Then the
approximation ratio is
\begin{displaymath}
\frac{\disc{\fr{X \cup Y,\DB}}{\epsilon}{}}{\disc{\fr{X,\DB}}{\epsilon}{}}
\Paren{\frac{\fr{X,\DB}-2\epsilon+\delta}{\fr{X,\DB}}}^{-1} \to
\frac{\fr{X,\DB}}{\fr{X,\DB}-2\epsilon}
\end{displaymath}  
when $\delta \to 0$.  If $\fr{X,\DB} \to 2\epsilon$, then the ratio
increases unboundedly.
\end{proof}

The worst case the relative error bounds for the discretization
function $\disc{}{\epsilon}{a}$ are the following.
\begin{example}[the worst case relative error bounds of $\disc{}{\epsilon}{a}$]
The smallest ratio is achieved when $\fr{X,\DB}=2k\epsilon$ and $\fr{X
\cup Y,\DB}=2k\epsilon-\delta$ for arbitrary small $\delta>0$ and some
$k \in \Set{1,\ldots,\Floor{1/\epsilon}}$. The ratio
\begin{displaymath}
\frac{(2k-1)\epsilon/(2k+1)\epsilon}{\Paren{2k\epsilon-\delta}/2\epsilon}
\end{displaymath}
is minimized by choosing $k=1$. Thus, the lower bound for the relative
error is $1/3$. 

The relative error cannot be bounded above since the frequencies
$\fr{X,\DB}=2\epsilon-\delta$ and $\fr{X \cup Y,\DB}=\delta$ with
discretizations $\disc{\fr{X,\DB}}{}{}=\disc{\fr{X \cup Y,\DB}}{}{}$
give the ratio
\begin{displaymath}
\frac{\epsilon/\epsilon}{\delta/\Paren{2\epsilon-\delta}}= 
\frac{2\epsilon}{\delta}-1
\to \infty
\end{displaymath}
when $\delta \to 0$ and $\epsilon>0$.
\exend \end{example}

The relative error for the accuracies of the association rules can be
bounded much better when discretizing by the discretization function
$\disc{}{\epsilon}{}$ having the approximation ratio guarantees
instead of the maximum absolute error guarantees.

\begin{theorem}
Let $\disc{}{\epsilon}{}$ be a discretization function with the
approximation ratio in the interval
$\IntC{\Paren{1-\epsilon},\Paren{1-\epsilon}^{-1}}$.  The
approximation ratio for the accuracy of the association rule
$\assoc{X}{Y}$ when the frequencies $\fr{X\cup Y,\DB}$ and
$\fr{X,\DB}$ are discretized by $\disc{}{\epsilon}{}$ is in the
interval
\begin{displaymath}
\IntC{\Paren{1-\epsilon}^{2},\Paren{1-\epsilon}^{-2}}.
\end{displaymath}
\end{theorem}
\begin{proof}
By choosing
\begin{displaymath}
\disc{\fr{X \cup Y,\DB}}{\epsilon}{}=\Paren{1-\epsilon}\fr{X \cup Y,\DB} 
\end{displaymath}
and
\begin{displaymath}
\disc{\fr{X ,\DB}}{\epsilon}{}=\Paren{1-\epsilon}^{-1}\fr{X,\DB}
\end{displaymath}
we get
\begin{displaymath}
\frac{\Paren{1-\epsilon}\fr{X \cup Y,\DB}}{\Paren{1-\epsilon}^{-1}\fr{X,\DB}}=
\Paren{1-\epsilon}^2\frac{\fr{X \cup Y,\DB}}{\fr{X,\DB}}.
\end{displaymath}
By choosing 
\begin{displaymath}
\disc{\fr{X \cup Y,\DB}}{\epsilon}{}=\Paren{1-\epsilon}\fr{X \cup Y,\DB}
\end{displaymath}
and
\begin{displaymath}
\disc{\fr{X ,\DB}}{\epsilon}{}=\Paren{1-\epsilon}^{-1}\fr{X,\DB} 
\end{displaymath}
we get
\begin{displaymath}
\frac{\Paren{1-\epsilon}^{-1}\fr{X \cup
Y,\DB}}{\Paren{1-\epsilon}\fr{X,\DB}}=
\Paren{1-\epsilon}^{-2}\frac{\fr{X \cup Y,\DB}}{\fr{X,\DB}}.
\end{displaymath}
It is easy to see that these are the worst case instances.
\end{proof}

Note that these bounds are tight also for the discretization function
$\disc{}{\epsilon}{r}$.

The discretization functions with the maximum absolute error
guarantees give also some guarantees for the approximation ratios of
accuracies.

\begin{theorem} \label{t:dirr}
Let $\disc{}{\epsilon}{}$ be a discretization function with the
approximation ratio in the interval
$\IntC{\Paren{1-\epsilon},\Paren{1-\epsilon}^{-1}}$.  Then the maximum
absolute error for the accuracy of the association rule $\assoc{X}{Y}$
when the frequencies $\fr{X\cup Y,\DB}$ and $\fr{X,\DB}$ are
discretized by $\disc{}{\epsilon}{}$ is at most
\begin{displaymath}
1-\Paren{1-\epsilon}^2=2\epsilon\Paren{1-\epsilon}.
\end{displaymath}
\end{theorem}
\begin{proof}
There are two extreme cases. First, the frequencies $\fr{X,\DB}$ and
$\fr{X \cup Y,\DB}$ can be almost equal but be discretized as far as
possible from each other, i.e.,
\begin{eqnarray*}
&&\Abs{\frac{\Paren{1-\epsilon}\fr{X \cup Y,\DB}}{\Paren{1-\epsilon}^{-1}\fr{X,\DB}}-\frac{\fr{X \cup Y,\DB}}{\fr{X,\DB}}} \\
&=&\Abs{\Paren{1-\epsilon}^2\frac{\fr{X \cup Y,\DB}}{\fr{X,\DB}}-\frac{\fr{X \cup Y,\DB}}{\fr{X,\DB}}} \\
&=&\Paren{1-\Paren{1-\epsilon}^2}
\frac{\fr{X \cup Y,\DB}}{\fr{X,\DB}}
\end{eqnarray*}
The maximum value is achieved by setting $\fr{X \cup
Y,\DB}=\fr{X,\DB}-\delta$ for arbitrary small $\delta>0$.

In the second case, the frequencies $\fr{X,\DB}$ and $\fr{X \cup
Y,\DB}$ are discretized to have the same value although they are as
apart from each other as possible. That is,
\begin{eqnarray*}
&&\Abs{\frac{\Paren{1-\epsilon}^{-1}\fr{X \cup Y,\DB}}{\Paren{1-\epsilon}\fr{X,\DB}}-\frac{\fr{X \cup Y,\DB}}{\fr{X,\DB}}} \\
&=&\Abs{\frac{\fr{X \cup Y,\DB}}{\Paren{1-\epsilon}^2\fr{X,\DB}}-\frac{\fr{X \cup Y,\DB}}{\fr{X,\DB}}} \\
&=&\Paren{\frac{1}{\Paren{1-\epsilon}^2}-1}\frac{\fr{X \cup Y,\DB}}{\fr{X,\DB}} \\
&=&\frac{1-\Paren{1-\epsilon}^2}{\Paren{1-\epsilon}^2}\frac{\fr{X \cup Y,\DB}}{\fr{X,\DB}}.
\end{eqnarray*}
However, in that case $\fr{X \cup Y,\DB}\leq
\Paren{1-\epsilon}^2\fr{X,\DB}$. Thus, the maximum absolute error is
again at most $1-\Paren{1-\epsilon}^2$.
\end{proof}

In this subsection we have seen that data-independent discretization
of frequencies with approximation guarantees can provide approximation
guarantees also for the accuracies of the association rules computed
from the discretized frequencies without any \emph{a priori}
information about the frequencies (especially when the frequencies are
discretized using a discretization function with the approximation
ratio guarantees).

\subsection{Data-Dependent Discretization}
In practice, taking the actual data into account usually improves the
performance of the approximation methods. Thus, it is natural to
consider also data-dependent discretization techniques. The problem of
discretizing frequencies by taking the actual frequencies into account
can be formulated as a computational problem as follows:
\begin{problem}[frequency discretization]
Given a finite subset $P$ of $\IntLO{0,1}$, a maximum error threshold
$\epsilon$ and a loss function $\error$, find a discretization
$\discretization$ for $P$ such that $\Abs{\disc{P}{}{}}$ is minimized
and the error $\err{P,\discretization}{}$ is at most
$\epsilon$.
\end{problem}

\begin{example}[frequency discretization]
Let the set $P \subset \IntLO{0,1}$ consist of points $1/10$, $3/10$,
$7/10$ and $9/10$, let the maximum error threshold $\epsilon$ be
$1/10$, and let the loss function $\error$ be the maximum absolute
error (Equation~\ref{eq:erra}).

Then the discretization function $\disc{}{}{}$ with smallest number of
discretization points and maximum absolute error at most $\epsilon$
is the mapping 
\begin{displaymath}
\disc{}{}{}=\Set{\frac{1}{10} \mapsto \frac{1}{5}, \frac{3}{10} \mapsto \frac{1}{5}, \frac{7}{10} \mapsto \frac{4}{5}, \frac{9}{10} \mapsto \frac{4}{5}}.
\end{displaymath}

If $\epsilon=1/9$ instead, then there are several mappings with the maximum absolute error at most $\epsilon$ and two discretization points. Namely,
all mappings
\begin{displaymath}
\disc{}{}{}=\Set{\frac{1}{10} \mapsto a, \frac{3}{10} \mapsto a, \frac{7}{10} \mapsto b, \frac{9}{10} \mapsto b}
\end{displaymath}
where $a \in \IntC{1/5-1/90,1/5+1/90}$ and $b \in
\IntC{4/5-1/90,4/5+1/90}$.  \exend \end{example}

In this subsection we derive sub-quadratic algorithms for discretizing
with respect to the maximum absolute error and polynomial-time
solutions for also many other classes of loss functions.

\subsubsection{Maximum absolute error}
A discretization of a point set $P \subseteq \IntLO{0,1}$ without
exceeding the maximum absolute error $\epsilon$ can be interpreted as
an interval cover of the point set $P$ with intervals of length
$2\epsilon$, i.e., a collection of length $2\epsilon$ sub-intervals of
$\IntC{0,1}$ that together cover all points in $P$.

A simple solution for the frequency discretization problem with the
loss function being the maximum absolute error is to repeatedly choose
the minimum uncovered point $d \in P$ and discretize all the
previously uncovered points of $P$ in the interval $\IntC{d,d+2\epsilon}$
to the value $d+\epsilon$. This is described as
Algorithm~\ref{a:Interval-Cover}.

\begin{algorithm}[h]
\caption{A straightforward algorithm for discretization with respect
to the maximum absolute error. \label{a:Interval-Cover}}
\begin{algorithmic}[1]
\Input{A finite set $P \subset \IntC{0,1}$ and a real value
$\epsilon \in \IntC{0,1}$.}
\Output{A discretization function $\discretization$ with
$\err{P,\discretization}{a} \leq \epsilon$.}
\Function{Interval-Cover}{$P,\epsilon$} 
\While{$P \neq \emptyset$} 
 \State $d \leftarrow \min P$ 
 \State $I \leftarrow \Set{x \in P : d \leq x \leq d+2\epsilon}$ 
 \ForAll{$x \in I$} 
  \State $\disc{x}{}{} \leftarrow d+\epsilon$ 
 \EndFor 
 \State $P \leftarrow P \setminus I$ 
\EndWhile 
\State \Return $\discretization$ 
\EndFunction
\end{algorithmic}
\end{algorithm}

\begin{theorem}
Algorithm~\ref{a:Interval-Cover} finds a discretization function
$\discretization$ such that the error $\err{P,\discretization}{a}$ is
at most $\epsilon$ and for all discretizations $\discretization'$ with
a smaller number of discretization points than $\Abs{\disc{P}{}{}}$
the error $\err{P,\discretization'}{a}$ is greater than $\epsilon$.
\end{theorem}
\begin{proof} 
The maximum absolute error is at most $\epsilon$ since all points are
covered by intervals of length $2\epsilon$ and the distance to the
center of any covering interval is at most $\epsilon$.

To see that a smaller number of discretization points would have a
larger error, let $x_1,\ldots,x_m$ be the discretization points of the
discretization $\discretization$ found by
Algorithm~\ref{a:Interval-Cover} for the point set $P$.  By
construction, there is a point $x_i-\epsilon \in P$ for each $1 \leq i
\leq m$. Furthermore, $\Abs{x_i-x_j}> 2\epsilon$ for all
discretization points $x_i$ and $x_j$ of $\discretization$ such that
$1 \leq i < j \leq m$, since otherwise the point $x_j-\epsilon \in P$
is contained in the interval $\IntC{x_i-\epsilon,x_i+\epsilon}$ or the
point $x_i-\epsilon \in P$ is contained in the interval
$\IntC{x_j-\epsilon,x_j+\epsilon}$.  Thus, no two points
$x_i-\epsilon,x_j-\epsilon \in P$ such that $1 \leq i<j\leq m$ can
share the same discretization point $x_k$ where $1 \leq k \leq m$.
\end{proof}

The straightforward implementation of Algorithm~\ref{a:Interval-Cover}
runs in time $\Oh{\Abs{P}^2}$. The bound is tight in the worst case as
shown by Example~\ref{ex:Interval-Cover}.

\begin{example}[The worst case running time of Algorithm~\ref{a:Interval-Cover}] \label{ex:Interval-Cover}
Let $P=\Set{1/\Abs{P},2/\Abs{P},\ldots,1-1/\Abs{P},1}$ and
$\epsilon<1/\Paren{2\Abs{P}}$. Then at each iteration only one point
is removed but all other points are inspected. There are $\Abs{P}$
iterations and the iteration $i$ takes time $\Oh{i}$. Thus, the total
time complexity is $\Oh{\Abs{P}^2}$.  \exend \end{example}

In the special case of $\epsilon$ being a constant, the time
complexity of the algorithm is linear in $\Abs{P}$ because each
iteration takes at most time $\Oh{\Abs{P}}$ and there can be at most
constant number of iterations: At each iteration, except possibly the
last one, at least length $2\epsilon$ subinterval of $\IntC{0,1}$ is
covered. Thus, the number of iterations can be bounded above by
$\Ceil{1/\Paren{2\epsilon}}=\Oh{1}$ and the total time needed is
$\Oh{\Abs{P}}$.

The worst case time complexity of the algorithm can be reduced to
$\Oh{\Abs{P}\log \Abs{P}}$ by constructing a heap for the point set
$P$. A minimum element in the heap can be found in constant time and
insertions and deletions to the heap can be done in time logarithmic
in $\Abs{P}$~\cite{b:knuth:3}.

The time complexity $\Oh{\Abs{P}\log\Abs{P}}$ is not optimal,
especially if preprocessing of the point set $P$ is allowed. For
example, if the set $P$ is represented as a sorted array, i.e., an
array $P$ such that $P[i] \leq P[j]$ for all $1 \leq i < j \leq
\Abs{P}$, then the problem can be solved in linear time in $\Abs{P}$
by Algorithm~\ref{a:Prefix-Cover}.

\begin{algorithm}[h]
\caption{A linear-time algorithm for discretizing a sorted point set
with respect to maximum absolute error. \label{a:Prefix-Cover}}
\begin{algorithmic}[1]
\Input{A finite set $P \subset \IntC{0,1}$ as an array in ascending
order and a real value $\epsilon \in \IntC{0,1}$.}
\Output{A discretization function $\discretization$ with
$\err{P,\discretization}{a} \leq \epsilon$.}
\Function{Prefix-Cover}{$P,\epsilon$} 
\For{$i=1,\ldots,\Abs{P}$}
 \If{$d<P[i]-\epsilon$}
  \State $d \leftarrow P[i]+\epsilon$
 \EndIf
 \State $\disc{P[i]}{}{} \leftarrow d$
\EndFor
\State \Return $\discretization$
\EndFunction
\end{algorithmic}
\end{algorithm}

The efficiency of Algorithm~\ref{a:Prefix-Cover} depends crucially on
the efficiency of sorting. In the worst case sorting real-valued
points takes time $\Oh{\Abs{P} \log \Abs{P}}$ but sometimes, for
example when the points are almost in order, the points can be sorted
faster. For example, the frequent itemset mining algorithm
\Apriori~\cite{i:agrawal96} finds the frequent itemsets in partially
descending order in their frequencies. Note that also the
generalization of the algorithm \Apriori, the levelwise algorithm
(Algorithm~\ref{a:Levelwise}) can easily be implemented in such a way
that it outputs frequent patterns in descending order in frequencies.

However, it is possible to find in time $\Oh{\Abs{P}}$ a
discretization function with maximum absolute error at most $\epsilon$
and the minimum number of discretization points, even if the points in
$P$ are not in ordered in some specific way in advance.  This can be
done by first discretizing the frequencies using the discretization
function $\disc{}{\epsilon}{a}$ (Equation~\ref{eq:disc:a}) and then
repairing the discretization. The high-level idea of the algorithm is
as follows:
\begin{enumerate}
\item
Put the points in $P$ into bins $0,1\ldots,\Floor{
1/\Paren{2\epsilon}}$ corresponding to intervals
$\IntLO{0,2\epsilon},\IntLO{2\epsilon,4\epsilon},\ldots,\IntLO{2\epsilon
\Floor{ 1/\Paren{2\epsilon}},1}$. Let $B$ be the set of bins such that
$B[i]$ corresponds to bin $i$.
\item
Find a minimal non-empty bin $i$ in $B$. (A non-empty bin $i$ is
called minimal if $i=0$ or the bin $i-1$ is empty.) 
\item
Find the smallest point $x$ in the bin $i$, replace the interval
corresponding to the bin $i$ by interval $\IntC{x,x+2\epsilon}$ and
move the points of the bin $i+1$ that are in the interval
$\IntC{x,x+2\epsilon}$ into the bin $i$.
\item
Remove bin $i$ from $B$.
\item
Go to step $2$ if there are still non-empty bins.
\end{enumerate}

The algorithm can be implemented to run in linear time in $\Abs{P}$:
The discretization to bins can be computed in time $\Oh{\Abs{P}}$
using a hash table for the set $B$~\cite{b:knuth:3}. A minimal
non-empty bin can be found in amortized constant time by processing
the consecutive runs of non-empty bins consecutively.

\begin{algorithm}[h!]
\caption{A linear-time algorithm for discretization with respect to
maximum absolute error. \label{a:Bin-Cover}}
\begin{algorithmic}[1]
\Input{A finite set $P \subset \IntC{0,1}$ and a real value
$\epsilon \in \IntC{0,1}$.}
\Output{A discretization function $\discretization$ with
$\err{P,\discretization}{a} \leq \epsilon$.}
\Function{Bin-Cover}{$P,\epsilon$}
\ForAll{$x \in P$}
 \State $i \leftarrow \Floor{ x/\Paren{2\epsilon}}$
 \State $B[i] \leftarrow B[i] \cup \Set{x}$
\EndFor
\ForAll{$B[i] \in B, B[i] \neq \emptyset$}
 \While{$i>0$ and $B[i-1]\neq\emptyset$}
  \State $i \leftarrow i-1$
  \State $d \leftarrow \min B[i]$
  \EndWhile
  \While{$B[i]\neq \emptyset$}
   \State $I \leftarrow \Set{x \in B[i] : d \leq x \leq d+2\epsilon}$
   \ForAll{$x \in I$}
    \State $\disc{x}{}{} \leftarrow d+\epsilon$
   \EndFor
   \State $B[i] \leftarrow B[i] \setminus I$
   \If{$\min B[i+1]<d+2\epsilon$}
    \State $i \leftarrow i+1$
   \Else
    \State $d \leftarrow \min B[i]$
   \EndIf
 \EndWhile
\EndFor
\State \Return $\discretization$
\EndFunction
\end{algorithmic}
\end{algorithm}

If the points in $P$ are given in an arbitrary order, then
Algorithm~\ref{a:Bin-Cover} is asymptotically optimal for minimizing
the number of discretization points with respect to the given maximum
absolute discretization error threshold $\epsilon$ as shown by
Theorem~\ref{t:optimaldisc}.

\begin{theorem} \label{t:optimaldisc}
No (deterministic) algorithm can find a discretization
$\discretization$ with the minimum number of discretization points
without inspecting all points in $P \subset \IntLO{0,1}$ when
$2\epsilon+\delta \leq 1$ for any $\delta>0$.
\end{theorem}
\begin{proof}
Let $P$ consist of points in the interval $\IntO{0,\delta}$ and
possibly the point $1$.  Furthermore, let the points examined by the
algorithm be in the interval $\IntO{0,\delta}$. Based on that
information, the algorithm cannot decide for sure whether or not the
point $1$ is in $P$.
\end{proof}

If the set $P$ is given in ascending or descending order, however,
then it is possible to find a set $\disc{P}{}{}$ of discretization
points of minimum cardinality among those that determine a
discretization of $P$ with the maximum absolute error at most
$\epsilon$, in time $\Oh{\Abs{\disc{P}{}{}} \log \Abs{P}}$ see
Algorithm~\ref{a:Log-Cover}. Although $\disc{P}{}{}$ is only an
implicit representation of the discretization function
$\discretization : P \to \disc{P}{}{}{}$, the discretization of any $x
\in P$ can be found in time $\Oh{\log \Abs{\disc{P}{}{}}}$ if the set
$\disc{P}{}{}$ is represented, e.g., as a sorted array.

\begin{algorithm}[h]
\caption{A sublinear-time algorithm for discretization a sorted point
set with respect to the maximum absolute error. \label{a:Log-Cover}}
\begin{algorithmic}[1]
\Input{A finite set $P \subseteq \IntC{0,1}$ as an array in ascending
order and a real value $\epsilon \in \IntC{0,1}$.}
\Output{A discretization points $\disc{P}{}{}$ with
$\err{P,\discretization}{a} \leq \epsilon$.}
\Function{Log-Cover}{$P,\epsilon$} 
\State $i \leftarrow 1$
\While{$i\leq\Abs{P}$}
 \State $d \leftarrow P[i]+\epsilon$
 \State $\disc{P}{}{} \leftarrow \disc{P}{}{} \cup \Set{d}$
 \State $j \leftarrow \Abs{P}+1$
 \While{$j>i+1$} 
  \State $k \leftarrow \Floor{\Paren{i+j}/2}$
  \If{$P[k]\leq d+\epsilon$}
   \State $i \leftarrow k$
  \Else
   \State $j \leftarrow k$
  \EndIf
 \EndWhile
 \State $i \leftarrow j$
\EndWhile
\State \Return $\disc{P}{}{}$
\EndFunction
\end{algorithmic}
\end{algorithm}

Note that the proposed techniques for discretizing with respect to the
maximum absolute error guarantees (i.e.,
Algorithms~\ref{a:Interval-Cover}, \ref{a:Prefix-Cover}, \ref{a:Bin-Cover}
and \ref{a:Log-Cover}) generalize to maximum error functions that are
strictly increasing transformations of the maximum absolute error
function. Furthermore, the algorithms can be modified to minimize
the maximum absolute error instead of the number of discretization
points by a simple application of binary search.

\subsubsection{Weighted sums of errors}
Sometimes it would be more natural to valuate the quality of
discretizations using a weighted sum
\begin{displaymath}
\sum_{x \in P} \wg{x}\err{x,\discretization}{}{}
\end{displaymath}
of errors $\err{x,\discretization}{}{}$ instead of the maximum error
$\max_{x \in P} \err{x,\discretization}{}{}$.  In that case, the
algorithms described previously in this chapter do not find the
optimal solutions. Fortunately, the problem can be solved optimally in
time polynomial in $\Abs{P}$ by dynamic programming; see
e.g.~\cite{a:fisher58,i:jagadish98}.

To describe the solution, we have to first define some notation.  Let
the point set $P$ be represented as an array in ascending order, i.e.,
$P[i]\leq P[j]$ for all $1 \leq i < j \leq \Abs{P}$, and let $P[i,j]$
denote the subarray $P[i] \ldots P[j]$.  The best discretization point
to represent the array $P[i,j]$ is denoted by $\mu_{i,j}$ and its
error by $\varepsilon_{i,j}$. The loss of the best discretization
$P[1,i]$ with $k$ discretization points with respect to the 
sum of errors is denoted by $\Delta^k_i$ and the $k-1$th
discretization point in that discretization is denoted by
$\omega^k_i$.

The optimal error for $P[1,i]$ using $k$ discretization points can be
defined by the following recursive formula:
\begin{displaymath}
\Delta^k_i = \left\{
\begin{array}{ll}
\varepsilon_{1,i} & \mbox{if } k=1 \mbox{ and} \\
\min_{k \leq j \leq i} \Set{\Delta^{k-1}_{j-1}+\varepsilon_{j,i}} \quad & \mbox{otherwise.}
\end{array} \right.
\end{displaymath}

The optimal sum-of-errors discretization by dynamic programming can be
divided into two subtasks:
\begin{enumerate}
\item
Compute the matrices $\mu$ of discretization points and $\varepsilon$
of their errors: $\mu_{i,j}$ is the discretization point for the
subset $P[i,j]$ and $\varepsilon_{i,j}$ is its error.
\item
Find the optimal discretizations for $P[1,i]$ with $k$ discretization
points for all $1\leq k \leq i \leq \Abs{P}$ from the matrices $\mu$
and $\varepsilon$ using dynamic programming.
\end{enumerate}

The optimal discretization function for $P$ can be found from any
matrix $\varepsilon \in \RN^{\Abs{P}\times\Abs{P}}$ of errors
and any matrix $\mu \in \RN^{\Abs{P}\times\Abs{P}}$ of
discretization points (although not all matrices $\varepsilon$ and
$\mu$ make sense nor are they computable).  For example, the matrices
can be given by an expert.

Simple examples of error and discretization point matrices computable
in polynomial time in $\Abs{P}$ are the matrices $\varepsilon$ and
$\mu$ for the weighted sums of absolute errors.  They can be computed
in time $\Oh{\Abs{P}^3}$ as described by
Algorithm~\ref{a:Valuate-Abs}. (Function \textproc{Median} computes
the weighted median of $P[i,j]$.)

\begin{algorithm}[h]
\caption{An algorithm to compute the loss and discretization matrices
$\varepsilon$ and $\mu$ for the point set $P$ and a weight function
$\weight$. \label{a:Valuate-Abs}}
\begin{algorithmic}[1]
\Input{A finite set $P \subset \IntC{0,1}$ and a weight function $\weight : P \to \RN$.}
\Output{Matrices $\varepsilon$ and $\mu$.}
\Function{Valuate-Abs}{$P,\weight$}
\For{$i=1,\ldots,\Abs{P}$}
 \For{$j=i,\ldots,\Abs{P}$}
  \State $\mu_{i,j} \leftarrow $ \Call{Median}{$P[i,j],\weight$}
  \State $\varepsilon_{i,j} \leftarrow 0$
  \For{$k=i,\ldots,j$}
   \State $\varepsilon_{i,j} \leftarrow \varepsilon_{i,j} + \wg{P[k]}\Abs{P[k]-\mu_{i,j}}$
  \EndFor
 \EndFor
\EndFor
\State \Return $\Tuple{\varepsilon,\mu}$
\EndFunction
\end{algorithmic}
\end{algorithm}

The discretization points $\mu_{i,j}$ and the errors
$\varepsilon_{i,j}$ of $P[i,j]$ for all $1\leq i \leq j \leq \Abs{P}$
can already be informative summaries of the set $P$. Besides of that,
it is possible to extract from the matrices $\varepsilon$ and $\mu$
the matrices $\Delta$ and $\omega$ corresponding to the partial sums
of errors and the discretizations.  This can be done by
Algorithm~\ref{a:Tabulator}. (The matrices $\Delta$ and $\omega$
determine the optimal discretizations for each number of
discretization points and each prefix $P[1,i]$ of $P$.)

\begin{algorithm}[h]
\caption{An algorithm to compute matrices $\Delta$ and $\omega$ from $P$, $\varepsilon$ and $\mu$. \label{a:Tabulator}}
\begin{algorithmic}[1]
\Input{A finite set $P \subset \IntC{0,1}$, and matrices $\varepsilon$ and $\mu$.}
\Output{Matrices $\Delta$ and $\omega$.}
\Function{Tabulator}{$P,\varepsilon,\mu$}
\ForAll{$i \in \Set{1,\ldots,\Abs{P}}$} \Comment{Initialize the errors $\Delta^k_i$.}
 \State $\Delta^1_i \leftarrow \varepsilon_{1,i}$
\EndFor
\ForAll{$k,i \in \Set{2,\ldots,\Abs{P}}, k \leq i$}
 \State $\Delta^k_i \leftarrow \infty$
\EndFor
\For{$k=1,\ldots,\Abs{P}$} \Comment {Find the best discretization of $P[1,i]$ with $k$ discretization points.}
 \State $\Delta' \leftarrow \infty$
 \ForAll{$j,i \in \Set{k,\ldots,\Abs{P}}, j \leq i$} 
  \If{$\Delta'<\Delta^k_i$}
   \State $\Delta^k_i \leftarrow \Delta'$
   \State $\omega^k_i \leftarrow j-1$
  \EndIf
 \EndFor
\EndFor
\State \Return $\Tuple{\Delta,\omega}$
\EndFunction
\end{algorithmic}
\end{algorithm}

The time complexity of Algorithm~\ref{a:Tabulator} is
$\Oh{\Abs{P}^3}$. The time consumption can be reduced to
$\Oh{k\Abs{P}^2}$ if we are interested only on discretizations with at
most $k$ discretization points. Furthermore, the method can be adapted
to other kinds of loss functions, too. For some loss functions, the
dynamic programming can be implemented with asymptotically better
efficiency guarantees~\cite{a:elomaa01,i:jagadish98}. There are
several ways to speed up the search in practice. For example, it is
not necessary to compute the parts of the matrices that are detected
to be not needed in the best solutions.

\begin{algorithm}[h!]
\caption{An algorithm to extract the best discretization of $k$
discretization points from the matrices $\Delta$ and
$\omega$. \label{a:Find-Discretization}}
\begin{algorithmic}[1]
\Input{A finite set $P \subset \IntC{0,1}$, matrices $\Delta$, $\mu$
and $\omega$, and an integer $k \in \Set{1,\ldots,\Abs{P}}$.}
\Output{The discretization $\discretization$ of $k$ discretization
points with the smallest error $\Delta^k_{\Abs{P}}$.}
\Function{Find-Discretization}{$P,\mu,\omega,k$}
\State $i \leftarrow \Abs{P}$
\For{$l=k,\ldots,1$}
 \For{$j=i,\ldots,\omega^l_i+1$}
  \State $\disc{P[i]}{}{} \leftarrow \mu_{\omega^j_i,i}$
 \EndFor
  \State $i \leftarrow \omega^l_i$
\EndFor
\State \Return $\discretization$
\EndFunction
\end{algorithmic}
\end{algorithm}

Although the matrices $\Delta$ and $\omega$ contain the information
about the optimal discretizations of all prefixes of $P$ for each
number of discretization points, usually the actual goal is to extract
the optimal discretizations from these matrices. 

The optimal discretizations of $k$ discretization points can be found
in time $\Oh{\Abs{P}}$ by Algorithm~\ref{a:Find-Discretization}.  It
can be adapted to find discretization with minimum number of
discretization points and the error less than $\epsilon$ in time
linear in $\Abs{P}$.
Note that if it is sufficient to obtain just the set $\disc{P}{}{}$ of
$k$ discretization points, then the task can be conducted in time
$\Oh{k}$ by Algorithm~\ref{a:Find-Discretization-Points}.

\begin{algorithm}[h!]
\caption{An algorithm to extract the best $k$ discretization points from the matrices $\Delta$ and
$\omega$. \label{a:Find-Discretization-Points}}
\begin{algorithmic}[1]
\Input{A finite set $P \subset \IntC{0,1}$, matrices $\Delta$, $\mu$
and $\omega$, and an integer $k \in \Set{1,\Abs{P}}$.}
\Output{The set $\disc{P}{}{}$ of $k$ discretization points with
points with the smallest error $\Delta^k_{\Abs{P}}$.}
\Function{Find-Discretization-Points}{$P,\mu,\omega,k$}
\State $\disc{P}{}{} \leftarrow \emptyset$
\State $i \leftarrow \Abs{P}$
\For{$l=k,\ldots,1$}
 \State $j \leftarrow \omega^l_i+1$
 \State $\disc{P}{}{} \leftarrow \disc{P}{}{} \cup \Set{\mu_{j,i}}$
 \State $i \leftarrow \omega^l_i$
\EndFor
\State \Return $\disc{P}{}{}$
\EndFunction
\end{algorithmic}
\end{algorithm}

Instead of finding the best discretization with a certain number of
discretization points, one could search for a hierarchical
discretization suggesting a good discretization of $k$ discretization
points for all values of $k$.

\begin{example}[hierarchical discretizations] \label{ex:hierdisc}
Let the point set $P$ be $\Set{0.1,0.2,0.5,0.6,0.9,1.0}$ and let us
consider hierarchical discretizations with respect to the maximum
absolute error. Two standard approaches to define hierarchical
clusterings are divisive (or top-down) and agglomerative (or
bottom-up) clusterings.

Divisive hierarchical clustering starts from the whole point set and
recursively divides it in such a way that the division always improves
the solution as much as possible. For example, the divisive clustering
of $P$ would be the following:

\begin{itemize}
\item
The first level of the clustering consists of only one cluster, namely
$\Set{0.1,0.2,0.5,0.6,0.9,1.0}$.
\item
The maximum absolute error is decreased as much as possible by
splitting the set into two parts $\Set{0.1,0.2,0.5}$ and
$\Set{0.6,0.9,1.0}$
\item
In the third level no split improves the maximum absolute
error. However, splitting $\Set{0.1,0.2,0.5}$ to $\Set{0.1,0.2}$ and
$\Set{0.5}$, or splitting $\Set{0.6,0.9,1.0}$ to $\Set{0.6}$ and
$\Set{0.9,1.0}$ decreases most the maximum absolute error for one of
the clusters with the maximum absolute error.
\item
The fourth level consists of the clusters $\Set{0.1,0.2}$,
$\Set{0.5}$, $\Set{0.6}$, and $\Set{0.9,1.0}$.
\item
In the fifth level we have again two equally good splitting
possibilities: $\Set{0.1,0.2}$ to $\Set{0.1}$ and $\Set{0.2}$, or
$\Set{0.9,1.0}$ to $\Set{0.9}$ and $\Set{1.0}$.
\item
The last level consists of singletons $\Set{0.1}$, $\Set{0.2}$,
$\Set{0.5}$, $\Set{0.6}$, $\Set{0.9}$, and $\Set{1.0}$.
\end{itemize}

Agglomerative hierarchical clustering starts from the singletons and
merges the clusters by minimizing the error introduced by the merges.
Thus, the agglomerative clustering of would be the following: First
level consists of singletons $\Set{0.1}$, $\Set{0.2}$, $\Set{0.5}$,
$\Set{0.6}$, $\Set{0.9}$, and $\Set{1.0}$. In the next three levels
$\Set{0.1}$ and $\Set{0.2}$, $\Set{0.5}$ and $\Set{0.6}$, and
$\Set{0.9}$ and $\Set{1.0}$ are merged in some order. Thus, the level
four consists of clusters $\Set{0.1,0.2}$, $\Set{0.5,0.6}$, and
$\Set{0.9,1.0}$. In the level five either $\Set{0.1,0.2}$ is merged
with $\Set{0.5,0.6}$, or $\Set{0.5,0.6}$ is merged with
$\Set{0.9,1.0}$. The last level consists of the set $P$.

It depends on the actual use of the discretized values which one of
these two approaches to hierarchical clustering is better.  \exend
\end{example}

In addition to standard divisive and agglomerative hierarchical
discretizations, it is possible to find hierarchical discretizations
that are optimal with respect to a given permutation $\pi :
\Set{1,\ldots,\Abs{P}} \to \Set{1,\ldots,\Abs{P}}$ in the following
sense: The discretization with $\pi(1)$ discretization points
has the minimum error among all discretizations with $\pi(1)$
discretization points.  The discretization with $\pi(2)$
discretization points is the one that has the minimum error among all
discretizations compatible with the discretization with $\pi(1)$
discretization points. In general, the discretization with $\pi(i)$
discretization points has the minimum error among the discretizations
with $\pi(i)$ discretization points that are compatible with the
chosen discretizations with $\pi(1),\pi(2),\ldots,\pi(i-1)$
discretization points.

The time complexity of the straightforward dynamic programming
implementation of this idea by modifying Algorithm~\ref{a:Tabulator} is
$\Oh{\Abs{P}^4}$.  Furthermore, for certain loss functions it is
possible to construct hierarchical discretizations that are close to
optimal for all values of the number of discretization points
simultaneously~\cite{i:dasgupta02}.

The discretizations could be applied to association rules instead of
frequent patterns. In that case, there are two values to be
discretized for each association rule: the frequency and the accuracy
of the rule. This can be generalized for patterns with $d$-dimensional
vectors of quality values. The problem is equivalent to clustering,
and thus in general, the problem is \NP-hard but many known
approximation algorithms for clustering can be
applied~\cite{i:badoiu02,i:delavega03,i:feder88,a:kanungo04,i:kumar04,a:kannan04}.

\section{Condensation by Discretization \label{s:conddisc}}

Discretization of frequencies can be used to simplify the collections
of frequent patterns. The high-level schema is the following:
\begin{enumerate}
\item
Discretize the frequencies of the frequent patterns.
\item
Find a condensed representation for the pattern collection with the
discretized frequencies.
\end{enumerate}

For example, the collection of closed frequent itemsets can be
approximated by the closed frequent itemsets with respect to
discretized frequencies.

\begin{example}[condensation by discretization and closed itemsets]
Let $\Items=\Set{1,\ldots,\Floor{(1-\sigma)n}}$, $\sigma \in \IntO{0,1}$ and
\begin{eqnarray*}
\DB&=&\Set{\Tuple{1,\Set{1}},\ldots,\Tuple{\Floor{(1-\sigma)n},\Set{\Floor{(1-\sigma)n}}}} \\
&&\cup \Set{\Tuple{\Floor{(1-\sigma)n}+i,\Items} : i \in \Set{1,\ldots,\Ceil{\sigma n}}}.
\end{eqnarray*}
Then 
\begin{displaymath}
\Closed{\sigma,\DB}=\Set{\Set{1},\ldots,\Set{n},\Items}
\end{displaymath}
with $\fr{\Items,\DB}=\Ceil{\sigma\Abs{\DB}}/\Abs{\DB}$ and
$\fr{\Set{A},\DB}=\Ceil{\sigma\Abs{\DB}+1}/\Abs{\DB}$ for each $A \in
\Items$.

If we allow error $1/\Abs{\DB}$ in the frequencies, then we can
discretize all frequencies of the non-empty $\sigma$-frequent closed
itemsets in $\DB$ to $\Ceil{\sigma\Abs{\DB}}/\Abs{\DB}$, i.e.,
$\Comp{\discretization}{\freq}{\emptyset,\DB}=1$ and
$\Comp{\discretization}{\freq}{X,\DB}=
\Ceil{\sigma\Abs{\DB}}/\Abs{\DB}$ for all other $X \subseteq
\Items$. 

Then the collection $\Closed{\sigma,\DB,\discretization}$ of
$\sigma$-frequent closed itemsets with respect to the discretization
$\discretization$ consists only of two itemsets $\emptyset$ and
$\Items$ with frequencies $1$ and $\Ceil{\sigma \Abs{\DB}}/\Abs{\DB}$.
\exend \end{example}

Note that if the original transaction database is available, then a
slightly similar approach to condense the collection of frequent
itemsets is to take a random sample of the transactions and compute
the closed frequent itemsets in the sample. This reduces the number of
closed itemsets but still results relatively good approximation for
the frequencies of the frequent
itemsets~\cite{i:mielikainen04:ssi,i:toivonen96}.  A major advantage
of computing the closed frequent itemsets in a sample of transactions
is that computing the closed frequent itemsets in the sample is
potentially much faster than computing the collection of (closed)
itemsets in the original data and discretizing the
frequencies. Disadvantages of this sampling approach are that the
outcome of the closed itemset mining from the sample is also a random
variable depending on the sample, and that the quality of the
approximation provided by the closed frequent itemsets in the sample
is at most as good as the quality of the optimal approximation
provided by the optimal discretization of the frequencies. Of course,
the sampling and discretization could be used in conjunction, by first
taking a relatively large sample of transactions for obtaining the
closed frequent itemsets efficiently and then discretizing the
frequencies of the closed frequent itemsets in the sample. This should
provide the computational efficiency and the approximation quality in
between of sampling and discretizing. We focus, however, solely on
discretizations.

\begin{realexample}[closed itemsets disappearing in the course completion database] \label{rex:clodisc}
Let us consider the collection $\Closed{\sigma,\DB}$ of closed
$0.20$-frequent itemsets in the course completion database (see
Subsection~\ref{ss:ds}). Recall (Example~\ref{rex:closed}) that the
number $\Abs{\Closed{\sigma,\DB}}$ of the closed $0.20$-frequent
itemsets in the course completion database is $2136$.

If the supports are discretized with the maximum absolute error $2$
(that is less than $0.1$ percent of the number of transactions in the
database), then the number of closed itemsets with respect to the
discretized supports is only $567$, i.e., less than $24$ percent of
$\Abs{\Closed{\sigma,\DB}}$.

In some parts of the itemset collection $\Closed{\sigma,\DB}$ the
reduction in the number of the closed itemsets can be even greater
than the average. For example, there are eight subsets of the itemset
$X=\Set{3,5,7,13,14,15,20}$ than are closed with respect to exact
supports but that have the same discretized support as $X$.  These
itemsets are shown in Table~\ref{t:rex:clodisc}.

\begin{table}[h!!] \centering
\caption{The itemsets with the same discretized support as
$\Set{3,5,7,13,14,15,20}$ in the course completion database. The
column are as follows: $\supp{X,\DB}$ is the exact support of the
itemset $X$ in the completion database $\DB$,
$\discretization^2_a(\supp{X,\DB})$ is $\supp{X,\DB}$ discretized with
maximum absolute error $2$, and $X \in \Closed{\sigma,\DB}$ is the
itemset $X$. \label{t:rex:clodisc}}
\begin{tabular}{|c|c|c|}
\hline
$\supp{X,\DB}$ & $\disc{\supp{X,\DB}}{2}{a}$ & $X \in \Closed{\sigma,\DB}$ \\
\hline
\hline
488 & 490 & $\{ 3,5,7,13,14,15,20 \}$  \\
489 & 490 & $\{ 3,5,7,13,15,20 \}$ \\
489 & 490 & $\{ 3,5,13,14,15,20 \}$ \\
490 & 490 & $\{ 3,5,13,15,20 \}$ \\
490 & 490 & $\{ 3,7,13,14,15,20 \}$ \\ 
491 & 490 & $\{ 3,13,14,15,20 \}$ \\
492 & 490 & $\{ 3,7,13,15,20 \}$ \\
492 & 490 & $\{ 5,7,13,14,15,20 \}$ \\
\hline
\end{tabular}
\end{table}
\exend \end{realexample}

The number of discretization points determines the quality of the
approximation: On one extreme --- a discretization using only one
discretization point --- the frequent itemsets that are closed with
respect to the discretized frequencies correspond to maximal frequent
itemsets.  When the number of discretization points increases, also
the number of closed frequent itemsets increase, the other extreme
case being the collection of frequent closed itemsets without any
discretization.

If the condensed representation depends on testing whether the
frequencies of the patterns are equal (such condensed representations
are, for example, the closed and the free patterns), then the number
of discretization points can be used as an estimate of the
effectiveness of the discretization. In addition to simplifying the
collections of frequent patterns, discretization can be used to make
the discovery of some patterns more efficient.

We evaluated the condensation abilities of discretizations by
discretizing the frequencies of the frequent itemsets in the Internet
Usage and IPUMS Census databases (see Subsection~\ref{ss:ds}), and
then computing which of the frequent itemsets are closed also with
respect to the discretized frequencies. (In these experiments, we
omitted the empty itemset from the itemset collections since its
frequency is always $1$.)

In the first series of experiments we were interested whether
data-dependent discretizations yield to smaller collections of closed
itemsets than their data-independent counterparts. We discretized the
frequencies using discretization function $\disc{}{\epsilon}{a}$
(Equation~\ref{eq:disc:a}) and the algorithm \textproc{Prefix-Cover}
(Algorithm~\ref{a:Prefix-Cover}) with different maximum absolute error
thresholds $\epsilon$ and removed the itemsets that were not closed
with respect to the discretized frequencies.

The results for Internet Usage database with the minimum frequency
threshold $0.05$ are shown in Table~\ref{t:disc:internet}.  The number
of the $0.05$-frequent itemsets, the number of the closed
$0.05$-frequent itemsets and the number of the maximal $0.05$-frequent
itemsets in Internet Usage database are $143391$, $141568$, and
$23441$, respectively.

\begin{table}[h] \center
\caption{The number of closed itemsets in the collection of
$0.05$-frequent itemsets in the Internet Usage database with
discretized frequencies for different maximum absolute error
guarantees.  The columns of the table are the maximum absolute error
$\epsilon$ allowed, the number of $\sigma$-frequent itemsets that are
closed with respect to the frequencies discretized using
Equation~\ref{eq:disc:a} and the number of $\sigma$-frequent itemsets
that are closed with respect to the frequencies discretized using
Algorithm~\ref{a:Prefix-Cover}.
\label{t:disc:internet}}
\begin{tabular}{|c|c|c|}
\hline
$\epsilon$ & fixed discretization & empirical discretization \\
\hline
\hline
$0.0010$ & $123426$ & $123104$ \\
$0.0050$ & $72211$ & $71765$ \\
$0.0100$ & $54489$ & $45944$ \\
$0.0200$ & $34536$ & $31836$ \\
$0.0400$ & $31587$ & $25845$ \\
$0.0600$ & $26087$ & $24399$ \\ 
$0.0800$ & $24479$ & $23916$ \\
$0.1000$ & $23960$ & $23705$ \\
\hline
\end{tabular}
\end{table}

\begin{table}[h!] \center
\caption{The number of closed itemsets in the collection of
$0.2$-frequent itemsets in the IPUMS Census database with discretized
frequencies for different maximum absolute error guarantees.  The
columns of the table have the same interpretation as the columns of
Table~\ref{t:disc:internet}. \label{t:disc:ipums}}
\begin{tabular}{|c|c|c|}
\hline
$\epsilon$ & fixed discretization & empirical discretization \\
\hline
\hline
$0.0010$ & $3226$ & $3242$ \\
$0.0050$ & $2362$ & $2375$ \\ 
$0.0100$ & $1776$ & $1772$ \\
$0.0200$ & $1223$ & $1225$ \\
$0.0400$ & $1014$ & $841$ \\
$0.0600$ & $932$ & $725$ \\
$0.0800$ & $711$ & $661$ \\
$0.1000$ & $627$ & $627$ \\
\hline
\end{tabular}
\end{table}

The results for IPUMS Census database with the minimum frequency
threshold $0.2$ are shown in Table~\ref{t:disc:ipums}. The results
were similar to other minimum frequency thresholds. The number of the
$0.2$-frequent itemsets, the number of the $0.2$ frequent closed
itemsets and the number of the maximal $0.2$-frequent itemsets in
IPUMS Census database are $86879$, $6689$, and $578$, respectively.
                                                                                
Clearly, the number of closed $\sigma$-frequent itemsets is an upper
bound and the number of maximal $\sigma$-frequent itemsets is a lower
bound for the number of frequent itemsets that are closed with respect
to the discretized frequencies.  The maximum absolute error is
minimized in the case of just one discretization point by choosing its
value to be the average of the maximum and the minimum frequencies.
The maximum absolute error for the best discretization with only one
discretization point for the $0.05$-frequent itemsets in the Internet
Usage database is $0.4261$. This is due to the fact that the highest
frequency in the collection of the $0.05$-frequent itemsets in the
Internet Usage database (excluding the empty itemset) is $0.9022$.
The maximum absolute error for the best discretization with one
discretization point for the $0.2$-frequent itemsets in the IPUMS
Census database is $0.4000$. That is, there is an itemset with
frequency equal to $0.2$ and an itemset with frequency equal to $1$ in
the collection of $0.2$-frequent itemsets in the IPUMS Census
database.

In addition to minimizing the maximum absolute error, we computed the
optimal discretizations with respect to the average absolute error
using dynamic programming (Algorithms~\ref{a:Valuate-Abs},
\ref{a:Tabulator} and \ref{a:Find-Discretization}).  In particular, we
computed the optimal discretizations for each possible number of
discretization points. The practical feasibility of the dynamic
programming discretization depends crucially on the number
$N=\Abs{\fr{\Frequent{\sigma,\DB},\DB}}$ of different frequencies as
its time complexity is $\Oh{N^3}$. Thus, the tests were conducted
using smaller collections of frequent itemsets than in the case of
discretization with respect to the maximum absolute error.

For the average absolute error, a uniform weighting over the frequent
itemsets were used. That is, the error of the discretization
$\discretization$ was
\begin{displaymath}
\frac{1}{\Abs{\Frequent{\sigma,\DB}}}\sum_{X \in \Frequent{\sigma,\DB}}
\Abs{\fr{X,\DB}-\disc{\fr{X,\DB}}{}{}}.
\end{displaymath}
 
\begin{figure} 
\includegraphics[width=\columnwidth]{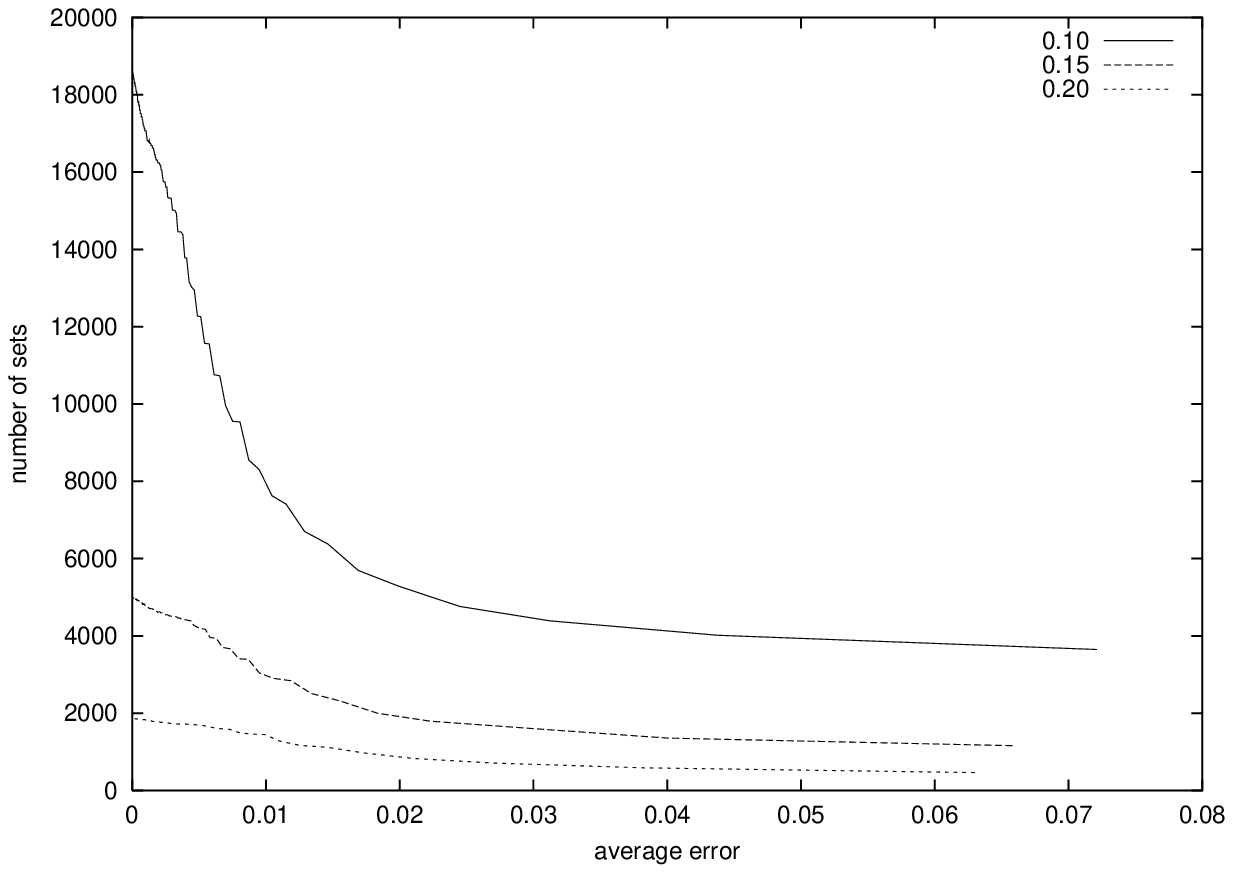}

\includegraphics[width=\columnwidth]{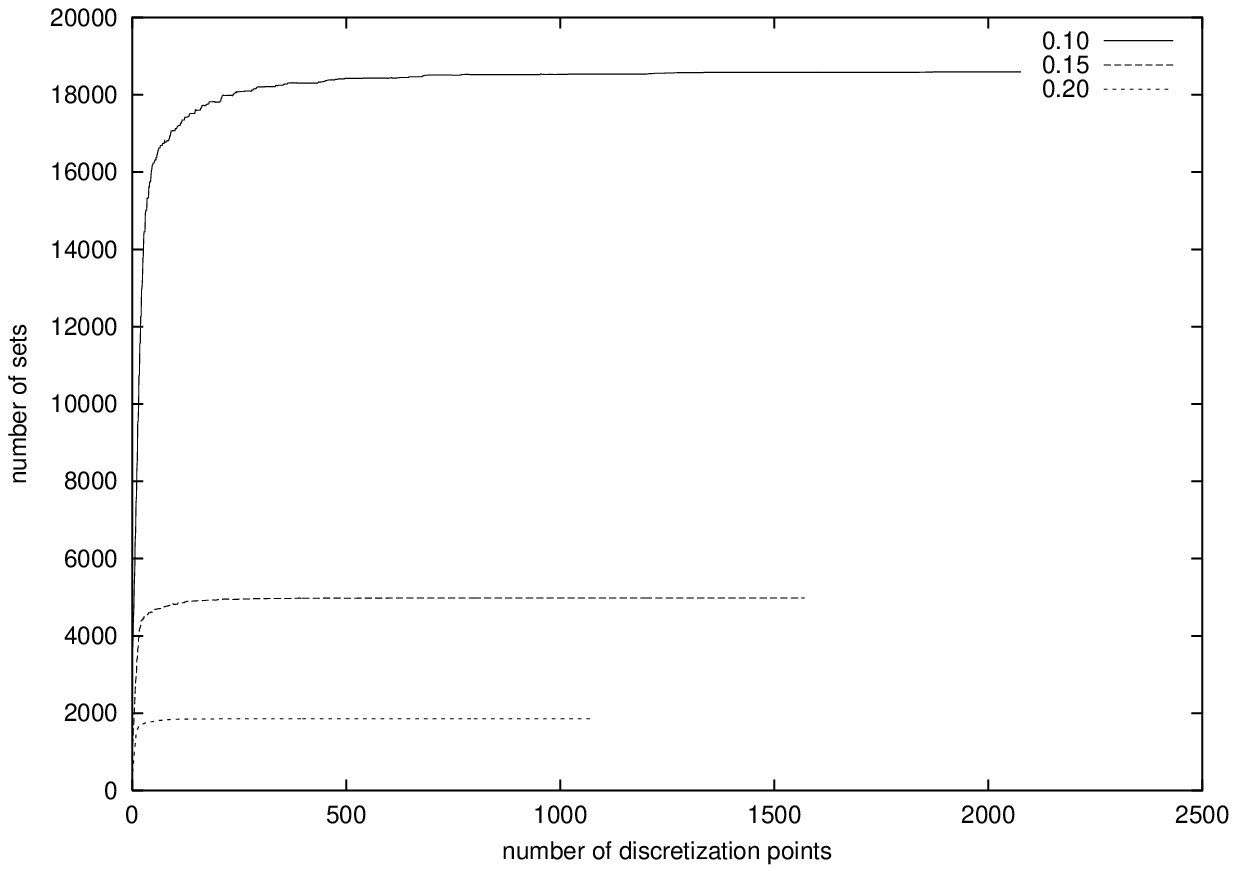}
\caption{The best average absolute error discretizations for Internet Usage data. \label{f:disc:internet}}
\end{figure}
                                                                                
\begin{figure}
\includegraphics[width=\columnwidth]{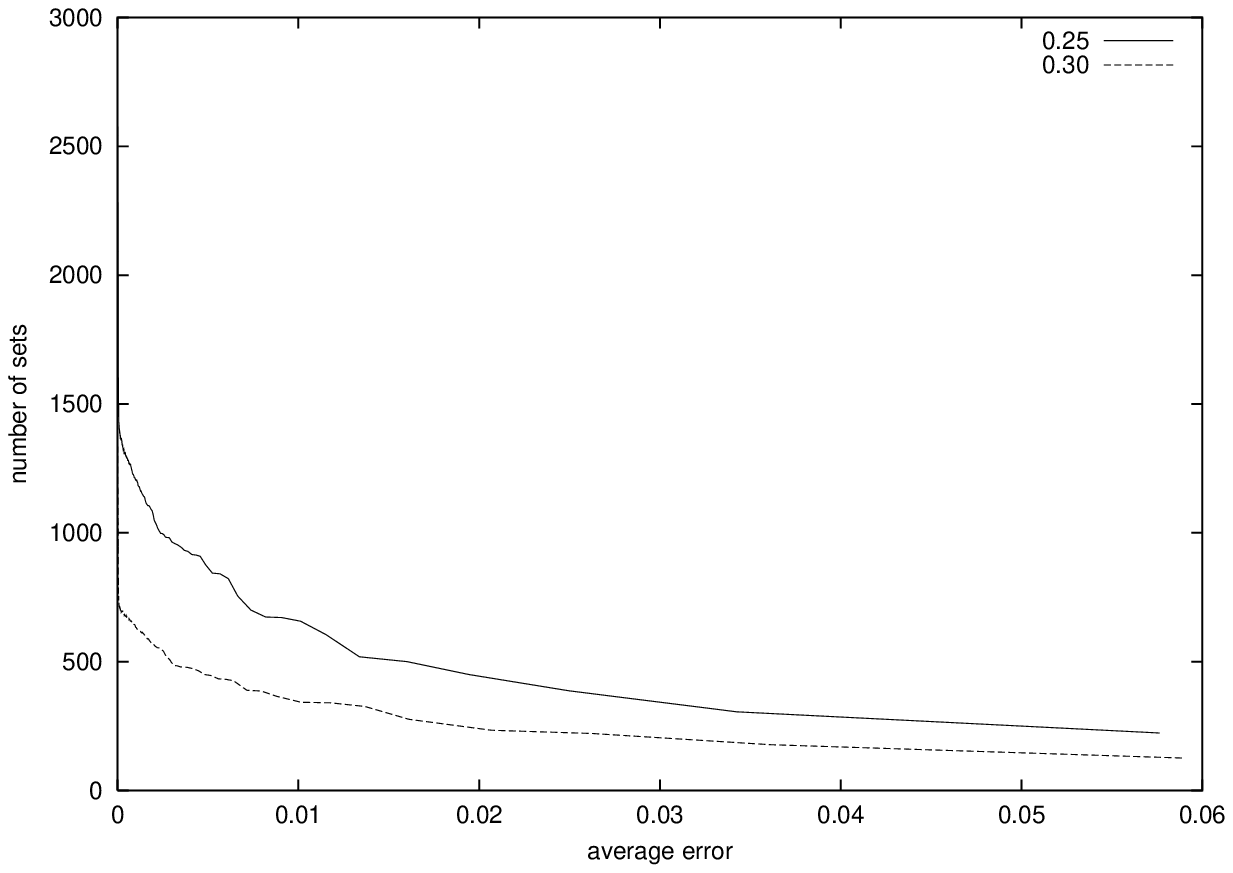}

\includegraphics[width=\columnwidth]{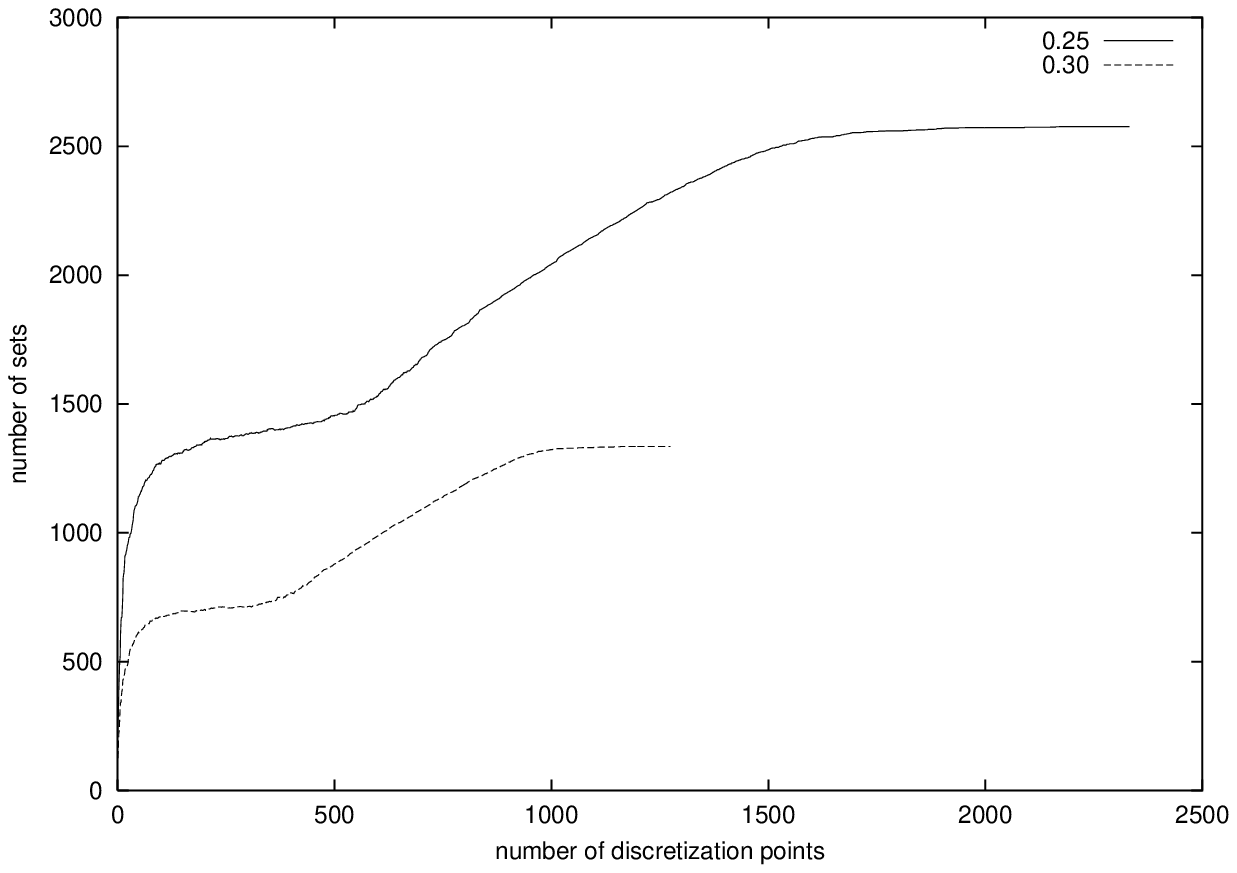}
\caption{The best average absolute error discretizations for IPUMS Census data. \label{f:disc:ipums}}
\end{figure}

The results are shown in Figure~\ref{f:disc:internet} and in
Figure~\ref{f:disc:ipums}. The figures can be interpreted as
follows. The labels of the curves are the minimum frequency thresholds
$\sigma$ for the collections of $\sigma$-frequent itemsets they
correspond to. The upper figures show the number of frequent itemsets
that are closed with respect to the discretized frequencies against
the average absolute error of the discretization.  The lower figures
show the number of the closed $\sigma$-frequent itemsets for
discretized frequencies against the number of discretization points.

On the whole, the results are encouraging, especially as the
discretizations do not exploit directly the structure of the pattern
collection but only the frequencies.  Although there are differences
between the results on different databases, it is possible to observe
that even with a quite small number of closed frequent itemsets and
discretization points, the frequencies of the frequent itemsets were
approximated adequately.

%% file: tradeoffs.tex

\chapter{Trade-offs between Size and Accuracy \label{c:tradeoffs}}


There are trade-offs between the understandability of the pattern
collection and its ability to describe the data at hand: 
\begin{itemize}
\item
If the pattern collection is small, then there is a chance that it could
eventually be understandable.
\item
If the pattern collection is large, then it might describe the data
underlying the pattern collection adequately.
\end{itemize}
Sometimes a very small collection of patterns can be both
understandable and accurate description the data. In general, however,
characterizing the data accurately requires many patterns assuming
that there are many different relevant data sets.

The trade-offs between understandability and accuracy have been
studied in pattern discovery mainly by comparing the cardinality of
the pattern collection to a quantitative measure of how well the
pattern collection describes the relevant aspects of the data.

Typically, one obtains smaller pattern collections by using
sufficiently high minimum quality value thresholds.  Finding a minimum
quality value threshold that captures most of the interesting and only
few uninteresting patterns is a challenging or even impossible task in
practice.

\begin{example}[discovering the backbone of a supermarket's profit]
Let $\DB$ be a transaction database of a supermarket containing
purchase events, the set $\Items$ of items consisting of the products
sold in that supermarket.  Furthermore, let $\wg{A}$ be the profit of
the item $A \in \Items$ and let $\wg{X}$ be the combined profit of the
items in the itemset $X$, i.e., $\wg{X}=\sum_{A \in X} \wg{A}$.

Suppose that we are interested in itemsets that fetch large portions
of the profit of the supermarket, i.e., the itemsets $X \subseteq
\Items$ with large \emph{(weighted) area} $\wg{X}\supp{X,\DB}$ in the
transaction database $\DB$. Then we have to face several problems, for
example the following ones.

First, there is no way to define a minimum frequency threshold that
would capture most of the itemsets fetching a large profit without
discovering many itemsets with less relevancy to the total profit of
the supermarket, although the support of the itemset is the only
data-dependent part of this interestingness measure.

Second, it is not clear what would be the right minimum area
threshold. For example, why should we choose $10000$ instead of $9999$
or vice versa? Intuitively this should not matter. However, even a
small change in the threshold might change the collection of
interesting patterns considerably.

Third, we could find out that we are actually more interested in some
other kinds of products, e.g., products with character and a weak
brand.  However, even realizing that these constraints are important
for a itemset being interesting might be very difficult from a large
collection of itemsets that contain also sufficiently many such
itemsets. Furthermore, weak brand can perhaps be detected based on the
discrepancy between the market value and the production costs of the
product but determining that a product has character is highly
subjective task without, e.g., extensive customer polls.

Thus, finding truly interesting patterns from data is often a
challenging, iterative and interactive process. \exend
\end{example}

To reduce the discrepancy between the size and the accuracy, several
condensed representations of pattern collections have been introduced.
(See Section~\ref{s:condensed} for more details.) They share, however,
the same fundamental difficulties as the other pattern collections: it
is difficult to find a small pattern collection that summarizes (the
relevant aspects of) the data well.  Overcoming these problems with
the size and the accuracy seems to be very difficult and they give
rise also to a crisp need for interactive exploration of pattern
collections and the trade-offs between the size and the accuracy.

If the whole pattern collection is too huge to comprehend, then a
natural solution to this problem is to consider only a subcollection
of patterns. There are a few properties that a good subcollection of a
pattern collection should fulfill. First, the subcollection should be
representative for the whole pattern collection. (This requirement is
based on the assumption that if the pattern collection describes the
data well, then also the representative subcollection should describe
the data quite well. The reason why the requirement is not defined
directly for the data, instead of the patterns, is that the data might
not be available or accessing it might be very expensive. Furthermore,
the methods described in this chapter can readily be adapted for
measuring the quality of the subcollection using the data instead of
the patterns.)  Second, the representative subcollection of $k$
patterns should not differ very much from the representative
subcollections of $k+1$ and $k-1$ patterns, i.e., the representative
subcollections should support interactive mining as it is presumably
highly non-trivial to guess the right value of $k$ immediately.

In this chapter we propose, as a solution to this problem, to order
the patterns in such a way that the $k$th pattern in the ordering
improves our estimate of the quality values of the patterns as much as
possible, given also the $k-1$ previous patterns in the ordering. Note
that this ensures that the representative subcollection of $k$
patterns does not differ much from the representative subcollections
of $k+1$ and $k-1$ patterns.

In addition to the pattern ordering problem, we study also the problem
of choosing the best $k$-subcollection of the patterns.  We show that
this problem is \NP{}-hard in general. However, for certain loss
functions and estimation methods, the optimal pattern ordering
provides a constant factor approximation for the best
$k$-subcollections for all values of $k$ \emph{simultaneously}. That
is, each length-$k$ prefix of the ordering is a subcollection that
describes the quality values of the patterns almost as well as the
best subcollection of cardinality $k$.

The feasibility of the method depends strongly on the loss function
and the estimation method at hand. To exemplify this, we describe
concrete instantiations of pattern orderings in two cases. First, we
use the pattern ordering to provide a refining representation of
frequent patterns. The representation is based on estimating the
frequencies of the patterns by the maximum of the frequencies of the
known superpatterns. Any prefix of this pattern ordering can be seen
as an approximation of closed frequent patterns. Second, we show how
transaction databases can be described as tilings. (Tiling is a
collection of tiles. A tile consists of an itemset and a set of
transaction identifiers of transactions that contain the itemset. We
use the fraction of the items of the database covered by the tiles as
the quality of the tiling.)

Finally, we empirically evaluate the suitability of pattern orderings
to serve as condensed representations in the case of the frequent
itemsets. More specifically, we estimate the frequencies of the
frequent itemsets using the maximum frequencies of the known
superitemsets and measure the loss by the average of the absolute
differences between the correct and the estimated frequencies.

This chapter is based mostly on the article ``The Pattern Ordering
Problem''~\cite{i:mielikainen03:pop}. (The example of tilings
described in Section~\ref{s:tilings} is related also to the article
``Tiling Databases''~\cite{i:geerts04}.)

\section{The Pattern Ordering Problem}
Most condensed representations of pattern collections consist of a
subcollection of the patterns such that the subcollection represents
the whole pattern collection well, often exactly. Representing the
whole pattern collection well by its subcollection depends on two
components.  

First, it depends on a function for estimating the
quality values of the patterns from the quality values of the patterns
in its subcollection, i.e., an estimation method
\begin{displaymath}
\estimation : \bigcup_{\SubPat \subseteq \Pat} \Pat \times
\IntC{0,1}^{\SubPat} \to \IntC{0,1}.
\end{displaymath}

\begin{example}[frequency estimation] \label{ex:indfreq}
A simple estimate for the frequency $\fr{X,\DB}$ of an itemset $X
\subseteq \Items$ is
\begin{equation} \label{eq:infreq}
\est{X,\Restrict{\freq}{\SubPat}}{\Items,\delta}
=\delta^{\Abs{\Items \setminus X}}
\prod_{A \in X, X \in \mathcal{S}} \fr{X,\DB}
\end{equation}
where $\mathcal{S} \subseteq 2^\Items$ and $\delta$ is the default
frequency for the items whose frequencies are not known.  This
estimation method assumes the independence of the items.

The downside of this estimation method is that it does not make use of
the other frequencies than the frequencies of the singleton
itemsets. Fortunately it can be generalized to exploit also other
frequencies. The idea of the generalization is to find a probability
distribution over the itemsets in the transactions that has the
\emph{maximum entropy} among the probability distributions compatible
with the frequency constraints. This estimation method has been
applied successfully in estimating the frequencies of itemsets based
on the frequencies of some other itemsets~\cite{a:pavlov03}.  \exend
\end{example}

\begin{example}[frequency estimation] \label{ex:maxfreq}
Another simple frequency estimate is
\begin{equation} \label{eq:maxfreq}
\est{X,\Restrict{\freq}{\SubPat}}{\mathit{Max}}
=\max \Set{\fr{Y,\DB} : Y \in \SubPat, Y \supseteq X}.
\end{equation}
Note that in the case of the closed itemsets
(Definition~\ref{d:closed}), the frequencies of the non-closed
itemsets are obtained using this rule.
\exend \end{example}

Second, the estimation is evaluated by a function that
measures the error of the estimation, i.e., a loss function
\begin{displaymath}
\error : \IntC{0,1}^{\Pat} \times \IntC{0,1}^{\Pat} \to \RN.
\end{displaymath}

\begin{example}[$L_p$ norms] \label{ex:lp}
One popular class of loss functions are $L_p$ norms
\begin{equation}
\err{\imeasure,\est{\cdot,\Restrict{\imeasure}{\SubPat}}{}}{L_p}=
\Paren{\sum_{x \in \Pat}
\Abs{\imeas{x}-\est{x,\Restrict{\imeasure}{\SubPat}}{}}^p}^{1/p}. \label{eq:lp}
\end{equation}
For example, if $p=2$ then the $L_p$ norm is the euclidean distance,
and if $p=1$, then it is the sum of absolute errors. The case where
$p=\infty$ corresponds to the maximum of the absolute errors.
\exend \end{example}

To simplify the considerations, we consider estimation methods and
loss functions as oracles, i.e., as functions that can be evaluated in
constant time regardless of their true computational complexity or
even computability. (Although the loss functions are often computable
in a reasonable time, there is not a necessity for that restriction in
the context of this chapter.)  With the aid of the estimation method
and the loss function, we can formulate the problem of finding a
subcollection that represents the whole pattern collection well as
follows:

\begin{problem}[the best $k$-subcollection patterns] \label{p:k-patterns}
Given a pattern collection $\Pat$, an interestingness measure
$\imeasure$, a positive integer $k$, an estimation function
$\estimation$, and a loss function $\error$, find the best
$k$-subcollection $\SubPat$ of $\Pat$, i.e., find a collection
$\SubPat \subseteq \Pat$ such that $\Abs{\SubPat} = k$ and for all
$k$-subcollections $\SubPat'$ of $\Pat$ hold
\begin{displaymath}
\err{\imeasure,\est{\cdot,\Restrict{\imeasure}{\SubPat}}{}}{}\leq
\err{\imeasure,\est{\cdot,\Restrict{\imeasure}{\SubPat'}}{}}{}.
\end{displaymath}
That is, $\SubPat$ has the smallest error among all $k$-subcollections
of $\Pat$.
\end{problem}

The problem of finding the best $k$-subcollection of patterns depends
on five parameters: the pattern collection $\Pat$, the interestingness
measure $\imeasure$, the estimation method $\estimation$, the loss
function $\error$ and the number $k$ of patterns allowed in the
subcollection.

\begin{example}[the best $k$-subcollection itemsets] \label{ex:k-itemsets}
Combining Example~\ref{ex:maxfreq} and Example~\ref{ex:lp} we get one
instance of Problem~\ref{p:k-patterns}:
\begin{itemize}
\item
The pattern collection $\Pat$ is a subcollection of the collection
$2^{\Items}$ of all subsets of the set $\Items$ of items. For example,
$\Pat$ could be the collection $\Frequent{\sigma,\DB}$ of the
$\sigma$-frequent itemsets in a given transaction database $\DB$.
\item
The interestingness measure $\imeasure$ is the frequency in the
transaction database $\DB$ and it is defined for all itemsets in the
pattern collection.
\item
The frequencies of the itemsets are estimated by taking the maximum of
the frequencies of known superitemsets of the itemset whose frequency
is under estimation, i.e., the estimation method $\estimation$ is as
defined by Equation~\ref{eq:maxfreq}.
\item
The loss in the estimation is measured by the maximum of the absolute
differences between the estimated and the correct frequencies of the
itemsets in $\Pat$. That corresponds to Equation~\ref{eq:lp} with
$p=\infty$.
\end{itemize}
These parameters together with the number $k$ (the maximum number of
patterns) form an instance of Problem~\ref{p:k-patterns}.
\exend \end{example}

Problem~\ref{p:k-patterns} is an optimization problem. It can easily
be transformed to a decision problem that asks whether there exists a
$k$-subcollection of $\Pat$ with the error at most $\epsilon$ instead
of looking for the $k$-subcollection with the smallest error.
Unfortunately, even a simple special case of the problem --- the
decision version of Example~\ref{ex:k-itemsets} --- is \NP{}-complete
as shown by Theorem~\ref{t:pop:NP}.

We show the \NP{}-hardness of Problem~\ref{p:k-patterns} by reduction
from Problem~\ref{p:mincover} which is known to be
\NP{}-complete~\cite{b:garey79}. (For more details in complexity
theory and \NP{}-completeness,
see~\cite{b:garey79,b:papadimitriou95}.)

\begin{problem}[minimum cover \cite{b:garey79}] \label{p:mincover}
Given a collection $C$ of subsets of a finite set $S$ and a positive
integer $k$, decide whether or not $C$ contains a cover of $S$ of size
$k$, i.e., whether or not there is a subset $C' \subseteq C$ with
$\Abs{C'}= k$ such that every element of $S$ belongs to at least one
member of $C'$.
\end{problem}

Note that we omit the empty itemset from the collection
$\Frequent{\sigma,\DB}$ in the proof of Theorem~\ref{t:pop:NP} to
simplify the reduction, since $\fr{\emptyset,\DB}=1$, i.e., it is
never necessary to estimate $\fr{\emptyset,\DB}$.
\begin{theorem} \label{t:pop:NP}
Given a collection $\Frequent{\sigma,\DB}$ of $\sigma$-frequent
itemsets in a transaction database $\DB$, a maximum error bound
$\epsilon$ and a cardinality bound $k$, it is \NP{}-complete to decide
whether or not there is a subcollection of $\Frequent{\sigma,\DB}'$
with the cardinality $k$ such that the maximum absolute error between
the correct frequency and the maximum of the frequencies of the
superitemsets in the subcollection is at most $\epsilon$. That is, it
is \NP{}-complete to decide whether or not there is a collection
$\Frequent{\sigma,\DB}' \subseteq \Frequent{\sigma,\DB}$ such that
$\Abs{\Frequent{\sigma,\DB}'}= k$ and
\begin{displaymath}
\max_{X \in \Frequent{\sigma,\DB}}\Set{\fr{X,\DB}-\max \Set{\fr{Y,\DB} : X \subseteq Y \in \Frequent{\sigma,\DB}'}} \leq \epsilon.
\end{displaymath}
\end{theorem}
\begin{proof}
The problem is in \NP{}\ since we can check in time polynomial in the
sum of the cardinalities of the itemsets in $\Frequent{\sigma,\DB}$
whether or not the maximum absolute error is at most $\epsilon$ for
all $X \in \Frequent{\sigma,\DB}$. That is, we can check in polynomial
time for each $X \in \Frequent{\sigma,\DB}$ and for each $Y \in
\Frequent{\sigma,\DB}'$ such that $X \subseteq Y$ whether
\begin{displaymath}
\Abs{\fr{X,\DB}-\fr{Y,\DB}} = \fr{X,\DB}-\fr{Y,\DB} \leq \epsilon.
\end{displaymath}

Let us now describe the reduction from Problem~\ref{p:mincover}. It is
easy to see that we can assume for each instance $\Tuple{S,C,k}$ of
Problem~\ref{p:mincover} that each element of $S$ is contained at
least in one set in $C$, no set in $C$ is contained in another set in
$C$, and the cardinality of each set in $C$ is greater than one.
Furthermore, we assume that the cardinalities of the sets in $C$ are
all at most three; the problem remains
\NP{}-complete~\cite{b:garey79}.

An instance $\Tuple{C,S,k}$ of the minimum cover problem is reduced to
an instance
$\Tuple{\Frequent{\sigma,\DB},\freq,\estimation,\error,k,\epsilon}$ as
follows.  The set $\Items$ of items is equal to the set $S$.  The
pattern collection $\Frequent{\sigma,\DB}$ consists of the sets in $C$
and all their non-empty subsets.  Thus, the cardinality of
$\Frequent{\sigma,\DB}$ is $\Oh{\Abs{S}^3}$, since we assumed that the
cardinality of the largest set in $C$ is three.  The transaction
database $\DB$ consists of one transaction for each set in $C$ and an
appropriate number of transactions that are singleton subsets of $S$
to ensure that $\fr{\Set{A},\DB}=\fr{\Set{B},\DB}>\fr{X,\DB}$ for all
$A,B \in S$ and $X \in C$.  (Thus, the minimum frequency threshold
$\sigma$ is $\Abs{\DB}^{-1}$.)

If we set $\epsilon=\fr{\Set{A},\DB}-\Abs{\DB}^{-1}$ for any element
$A$ in $S$, then there is a set $C' \subset C$ such that $\Abs{C'}
= k$ if and only if
\begin{displaymath}
\max_{X \in \Frequent{\sigma,\DB}}\Set{\fr{X,\DB}-\max \Set{\fr{Y,\DB} : X \subseteq Y \in \Frequent{\sigma,\DB}'}} \leq \epsilon
\end{displaymath}
holds for the same collection $C'=\Frequent{\sigma,\DB}' \subseteq
\Frequent{\sigma,\DB}$ with respect to the transaction database $\DB$.
(Note that without loss of generality, we can assume that
$\Frequent{\sigma,\DB}' \subseteq \Maximal{\sigma,\DB}=C$.)  Thus, the
problem is \NP{}-hard, too.
\end{proof}

Thus, the decision version of the special case of Problem~\ref{p:k-patterns}
as described by Example~\ref{ex:k-itemsets} is \NP{}-complete by
Theorem~\ref{t:pop:NP}. Thus, so is Problem~\ref{p:k-patterns} itself.
(For an alternative example of such a special case of
Problem~\ref{p:k-patterns} shown to be \NP{}-complete,
see~\cite{i:afrati04}.)

Furthermore, the proof of Theorem~\ref{t:pop:NP} implies also the
following inapproximability result for the optimization version of the
problem. (For more details in approximability,
see~\cite{b:ausiello99}.)
\begin{theorem}
Given a collection $\Pat$ of itemsets in a transaction database $\DB$,
their frequencies, the estimation method $\estimation_{\mathit{Max}}$ as
defined by Equation~\ref{eq:maxfreq} and loss function
$\err{\imeasure,\est{\cdot,\Restrict{\imeasure}{\SubPat}}{}}{L_\infty}$
as defined by Equation~\ref{eq:lp}, it is \NP{}-hard to find a
subcollection $\SubPat$ of $\Pat$ such
that
\begin{displaymath}
\err{\Restrict{\freq}{\Pat},\est{\cdot,\Restrict{\freq}{\SubPat}}{\mathit{Max}}}{L_\infty} \leq \epsilon
\end{displaymath}
and the cardinality of $\SubPat$ being within a factor $c \log
\Abs{\Items}$ (for some constant $c>0$) from the cardinality of the smallest
subcollection of $\Pat$ with error at most $\epsilon$.
\end{theorem}
\begin{proof}
The reduction in the proof of Theorem~\ref{t:pop:NP} shows that the
problem is \APX{}-hard~\cite{b:ausiello99,a:papadimitriou91}.  

If the collection $\Frequent{\sigma,\DB}$ is replaced by the
collection $\Pat=C \cup \Set{\Set{A} : A \in S}$, then we can get rid
of the cardinality constraints for the sets in $C$ while still
maintaining the itemset collection $\Pat$ being of polynomial size in
the size of the input $\Tuple{C,S,k}$.  This gives us stronger
inapproximability results. Namely, it is \NP{}-hard to find a set
cover $C' \subseteq C$ of the cardinality within a logarithmic factor
$c\log \Abs{S}$ (for some constant $c>0$) from the smallest set cover
of $S$ in $C$~\cite{b:ausiello99,i:raz97}.

If we could find a collection $\SubPat \subseteq \Pat$ of the
cardinality $k$ and the error at most $\epsilon$, then that collection
could also be a set cover of $S$ of the cardinality $k$.
\end{proof}

Even if there was a polynomial-time solution for
Problem~\ref{p:k-patterns}, it is not clear whether it is the right
problem to solve after all. A major disadvantage of the problem is
that it does not take into account the requirement that the solution
consisting of $k$ patterns should be close to the solutions consisting
of $k+1$ and $k-1$ patterns.  In general, it would be desirable that
the solutions of all cardinalities would be somewhat similar.

One approach to ensure that is to order the patterns somehow and
consider each length-$k$ prefix of the ordering as the representative
$k$-subcollection of the patterns. The ordering should be such that
the prefixes of the ordering are good representative subcollections of
(the quality values of) the pattern collection. For example, the
patterns could be ordered based on how well the prefixes of the
ordering describe the collection.

\begin{problem}[pattern ordering] \label{p:pattern ordering}
Given a pattern collection $\Pat$, an interestingness measure
$\imeasure$, an estimation function $\estimation$ and a loss function
$\error$, find an ordering $p_1,\ldots,p_{\Abs{\Pat}}$ of the patterns
such that
\begin{equation} \label{eq:ordering}
\err{\imeasure,\est{\cdot,\Restrict{\imeasure}{\Set{p_1,\ldots,p_i}}}{}}{} \leq
\err{\imeasure,\est{\cdot,\Restrict{\imeasure}{\Set{p_1,\ldots,p_{i-1},p_j}}}{}}{}
\end{equation}
for all $i \in\Set{1,\ldots,\Abs{\Pat}}$ and $j \in \Set{i,\ldots,\Abs{\Pat} }$.
\end{problem}

The pattern ordering can be seen as a refining approximation of the
pattern collection: the first pattern in the ordering describes the
pattern collection at least as well as any other pattern in the
collection, the second pattern is the the best choice if the first
pattern is already chosen to the representation. In general, the $k$th
pattern in the ordering is the best choice to improve the estimate
given the first $k-1$ patterns in the ordering. 

\begin{algorithm}
\caption{The pattern ordering algorithm. \label{a:Order-Patterns}}
\begin{algorithmic}[1]
\Input{The collection $\Pat$ of patterns, the interestingness measure
$\imeasure$, the estimation method $\estimation$, and the loss
function $\error$.}
\Output{The optimal pattern ordering as defined by
Equation~\ref{eq:ordering} and the loss
$\varepsilon_i=\err{\imeasure,\est{\cdot,\Restrict{\imeasure}{\Pat_i}}{}}{}$ for
each $i$-prefix $\Pat_i$ of the pattern ordering.}
\Function{Order-Patterns}{$\Pat,\imeasure,\estimation,\error$}
\State $\Pat_0 \leftarrow \emptyset$
\For{$i=0,\ldots,\Abs{\Pat}-1$}
 \State $p_{i+1} \leftarrow \arg \min_{p \in \Pat \setminus \Pat_i} \Set{\err{\imeasure,\est{\cdot,\Restrict{\imeasure}{\Pat_i \cup \Set{p}}}{}}{}}$
 \State $\Pat_{i+1} \leftarrow \Pat_i \cup \Set{p_{i+1}}$
 \State $\varepsilon_{i+1} \leftarrow \err{\imeasure,\est{\cdot,\Restrict{\imeasure}{\Pat_{i+1}}}{}}{}$
\EndFor
\State \textbf{return } $\Tuple{\Tuple{p_1,\ldots,p_{\Abs{\Pat}}},\Tuple{\varepsilon_1,\ldots,\varepsilon_{\Abs{\Pat}}}}$
\EndFunction
\end{algorithmic}
\end{algorithm}

The pattern ordering and the estimation errors for all prefixes of the
ordering can be computed efficiently by
Algorithm~\ref{a:Order-Patterns}.  The running time of the algorithm
depends crucially on the complexity of evaluating the expression
$\err{\imeas{\Pat},\est{\Pat,\Restrict{\imeasure}{\Pat_i \cup
\Set{p}}}{}}{}$ for each pattern $p \in \Pat \setminus \Pat_i$ and for
all $i=0,\ldots,\Abs{\Pat}-1$.  If $\Maximum{\Pat}$ is the maximum
time complexity of finding the pattern $p_{i+1}$ that improves the
prefix $\Pat_i$ as much as possible with respect to the estimation
method and the loss function, then the time complexity of
Algorithm~\ref{a:Order-Patterns} is bounded above by
$\Oh{\Abs{\Pat}\Maximum{\Pat}}$. Note that the algorithm requires at
most $\Oh{\Abs{\Pat}^2}$ loss function evaluations since there are
$\Oh{\Abs{\Pat}}$ possible patterns to be the $i$th pattern in the
ordering.

\begin{example}[on the efficiency of Algorithm~\ref{a:Order-Patterns}] \label{ex:order-patterns}
Let the estimation method be
\begin{displaymath}
\est{p,\Restrict{\imeasure}{\SubPat}}{\mathit{simple}}= \left\{
\begin{array}{ll}
\imeas{p} \quad & \mbox{if } p \in \SubPat \mbox{ and} \\
0 & \mbox{otherwise},
\end{array}\right.
\end{displaymath}
i.e., let the quality values be zero unless explicitly given, and let
the loss be the sum of the differences
\begin{displaymath}
\imeas{p}-\est{p,\Restrict{\imeasure}{\SubPat}}{\mathit{simple}}=\left\{
\begin{array}{ll}
0 \quad & \mbox{if } p \in \SubPat \mbox{ and} \\
\imeas{p} & \mbox{otherwise}.
\end{array}\right.
\end{displaymath}

Then finding the pattern that improves to solution the most can be
found in time logarithmic in $\Abs{\Pat}$ by using a
heap~\cite{b:knuth:3}. More specifically, each quality value of a
pattern in $\Pat$ is put into the heap in time
$\Oh{\Abs{\Pat}\log\Abs{\Pat}}$. The best pattern can be found in each
iteration by picking the pattern with highest quality value in the
heap. Thus, the total running time of the algorithm is then
$\Oh{\Abs{\Pat} \log \Abs{\Pat}}$. (Note that the optimal pattern
ordering could be obtained in this case also by sorting the patterns
with respect to their quality values.)  \exend \end{example}

The patterns could be ordered also by starting with the whole pattern
collection $\Pat$ and repeatedly removing from the collection the
pattern whose omission increases the error least, rather than starting
with an empty collection and adding the pattern that decreases the
error most. 

If the pattern ordering and the errors for all of its prefixes are
computed (as Algorithm~\ref{a:Order-Patterns} does), then the user can
very efficiently explore the trade-offs between the size and the
accuracy: If the number of patterns is overwhelming, then the user can
consider shorter prefixes of the pattern ordering. If the accuracy of
the estimates is not high enough, then the user can add more patterns
to the prefix.

Furthermore, this exploration can be done very efficiently. Finding
the prefix of length $k$ can always be implemented to run in constant
time by representing the pattern ordering as an array of patterns.
The shortest prefix with error at most a given threshold $\epsilon$
can be found in time $\Oh{\Abs{\Pat}}$ by scanning the array of
patterns sequentially. Similarly, the prefix of length at most $k$
with the smallest error can be found in time linear in $\Abs{\Pat}$.
If the loss function is nonincreasing, i.e., it is such that
\begin{displaymath}
\est{\cdot,\Restrict{\imeasure}{\SubPat}}{} \leq
\est{\cdot,\Restrict{\imeasure}{\SubPat \setminus \Set{p}}}{}
\end{displaymath}
for each $p \in \SubPat$ and each $\SubPat \subseteq \Pat$, then the
time consumption of these tasks can be reduced to $\Oh{\log
\Abs{\Pat}}$ by a simple application of binary search.

In addition to efficient exploration of trade-offs between the size and
the accuracy, the pattern ordering can shed some light to the
relationships between the patterns in the collections. For example,
the prefixes of the pattern ordering suggest which patterns are
complementary to each other and show which improve the quality value
estimation.

\section{Approximating the Best $k$-Subcollection of Patterns}
On one hand, the problem of finding the best $k$-subcollection of
patterns is \NP{}-hard as shown by Theorem~\ref{t:pop:NP}.  Thus,
there is not much hope for polynomial-time algorithms for finding the
best $k$-subcollection in general. On the other hand, the optimal
pattern ordering can be found by Algorithm~\ref{a:Order-Patterns}.
Furthermore, the greedy procedure (of which
Algorithm~\ref{a:Order-Patterns} is one example) has been recognized
to provide efficiently exact or approximate solutions for a wide
variety of other
problems~\cite{a:feige98:setcover,a:guha99,a:helman93,i:kempe03}.
Actually, Algorithm~\ref{a:Order-Patterns} provides the optimal
solution for some special cases.  For example, the prefixes of the
optimal pattern ordering for Example~\ref{ex:order-patterns} are also
the best subcollections. Furthermore, the optimal pattern ordering
always determines the best pattern to describe the quality values of
the whole collection. Unfortunately it does not provide necessarily
the optimal solution for an arbitrary value of $k$.

\begin{example}[the suboptimality of the optimal pattern ordering] \label{ex:suboptimal}
Let the pattern collection $\Pat$ be $2^{\Set{A,B,C}}$ and the
interestingness measure be the support.  Let the support of the
itemset $\Set{A,B,C}$ be $1$ and the other supports be $3$.
Furthermore, let the estimation method be as defined by
Equation~\ref{eq:maxfreq} and let the loss function be the euclidean
distance, i.e., Equation~\ref{eq:lp} with $p=2$.

Then the initial loss is $55$.  The best $3$-subcollection consists of
itemsets $\Set{A,B}$, $\Set{A,C}$ and $\Set{B,C}$ with the loss $1$
whereas Algorithm~\ref{a:Order-Patterns} chooses the itemset
$\Set{A,B,C}$ instead of one of the $2$-itemsets, resulting the loss
$4$.  The decreases of losses are $54$ and $51$, respectively.
\exend \end{example}

There is still a possibility, however, that the optimal pattern
ordering provides reasonable approximations for at least some
$k$-subcollections, loss functions and estimation methods.

In fact, under certain assumptions about the estimation method
$\estimation$ and the loss function $\error$, it is possible to show
that each $k$-prefix of the pattern ordering is within a constant
factor from the corresponding best $k$-subcollection of patterns in
$\Pat$ for all $k=1,\ldots,\Abs{\Pat}$ simultaneously.  

More specifically, if the estimation method $\estimation$ and the loss
function $\error$ together satisfy certain conditions, then for each
$k$-prefix $\Pat_k$ of the pattern ordering the decrease of loss
\begin{displaymath}
\err{\imeasure,\est{\cdot,\Restrict{\imeasure}{\emptyset}}{}}{}-\err{\imeasure,\est{\cdot,\Restrict{\imeasure}{\Pat_k}}{}}{}
\end{displaymath}
is within the factor $1-1/e \geq 0.6321$ from the maximum decrease of
loss
\begin{displaymath}
\err{\imeasure,\est{\cdot,\Restrict{\imeasure}{\emptyset}}{}}{}-\err{\imeasure,\est{\cdot,\Restrict{\imeasure}{\Pat^*_k}}{}}{}
\end{displaymath}
for any $k$-subcollection of $\Pat$, i.e.,
\begin{displaymath}
\frac{\err{\imeasure,\est{\cdot,\Restrict{\imeasure}{\emptyset}}{}}{}-\err{\imeasure,\est{\cdot,\Restrict{\imeasure}{\Pat_k}}{}}{}}{\err{\imeasure,\est{\cdot,\Restrict{\imeasure}{\emptyset}}{}}{}-\err{\imeasure,\est{\cdot,\Restrict{\imeasure}{\Pat^*_k}}{}}{}} \geq \frac{e-1}{e}
\end{displaymath}
for all $k \in \Set{1,\ldots,\Abs{\Pat}}$.

To simplify the notation, we use the following shorthands:
\begin{eqnarray*}
\varepsilon_i &=& \err{\imeasure,\est{\cdot,\Restrict{\imeasure}{\Pat_i}}{}}{} \\
\varepsilon^*_i &=& \err{\imeasure,\est{\cdot,\Restrict{\imeasure}{\Pat^*_i}}{}}{} \\
\delta_i &=& \err{\imeasure,\est{\cdot,\Restrict{\imeasure}{\emptyset}}{}}{}-\err{\imeasure,\est{\cdot,\Restrict{\imeasure}{\Pat_i}}{}}{} \\
\delta^*_i &=& \err{\imeasure,\est{\cdot,\Restrict{\imeasure}{\emptyset}}{}}{}-\err{\imeasure,\est{\cdot,\Restrict{\imeasure}{\Pat^*_i}}{}}{}
\end{eqnarray*}
The pattern collection $\Pat$, the interestingness measure
$\imeasure$, the estimation method $\estimation$ and the loss function
$\error$ are assumed to be clear from the context.

First we show that if the loss decreases sufficiently from the
$i-1$-prefix to the $i$-prefix for all $i=1,\ldots,\Abs{\Pat}$,
then $\delta_k \geq \Paren{1-1/e}\delta^*_k$ holds for all
$k=1,\ldots,\Abs{\Pat}$.

\begin{lemma} \label{l:pop:approx1}
If
\begin{equation}
\delta_i-\delta_{i-1}
\geq
\frac{1}{k}\Paren{\delta^*_k-\delta_{i-1}}
\label{eq:improvement}
\end{equation}
holds for all $i$ and $k$ with $1 \leq i \leq k \leq \Abs{\Pat}$ then
\begin{displaymath}
\delta_k \geq \Paren{1-\frac{1}{e}}\delta^*_k
\end{displaymath}
for all $k=1,\ldots,\Abs{\Pat}$.
\end{lemma}
\begin{proof}
From Equation~\ref{eq:improvement} we get
\begin{eqnarray*}
\delta_i & \geq & \frac{1}{k}\delta^*_k + \Paren{1-\frac{1}{k}} \delta_{i-1} \\
& \geq & \frac{1}{k}\delta^*_k \sum_{j=0}^i\left(1-\frac{1}{k}\right)^j \\
& = & \frac{1}{k}\delta^*_k\frac{\left(1-1/k\right)^i-1}{\left(1-1/k\right)-1} \\
& = & \Paren{1-\Paren{1-\frac{1}{k}}^i} \delta^*_k
\end{eqnarray*}
since by definition $\delta_0=\varepsilon_0-\varepsilon_0=0$.

Thus,
\begin{displaymath}
\delta_k \geq
\Paren{1-\Paren{1-\frac{1}{k}}^k} \delta^*_k \geq \Paren{1-\frac{1}{e}} \delta^*_k
\end{displaymath}
as claimed.
\end{proof}

The approximation with respect to the optimal loss is not so easy. In
fact, the optimal pattern ordering does not provide any approximation
ratio guarantees in general: there might be a collection $\Pat^*_k$ of
$k$ patterns that provide zero loss estimation of $\imeasure$ but
still the $k$-subcollection chosen by Algorithm~\ref{a:Order-Patterns}
can have non-zero loss.  (Note that also in the
Example~\ref{ex:suboptimal} the ratio of losses is $4$ whereas the
ratio between the decreases of losses is $17/18$.) Still, we can
transform Lemma~\ref{l:pop:approx1} to give bounds for the loss
instead of the decrease of the loss.

\begin{lemma}  \label{l:pop:approx2}
If
\begin{displaymath}
\varepsilon_{i-1}-\varepsilon_i \geq \frac{1}{k}\Paren{\varepsilon_{i-1}-\varepsilon^*_k}
\end{displaymath}
for all $i$ and $k$ with $1 \leq i \leq k \leq \Abs{\Pat}$ then also
\begin{displaymath}
\varepsilon_k \leq \Paren{1-\frac{1}{e}}\varepsilon^*_k+\frac{1}{e}\varepsilon_0
\end{displaymath}
holds for all $k=1,\ldots,\Abs{\Pat}$.
\end{lemma}
\begin{proof}
First note that
\begin{displaymath}
\varepsilon_{i-1}-\varepsilon_i \geq \frac{1}{k}\Paren{\varepsilon_{i-1}-\varepsilon^*_k} \iff
\delta_{i}-\delta_{i-1} \geq \frac{1}{k}\Paren{\delta^*_k-\delta_{i-1}} 
\end{displaymath}
and second that
\begin{displaymath}
\varepsilon_k \leq \Paren{1-\frac{1}{e}}\varepsilon^*_k + \frac{1}{e}\varepsilon_0
\iff
\delta_k \geq \Paren{1-\frac{1}{e}}\delta^*_k.
\end{displaymath}
Thus, Lemma~\ref{l:pop:approx1} gives the claimed result.
\end{proof}

The bound given by Lemma~\ref{l:pop:approx2} is considerably weaker
than the bound given by Lemma~\ref{l:pop:approx1} due to the additive
term of a constant fraction of the initial error, i.e., the error of
our initial assumption about the quality values.

Still, the prefixes of the optimal pattern ordering serve as good
representative $k$-subcollections of $\Pat$ for all values of $k$
simultaneously, in addition to being a refining description of the
quality values of the pattern collection.

\section{Approximating the Quality Values}

As a more concrete illustration of the approximation abilities of the
pattern orderings, in this section we shall consider the orderings of
patterns in downward closed collections $\Pat$ with anti-monotone
interestingness measures when the quality value of a pattern is
estimated to be the maximum of the quality values of its known
superpatterns (the collection $\SubPat$), i.e.,
\begin{equation} \label{eq:maxquality}
\est{p,\Restrict{\imeasure}{\SubPat}}{\mathit{Max}}= \max \Set{\imeas{p'} : p \preceq p' \in \SubPat}.
\end{equation}

Note that this estimation method was used also in
Example~\ref{ex:maxfreq}.  The next results show that the estimation
method $\est{}{\mathit{Max}}$ gives the correct quality values for all patterns
in $\Pat$ exactly when the subcollection used in the estimation
contains all closed patterns in the collection $\Pat$.
\begin{theorem} \label{t:pop:closed}
The collection $\Cl{\Pat}$ of the closed patterns in $\Pat$
is the smallest subcollection of $\Pat$ such that
\begin{displaymath}
\imeas{p}=\est{p,\Restrict{\imeasure}{\Cl{\Pat}}}{\mathit{Max}}
\end{displaymath}
for all $p \in \Pat$.
\end{theorem}
\begin{proof}
By definition, for each pattern $p \in \Pat$ there is a pattern $p'
\in \Cl{\Pat}$ such that $p \preceq p'$ and $\imeas{p}=\imeas{p'}$. As
we assume that the interestingness measure $\imeasure$ is
anti-monotone, taking the maximum quality value of the superpatterns
$p' \in \cl{\Pat}$ of a pattern $p \in \Pat$ determines to quality
value of $p$ correctly. Thus, $\Restrict{\imeasure}{\Cl{\Pat}}$ is
sufficient to determine $\Restrict{\imeasure}{\Pat}$.

To see that all closed patterns in $\Pat$ are needed, notice that the
quality value of a pattern $p \in \Cl{\Pat}$ is greater than any of
the quality values of its proper superpatterns. Thus, the quality
values of the patterns in $\Cl{\Pat}$ as
$\Restrict{\imeasure}{\Cl{\Pat}}$ is needed to determine even
$\Restrict{\imeasure}{\Cl{\Pat}}$ using the estimation method
$\est{}{\mathit{Max}}$.
\end{proof}

Thus, the problem of finding the smallest subcollection $\SubPat$ of
$\Pat$ such that
$\err{\imeas{p},\est{p,\Restrict{\imeasure}{\SubPat}}{\mathit{Max}}}{}{}=0$
with respect to the estimation method $\est{}{\mathit{Max}}$ and any reasonable
loss function $\error$ (i.e., a loss function such that only the
correct estimation of $\Restrict{\imeasure}{\Pat}$ has zero loss and
such that the loss can be evaluated efficiently for any $p \in \Pat$)
can be solved efficiently by Algorithm~\ref{a:Closed-Patterns}.

If some error is allowed, then the complexity of the solution depends
also on the loss function. Let us first consider the maximum absolute
error
\begin{displaymath}
\err{\imeasure,\est{\cdot,\Restrict{\imeasure}{\SubPat}}{\mathit{Max}}}{\mathit{Max}}
= \max_{p \in \Pat} \Set{ \imeas{p}-\est{p,\Restrict{\imeasure}{\SubPat}}{\mathit{Max}} },
\end{displaymath}
i.e., the loss defined by Equation~\ref{eq:lp} with $p=\infty$.

By Theorem~\ref{t:pop:NP}, the problem of finding the best
$k$-subcollection of $\Pat$ with the loss at most $\epsilon$ is
\NP{}-hard even when $\Pat=\Frequent{\sigma,\DB}$ and
$\imeasure=\freq$. The maximum absolute error is not very informative
loss function since it does not take into account the number of
patterns with error exceeding the maximum error threshold $\epsilon$.
Still, it makes a difference whether there is one or one million
patterns exceeding the maximum absolute difference threshold.

If the loss function is the number of frequencies that are not
estimated with the absolute error at most $\epsilon$, i.e.,
\begin{equation} \label{eq:maxepsilonerror}
\err{\imeasure,\est{\cdot,\Restrict{\imeasure}{\SubPat}}{\mathit{Max}}}{Max,\epsilon}
= \Abs{\Set{p \in \Pat : \Abs{\imeas{p}-\est{p,\Restrict{\imeasure}{\SubPat}}{\mathit{Max}}}>\epsilon}}
\end{equation}
then the problem can be modeled as a special case of the maximum
$k$-coverage problem (Problem~\ref{p:k-coverage}).
\begin{problem}[maximum $k$-coverage \cite{b:ausiello99}] \label{p:k-coverage}
Given a collection $C$ of subsets of a finite set $S$ and a positive
integer $k$, find a $k$-subcollection $C'$ of $C$ with the largest
coverage of $S$, i.e., the collection $C' \subseteq C$ of cardinality
$k$ that maximizes the cardinality of $S \setminus \Paren{\bigcup_{X
\in C'} X}$.
\end{problem}

\begin{theorem}
Let the estimation method be $\estimation_{\mathit{Max}}$
(Equation~\ref{eq:maxquality}) and the loss function be
$\error_{Max,\epsilon}$ (Equation~\ref{eq:maxepsilonerror}).  Then
Problem~\ref{p:k-patterns} is a special case of
Problem~\ref{p:k-coverage}.
\end{theorem}
\begin{proof}
The reduction from an instance
$\Tuple{\Pat,\imeasure,\estimation_{\mathit{Max}},\error_{Max,\epsilon},k}$ of
Problem~\ref{p:k-patterns} to an instance $\Tuple{C,S,k}$ of
Problem~\ref{p:k-coverage} is straightforward.  The set $S$ consists
of all patterns in $\Pat$, and $C$ consists of sets $\Set{ p' \in \Pat
: p' \preceq p, \imeas{p'}-\imeas{p}\leq \epsilon}$ for each $p \in
\Pat$.
\end{proof}

If the sum of errors is used instead of the maximum absolute error,
the following approximation bounds can be guaranteed:
\begin{theorem}
For the length-$k$ prefix $\Pat_k$ of the optimal solution for the
pattern ordering problem of the pattern collection $\Pat$ and the best
$k$-subcollection $\Pat^*_k$ of $\Pat$, we have
\begin{displaymath}
\delta_k \geq \Paren{1-\frac{1}{e}}\delta^*_k
\end{displaymath}
for the estimation method $\est{}{\mathit{Max}}$
(Equation~\ref{eq:maxquality}) and for any loss function
\begin{equation}
\err{\imeasure,\est{\cdot,\Restrict{\imeasure}{\SubPat}}{\mathit{Max}}}{\funct}=\sum_{p \in \Pat} \fun{\imeas{p}-\est{p,\Restrict{\imeasure}{\SubPat}}{\mathit{Max}}}
\end{equation}
where $\funct$ is an increasing function.
\end{theorem}
\begin{proof}
Using Lemma~\ref{l:pop:approx1}, it is sufficient to show that
Equation~\ref{eq:improvement} holds.

Let $p_1,\ldots,p_{\Abs{\Pat}}$ be to ordering of the patterns in
$\Pat$ given by Algorithm~\ref{a:Order-Patterns} and let
$\Pat_i=\Set{p_1,\ldots,p_i}$.  The pattern collection $\Pat$ can be
partitioned into $k$ groups $\Pat_{p},p \in \Pat^*_k$, as follows:
\begin{displaymath}
\Pat_{p_i}=\Set{p \preceq p_i : i=\min\Set{j \in \Set{1,\ldots,\Abs{\Pat}} : \imeas{p_j}=\est{p,\Restrict{\imeasure}{\Pat^*_k}}{\mathit{Max}}}}.
\end{displaymath}

Note that
\begin{displaymath}
\err{\imeasure,\est{\cdot,\Restrict{\imeasure}{\SubPat}}{\mathit{Max}}}{\funct}
\geq \err{\imeasure,\est{\cdot,\Restrict{\imeasure}{\SubPat \cup \SubPat'}}{\mathit{Max}}}{\funct}
\end{displaymath}
for all $\SubPat, \SubPat' \subseteq \Pat$.
This implies that also
\begin{displaymath}
\err{\imeasure,\est{\cdot,\Restrict{\imeasure}{\Pat_i}}{\mathit{Max}}}{\funct}
\geq \err{\imeasure,\est{\cdot,\Restrict{\imeasure}{\Pat_i \cup \Pat^*_k}}{\mathit{Max}}}{\funct}
\end{displaymath}
for all $i,k\in \Set{1,\ldots,\Abs{\Pat}}$.

For any $i,k \in \Set{1,\ldots,\Abs{\Pat}}$, the decrease of loss $\delta^*_k-\delta_{i-1}$ can be written as
\begin{displaymath}
\Paren{\sum_{p \in \Pat} \fun{\imeas{p}-\est{p,\Restrict{\imeasure}{\Pat_{i-1} \cup \Pat^*_k}}{\mathit{Max}}}}-\delta_{i-1}.
\end{displaymath}
and it can be further partitioned into sums
\begin{displaymath}
\sum_{p' \in \Pat_p} \Paren{\fun{\imeas{p'}-\est{p',\Restrict{\imeasure}{\Pat_{i-1} \cup \Set{p}}}{\mathit{Max}}}-
\fun{\imeas{p'}-\est{p',\Restrict{\imeasure}{\Pat_{i-1}}}{\mathit{Max}}}}
\end{displaymath}
for each $p \in \Pat^*_k$.  At least one of those sums must be at
least $1/k$-fraction of $\delta^*_k-\delta_{i-1}$. Thus, the claim
holds.
\end{proof}

Furthermore, the search for the optimal pattern ordering using the
estimation method $\est{}{\mathit{Max}}$ (Equation~\ref{eq:maxquality}) can be
speeded up by considering, without loss of generality, only the closed
patterns:
\begin{theorem}
For all loss functions $\error$ and all subcollections $\SubPat$ of
the pattern collection $\Pat$ we have
\begin{displaymath}
\err{\imeasure,\est{\cdot,\Restrict{\imeasure}{\SubPat}}{\mathit{Max}}}{} =
\err{\imeasure,\est{\cdot,\Restrict{\imeasure}{\Set{\mathit{cl}(p) : p \in \SubPat}}}{\mathit{Max}}}{}
\end{displaymath}
\end{theorem}
\begin{proof}
Any pattern $p \in \Pat$ can be replaced by its closure $\cl{p,\preceq,\imeasure}$ since
$\imeas{p}=\imeas{\cl{p}}$. Furthermore, if $p' \preceq p$ then $p' \preceq \cl{p}$
for all $p,p' \in \Pat$.
\end{proof}

\begin{realexample}[ordering $0.20$-frequent itemsets in the course completion database] \label{rex:freqest}
Let us examine how the estimation method $\est{}{\mathit{Max}}$
orders the $0.20$-frequent itemsets in
the course completion database (see Subsection~\ref{ss:ds}) when the loss
function is the average of the absolute differences.

\begin{figure}[h!]
\includegraphics[width=\textwidth]{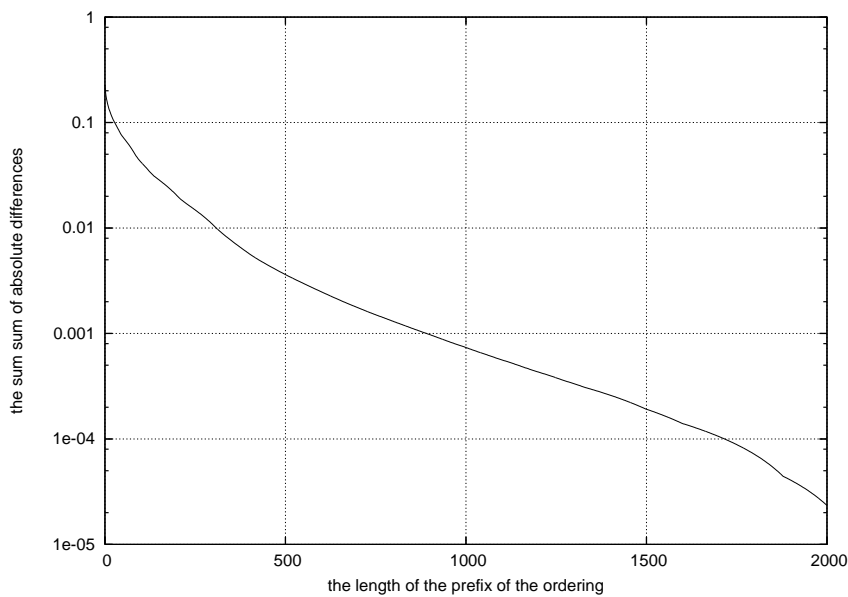}
\caption{The average of the absolute differences between the frequencies
estimated by Equation~\ref{eq:maxquality} and the correct frequencies
for each prefix of the pattern ordering.  \label{f:rex:freqest:error}}
\end{figure}

\begin{figure}[h!]
\includegraphics[width=\columnwidth]{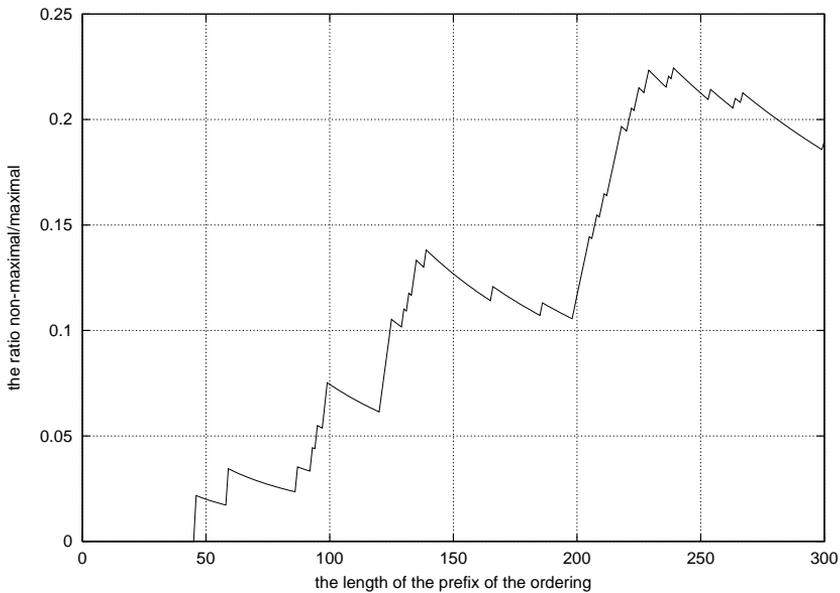}
\caption{The ratio between the non-maximal and the maximal itemsets in
the prefixes of the pattern ordering.\label{f:rex:freqest:ratio}}
\end{figure}

\begin{figure}[hp!]
\includegraphics[width=\columnwidth]{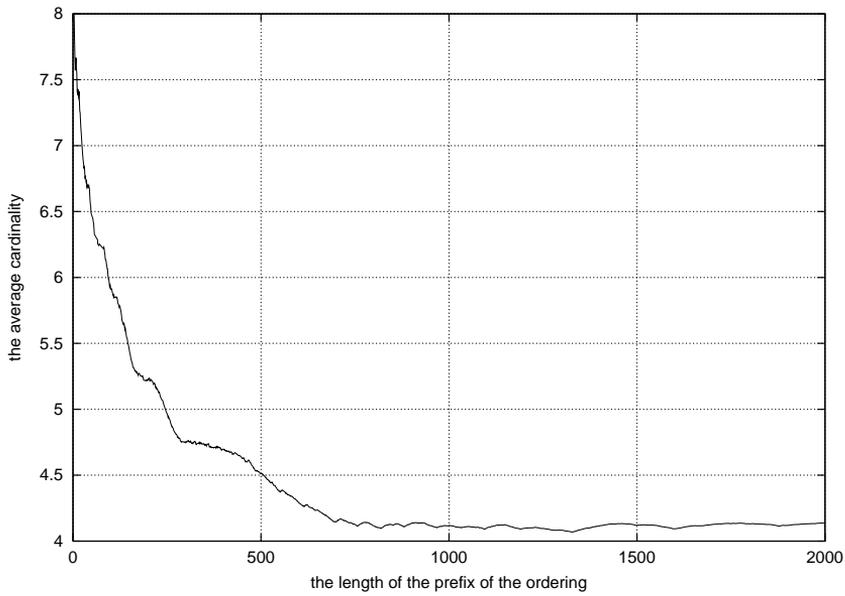}
\caption{The average cardinality of the itemsets. \label{f:rex:freqest:avgc}}
\end{figure}

\begin{figure}[hp!]
\includegraphics[width=\textwidth]{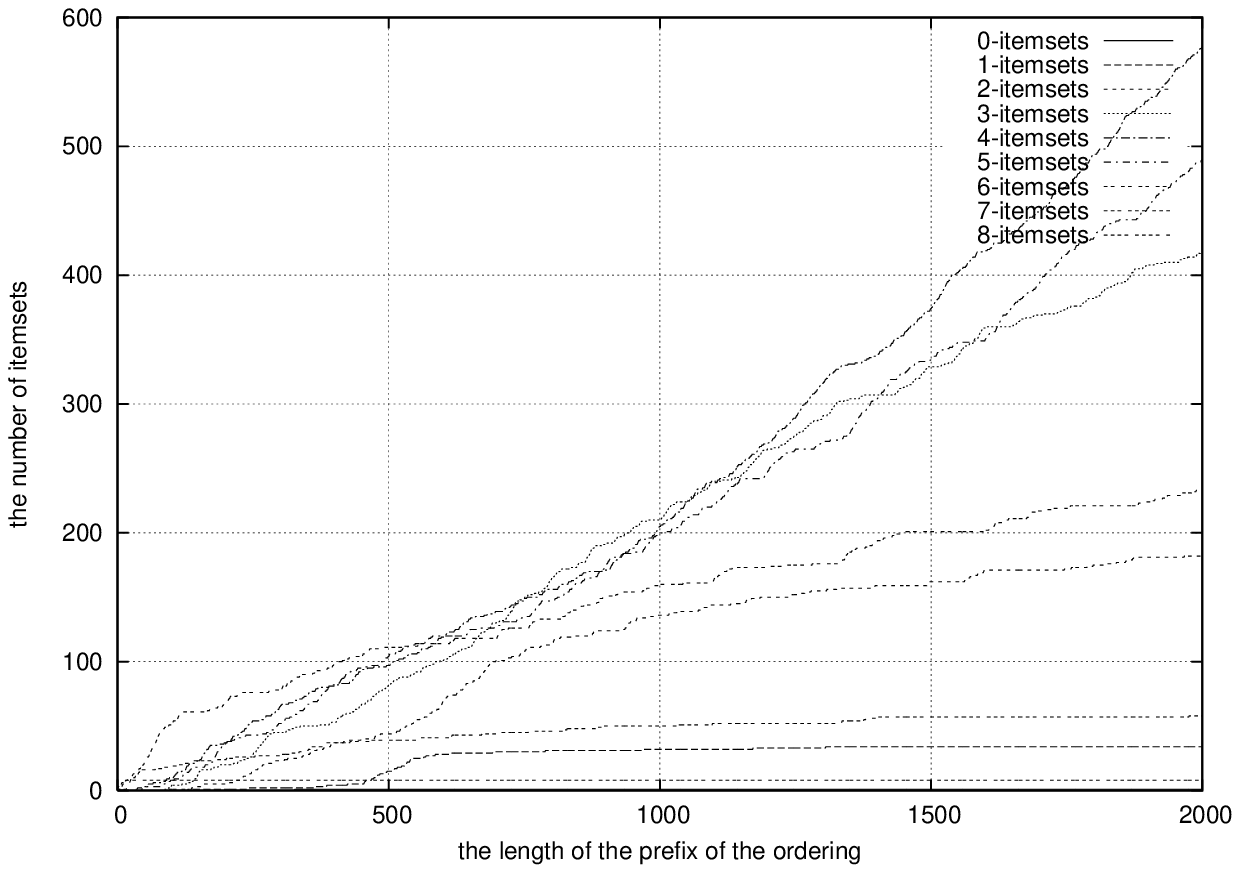}
\caption{The number of itemsets of each cardinality. \label{f:rex:freqest:ordernumbers}}
\end{figure}

The averages of the absolute differences in the frequency estimates
for all prefixes up to length 2000 of the pattern ordering are shown
in Figure~\ref{f:rex:freqest:error}. It can be noticed that the error
decreases quite quickly.

The decrease of the error does not tell much about the other
properties of the pattern ordering. As the estimation method is taking
the maximum of the frequencies of the known superitemsets, it is
natural to ask whether the first itemsets in the ordering are
maximal. Recall that the number of maximal $0.20$-frequent itemsets is
$253$ and the number of closed $0.20$-frequent itemsets is $2136$. The
first $46$ itemsets in the ordering are maximal but after that there
are also itemsets that are not maximal in the collection
$0.20$-frequent itemsets in the course completion database. The last
maximal itemset appears as the $300$th itemset in the ordering. The
interesting part of ratios between the non-maximal and the maximal
itemsets in the prefixes of the ordering is illustrated in
Figure~\ref{f:rex:freqest:ratio}. (Note that after the $300$th itemset
in the ordering, the ratio changes by an additive term $1/253$ for
each itemset.)

One explanation for the relatively large number of maximal itemsets in
the beginning of the ordering is that the initial estimate for all
frequencies is $0$, whereas the frequency of each maximal itemset is
in our case at least $0.20$. Furthermore, the first maximal itemsets
are quite large (thus containing a large number of at least
$0.20$-frequent itemsets) and the majority of the frequencies of the
$0.20$-frequent itemsets are within $0.025$ to $0.20$. Still, the
itemset ordering differs considerably from first listing all maximal
$0.20$-frequent itemsets and then all other $0.20$-frequent itemsets.

For a more refined view to the ordering, the average cardinality of
the itemsets in each prefix is shown in
Figure~\ref{f:rex:freqest:avgc} and the number of itemsets of each
cardinality in each prefix is shown in
Figure~\ref{f:rex:freqest:ordernumbers}.

The average cardinality of the itemsets in the prefixes drops quite
quickly close to the global average cardinality. That is, after the
initial major corrections in the frequencies (i.e., listing some of
the largest maximal itemsets) there are both small and large itemsets
in the ordering. Furthermore, the itemsets of all cardinalities are
listed quite much in the same relative speed.  Thus, based on these
statistics, the method seems to provide some added value compared to
listing the itemsets levelwise from the largest to the smallest
cardinality as well as listing the itemsets in the order of increasing
frequency. That is, the method seems to list one itemset here and
another there, giving a refining view to the itemset collection, as
hoped.  \exend \end{realexample}

\section{Tiling Databases \label{s:tilings}}
In this section we illustrate the use of pattern orderings as refining
description of data, transaction databases in particular.

A transaction database $\DB$ can be seen as an $n \times m$ binary
matrix $M_\DB$ such that
\begin{displaymath}
M_\DB[i,A]=\left\{
\begin{array}{ll}
1 \quad & \mbox{if } A \in X \mbox{ for some } \Tuple{i,X} \in \DB \\
0 & \mbox{otherwise.}
\end{array}\right.
\end{displaymath}

Viewing transaction databases as binary matrices suggests also pattern
classes and interestingness measures different from itemsets and
frequencies.

For example, it is not clear why itemsets (i.e., sets of column
indices) would be especially suitable for describing binary
matrices. Instead of sets of column indices, it could be more natural
to describe the matrices by their monochromatic
submatrices. Furthermore, as the transaction databases are often
sparse, we shall focus on submatrices full of ones, i.e.,
tiles~\cite{i:geerts04}, also known as
bi-sets~\cite{i:besson04:concepts}, and closely related to formal
concepts~\cite{b:ganter99}. (Some other approaches to take also the
transaction identifiers into account to choose a representative
collection of itemsets are described
in~\cite{i:toivonen95,i:wang04:summary}.)  As a quality measure we
shall consider the areas of the tiles.

\begin{definition}[tiles, tilings and their area]
Let $\DB$ be a transaction database over $\Items$.

A \emph{tile} $\tile{C,X}$ is a set $C \times X$ such that $C
\subseteq \tid{\DB}$ and $X \subseteq \Items$. The sets $C$ and $X$
can be omitted when they are not of importance. A tile $\tile{C,X}$ is
\emph{contained} in $\DB$ if for each $\Tuple{i,A} \in \tile{C,X}$ there is a
transaction $\Tuple{i,Y} \in \DB$ such that $A \in Y$ (and thus $X
\subseteq Y$, too).

A tile $\tile{C,X}$ is \emph{maximal} in $\DB$ if it is contained in
$\DB$ and none of its supertiles is contained in $\DB$, i.e., a tile
$\tile{C,X}$ is maximal in $\DB$ if
$\tile{C,X}=\tile{\cover{X,\DB},\cl{X,\DB}}$.

The \emph{area} of a tile $\tile{C,X}$ is
\begin{displaymath}
\area{\tile{C,X}}=\Abs{\tile{C,X}}=\Abs{C}\Abs{X}.
\end{displaymath}
The area of an itemset $X$ in $\DB$ is the same as the area of the
tile $\tile{\cover{X,\DB},X}$.

A \emph{tiling} $\Tiling$ is a collection of tiles. A tiling is
\emph{contained} in $\DB$ if all tiles in the tiling are in $\DB$.

The \emph{area} of a tiling $\Tiling$ is
\begin{displaymath}
\area{\Tiling}=\Abs{\bigcup_{\tau \in \Tiling} \tau}.
\end{displaymath}

The closure of a tiling $\Tiling$ in $\DB$ is
\begin{displaymath}
\cl{\Tiling,\DB}=\Set{\tile{\cover{X,\DB},\cl{X,\DB}} : \tile{C,X} \in
\Tiling}.
\end{displaymath}
\end{definition}

The motivation to consider the area of tilings is the following. As
the transaction databases $\DB$ is typically sparse, it might be a
good idea to describe $\DB$ by indicating where are the ones in the
binary matrix by the row and the column indices of submatrices full of
ones. Thus, the quality of a tile or a tiling is measured by the
number of ones covered by it whereas the goal in frequent itemset
mining is to find as high tiles as possible.

Tiles and tilings are most suitable for pattern ordering since the
area of a tiling determines a natural loss function for tilings. 

More specifically, the task of tiling transaction databases can be
formulated as a special case of Problem~\ref{p:pattern ordering}. The
pattern collection $\Pat$ is the collection of tiles in $\DB$. The
interestingness measure $\imeasure$ is the area of the tile. The
estimation method does not depend on the areas of the known tiles but
just the tiles. The known tiles are sufficient to determine the areas
of all subtiles of them.  (Similarly, in the case of frequent
itemsets, a subcollection of frequent itemsets would be sufficient to
determine the cardinalities of those frequent itemsets and their
subitemsets.) A loss function $\error$ can be, e.g., the number of
ones not yet covered, i.e.,
\begin{displaymath}
\err{\DB,\Tiling}{\mathit{area}}=
\Abs{\Set{\Tuple{i,A} : \Tuple{i,X} \in \DB, A \in
X}}-\area{\Tiling}.
\end{displaymath}
Thus, the number of transaction databases with $\Abs{\DB}$
transactions over $\Items$ that are compatible with the tiling
$\Tiling$ is at most
\begin{equation} \label{eq:ubdb}
\Choose{\Abs{\DB}\Abs{\Items}-\area{\Tiling}}{\Abs{\Set{\Tuple{i,A} : \Tuple{i,X} \in \DB, A \in
X}}-\area{\Tiling}}.
\end{equation}

Note that we use the transaction database $\DB$ and the tiling
$\Tiling$ as the parameters of the loss functions, instead of the area
function for tiles in $\DB$ and the area function for tiles in
$\Tiling$, since they have the same information as the area function
for all tiles in $\DB$ and the area function for the tiles in the
tiling $\Tiling$.

\begin{proposition}
Each $k$-prefix of the best ordering of tiles in $\DB$ with respect to
the loss function $\error_{\mathit{area}}$ defines a tiling
$\Tiling_k$ that has area within a factor $1-1/e$ from the best
$k$-tiling $\Tiling^*_k$ in $\DB$.
\end{proposition}
\begin{proof}
Based on Lemma~\ref{l:pop:approx1}, it is sufficient to show that
Equation~\ref{eq:improvement} holds, i.e., that
\begin{displaymath}
\area{\Tiling_i}-\area{\Tiling_{i-1}}\geq
\frac{1}{k}\Paren{\area{\Tiling^*_k}-\area{\Tiling_{i-1}}}
\end{displaymath}
for all $i$ and $k$ such that $1 \leq i \leq k$.

Let $\Tiling^*_k=\tau_{j_1},\ldots,\tau_{j_k}$.  There must be on
$\tau^*_i \in \Tiling^*_k$ such that
\begin{displaymath}
\area{\Tiling_{i-1} \cup \Set{\tau^*_i}}- \area{\Tiling_{i-1}} \geq
\frac{1}{k}\Paren{\area{\Tiling^*_k}-\area{\Tiling_{i-1}}}
\end{displaymath}
since
\begin{displaymath}
\area{\Tiling^*_k} \leq
\sum_{\tau \in \Tiling^*_k} \area{\tau}.
\end{displaymath}

Thus, there is a tile $\tau_i$ in $\DB$ but not in $\Tiling_{i-1}$
such that $\area{\Tiling_{i-1} \cup \Set{\tau_i}} \geq
\area{\Tiling_{i-1} \cup \Set{\tau^*_i}}$, i.e., the claim holds.
\end{proof}

Hence, each prefix of the ordering of tiles in $\DB$ gives a good
approximation for the best tiling of the same cardinality. All we have
to do is to find the tiles in $\DB$.

The first obstacle for finding the tiles in $\DB$ is that the number
of tiles can be very large. A slight relief is that we can restrict
our focus to maximal tiles in $\DB$ instead of all tiles.
\begin{proposition}
Replacing the tiles $\tile{C,X}$ of a tiling $\Tiling$ in $\DB$ by the
maximal tiles $\tile{\cover{X,\DB},\cl{X,\DB}}$ in $\DB$ does not
decrease the area of the tiling.
\end{proposition}
\begin{proof}
This is immediate since $\tile{C,X} \subseteq \tile{\cover{X,\DB},\cl{X,\DB}}$ 
for each tile $\tile{C,X}\in \Tiling$
contained in $\DB$, and thus
\begin{displaymath}
\bigcup_{\tile{C,X} \in \Tiling} \tile{C,X} \subseteq \bigcup_{\tile{C,X} \in \Tiling} \tile{\cover{X,\DB},\cl{X,\DB}}
\end{displaymath}
which implies that $\area{\Tiling} \leq \area{\cl{\Tiling,\DB}}$.
\end{proof}

The number of maximal tiles in $\DB$ could still be prohibitive.  The
number of maximal tiles to be considered can be decreased by finding
only large maximal tiles, i.e., the maximal tiles with area at least
some threshold.  Unfortunately, even the problem of finding the
largest tile is \NP{}-hard~\cite{i:geerts04,a:peeters03}, but there
are methods that can find the large maximal tiles in
practice~\cite{i:besson04:concepts,i:geerts04}.

Nevertheless, ordering the large maximal tiles is not the same as
ordering all maximal tiles. Although the number of all maximal tiles
might be prohibitive, it would possible construct any prefix of the
optimal ordering of the maximal tiles in $\DB$ if we could find for
any prefix $\Tiling_i$ of the ordering the tile $\tau_{i+1}$ in $\DB$
that maximizes $\area{\Tiling_i \cup \Set{ \tau_{i+1}}}$. Clearly,
also this problem is \NP{}-hard but in practice such tiles can be
found reasonably efficiently~\cite{i:geerts04}.

\begin{realexample}[Tiling the course completion database] \label{rex:tiling}
Let us consider the course completion database (see
Subsection~\ref{ss:ds}).

We computed the greedy tilings using Algorithm~\ref{a:Order-Patterns}
and an algorithm for discovering the tile that improves the current
tiling as much as possible.  We compared the greedy tiling to the
tiling obtained by ordering all frequent and maximal frequent itemsets
by their frequencies.  The greedy tiling is able to describe the
database quite adequately: the $34$ first tiles (shown in
Table~\ref{t:rex:tiling:cover}) in the tiling cover $43.85$ percent
($28570/65152$-fraction) of the ones in the databases. As a
comparison, the $34$ most frequent itemsets and the $34$ most frequent
closed itemsets (shown in Table~\ref{t:rex:tiling:frequent}) cover
just $19.13$ percent ($12462/65152$-fraction) of the
database. Furthermore, already the $49$ first tiles in the greedy
tiling cover more than half of the ones in the database.

The relatively weak performance of frequent itemsets can be explained
by the fact that frequent itemsets do not care about the other
frequent itemsets and also the interaction between closed itemsets is
very modest. Furthermore, all of the $34$ most frequent itemsets are
quite small, the three largest of them consisting of only three items,
whereas 22nd tile\footnote{The tile has the largest itemset within the
34 first tiles in the tiling and it consists almost solely of courses
offered by the faculty of law. The group of students inducing the tile
seem to comprise former computer science students who wanted to become
lawyers and a couple of law students that have studied a couple of
courses at the department of computer science.} contains $23$ items.

\begin{table} \centering
\caption{The $34$ first itemset in the greedy tiling of the course completion database.\label{t:rex:tiling:cover}}
\begin{tabular}{|c|c|c|}
\hline
$\supp{X,\DB}$ & $\area{X,\DB}$ & $X$ \\
\hline
\hline
$411$ & $4521$ & $\{0,2,3,5,6,7,12,13,14,15,20\}$ \\
$1345$ & $2690$ & $\{0,1\}$ \\
$765$ & $3060$ & $\{2,3,4,5\}$ \\
$367$ & $1835$ & $\{2,21,22,23,24\}$ \\
$418$ & $1672$ & $\{7,9,17,31\}$ \\
$599$ & $1198$ & $\{8,11\}$ \\
$513$ & $1026$ & $\{16,32\}$ \\
$327$ & $1635$ & $\{7,14,18,20,27\}$ \\
$706$ & $2118$ & $\{0,3,10\}$ \\
$357$ & $1428$ & $\{6,7,17,19\}$ \\
$405$ & $1215$ & $\{0,24,29\}$ \\
$362$ & $1810$ & $\{7,9,13,15,30\}$ \\
$197$ & $985$ & $\{2,19,33,34,45\}$ \\
$296$ & $1184$ & $\{2,3,25,28\}$ \\
$422$ & $844$ & $\{18,26\}$ \\
$166$ & $830$ & $\{21,23,37,43,48\}$ \\
$269$ & $538$ & $\{36,52\}$ \\
$393$ & $786$ & $\{3,35\}$ \\
$329$ & $1645$ & $\{6,7,13,15,41\}$ \\
$221$ & $442$ & $\{40,60\}$ \\
$735$ & $2205$ & $\{2,5,12\}$ \\
$20$ & $460$ & $\{1,11,162,166,175,177,189,191,$ \\
& & $204,206,208,209,216,219,223,226,$ \\
& & $229,233,249,257,258,260,272\}$ \\
$294$ & $882$ & $\{14,20,38\}$ \\
$410$ & $410$ & $\{39\}$ \\
$852$ & $1704$ & $\{1,8\}$ \\
$313$ & $939$ & $\{7,9,44\}$ \\
$193$ & $386$ & $\{42,49\}$ \\
$577$ & $1154$ & $\{21,23\}$ \\
$649$ & $649$ & $\{25\}$ \\
$1069$ & $1069$ & $\{6\}$ \\
$939$ & $1878$ & $\{0,4\}$ \\
$336$ & $336$ & $\{46\}$ \\
$264$ & $528$ & $\{19,50\}$ \\
$328$ & $328$ & $\{47\}$ \\
\hline
\end{tabular}
\end{table}

\begin{table} \centering
\caption{The $34$ most frequent (closed) itemsets
in the course completion database.\label{t:rex:tiling:frequent}}
\begin{tabular}{|c|c|c|}
\hline
$\supp{X,\DB}$ & $\area{X,\DB}$ & $X$ \\
\hline
\hline
$2405$ & $0$ & $\emptyset$ \\
$2076$ & $2076$ & $\{0\}$ \\
$1547$ & $1547$ & $\{1\}$ \\
$1498$ & $1498$ & $\{2\}$ \\
$1345$ & $2690$ & $\{0,1\}$ \\
$1293$ & $2586$ & $\{0,2\}$ \\
$1209$ & $1209$ & $\{3\}$ \\
$1098$ & $2196$ & $\{2,3\}$ \\
$1081$ & $1081$ & $\{4\}$ \\
$1071$ & $1071$ & $\{5\}$ \\
$1069$ & $1069$ & $\{6\}$ \\
$1060$ & $1060$ & $\{7\}$ \\
$1057$ & $2114$ & $\{1,2\}$ \\
$1052$ & $2104$ & $\{0,3\}$ \\
$1004$ & $2008$ & $\{3,5\}$ \\
$992$ & $1984$ & $\{2,5\}$ \\
$983$ & $1966$ & $\{2,7\}$ \\
$971$ & $1942$ & $\{2,6\}$ \\
$960$ & $2880$ & $\{2,3,5\}$ \\
$958$ & $2874$ & $\{0,2,3\}$ \\
$943$ & $1886$ & $\{0,5\}$ \\
$939$ & $1878$ & $\{0,4\}$ \\
$931$ & $931$ & $\{8\}$ \\
$924$ & $1848$ & $\{0,7\}$ \\
$921$ & $1842$ & $\{0,6\}$ \\
$920$ & $920$ & $\{9\}$ \\
$915$ & $2745$ & $\{0,1,2\}$ \\
$911$ & $1822$ & $\{1,3\}$ \\
$896$ & $1792$ & $\{6,7\}$ \\
$887$ & $2661$ & $\{0,3,5\}$ \\
$880$ & $1760$ & $\{3,4\}$ \\
$875$ & $2625$ & $\{0,2,5\}$ \\
$870$ & $1740$ & $\{2,4\}$ \\
$862$ & $2586$ & $\{0,2,7\}$ \\
\hline
\end{tabular}
\end{table}

\begin{table} \centering
\caption{The $34$ maximal itemsets with minimum support threshold $700$
in the course completion database.\label{t:rex:tiling:maximal}}
\begin{tabular}{|c|c|c|}
\hline
$\supp{X,\DB}$ & $\area{X,\DB}$ & $X$ \\
\hline
\hline
$748$ & $748$ & $\{15\}$ \\
$744$ & $744$ & $\{16\}$ \\
$733$ & $733$ & $\{17\}$ \\
$707$ & $707$ & $\{18\}$ \\
$709$ & $1418$ & $\{0,11\}$ \\
$700$ & $1400$ & $\{7,13\}$ \\
$721$ & $1442$ & $\{7,14\}$ \\
$732$ & $2196$ & $\{0,1,4\}$ \\
$730$ & $2190$ & $\{0,1,5\}$ \\
$712$ & $2136$ & $\{0,1,7\}$ \\
$741$ & $2223$ & $\{0,1,8\}$ \\
$749$ & $2247$ & $\{0,2,9\}$ \\
$706$ & $2118$ & $\{0,3,10\}$ \\
$721$ & $2163$ & $\{1,2,6\}$ \\
$755$ & $2265$ & $\{1,2,7\}$ \\
$750$ & $2250$ & $\{2,3,6\}$ \\
$738$ & $2214$ & $\{2,3,9\}$ \\
$724$ & $2172$ & $\{2,3,10\}$ \\
$705$ & $2115$ & $\{2,5,6\}$ \\
$716$ & $2148$ & $\{2,5,10\}$ \\
$726$ & $2178$ & $\{2,7,9\}$ \\
$705$ & $2115$ & $\{3,5,6\}$ \\
$720$ & $2160$ & $\{3,5,10\}$ \\
$704$ & $2112$ & $\{3,6,7\}$ \\
$737$ & $2948$ & $\{0,1,2,3\}$ \\
$722$ & $2888$ & $\{0,2,3,4\}$ \\
$849$ & $3396$ & $\{0,2,3,5\}$ \\
$741$ & $2964$ & $\{0,2,3,7\}$ \\
$729$ & $2916$ & $\{0,2,6,7\}$ \\
$706$ & $2824$ & $\{0,3,4,5\}$ \\
$749$ & $2996$ & $\{1,2,3,5\}$ \\
$765$ & $3060$ & $\{2,3,4,5\}$ \\
$757$ & $3028$ & $\{2,3,5,7\}$ \\
$727$ & $2908$ & $\{2,3,5,12\}$ \\
\hline
\end{tabular}
\end{table}

A slightly better performance can obtained with $34$ maximal itemsets
(shown in Table~\ref{t:rex:tiling:maximal}): the $34$ maximal itemsets
determine a tiling that covers $26.64$ percent
($17356/65152$-fraction) of the database.  (The $34$ maximal itemsets
were obtained by choosing the minimum support threshold to be
$700$. This is also the origin of choosing the value $34$ as the
number of illustrated itemsets.)  Maximal itemsets depend more on each
other since the maximal itemsets form an antichain
(see Chapter~\ref{c:chains} for more details).

It can be argued that we could afford a slightly larger number of
frequent itemsets since they are in some sense simpler than the
tiles. We tested this with the collections of the closed
$0.20$-frequent itemsets (2136 itemsets) and the maximal
$0.20$-frequent itemsets (253 itemsets) which have been used in
previous real examples. They cover $43.80$ percent
($28535/65152$-fraction) and $41.12$ percent ($26789/65152$-fraction),
respectively.  That is still less than the $34$ first tiles in the
greedy tiling.
\exend \end{realexample}

Thus, sometimes the pattern ordering can be computed incrementally
although generating the whole pattern collection would be infeasible.
Still, there are many ways how the tilings could be improved. 

First, the definitions of tiles and tilings could be adapted also to
submatrices full of zeros since a submatrix full of zeros is
equivalent to a submatrix full of ones in the binary matrix where all
bits are flipped.

Second, the complexity of describing of a particular tile could be
taken in to account. Assuming no additional information about the
matrix, a $\tile{C,X}$ in $\DB$ can be described using
\begin{displaymath}
\Abs{C}\log \Abs{\DB}+\Abs{X}\log \Abs{\Items}
\end{displaymath}
bits. However, taking into account the encoding costs of the tiles, it
is not sufficient to consider only maximal tiles.
\begin{example}[Maximal tiles with costs are not optimal]
Let the the transaction database $\DB$ consist of two transactions:
$\Tuple{1,\Set{A}}$ and $\Tuple{2,\Set{A,B}}$. Then the maximal tiles
describing $\DB$ are $\Set{\Tuple{1,A},\Tuple{2,A}}$ and
$\Set{\Tuple{2,A},\Tuple{2,B}}$ whereas tiles $\Set{\Tuple{1,A}}$ and
$\Set{\Tuple{2,A},\Tuple{2,B}}$ would be sufficient and slightly
cheaper, too.
\exend \end{example}

Third, the bound for the number of databases given by
Equation~\ref{eq:ubdb} does not take into account the fact that the
tiles in the tiling are maximal.  Let $\tid{\tau}$ and $\Items_\tau$
be the transaction identifiers and the items in a tile $\tau$,
respectively. The maximality of the tile $\tau$ restricts the
collection of compatible databases $\DB$ as follows.  The tile $\tau$
must be in the compatible database $\DB$.  For each transaction
identifier $i \in \tid{\DB} \setminus \tid{\tau}$ there must be an
item $A \in \Items_\tau$ such that $\Tuple{i,X} \in \DB$ does not
contain $A$. For each item $\Items \setminus \Items_{\tau}$ there must
be a transaction identifier $i \in \tid{\tau}$ such that $\Tuple{i,X}
\in \DB$ does contain $A$. The collection of the transaction databases
compatible with a tiling $\Tiling$ is the intersection of the
collections of the transaction databases compatible with each tile in
$\Tiling$.

\section{Condensation by Pattern Ordering}
We evaluated the ability of the pattern ordering approach to condense
the collections of the $\sigma$-frequent itemsets using the estimation
method
\begin{displaymath}
\est{X,\Restrict{\freq}{\Frequent{\sigma,\DB}'}}{\mathit{Max}}= \max
\Set{\fr{Y,\DB} : X \supseteq Y \in \Frequent{\sigma,\DB}'}.
\end{displaymath}
where $\Frequent{\sigma,\DB}'$ is the subcollection of
$\sigma$-frequent itemsets for which the frequencies are known (see
also Example~\ref{ex:maxfreq}).  The loss function used in the
experiments was the average absolute error with uniform distribution
over the itemset collection, i.e.,
\begin{eqnarray*}
&&\err{\Restrict{\freq}{\Fr{\sigma,\DB}},\est{\cdot,\Restrict{\freq}{\Frequent{\sigma,\DB}'}}{\mathit{Max}}}{}= \\
&&\frac{1}{\Abs{\Frequent{\sigma,\DB}}}\sum_{X \in \Frequent{\sigma,\DB}} \Paren{\fr{X,\DB}-\max_{X \subseteq Y \in \Frequent{\sigma,\DB}'} \fr{Y,\DB}}.
\end{eqnarray*}

\begin{figure}[p]
\centering
\includegraphics[width=\textwidth]{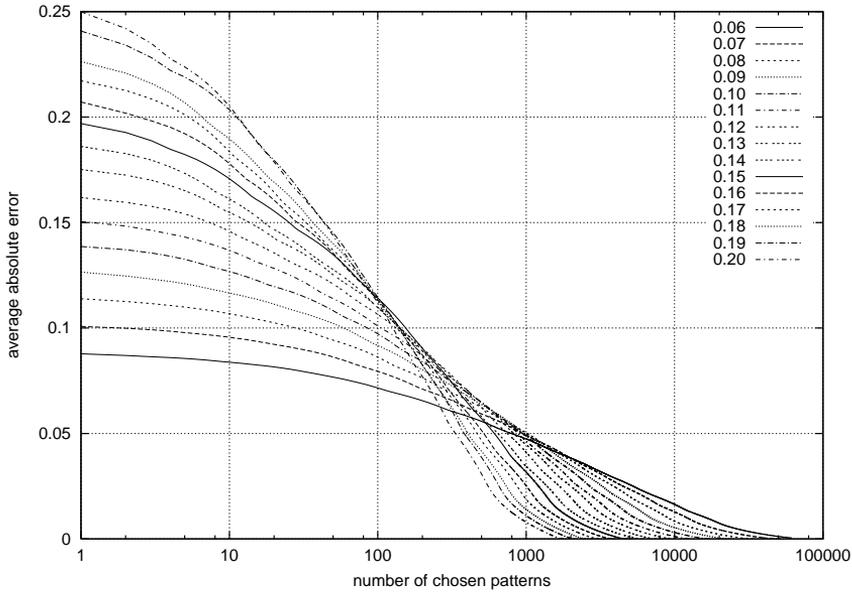}
\caption{Pattern orderings for Internet Usage data. \label{f:pop:internet}}
\end{figure}

\begin{figure}[p]
\centering
\includegraphics[width=\textwidth]{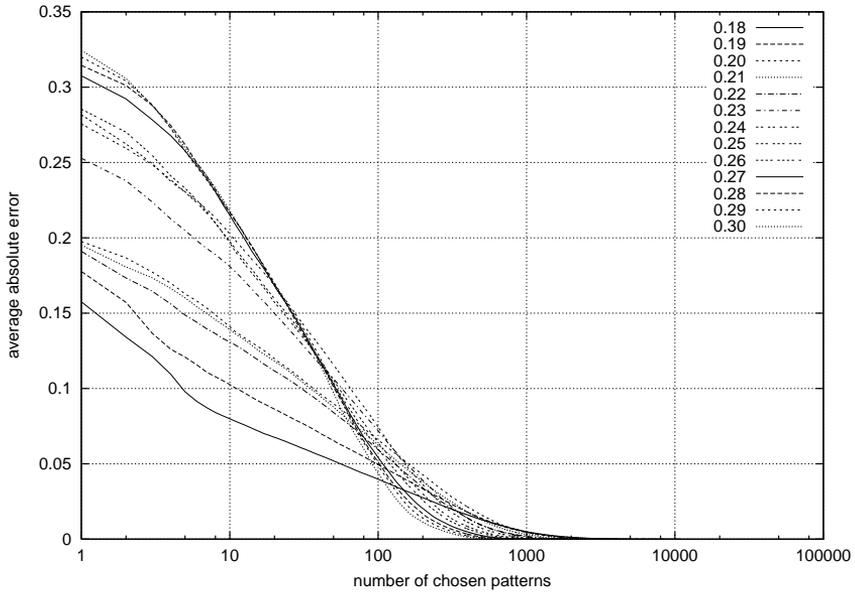}
\caption{Pattern orderings for IPUMS Census data. \label{f:pop:ipums}}
\end{figure}

The pattern orderings were found by the algorithm
\textproc{Order-Patterns} (Algorithm~\ref{a:Order-Patterns}).  Then we
computed the pattern orderings for $\sigma$-frequent itemsets in the
transaction databases Internet Usage and IPUMS Census for several
different minimum frequency thresholds $\sigma \in \IntC{0,1}$.  The
results are shown in Figure~\ref{f:pop:internet} and in
Table~\ref{t:pop:internet} for Internet Usage and in
Figure~\ref{f:pop:ipums} and in Table~\ref{t:pop:ipums} for IPUMS
Census.

\begin{table}[p] \centering
\begin{tabular}{|c|c|c|c|c|c|c|c|c|}
\hline
$\sigma$ & {\small $\Abs{\Frequent{\sigma,\DB}}$} & {\small $0$} & {\small $0.001$} & {\small $0.005$} & {\small $0.01$} & {\small $0.02$} & {\small $0.04$} & {\small $0.08$} \\
\hline
\hline
0.20 & 1856 & 1856 & 1574 & 1190 & 900 & 619 & 418 & 188 \\
0.19 & 2228 & 2228 & 1870 & 1396 & 1052 & 728 & 486 & 212 \\
0.18 & 2667 & 2667 & 2217 & 1625 & 1206 & 820 & 522 & 211 \\
0.17 & 3246 & 3246 & 2672 & 1925 & 1421 & 970 & 597 & 231 \\
0.16 & 4013 & 4013 & 3254 & 2295 & 1671 & 1132 & 655 & 242 \\
0.15 & 4983 & 4983 & 3994 & 2764 & 1995 & 1377 & 775 & 270 \\
0.14 & 6291 & 6290 & 4955 & 3339 & 2362 & 1602 & 860 & 261 \\
0.13 & 8000 & 7998 & 6208 & 4093 & 2881 & 1972 & 1034 & 281 \\
0.12 & 10476 & 10472 & 7970 & 5118 & 3562 & 2414 & 1189 & 289 \\
0.11 & 13813 & 13802 & 10267 & 6352 & 4305 & 2804 & 1284 & 264 \\
0.10 & 18615 & 18594 & 13468 & 8068 & 5409 & 3395 & 1423 & 245 \\
0.09 & 25729 & 25686 & 18035 & 10399 & 6920 & 4094 & 1587 & 203 \\
0.08 & 36812 & 36714 & 24870 & 13681 & 9032 & 5008 & 1708 & 153 \\
0.07 & 54793 & 54550 & 35441 & 18477 & 12147 & 6276 & 1803 & 95 \\
0.06 & 85492 & 84873 & 52295 & 25595 & 16376 & 7568 & 1747 & 29 \\
\hline
\end{tabular}
\caption{Pattern orderings for Internet Usage data. \label{t:pop:internet}}
\end{table}

\begin{table} \centering
\begin{tabular}{|c|c|c|c|c|c|c|c|c|}
\hline
$\sigma$ & {\small $\Abs{\Frequent{\sigma,\DB}}$} & {\small $0$} & {\small $0.001$} & {\small $0.005$} & {\small $0.01$} & {\small $0.02$} & {\small $0.04$} & {\small $0.08$} \\
\hline
\hline
0.30 & 8205 & 1335 & 444 & 285 & 212 & 153 & 107 & 61 \\
0.29 & 9641 & 1505 & 496 & 317 & 236 & 167 & 116 & 65 \\
0.28 & 11443 & 1696 & 551 & 351 & 260 & 184 & 120 & 66 \\
0.27 & 13843 & 1948 & 624 & 395 & 292 & 203 & 128 & 68 \\
0.26 & 17503 & 2293 & 725 & 456 & 338 & 233 & 147 & 71 \\
0.25 & 20023 & 2577 & 810 & 502 & 369 & 256 & 161 & 77 \\
0.24 & 23903 & 3006 & 944 & 583 & 427 & 293 & 185 & 92 \\
0.23 & 31791 & 3590 & 1093 & 661 & 477 & 328 & 196 & 85 \\
0.22 & 53203 & 4271 & 1194 & 678 & 481 & 316 & 171 & 57 \\
0.21 & 64731 & 5246 & 1454 & 813 & 573 & 372 & 189 & 62 \\
0.20 & 86879 & 6689 & 1771 & 949 & 661 & 424 & 218 & 67 \\
0.19 & 151909 & 8524 & 1974 & 953 & 628 & 363 & 151 & 27 \\
0.18 & 250441 & 10899 & 2212 & 992 & 625 & 312 & 99 & 10 \\
\hline
\end{tabular}
\caption{Pattern orderings for IPUMS Census data. \label{t:pop:ipums}}
\end{table}

In the figures the axes are the following.  The $x$-axis corresponds
to the length of the prefix of the pattern ordering and the $y$-axis
is corresponds to the average absolute error of the frequency
estimation from the corresponding prefix. The labels of the curves
express the minimum frequency thresholds of the corresponding frequent
itemset collections.

The tables can be interpreted as follows.  The columns $\sigma$ and
$\Abs{\Frequent{\sigma,\DB}}$ correspond to the minimum frequency
threshold $\sigma$ and the number of $\sigma$-frequent itemsets. The
rest of the columns $0$, $0.001$, $0.005$, $0.01$, $0.02$, $0.04$ and
$0.08$ correspond to the number of itemsets in the shortest prefix
with the loss at most $0$, $0.001$, $0.005$, $0.01$, $0.02$, $0.04$
and $0.08$, respectively. (Note that the column $0$ corresponds to the
number of closed frequent itemsets by Theorem~\ref{t:pop:closed}.)

The results show that already relatively short prefixes of the pattern
orderings provide frequency estimates with high accuracy. The
inversions of the orders of the error curves in
Figure~\ref{f:pop:internet} and in Figure~\ref{f:pop:ipums} are due to
the used combination of the estimation method and the loss functions:
On one hand the average absolute error is lower for frequent itemset
collections with lower minimum frequency threshold for the frequency
estimation without any data since initially all frequency estimates of
the frequent itemsets are zero.  On the other hand the frequencies can
be estimated correctly from the closed frequent itemsets and the
number of closed frequent itemsets is smaller for higher minimum
frequency thresholds.

%% file: chains.tex

\chapter{Exploiting Partial Orders of Pattern Collections \label{c:chains}}


Large collections of interesting patterns can be very difficult to
understand and it can be too expensive even to manipulate all
patterns. Because of these difficulties, recently a large portion of
pattern discovery research has been focused on inventing condensed
representations for pattern collections. (See
Section~\ref{s:condensed} for more details.)

Most of the condensed representations are based on relatively local
properties of the pattern collections: the patterns in the condensed
representations are typically chosen solely based on small
neighborhoods in the original pattern collection regardless of which
of the patterns are deemed to be redundant and which are chosen to the
condensed representation.

Two notable exceptions to this are the condensed frequent pattern
bases~\cite{i:pei02} and non-derivable
itemsets~\cite{i:calders02}. Still, even these condensed
representations have certain drawbacks and limitations.

The construction of condensed frequent patterns bases is based on a
greedy strategy: The patterns are pruned from minimal to maximal or
vice versa. A pattern is deemed to be redundant (i.e., not being in
the pattern base) if its frequency is close enough to the frequency of
some already found irredundant pattern that is its super- or
subpattern, depending on the processing direction of the pattern
collection. Alas, also the condensed frequent pattern bases can be
interpreted to be dependent only on the local neighborhoods of the
patterns, although the neighborhoods are determined by the frequencies
rather than only by the structure of the underlying pattern class.

The non-derivable itemsets (Definition~\ref{d:ndi}) take into account
more global properties of the pattern collection. Namely, the
irredundancy (i.e., non-derivability) of an itemset with respect to
derivability depends on the frequencies of its all subitemsets.  
However, the irredundancy of the itemset in the case of non-derivable
itemsets is determined using inclusion-exclusion truncations. Although
the non-derivable itemsets can be superficially understood as the
itemsets whose frequencies cannot be derived exactly from the
frequencies of their subitemsets, it is not so easy to see immediately
which aspects of the itemset collection and the frequencies of the
itemsets one particular non-derivable itemset represents, i.e., to see
the essence of the upper and the lower bounds of the itemsets for the
underlying transaction database.

The pattern collections have also other structure than the quality
values. (In fact, not all pattern collections have quality values at
all. For example, the interesting patterns could be determined by an
oracle that is not willing to say anything else than whether or not a
pattern is fascinating.) In particular, virtually all pattern
collections adhere some non-trivial partial order
(Definition~\ref{d:partial}).

The goal of this chapter is to make pattern collections more
understandable and concise by exploiting the partial orders of the
collections. We use the partial orders to partition a given pattern
collection to subcollections of (in)comparable patterns, i.e., to
(anti)chains. In addition to clustering the patterns in the collection
into smaller groups using the partial order of the pattern class, we
show that the chaining of patterns can also condense the pattern
collection: for many pattern classes each chain representing possibly
several patterns can be represented as only a slightly more complex
pattern than each of the patterns in the chain.

In this chapter, we propose the idea of (anti)chaining patterns,
illustrate its usefulness and potential pitfalls, and discuss the
computational aspects of the chaining massive pattern
collections. Furthermore, we explain how, for some pattern classes,
each chain can represented as one slightly more complex pattern than
the patterns in the underlying pattern collection.

This chapter is based on the article ``Chaining
Patterns''~\cite{i:mielikainen03:chains}.

\section{Exploiting the Structure}

The collections of interesting patterns (and the underlying pattern
classes, too) have usually some structure.

\begin{example}[structuring the collection of itemsets by frequencies]
The collection $2^{\Items}$ of all itemsets can be structured based on
their frequencies: every subset of a frequent itemset is frequent and
every superset of an infrequent itemset is infrequent. Thus, for each
minimum frequency threshold $\sigma \in \IntC{0,1}$, a given
transaction database $\DB$ determines a partition
\begin{displaymath}
\Tuple{\Frequent{\sigma,\DB},2^\Items \setminus \Frequent{\sigma,\DB}}
\end{displaymath}
of the itemset collection $2^\Items$.
\exend \end{example}

The downward closed collections of frequent itemsets are examples of
data-dependent structures of pattern collections. The pattern
collections have also some data-independent structure. Maybe the most
typical data-independent structure in a pattern collection is a
partial order over the patterns.

\begin{example}[set inclusion as a partial order over itemsets]
Let the pattern class be again $2^\Items$. A natural partial order for
itemsets is the partial order determined by the set inclusion
relation:
\begin{displaymath}
X \prec Y \iff X \subset Y
\end{displaymath}
for all $X,Y\subseteq \Items$.
\exend \end{example}

A partial order where no two patterns are comparable, i.e., an empty
partial order, is called a \emph{trivial} partial order. For example,
any partial order restricted to maximal or minimal patterns is
trivial. A trivial partial order is the least informative partial
order in the sense that it does not relate the patterns to each other
at all.

Besides of merely detecting the structure in the pattern collection,
the found structure can sometimes be further exploited. For example,
the frequent itemsets can be stored into an \emph{itemset tree} by
defining a total order for over $\Items$. In an itemset tree, each
itemset corresponds to a path from root to some node the labels of the
edges being the items of the itemsets in ascending order. (Itemset
trees are known also by several other names,
see~\cite{a:agarwal01,i:agrawal96,a:han04,a:zaki00}.)

\begin{example}[an itemset tree] \label{ex:itemsettree}
Let the itemset collection consist of itemsets $\emptyset$, $\Set{A}$,
$\Set{A,B,C}$, $\Set{A,B,D}$, $\Set{A,C}$, $\Set{B}$, $\Set{B,C,D}$,
and $\Set{B,D}$. The itemset tree representing this itemset collection
is shown as Figure~\ref{f:itemsettree}.

\begin{figure}[h!]
\includegraphics[width=\textwidth]{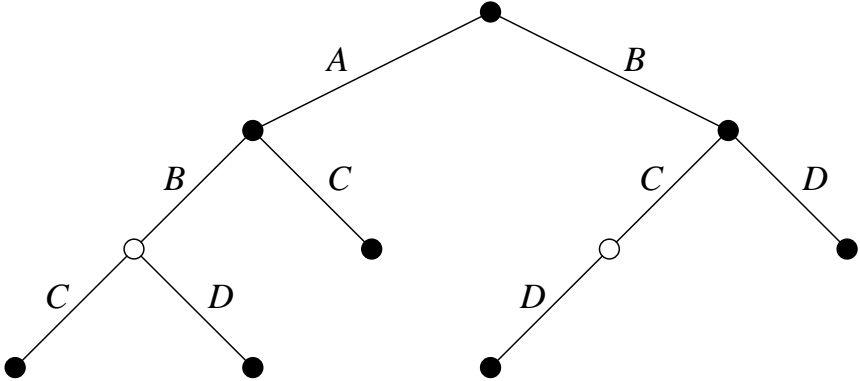}
\caption{An itemset tree representing the itemset collection of
Example~\ref{ex:itemsettree}.  Each itemset can be seen in the tree as
a path from the root to a solid node. \label{f:itemsettree}}
\end{figure}
\exend \end{example}

Representing an itemset collection as an itemset tree can save space
and support efficient quality value queries. The quality value of an
itemset $X$ can be retrieved (or decided that it is not in the itemset
tree) in time $\Oh{\Abs{X}}$. (Time and space complexities similar to
itemset tries can be obtained also by refining the itemset trees to
automata~\cite{i:mielikainen04:automata}.)  Unfortunately, the
structure of itemset trees is strongly dependent on the ordering of
the items: there are not always natural orderings for the items and an
arbitrary ordering can induce artificial structure to the itemset tree
that hides the essence of the pattern collection.

The exploitation of the partial order structure of a pattern
collection somehow might still be beneficial although, for example, it
is not clear whether the itemset tree makes a partial order of an
itemset collection more understandable or even more obscure from the
human point of view.  A simple approach to reduce the obscurity of the
itemset trees is to construct for each itemset collection a random
forest of itemset trees where each itemset tree represents the itemset
collection using some random ordering of the items. (These random
forests should not be confused with the random forests of Leo
Breiman~\cite{a:breiman01}.)  Unfortunately, the ordering of the items
is still present in each of the itemset trees. Fortunately, there are
structures in partial orders that do not depend on anything else than
the partial order. Two important examples of such structures are
\emph{chains} and \emph{antichains}.

\begin{definition}[chains and antichains]
A subset $\Chain$ of a partially ordered set $\Pat$ is called a
\emph{chain} if and only if all elements in $\Chain$ are comparable
with each other, i.e., $p \preceq p'$ or $p' \preceq p$ holds for all
$p,p' \in \Chain$.

The \emph{rank} of a pattern $p$ in chain $\Chain$, denoted by
$\rk{p,\Chain}$, is the number of elements that has to be removed from
the chain before $p$ is the minimal pattern in the chain.

A subset $\AntiChain$ of a partially ordered set $\Pat$ is called an
\emph{antichain} if and only if all elements in $\Chain$ are
incomparable with each other, i.e., $p \prec p'$ holds for no $p,p'
\in \Chain$.
\end{definition}

\begin{example}[chains and antichains]
The itemset collection
\begin{displaymath}
\Set{\Set{A,B,C,D,E,F},\Set{A,C,E},\Set{C,E},\Set{C}}
\end{displaymath}
is a chain with respect to the set inclusion relation since all
itemsets in the collections are comparable with each other.
Similarly, the itemset collection $\Set{\Set{A,B} \Set{A,C}}$ is an
antichain.
The itemset collection $\Set{\Set{A,B},\Set{A},\Set{B}}$ is not a
chain nor an antichain: the collection is not a chain since $\Set{A}$
and $\Set{B}$ are not comparable, and it is not an antichain since
$\Set{A,B}$ is comparable with $\Set{A}$ and $\Set{B}$.
\exend \end{example}

A chain or an antichain in a pattern collection can be understood more
easily than the whole pattern collection, since each pattern either
does or does not have relationship with each other pattern in the
chain or the antichain, respectively. Thus, a natural approach to make
a pattern collection more digestible using a partial order structure
is to partition the pattern collection into chains or antichains.

\begin{definition}[chain and antichain partitions]
A \emph{chain partition} (an \emph{antichain partition}) of a
partially ordered set $\Pat$ is a partition of the set $\Pat$ into
disjoint chains $\Chain_1,\ldots,\Chain_m$ (disjoint antichains
$\AntiChain_1,\ldots,\AntiChain_m$).

A chain partition (an antichain partition) of $\Pat$ is \emph{minimum}
if and only if there is no chain partition (no antichain partition) of
$\Pat$ consisting of a smaller number of chains (antichains).

A chain partition (an antichain partition) of $\Pat$ is \emph{minimal}
if and only if there are no two chains $\Chain_i$ and $\Chain_j$
(antichains $\AntiChain_i$ and $\AntiChain_j$) in the chain partition
(the antichain partition) such that their union is a chain (an
antichain).
\end{definition}

A chain or an antichain partition can be interpreted as a structural
clustering of the patterns. Each chain represents a collection of
comparable patterns and each antichain a collection of incomparable
ones, i.e., a chain consists of structurally similar patterns whereas
an antichain can be seen as a representative collection of patterns.

The minimum chain partition is not necessarily unique. The lack of
uniqueness is not only a problem because of the exploratory nature of
data mining.  Different partitions highlight different aspects of the
pattern collection which can clearly be beneficial when one is trying
to understand the pattern collection (and the underlying data set).

The maximum number of chains (antichains) in a chain partition (an
antichain partition) of a pattern collection $\Pat$ is $\Abs{\Pat}$
since each pattern $p \in \Pat$ as a singleton set $\Set{p}$ is a
chain and an antichain simultaneously. The minimum number of chains in
a chain partition is at least the cardinality of the largest antichain
in $\Pat$ since no two distinct patterns in the largest antichain can
be in the same chain. This inequality can be shown to be actually an
equality and the result is known as \emph{Dilworth's Theorem}:
\begin{theorem}[Dilworth's Theorem \cite{b:jukna01}] \label{t:dilworth}
A partially ordered set $\Pat$ can be partitioned into $m$ chains if
and only if the largest antichain in $\Pat$ is of cardinality at most
$m$.
\end{theorem}

\begin{example}[bounding the number of chains from below] \label{ex:chainlower}
The maximal patterns in a pattern collections form an
antichain. Thus, the number of maximal patterns is a lower bound for
the number of chains in the minimum chain partition.
\exend \end{example}

Similarly to bounding the minimum chain partitions by maximum
antichains, it is possible to bound the minimum number of antichains
needed to cover all patterns in $\Pat$ by the cardinality of the
maximum chain in $\Pat$:
\begin{theorem}[\cite{b:schrijver03}] \label{t:antichains}
The number of antichains in a minimum antichain partition of a
partially ordered set $\Pat$ is equal to the cardinality of a maximum
chain in $\Pat$.
\end{theorem}

\section{Extracting Chains and Antichains}

The problem of finding the minimum chain partition for a partially
ordered pattern collection $\Pat$ can be formulated as follows:
\begin{problem}[minimum chain partition]
Given a pattern collection $\Pat$ and a partial order $\prec$ over
$\Pat$, find a partition of $\Pat$ into the minimum number of chains
$\Chain_1,\ldots,\Chain_m$.
\end{problem}

The minimum chain partition can be found efficiently by finding a
maximum matching in a bipartite graph~\cite{b:lovasz86}. The maximum
bipartite matching problem is the following~\cite{b:schrijver03}:
\begin{problem}[maximum bipartite matching] \label{p:bimatch}
Given a bipartite graph $\Tuple{U,V,E}$ where $U$ and $V$ are sets of
vertices and $E$ is a set of edges between $U$ and $V$, i.e., a set of
pairs in $U \times V$, find a maximum bipartite matching $M \subseteq
E$, i.e., find a largest subset $M$ of $E$ such that
\begin{displaymath}
\degr{u,M}=\Abs{ \Set{e \in M : (u,v)=e \mbox { for some } v \in V}}
\leq 1
\end{displaymath}
for all $u \in U$ and
\begin{displaymath}
\degr{v,M}=\Abs{ \Set{e \in M : (u,v)=e \mbox { for some } u \in U}}
\leq 1
\end{displaymath}
for all $v \in V$.
\end{problem}

The matching is computed in a bipartite graph consisting two copies
$\Pat$ and $\Pat'$ of the pattern collection $\Pat$ and the partial
order $\prec'$ as edges between $\Pat$ and $\Pat'$ corresponding to
the partial order $\prec$.  Thus, the bipartite graph representation
of the pattern collection $\Pat$ is a triplet
$\Tuple{\Pat,\Pat',\prec'}$.

\begin{proposition} \label{t:chains}
The a matching $M$ in a bipartite graph $\Tuple{\Pat,\Pat',\prec'}$
determines a chain partition.
The number of unmatched vertices in $\Pat$ (or equivalently in $\Pat'$) 
is equal to the number of chains.
\end{proposition}
\begin{proof}
Let us consider the partially ordered set $\Pat$ as a graph
$\Tuple{\Pat,\prec}$ A matching $M \subseteq \prec'$ ensures that the
in-degree and out-degree of each $p \in \Pat$ is at most one.  Thus,
the set $M$ partitions the graph $\Tuple{\Pat,\prec}$ to paths. By
transitivity of partial orders, each path is a chain.

The number of unmatched patterns in $\Pat$ correspond to the minimal
patterns of the chains. Each unmatched pattern $p \in \Pat$ is a
minimal pattern in some chain and if a patterns $p$ is minimal pattern
in some chain then it is unmatched. As each chain contains exactly one
minimal pattern, the number of unmatched patterns in $\Pat$ is equal
to the number of chains in the chain partition corresponding to the
matching $M$.
\end{proof}

Due to Proposition~\ref{t:chains}, the number of chains is minimized
when the cardinality of the matching is maximized. The chain partition
can be extracted from the matching $M$ in time linear in the
cardinality of $\Pat$. The partition of a partially ordered pattern
collection into the minimum number of chains can be computed as described
by Algorithm~\ref{a:Partition-into-Chains}.

\begin{algorithm}[h]
\caption{A minimum chain partition. \label{a:Partition-into-Chains}}
\begin{algorithmic}[1]
\Input{A pattern collection $\Pat$ and a partial order $\prec$ over
$\Pat$.}
\Output{Partition of $\Pat$ into the minimum number $m$ of chains
$\Chain_1,\ldots,\Chain_m$.}
\Function{Partition-into-Chains}{$\Pat,\prec$}
\State $M \leftarrow$ \Call{Maximum-Matching}{$\Pat,\Pat,\prec$}
\State $m \leftarrow 0$
\ForAll{$p \in \Pat$}
 \State $\prv{p} \leftarrow p$
 \State $\nxt{p} \leftarrow p$
\EndFor
\ForAll{$\Tuple{p,p'} \in M$}
 \State $\nxt{p} \leftarrow p'$
 \State $\prv{p'} \leftarrow p$
\EndFor
\ForAll{$p \in \Pat, p=\prv{p}$}
 \State $m \leftarrow m+1$
 \State $\Chain_m\leftarrow \Set{p}$
 \While{$p \neq \nxt{p}$}
  \State $p \leftarrow \nxt{p}$
  \State $\Chain_m \leftarrow \Chain_m \cup \Set{p}$
 \EndWhile
\EndFor
\State \textbf{return} $\Tuple{\Chain_1,\ldots,\Chain_m}$
\EndFunction
\end{algorithmic}
\end{algorithm}

A maximum matching $M$ in a bipartite graph $\Tuple{U,V,E}$ can be
found in time $\Oh{\sqrt{\min
\Set{\Abs{U},\Abs{V}}}\Abs{E}}$~\cite{a:galil86}.  Thus, if the
partial order $\prec$ is known explicitly, then the minimum chain
partition ca be found in time $\Oh{\sqrt{\Abs{\Pat}}\Abs{\prec}}$
which can be bounded above by $\Oh{\Abs{\Pat}^{5/2}}$ since the
cardinality of $\prec$ is at most $\Abs{\Pat}\Paren{\Abs{\Pat}-1}/2$.

The idea of partitioning the graph $\Tuple{\Pat,\prec}$ into the
minimum number of disjoint paths can be generalized to partitioning it
into disjoint degree-constrained subgraphs with maximum number of
matched edges by finding a maximum bipartite $\mathit{b}$-match\-ing
instead of a maximum (ordinary) bipartite matching matching.  The
maximum bipartite $b$-matching differs from the maximum bipartite
matching (Problem~\ref{p:bimatch}) only by the degree
constraints. Namely, each vertex in $v \in U \cup V$ has a positive
integer $\mathit{b}(v)$ constraining the maximum degree of the vertex:
the degree $\degr{v,M}$ of $v$ in the matching $M$ can be at most
$\mathit{b}(v)$. Thus the maximum bipartite matching is a special case
of the maximum bipartite $\mathit{b}$-matching with $\mathit{b}(v)=1$
for all $v \in U \cup V$.

If there is a weight function $\weight : \prec \to \mathbb{R}$, then
the graph $\Tuple{\Pat,\prec}$ can be partitioned also into disjoint
paths with maximum total weight. That is, the pattern can be
partitioned into disjoint chains in such a way that the sum of the
weights of consecutive patterns in the chains is maximized. This can
be done by finding a maximum weight bipartite matching that differs
from the maximum bipartite matching (Problem~\ref{p:bimatch}) by the
objective function: instead of maximizing the cardinality $\Abs{M}$ of
the matching $M$, the weight $\sum_{e \in M} \wg{e}$ of the edges in
the matching $M$ is maximized.

However, there are two traits in partitioning the partially ordered
pattern collections into chains: pattern collections are often
enormously large and the partial order over the collection might be
known only implicitly.

Due to the problem of pattern collections being massive, finding
the maximum bipartite matching in time
$\Oh{\sqrt{\Abs{\Pat}}\Abs{\prec}}$ can be too slow. This problem can
be overcome by finding a maximal matching instead of a maximum
matching. A maximal matching in $\Tuple{\Pat,\Pat',\prec'}$ can be
found in time $\Oh{\Abs{\Pat}+\Abs{\prec}}$ as shown by
Algorithm~\ref{a:Maximal-Matching}.

\begin{algorithm} [h]
\caption{A greedy algorithm for finding a maximal matching in a bipartite graph $\Tuple{\Pat,\Pat',\prec'}$. \label{a:Maximal-Matching}}
\begin{algorithmic}[1]
\Input{A bipartite graph $\Tuple{\Pat,\Pat',\prec'}$.}
\Output{A maximal matching $M$ in the graph.}
\Function{Maximal-Matching}{$\Pat,\Pat',\prec'$}
\ForAll{$p \in \Pat$}
\State $\prv{p} \leftarrow p$
\State $\nxt{p} \leftarrow p$
\EndFor
\ForAll{$\Tuple{p,p'} \in \prec'$}
\If{$p=\nxt{p}$ and $p'=\prv{p'}$}
\State $\nxt{p} \leftarrow p'$
\State $\nxt{p'} \leftarrow p$
\EndIf
\EndFor
\State $M \leftarrow \emptyset$
\ForAll{$p \in \Pat,p \neq \nxt{p}$}
\State $M \leftarrow M \cup \Set{ \Tuple{p,\nxt{p}}}$
\State $\nxt{p} \leftarrow p$
\EndFor
\State \textbf{return} $M$
\EndFunction
\end{algorithmic}
\end{algorithm}

It is easy to see that the cardinality of a maximal matching is at
least half of the cardinality of the maximum matching in the same
graph. Unfortunately, this does not imply any non-trivial
approximation quality guarantees for the corresponding chain
partitions.

\begin{example}[minimum and minimal chain partitions by maximum and maximal matchings] \label{ex:maxmatch}
Let us consider the pattern collection
$\Set{1,2,\ldots,2n}$ with partial order 
\begin{displaymath}
\prec=\Set{\Tuple{i,j} : i<j}.
\end{displaymath}
The maximum matching
\begin{displaymath}
\Set{\Tuple{1,2},\Tuple{2,3},\ldots,\Tuple{2n-1,2n}}
\end{displaymath}
determines only one chain
\begin{displaymath}
\Chain=\Set{1,2,\ldots,2n}
\end{displaymath}
whereas the worst maximal matching
\begin{displaymath}
\Set{\Tuple{1,2n},\Tuple{2,2n-1},\ldots,\Tuple{n,n+1}}
\end{displaymath}
determines $n$ chains
\begin{displaymath}
\Chain_1=\Set{1,2n},\Chain_2=\Set{2,2n-1},\ldots,\Chain_n=\Set{n,n+1}.
\end{displaymath}

Thus, in the worst case the chain partition found by maximal matching
is $\Abs{\Pat}/2$ times worse than the optimal chain partition found
by maximum matching.
\exend \end{example}

The quality of the maximal matching, i.e., the quality of the minimal
chain partition can be improved by finding a total order conforming
the partial order. If the partial order is known explicitly, then a
total order conforming it can be found by topological sorting in time
$\Oh{\Abs{\prec}+\Abs{\Pat}}$. Sometimes there is a total order that
can be computed without even knowing the partial order explicitly. For
example, frequent itemsets can be sorted with respect to their
cardinalities. This kind of ordering can reduce the number of chains
in the chain partition found by maximal matchings considerably. The
amount of the improvement depends on how well the total order is able
to capture the essence of the partial order (whatever it might be).

\begin{example}[improving maximal matching using a total order]
Let us consider the pattern collection and the partial order of
Example~\ref{ex:maxmatch}.  If the patterns in the collection
$\Set{1,2,\ldots,2n}$ are ordered in ascending or in descending order,
then the maximal matching agrees with the maximum matching, i.e., the
minimal chain partition agrees with the minimum chain partition.
\exend \end{example}

If the partial order is given implicitly, as a function that can be
evaluated for any pair of patterns $p,p' \in \Pat$, then the explicit
construction of the partial order relation $\prec$ might itself be a
major bottleneck of the chaining of the patterns.  The brute force
construction of the partial order $\prec$, i.e., testing of all pairs
of patterns in $\Pat$ requires $\Oh{\Abs{\Pat}^2}$ comparisons. In the
worst case this upper bound is tight.
\begin{example}[the number of comparison in the worst case]
Let the pattern collection $\Pat$ be an antichain with respect to a
partial order $\prec$. Then all patterns in $\Pat$ must be compared
with all other patterns in $\Pat$ in order to construct $\prec$
explicitly, i.e., to ensure that all patterns in $\Pat$ are
incomparable with each other and thus that $\Pat$ indeed is an
antichain.  \exend
\end{example}

\begin{algorithm}
\caption{An algorithm to find a minimal chain partition. \label{a:Minimal-Partition-into-Chains}}
\begin{algorithmic}[1]
\Input{A pattern collection $\Pat$ and a partial order $\prec$ over
$\Pat$.}
\Output{A minimal chain partition $\Chain_1,\ldots,\Chain_m$ of the
pattern collection $\Pat$.}
\Function{Minimal-Partition-into-Chains}{$\Pat,\prec$}
\State $m \leftarrow 0$
\ForAll{$p \in \Pat$}
 \State $\prv{p} \leftarrow p$
 \State $\nxt{p} \leftarrow p$
 \State $i \leftarrow 1$ 
 \While{$i \leq m$ and $p=\prv{p}=\nxt{p}$}
  \If{$p \prec \min \Chain_i$}
   \State $\nxt{p} \leftarrow \min \Chain_i$
   \State $\prv{\min \Chain_i} \leftarrow p$
  \ElsIf{$\max \Chain_i \prec p$}
   \State $\prv{p} \leftarrow \max \Chain_i$
   \State $\nxt{\max \Chain_i} \leftarrow p$
  \EndIf  
  \State $p' \leftarrow \prv{\max \Set{p'' \in \Chain_i : p \prec p''}}$
  \If{$p=\prv{p}=\nxt{p}$ and $p' \prec p$}
   \State $\nxt{p} \leftarrow \nxt{p'}$
   \State $\prv{p} \leftarrow p'$
   \State $\nxt{p'} \leftarrow p$
   \State $\prv{\nxt{p}} \leftarrow p$
  \EndIf
  \If{$p \neq \prv{p}$ or $p \neq \nxt{p}$}
   \State $\Chain_i \leftarrow \Chain_i \cup \Set{p}$
  \EndIf
  \State $i \leftarrow i+1$
 \EndWhile
 \If{$p=\prv{p}=\nxt{p}$}
  \State $m \leftarrow m+1$
  \State $\Chain_m \leftarrow \Set{p}$
 \EndIf
\EndFor
\State \Return $\Tuple{\Chain_1,\ldots,\Chain_m}$
\EndFunction
\end{algorithmic}
\end{algorithm}

Fortunately, the partial order relations have two useful properties
that can be exploited in the construction of $\prec$, namely
transitivity and antisymmetry holding for any partial order relation
$\preceq$. Due to transitivity, $p \preceq p'$ and $p' \preceq p''$
together imply $p \preceq p''$, and antisymmetry guarantees that the
graph $\Tuple{\Pat,\prec}$ is acyclic. The partial order $\preceq$ can
be computed also as a side product of the construction of a chain
partition as shown for minimal chain partitions by
Algorithm~\ref{a:Minimal-Partition-into-Chains}.

Although Algorithm~\ref{a:Minimal-Partition-into-Chains} needs time
$\Oh{\Abs{\Pat}^2}$ in the worst case, $\Abs{\Chain_i}$ comparisons
are always sufficient to decide whether a pattern $p \in \Pat$ can be
added to $\Chain_i$. Furthermore, the number of comparison can be
reduced to $1+\Floor{\log_2 \Abs{\Chain_i}}$ comparisons if $\Chain_i$
is represented as, e.g., a search tree instead of a linked list. The
number of comparisons can be reduced also by reusing already evaluated
comparisons and transitivity. Furthermore, there are several other
strategies to construct the partial order relation $\prec$.  The
usefulness of different strategies depends on the cost of evaluating
the comparisons and the actual partial order. Thus, it seems that
choosing the best strategy for constructing the partial order has to
be estimated experimentally in general.

Another partition of a pattern collection based on a partial order is
an antichain partition.  The problem of finding a minimum antichain
partition of a partially ordered pattern collection $\Pat$ can be
formulated as follows:
\begin{problem}[minimum antichain partition] \label{p:antichains}
Given a pattern collection $\Pat$ and a partial order $\prec$ over
$\Pat$, find a partition of $\Pat$ into the minimum number of
antichains $\AntiChain_1,\ldots,\AntiChain_m$.
\end{problem}

Solving Problem~\ref{p:antichains} is relatively easy based on
Theorem~\ref{t:antichains}.
Algorithm~\ref{a:Partition-into-Antichains} solves the problem in the
case of arbitrary pattern collections.
\begin{algorithm}
\caption{A minimum antichain partition. \label{a:Partition-into-Antichains}}
\begin{algorithmic}[1]
\Input{A pattern collection $\Pat$ and a partial order $\prec$ over
$\Pat$.}
\Output{Partition of $\Pat$ into the minimum number $m$ of antichains
$\AntiChain_1,\ldots,\AntiChain_m$.}
\Function{Partition-into-Antichains}{$\Pat,\prec$}
\State $\Pat' \leftarrow \Pat$
\State $m \leftarrow 0$
\While{$\Pat' \neq \emptyset$}
\State $m \leftarrow m+1$
\State $\AntiChain_m \leftarrow \Set{ p \in \Pat' : p \preceq p' \in \Pat' \Rightarrow p=p'}$
\State $\Pat \leftarrow \Pat' \setminus \AntiChain_m$
\EndWhile
\State \textbf{return} $\Tuple{\AntiChain_1,\ldots,\AntiChain_m}$
\EndFunction
\end{algorithmic}
\end{algorithm}

In many cases the minimum antichain partition can be found even more
easily. For example, the minimum antichain partition of
$\sigma$-frequent itemsets can be computed in time linear in the sum
of cardinalities of the $\sigma$-frequent itemsets: The length $m$ of
the longest chain in $\Frequent{\sigma,\DB}$ is one greater than the
cardinality of the largest itemset in the collection. Thus, the
collection $\Frequent{\sigma,\DB}$ can be partitioned into $m$
antichains $\AntiChain_1,\ldots,\AntiChain_m$ containing all
$\sigma$-frequent itemsets of cardinalities $0,\ldots,m-1$,
respectively. Clearly, this partition can be constructed in time
linear in $\sum_{X \in \Frequent{\sigma,\DB}} \Abs{X}$ by maintaining
$m$ lists of patterns.

\section{Condensation by Chaining Patterns}

A chain partition of a pattern collection can be more than a mere
structural clustering if the collection has more structure than a
partial order. One example of such a pattern collection is a
transaction database without its transaction identifiers.

\begin{example}[itemset chains] \label{ex:ichain1}
Let us consider the transaction database $\DB$ shown as Table~\ref{t:ex:ichain1}.

\begin{table}[h!] \centering
\caption{The transaction identifiers and the itemsets of the transactions in $\DB$. \label{t:ex:ichain1}}
\begin{tabular}{|c|c|}
\hline
$\mathit{tid}$ & $X$ \\
\hline
\hline
$1$ & $\Set{1}$ \\
$2$ & $\Set{2}$ \\
$3$ & $\Set{2}$ \\
$4$ & $\Set{2}$ \\
$5$ & $\Set{2}$ \\
$6$ & $\Set{2}$ \\
$7$ & $\Set{1,3}$ \\
$8$ & $\Set{1,3}$ \\
\hline
\end{tabular}
\begin{tabular}{|c|c|}
\hline
$\mathit{tid}$ & $X$ \\
\hline
\hline
$9$ & $\Set{1,3}$ \\
$10$ & $\Set{2,4}$ \\
$11$ & $\Set{2,4}$ \\
$12$ & $\Set{2,4}$ \\
$13$ & $\Set{2,4}$ \\
$14$ & $\Set{1,2,3}$ \\
$15$ & $\Set{1,2,3}$ \\
$16$ & $\Set{1,2,4}$ \\
\hline
\end{tabular}
\end{table}

Note that the transaction database $\DB$ could be represented also as
a collection of weighted itemsets, i.e., as a collection
\begin{displaymath}
\Set{\Set{1},\Set{2},\Set{1,3},\Set{2,4},\Set{1,2,3},\Set{1,2,4}}=\Set{1,2,13,23,123,124}
\end{displaymath}
of itemsets together with a weight function
\begin{displaymath}
\weight=\Set{1 \mapsto 1, 2 \mapsto 5,13 \mapsto 3,24 \mapsto 4,123 \mapsto 2, 124 \mapsto 1}.
\end{displaymath}

The collection of itemsets representing the transaction database $\DB$
can be partitioned into two chains $\Chain_1=\Set{1,13,123}$ and
$\Chain_2=\Set{2,24,124}$.
\exend \end{example}

Each chain $\Chain$ of itemsets can be written as a one itemset $X$ by
adding to each item in the itemsets of $\Chain$ the information about
the minimum rank of the itemset in $\Chain$ containing that item. That
is, a chain 
$\Chain=\Set{X_1,\ldots,X_n}$ such that $X_1 \subset \ldots \subset X_n=A_1\ldots
A_m$
can be written as
\begin{displaymath}
\Chain=\Set{A_{1}^{\rk{A_1,\Chain}},\ldots,A_{m}^{\rk{A_m,\Chain}}} =
A_{1}^{\rk{A_1,\Chain}}\ldots A_{m}^{\rk{A_m,\Chain}}
\end{displaymath}
where
\begin{displaymath}
\rk{A,\Chain}=\min_{A \in X \in \Chain} \rk{X,\Chain}
\end{displaymath}
for any item $A$. (Note that the superscript corresponding to the
ranks serve also as separators of the items, i.e., no other separators
such as commas are needed.)  Furthermore, if there are several items
$A_i, i \in I=\Set{i_1,\ldots,i_{\Abs{I}}}$, with the same rank
$\rk{A_i,\Chain}=k$, then we can write $\Set{A_i : i \in
I}^k=\Set{A_{i_1},\ldots,A_{i_{\Abs{I}}}}^k$ instead of
$A_{i_1}^k\ldots A_{i_{\Abs{I}}}^k$.  The ranks can even be omitted in
that case if the items are ordered by their ranks.

The quality values of the itemsets in the chain $\Chain$ can be
expressed as a vector of length $\Abs{\Chain}$ where $i$th position of
the vector is quality value of the itemset with rank $i-1$ in the
chain. Also, if the interestingness measure is known to be strictly
increasing or strictly decreasing with respect to the partial order,
then the ranks can be replaced by the quality values of the itemsets.

\begin{example}[representing the itemset chains]
The itemset chains $\Chain_1$ and $\Chain_2$ of
Example~\ref{ex:ichain1} can be written as 
\begin{displaymath}
\Chain_1=1^02^23^1
\end{displaymath}
and
\begin{displaymath}
\Chain_2=1^22^04^1
\end{displaymath}
where the superscripts are the ranks.  The whole transaction database
(neglecting the actual transaction identifiers) is determined if also
the weight vectors
\begin{displaymath}
\wg{\Chain_1}=\Tuple{1,3,2}
\end{displaymath}
and 
\begin{displaymath}
\wg{\Chain_2}=\Tuple{5,4,1}
\end{displaymath}
associated to the chains $\Chain_1$ and $\Chain_2$ are given.
\exend \end{example}

From a chain represented as an itemset augmented with the ranks of
items in the chain, it is possible to construct the original chain.
Namely, a rank-$k$ itemset of the chain
\begin{displaymath}
\Chain=A_{1}^{\rk{A_1,\Chain}}\ldots A_{m}^{\rk{A_m,\Chain}} 
\end{displaymath}
is
\begin{displaymath}
\Set{ A_i : \rk{A_i,\Chain} \leq k, 1 \leq i \leq m}.
\end{displaymath}

This approach to represent pattern chains can be adapted to a wide
variety of different pattern classes such sequences and graphs.
Besides of making the pattern collection more compactly representable
and hopefully more understandable, this approach can also compress the
pattern collections.

\begin{example}[condensation by itemset chains]
Let an itemset collection consists
of itemsets
\begin{displaymath}
\Set{0},\Set{0,1},\ldots,\Set{0,\ldots,n-1},\Set{1,\ldots,n},\Set{2,\ldots,n},\ldots,\Set{n}.
\end{displaymath}
The collection can be partitioned to two chains
\begin{displaymath}
\Chain_1=\Set{\Set{0},\Set{0,1},\ldots,\Set{0,\ldots,n-1}}
\end{displaymath}
and
\begin{displaymath}
\Chain_2=\Set{\Set{1,\ldots,n},\Set{2,\ldots,n},\Set{n}}.
\end{displaymath}
The size of each chain is $\Th{n^2}$ items if they are represented
explicitly but only $\Oh{n\log n}$ items if represented as itemsets
augmented with the item ranks, i.e., as
\begin{displaymath}
\Chain_1=\Set{0^0,1^1,\ldots,\Paren{n-1}^{n-1}}=0^01^1\ldots \Paren{n-1}^{n-1}
\end{displaymath}
and
\begin{displaymath}
\Chain_2=\Set{1^{n-1},2^{n-2},\ldots,n^0}=1^{n-1}2^{n-2}\ldots n^0.
\end{displaymath}
\exend \end{example}

\begin{realexample}[chain and antichain partitions in the course completion database] \label{rex:chains}
To illustrate chain and antichain partitions, let us again consider
the course completion database (see Subsection~\ref{ss:ds}) and especially
the $0.20$-frequent closed itemsets in it (see Example~\ref{rex:closed}).

By Dilworth's Theorem (Theorem~\ref{t:dilworth}), each antichain in the
collection gives a lower bound for the minimum number of chains in any
chain partition of the collection. As itemsets of each cardinality
form an antichain, we know (see Table~\ref{t:rex:closed}) that there
are at least $638$ chains in any chain partition of the collection of
$0.20$-frequent closed itemsets in the course completion database. 

The minimum number of chains in the collection is slightly higher,
namely $735$. (That can be computed by summing the values of the
second column of Table~\ref{t:rex:chains:distr} representing the
numbers of chains of different lengths.) The mode and median lengths
of the chains are both three.

\begin{table}[h] \centering
\caption{The number of chains of all non-zero lengths in the minimum
chain partition of the $0.20$-frequent closed itemsets in the course
completion database. \label{t:rex:chains:distr}}
\begin{tabular}{|c|c|}
\hline
the length of chain & the number of chains \\
\hline 
\hline
$1$ & $63$ \\
$2$ & $189$ \\
$3$ & $277$ \\
$4$ & $168$ \\
$5$ & $36$ \\
$6$ & $2$ \\
\hline
\end{tabular}
\end{table}

Ten longest chains are shown in Table~\ref{t:rex:chains:top10}.  (The
eight chains of length five are chosen arbitrarily from the $36$
chains of length five.) The columns of the table are follows. The
column $\Abs{\Chain}$ corresponds to the lengths of the chains, the
column $\Chain$ to the chains, and the column $\supp{\Chain,\DB}$ to
the vectors representing the supports of the itemsets in the chain.

\begin{table}[h] \centering
\caption{Ten longest chains in the minimum chain partition of the
$0.20$-frequent closed itemsets in the course completion
database. \label{t:rex:chains:top10}}
\begin{tabular}{|c|c|c|}
\hline
$\Abs{\Chain}$ & $\Chain$ & $\supp{\Chain,\DB}$ \\
\hline
\hline
$6$ & $12^{0}2^{1}15^{2}13^{3}0^{4}\Set{3,5}^{5}$ & $\Tuple{763, 739, 616, 558, 523, 520}$ \\
$6$ & $7^{0}10^{1}3^{2}5^{3}13^{4}2^{5}$ & $\Tuple{1060, 570, 565, 559, 501, 496}$ \\
$5$ & $\Set{6,13}^{0}15^{1}2^{2}12^{3}\Set{3,5}^{4}$ & $\Tuple{625, 579, 528, 507, 504}$ \\
$5$ & $\Set{0,12}^{0}6^{1}13^{2}7^{3}\Set{2,3,5}^{4}$ & $\Tuple{690, 569, 500, 495, 481}$ \\
$5$ & $15^{0}0^{1}1^{2}5^{3}3^{4}$ & $\Tuple{748, 678, 539, 497, 491}$ \\
$5$ & $\Set{2,5}^{0}15^{1}0^{2}7^{3}\Set{6,12}^{4}$ & $\Tuple{992, 666, 608, 575, 499}$ \\
$5$ & $0^{0}10^{1}2^{2}1^{3}3^{4}$ & $\Tuple{2076, 788, 692, 526, 510}$ \\
$5$ & $\Set{2,3}^{0}6^{1}12^{2}7^{3}13^{4}$ & $\Tuple{1098, 750, 601, 574, 515}$ \\
$5$ & $1^{0}9^{1}5^{2}0^{3}2^{4}$ & $\Tuple{1587, 684, 547, 489, 485}$ \\
$5$ & $\Set{3,15}^{0}2^{1}13^{2}6^{3}5^{4}$ & $\Tuple{675, 668, 587, 527, 523}$ \\
\hline
\end{tabular}
\end{table}

The chains in Table~\ref{t:rex:chains:top10} show one major problem of
chaining by (unweighted) bipartite matching: the quality values can
differ quite much inside one chain. This problem can be slightly
diminished by using weighted bipartite matching where the weight of
the edge depends on how much the quality values of the corresponding
itemsets differ from each other. This ensures only that the sum of the
differences of the quality values of consecutive itemsets in the
chains is minimized Thus, in long chains the minimum and the maximum
quality values can still differ considerably. A more heuristic
approach would be to further partition the obtained chains in such a
way that the quality values of any two itemsets in the same chain do
not differ too much from each other. Such partitions can be computed
efficiently for several loss functions using the techniques described
in Chapter~\ref{c:views}.  The minimality of the chain partition,
however, is sacrificed when the chains in the partition are further
partitioned.

A simple minimum antichain partition of the collection of
$0.20$-frequent itemsets in the course completion database is the
partition of the itemsets by their cardinalities (see
Table~\ref{t:rex:frequent}). Especially, the $0.20$-frequent items
(Table~\ref{t:freqitems}) form an antichain in the collection of
$0.20$-frequent itemsets in the database. The frequent items can be
considered as a simple summary of the collection of all frequent
itemsets and the underlying transaction database, too.

Also the antichains can contain itemsets with very different quality
values. Again, this problem can be diminished by further
partitioning each antichain using the quality values of the patterns.
\exend \end{realexample}

We evaluated the condensation abilities of pattern chaining
experimentally by chaining closed $\sigma$-frequent itemsets of the
IPUMS Census and Internet Usage databases for several different
minimum frequency thresholds $\sigma \in \IntC{0,1}$.  We chained the
itemsets optimally by finding a maximum bipartite matching in the
corresponding bipartite graph
(Algorithm~\ref{a:Partition-into-Chains}) and in a greedy manner
(Algorithm~\ref{a:Minimal-Partition-into-Chains}) when the itemsets
were ordered by their cardinalities.

As noticed in Example~\ref{ex:chainlower}, the number of chains is
bounded above by the cardinality of the pattern collection and below
by the number of maximal patterns in the collections. In the case of
closed $\sigma$-frequent itemsets this means that the number of chains
is never greater than the number of closed $\sigma$-frequent itemsets
and never smaller than the number of maximal $\sigma$-frequent
itemsets. Furthermore, the lower bound given by the maximal itemsets
might not be very tight:

\begin{example}[slackness of lower bounds determined by maximal itemsets]
If the collection of closed $\sigma$-frequent itemsets in $\DB$ is
\begin{displaymath}
\Closed{\sigma,\DB}=2^\Items=\Set{X \subseteq \Items}
\end{displaymath}
then the collection of maximal $\sigma$-frequent itemsets in $\DB$ is
\begin{displaymath}
\Maximal{\sigma,\DB}=\Set{\Items}
\end{displaymath}
but largest antichain $\AntiChain$ in $\Closed{\sigma,\DB}$ consists
of all itemsets of cardinality $\Floor{ \Abs{\Items}/2}$. Thus, the
cardinality of $\AntiChain$ is
\begin{displaymath}
\Abs{\AntiChain}=\Choose{\Abs{\Items}}{\Floor{\Abs{\Items}/2}}.
\end{displaymath}
\exend \end{example}

\begin{figure}[p] \centering
\includegraphics[width=\textwidth]{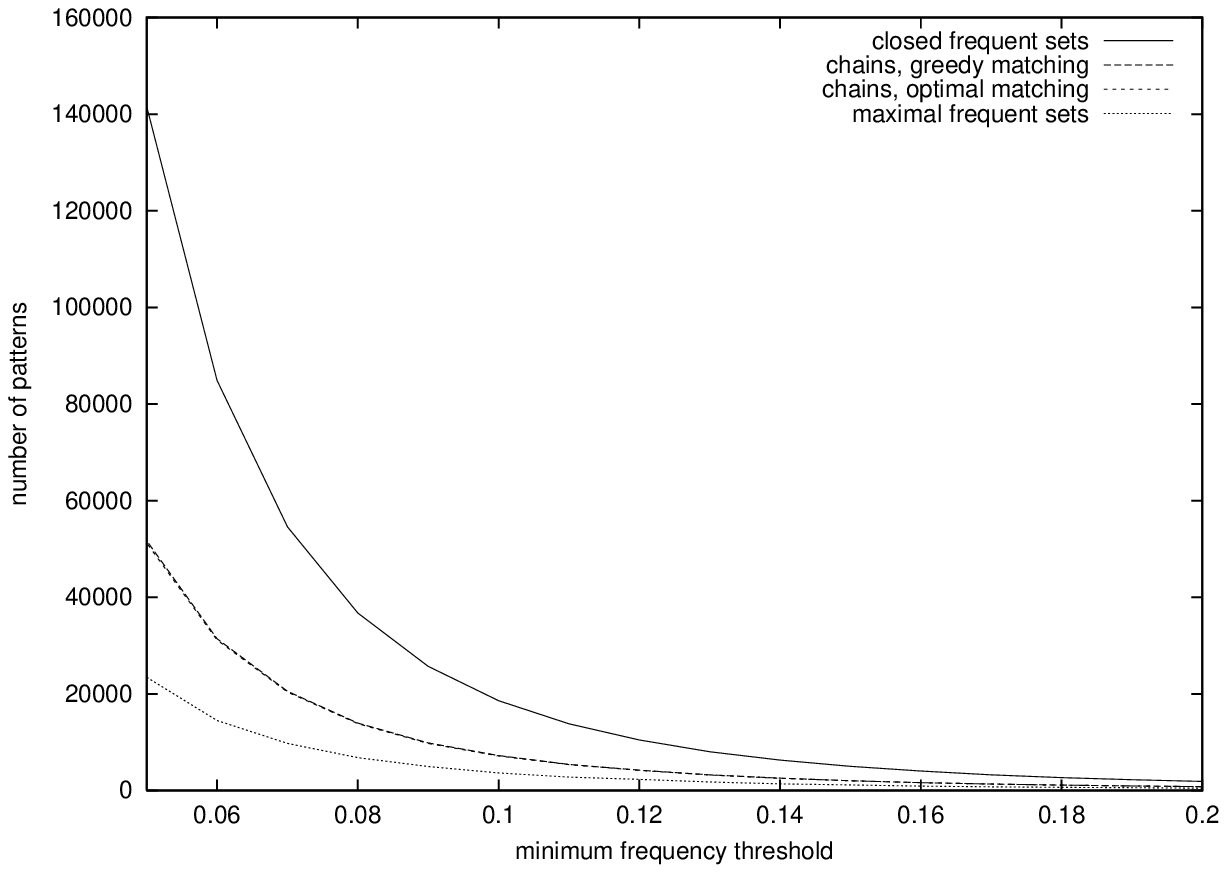}
\includegraphics[width=\textwidth]{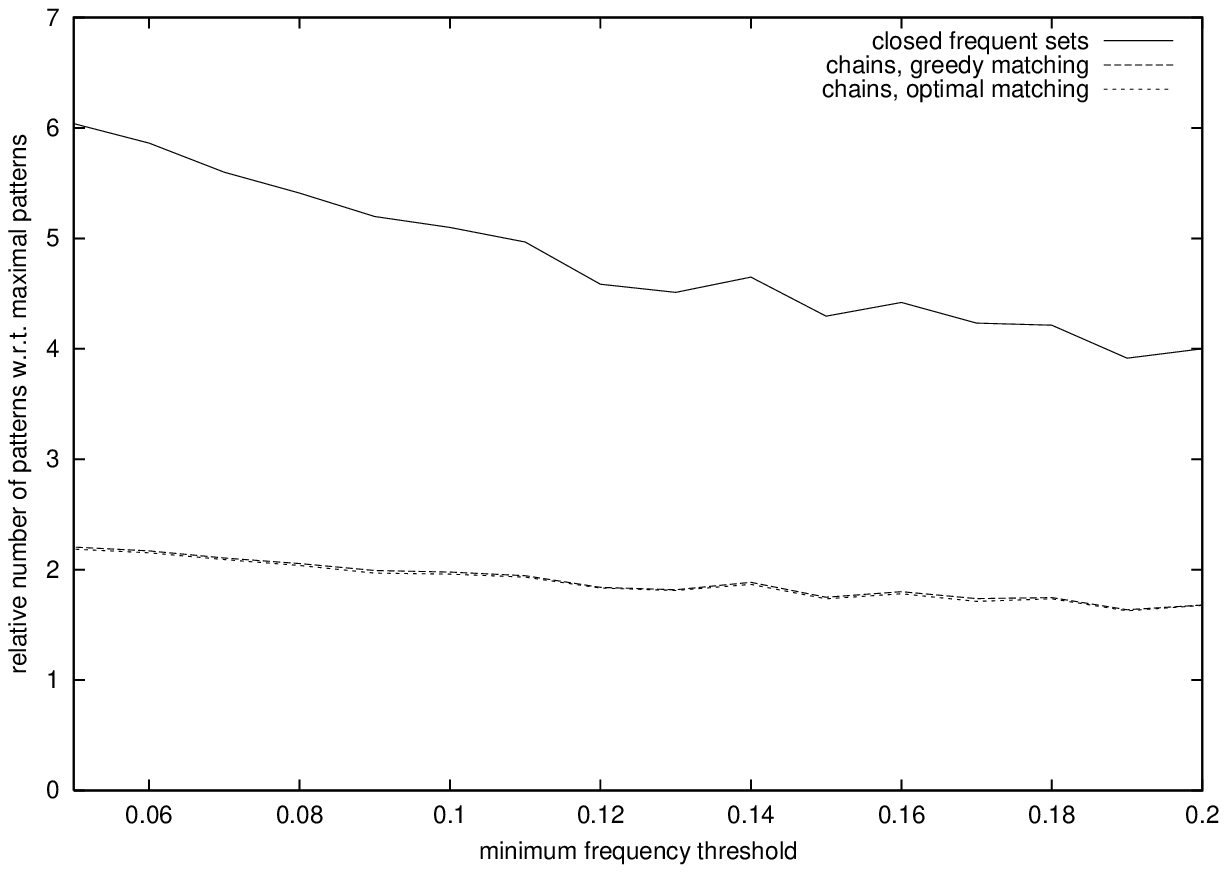}
\caption{Pattern chains in Internet Usage data. \label{f:chains:internet}}
\end{figure}

\begin{figure}[p] \centering
\includegraphics[width=\textwidth]{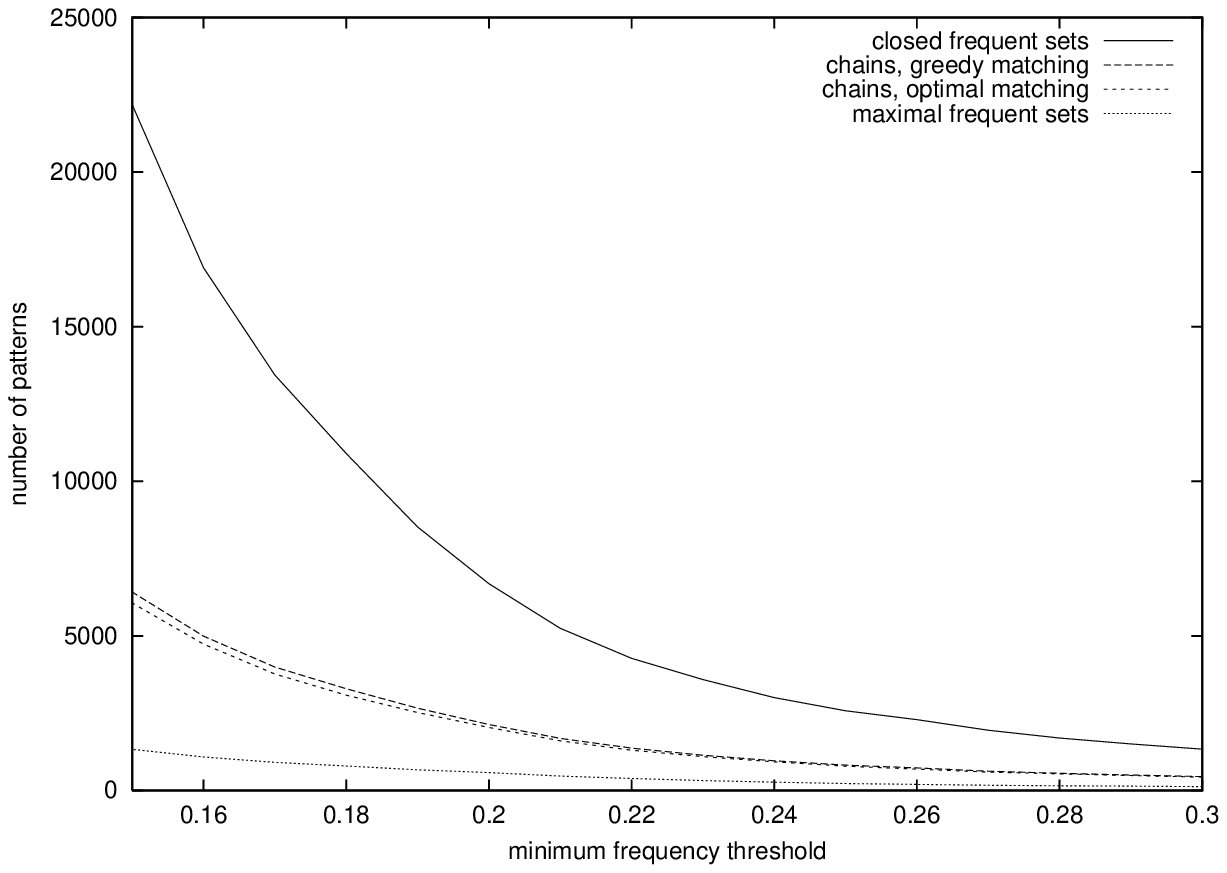}
\includegraphics[width=\textwidth]{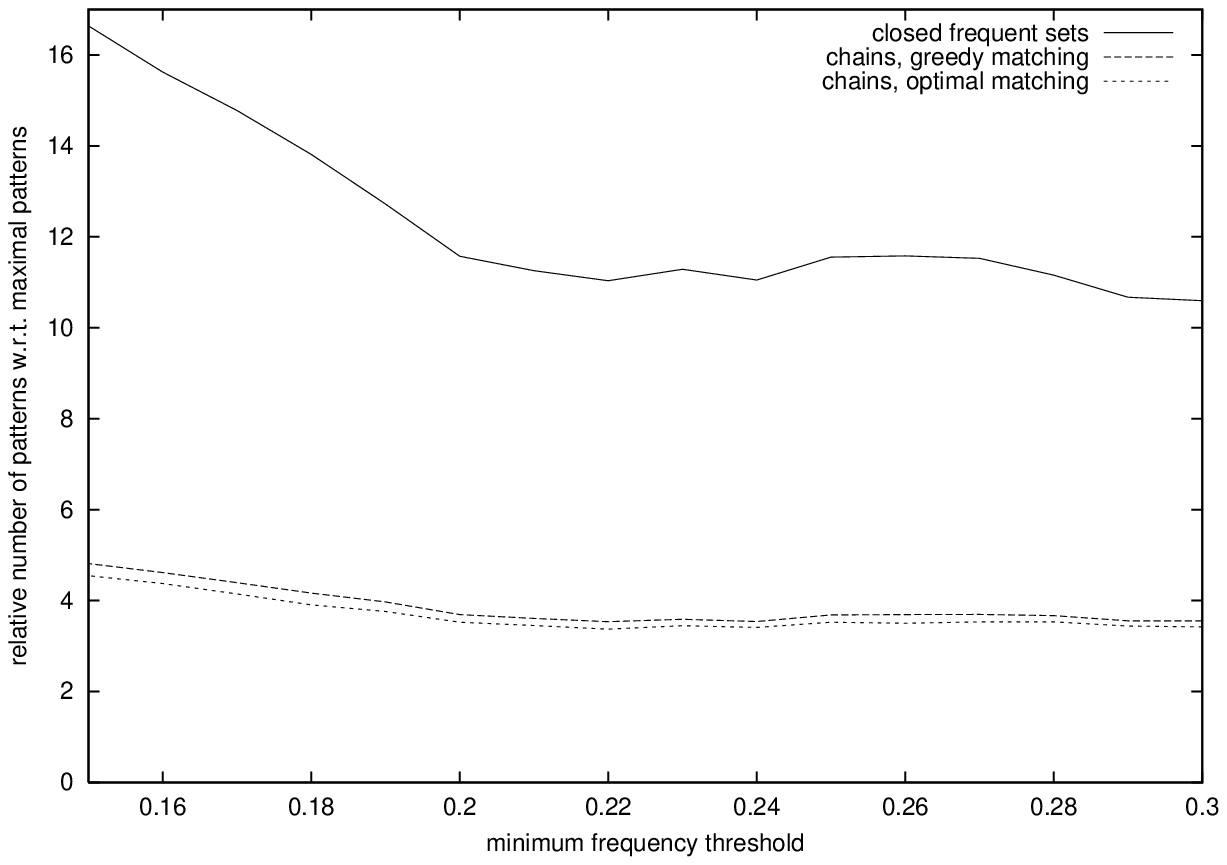}
\caption{Pattern chains IPUMS Census data. \label{f:chains:ipums}}
\end{figure}

The chaining of closed $\sigma$-frequent itemsets was computed for
many different minimum frequency thresholds $\sigma \in
\IntC{0,1}$. The results are shown in Figure~\ref{f:chains:internet}
and in Figure~\ref{f:chains:ipums}. The upper figures show the minimum
frequency thresholds against the number of patterns. Each curve
corresponds to some class of patterns expressed by the label of the
curve. The lower figures show the minimum frequency thresholds against
the relative number of closed frequent itemsets and chains with
respect to the number of maximal frequent itemsets.

The number of chains in experiments were smaller than the number of
closed frequent itemsets. Thus, the idea of finding a minimum chain
partition seems to be useful for condensation. It is also worth to
remember that the fundamental assumption in frequent itemset mining is
that not very large itemsets are frequent since also all subitemsets
of the frequent itemsets are frequent. This implies that the chains
with respect to the partial order relation subset inclusion cannot be
very long as the length of the longest chain in the frequent itemset
collection is the cardinality of the largest frequent itemset. This
observation makes the results even more satisfactory.

All the more interesting results were obtained when comparing the
minimal and the minimum chain partitions: the greedy heuristic
produced almost as small chain partitions as the computationally much
more demanding approach based on maximum bipartite matchings. (Similar
results were obtained also with all other transaction databases we
experimented.)  It is not clear, however, whether the quality of the
maximal matchings is specific to closed frequent itemsets or if the
results generalize to some other pattern collections as well.

%% file: profiles.tex

\chapter{Relating Patterns by Their Change Profiles \label{c:profiles}}


To make pattern collections more understandable, it would often be
useful to relate the patterns to each other. In Chapter~\ref{c:chains}
the relationships between patterns were determined by a partial order
over the pattern collection. The patterns can be related to each other
also by their quality values.  For example, absolute or relative
differences between the quality values of the patterns could be used
to measure their (dis)similarity. It is not immediate, however,
whether comparing the quality values of two patterns actually tells
much about their similarity.

An alternative approach is to measure the similarity between two
patterns based on how they relate to other patterns. That is, the
patterns are considered similar if they are related to other patterns
similarly. This approach depends strongly on what it means to be
related to other patterns.  A simple solution is to consider how the
quality value of the pattern has to be modified in order to obtain the
quality values of its super- and subpatterns.
\begin{example}[modifying quality values]
Two simplest examples of modifications of $\imeas{p}$ to $\imeas{p'}$
are multiplying the quality value $\imeas{p}$ by the value
$\imeas{p'}/\imeas{p}$, and adding to the quality value $\imeas{p}$
the value $\imeas{p'}-\imeas{p}$.

In this chapter we restrict the modifications to the first case, i.e.,
modifying the quality value $\imeas{p'}$ of a pattern $p' \in \Pat$
from the quality value $\imeas{p}$ of a pattern $p \in \Pat$ by
multiplying $\imeas{p}$ by $\imeas{p'}/\imeas{p}$.  \exend
\end{example}

These modifications for one pattern can be combined as a mapping from
the patterns to modifications.  This mapping for a pattern $p$ is
called a \emph{change profile} $\ch{}{p}{}$ of the pattern $p$ and
each value $\ch{p'}{p}{}$ is called the \emph{change} of $p$ with
respect to $p' \in \Pat$. To simplify the considerations, the change
profile $\ch{}{p}{}$ is divided into two parts (adapting the
terminology of~\cite{a:mitchell82,a:mannila97:levelwise}): the
\emph{specializing} change profile $\ch{}{p}{s}$ describes the changes
to the superpatterns and the \emph{generalizing} change profile
$\ch{}{p}{s}$ describes the changes to the subpatterns.  When the type
of the change profile is not of importance, a change profile of $X$ is
denoted by $\ch{}{X}{}$.

\begin{example}[specializing change profiles for itemsets] \label{ex:ch}
Let us consider the transaction database $\DB$ shown as
Table~\ref{t:ex:ch:db}.
\begin{table}[h!] \centering
\caption{The transaction identifiers and the itemsets of the
transactions in $\DB$. \label{t:ex:ch:db}}
\begin{tabular}{|c|c|}
\hline
$\mathit{tid}$ & $X$ \\
\hline
\hline
$1$ & $\Set{A}$ \\
$2$ & $\Set{A,C}$ \\
$3$ & $\Set{A,B,C}$ \\
$4$ & $\Set{B,C}$ \\
\hline
\end{tabular}
\end{table}

The collection of $1/4$-frequent itemsets in $\DB$ and their
frequencies are shown in Table~\ref{t:ex:ch:frequent}.
\begin{table}[h!] \centering
\caption{The frequent itemsets and their frequencies in
$\DB$. \label{t:ex:ch:frequent}}
\begin{tabular}{|c|c|}
\hline
$X$ & $\fr{X,\DB}$\\
\hline
\hline
$\emptyset$ & $1$ \\
$A$ & $3/4$ \\
$B$ & $1/2$ \\
$C$ & $3/4$ \\
$AB$ & $1/4$ \\
$AC$ & $1/2$ \\
$ABC$ & $1/4$ \\
\hline
\end{tabular}
\end{table}

For the itemsets and the frequencies, the changes in the specializing
change profiles are of form
\begin{displaymath}
\ch{Y}{X}{s}=\frac{\fr{X \cup Y,\DB}}{\fr{X,\DB}}.
\end{displaymath}

Thus, the specializing change profiles of the singleton itemsets $A$,
$B$ and $C$ of the $1/4$-frequent itemsets in $\DB$ are determined by
the changes
\begin{eqnarray*}
\ch{}{A}{s} &=& \Set{B \mapsto \frac{1}{3}, C \mapsto \frac{2}{3},BC
\mapsto \frac{1}{3}}= \Set{B,BC \mapsto \frac{1}{3}, C \mapsto
\frac{2}{3}}, \\
\ch{}{B}{s}&=& \Set{A \mapsto \frac{1}{2}, C \mapsto 1, AC \mapsto
\frac{1}{2}}= \Set{A,AC \mapsto \frac{1}{2}, C \mapsto 1}
\mbox{ and} \\
\ch{}{C}{s}&=&\Set{A \mapsto \frac{2}{3}, B \mapsto \frac{2}{3}, AB
\mapsto \frac{1}{3}}=\Set{A,B \mapsto \frac{2}{3}, AB \mapsto
\frac{1}{3}}.
\end{eqnarray*}
\exend \end{example}

The change profiles attempt to reach from a local description of data,
i.e., a pattern collection, to more global view, i.e., to
relationships between the patterns in the collection. The change
profiles can be used to define similarity measures between the
patterns, to score the patterns and also in the condensed
representations of pattern collections.

In this chapter, we introduce the concept of change profiles, a new
representation of pattern collections that pursues to bridge the gap
between local and global descriptions of data. We describe several
variants of change profiles and study their properties.  We consider
different approaches to cluster change profiles and show that they are
\NP-hard and inapproximable for a wide variety of dissimilarity
functions for change profiles, but that in practice change profiles
can be used to provide reasonable clusterings. Furthermore, we suggest
representing a pattern collection using approximate change profiles
and propose algorithms to estimate the quality values from the
approximate change profiles.

This chapter is based on the article ``Change
Profiles''~\cite{i:mielikainen03:profiles}.  In the remaining of the
chapter we shall focus on frequent itemsets; change profiles can
readily be generalized to arbitrary pattern collections with a partial
order.

\section{From Association Rules to Change Profiles}
The frequency $\fr{X,\DB}$ of a frequent itemset $X$ in a transaction
database $\DB$ can be interpreted as the probability $\Prob{X}$ of the
event ``a transaction drawn randomly from the transaction database
$\DB$ contains itemset $X$'' and the accuracy $\acc{\assoc{X}{Y},\DB}$
of an association rule $\assoc{X}{Y}$ as the conditional probability
$\Prob{Y|X}$. Thus, each association rule $\assoc{X}{Y}$ describes one
relationship of the itemset $X$ to other itemsets. (Empirical
conditional probabilities of also different kinds of events have been
studied in data mining under the name of
\emph{cubegrades}~\cite{a:imielinski02}.)

A more global view of the relationships between the frequent itemset
$X$ and other frequent itemsets can be obtained by combining the
association rules $\assoc{X}{Y}$ with common body into a mapping from
the frequent itemsets to the interval $\IntC{0,1}$. This mapping
is called a \emph{specializing change profile}:

\begin{definition}[specializing change profiles] \label{d:chs}
A specializing change profile of a $\sigma$-frequent itemset $X$ in
$\DB$ is a mapping
\begin{displaymath}
\ch{}{X}{s} : \Set{Y \subseteq \Items : X \cup Y \in \Frequent{\sigma,\DB}} \to \IntC{0,1}
\end{displaymath}
consisting the accuracies of the $\sigma$-frequent rules
$\assoc{X}{Y}$ in $\DB$, i.e.,
\begin{displaymath}
\ch{Y}{X}{s}=\frac{\fr{X \cup Y,\DB}}{\fr{X,\DB}}
\end{displaymath}
where $X \cup Y \in \Frequent{\sigma,\DB}$.
\end{definition}

A specializing change profile $\ch{}{X}{s}$ can be interpreted as the
conditional probability $\Prob{\mathbf{Y}|X}$ where $\mathbf{Y}$ is a
random variable. 

\begin{example}[specializing change profiles] \label{ex:chs}
Let us consider the collection $\Frequent{\sigma,\DB}$ of the
$\sigma$-frequent itemsets in $\DB$ with $\sigma=1/4$ where $\DB$ is
as shown in Table~\ref{t:ex:ch:db}, and the $\sigma$-frequent itemsets
and their frequencies as shown in Table~\ref{t:ex:ch:frequent}.  Then
the specializing change profiles of $\Frequent{\sigma,\DB}$ are
\begin{eqnarray*}
\ch{}{\emptyset}{s}&=&\Set{\emptyset \mapsto 1, A,C \mapsto \frac{3}{4}, B,AC,BC \mapsto \frac{1}{2}, AB,ABC \mapsto \frac{1}{4}}, \\
\ch{}{A}{s}&=&\Set{\emptyset,A \mapsto 1, C,AC \mapsto \frac{2}{3}, B,AB,BC,ABC \mapsto \frac{1}{3}}, \\
\ch{}{B}{s}&=&\Set{\emptyset,B,C,BC \mapsto 1, A,AB,AC,ABC \mapsto \frac{1}{2}}, \\
\ch{}{C}{s}&=&\Set{\emptyset,C \mapsto 1, A,B,AC,BC \mapsto \frac{2}{3}, AB,ABC \mapsto \frac{1}{3}}, \\
\ch{}{AB}{s}&=&\Set{\emptyset,A,B,C,AB,AC,BC,ABC \mapsto 1}, \\
\ch{}{AC}{s}&=&\Set{\emptyset,A,C,AC \mapsto 1, B,AB,BC,ABC \mapsto \frac{1}{2}}, \\
\ch{}{BC}{s}&=&\Set{\emptyset,B,C,BC \mapsto 1, A,AB,AC,ABC \mapsto \frac{1}{2}} \mbox{ and} \\
\ch{}{ABC}{s}&=&\Set{\emptyset,A,B,C,AB,AC,BC,ABC \mapsto 1}.
\end{eqnarray*}
\exend \end{example}

Similarly to the specializing change profiles, we can define a change
profile to describe how the frequency of a $\sigma$-frequent itemset
$X$ changes when some items are removed from it. A change profile of
this kind is called a \emph{generalizing change profile}:

\begin{definition}[generalizing change profiles] \label{d:chg}
A generalizing change profile of a $\sigma$-frequent itemset $X$ in
$\DB$ is a mapping
\begin{displaymath}
\ch{}{X}{g}:\Frequent{\sigma,\DB} \to \IntC{1,\frac{1}{\sigma}}
\end{displaymath}
consisting of the inverse accuracies of the frequent rules $\assoc{X\setminus Y}{X}$, i.e.,
\begin{displaymath}
\ch{Y}{X}{g}=\frac{\fr{X\setminus Y,\DB}}{\fr{X,\DB}}.
\end{displaymath}
where $Y \subseteq \Items$.
\end{definition}

The generalizing change profile $\ch{}{X}{g}$ corresponds to the
mapping $1/\Prob{X|X \setminus \mathbf{Y}}$ where $\mathbf{Y}$ is a
random variable.

\begin{example}[generalizing change profiles] \label{ex:chg}
Let us consider the collection $\Frequent{\sigma,\DB}$ of the
$\sigma$-frequent itemsets in $\DB$ with $\sigma=1/4$ where $\DB$ is
as shown in Table~\ref{t:ex:ch:db}, and the $\sigma$-frequent itemsets
and their frequencies as shown in Table~\ref{t:ex:ch:frequent}.  Then
the generalizing change profiles in $\Frequent{\sigma,\DB}$ are
\begin{eqnarray*}
\ch{}{\emptyset}{g}&=&\Set{\emptyset,A,B,C,AB,AC,BC,ABC \mapsto 1}, \\
\ch{}{A}{g}&=&\Set{\emptyset,B,C,BC \mapsto 1, A, AB, AC, ABC \mapsto \frac{4}{3}}, \\
\ch{}{B}{g}&=&\Set{\emptyset,A,C,AC \mapsto 1, B, AB, BC, ABC \mapsto 2}, \\
\ch{}{C}{g}&=&\Set{\emptyset,A,B,AB \mapsto 1, C, AC, BC, ABC \mapsto \frac{4}{3}}, \\
\ch{}{AB}{g}&=&\Set{\emptyset,C \mapsto 1, A,AC \mapsto 2, B,BC \mapsto 3, AB,ABC \mapsto 4}, \\
\ch{}{AC}{g}&=&\Set{\emptyset,B \mapsto 1, A,C,AB,BC \mapsto \frac{3}{2}, AC,ABC \mapsto 2}, \\
\ch{}{BC}{g}&=&\Set{\emptyset, A,C,AC \mapsto 1, B,AB \mapsto \frac{3}{2}, BC,ABC \mapsto 2} \mbox{ and} \\
\ch{}{ABC}{g}&=&\Set{\emptyset,C \mapsto 1, A,B,AC \mapsto 2, BC \mapsto 3, AB,ABC \mapsto 4}.
\end{eqnarray*}
\exend \end{example}

Each specializing and generalizing change profile $\ch{}{X}{s}$ and
$\ch{}{X}{g}$ describe upper and lower neighborhoods
\begin{displaymath}
\Nbr{X}{s}=\Set{ X \cup Y \in \Frequent{\sigma,\DB} : Y \subseteq \Items}
\end{displaymath}
and 
\begin{displaymath}
\Nbr{X}{g}=\Set{ X \setminus Y \in \Frequent{\sigma,\DB} : Y \subseteq \Items}
\end{displaymath}
of the frequent itemset $X$ in the collection $\Frequent{\sigma,\DB}$,
respectively.  The neighborhood
\begin{displaymath}
\Nbr{X}{}=\Nbr{X}{s} \cup \Nbr{X}{g}
\end{displaymath}
of $X$ consists of the frequent itemsets $Y \in \Nbr{X}{s}$ that
contain the frequent itemset $X$ and the frequent itemsets $ X
\setminus Y \in \Nbr{X}{g}$ that are contained in $X$, i.e., the
frequent itemsets that are comparable with $X$.

As seen in Example~\ref{ex:chs} and Example~\ref{ex:chg}, the change profiles
(Definition~\ref{d:chs} and Definition~\ref{d:chg}) are often highly
redundant. This is due to the following properties of itemsets:
\begin{observation} \label{obs:itemsets}
Let $X,Y \subseteq \Items$. Then
\begin{displaymath}
X \cup Y=X \cup \Paren{Y \setminus X}
\end{displaymath}
and
\begin{displaymath}
X \setminus Y=X \setminus \Paren{Y \cap X}.
\end{displaymath}
\end{observation}

The number of defined values of the change profile is reduced (without
losing any information) considerably by exploiting
Observation~\ref{obs:itemsets}.

\begin{example}[redundancy in change profiles] \label{ex:redundancy}
Let $X$ be a frequent itemset with only one frequent superitemset $X
\cup \Set{A}$ where $A \notin X$.  There are $2^{\Abs{X}+1}$
subitemsets of $X \cup \Set{A}$.  The first equation in
Observation~\ref{obs:itemsets} implies that frequency of $X \cup Y$ is
equal to the frequency of $X$ if $Y \subseteq X$. Thus, the
specializing changes $\ch{Y}{X}{s}=1$ for all $Y \subseteq X$ can be
neglected.  Furthermore, the specializing changes $\ch{Y \cup
\Set{A}}{X}{s}$ are equal for all $Y \subseteq X$ and it is sufficient
to store just the specializing change $\ch{\Set{A}}{X}{s}$.  This
reduces the size of the specializing change profile of $X$ by factor
$2^{\Abs{X}+1}$.

Let $X$ be an arbitrary frequent itemset and let
$\Frequent{\sigma,\DB}$. Based on the second equation of
Observation~\ref{obs:itemsets}, there is no need to store changes
$\ch{Y}{X}{g}=1$ for $Y \subseteq \Items$ such that $Y \not \subseteq
X$.  This reduces the number of of changes in the generalizing change
profile of $X$ by factor $2^{\Abs{\Items} -\Abs{X}}=2^{\Abs{\Items
\setminus X}}$.  \exend
\end{example}

The change profiles with redundancy reduced as in
Example~\ref{ex:redundancy} are called \emph{concise change profiles}:

\begin{definition}[concise specializing change profiles]
A concise specializing change profile $\cch{}{X}{s}$ is a restriction
of a specializing change profile $\ch{}{X}{s}$ to itemsets $Y$ such
that $X \cap Y=\emptyset$ and $X \cup Y \in \Frequent{\sigma,\DB}$.
\end{definition}

\begin{definition}[concise generalizing change profiles]
A concise generalizing change profile $\cch{}{X}{g}$ is a restriction
of a generalizing change profile $\ch{}{X}{g}$ to itemsets $Y$ such
that $Y \subseteq X$.
\end{definition}

\begin{example}[concise change profiles] \label{ex:cch}
Let us consider the collection $\Frequent{\sigma,\DB}$ of the
$\sigma$-frequent itemsets in $\DB$ with $\sigma=1/4$ where $\DB$ is
as shown in Table~\ref{t:ex:ch:db}, and the $\sigma$-frequent itemsets
and their frequencies as shown in Table~\ref{t:ex:ch:frequent}.  The
concise specializing change profiles of $\Frequent{\sigma,\DB}$ are
\begin{eqnarray*}
\cch{}{\emptyset}{s}&=&\Set{\emptyset \mapsto 1, A,C \mapsto \frac{3}{4}, B,AC,BC \mapsto \frac{1}{2}, AB,ABC \mapsto \frac{1}{4}}, \\
\cch{}{A}{s}&=&\Set{\emptyset \mapsto 1, C \mapsto \frac{2}{3}, B,BC \mapsto \frac{1}{3}}, \\
\cch{}{B}{s}&=&\Set{\emptyset,C \mapsto 1, A,AC \mapsto \frac{1}{2}}, \\
\cch{}{C}{s}&=&\Set{\emptyset \mapsto 1, A,B \mapsto \frac{2}{3}, AB \mapsto \frac{1}{3}}, \\
\cch{}{AB}{s}&=&\Set{\emptyset,C \mapsto 1}, \\
\cch{}{AC}{s}&=&\Set{\emptyset \mapsto 1, B \mapsto \frac{1}{2}}, \\
\cch{}{BC}{s}&=&\Set{\emptyset \mapsto 1, A \mapsto \frac{1}{2}} \mbox{ and} \\
\cch{}{ABC}{s}&=&\Set{\emptyset \mapsto 1}
\end{eqnarray*}
and the concise generalizing change profiles of
$\Frequent{\sigma,\DB}$ are
\begin{eqnarray*}
\cch{}{\emptyset}{g}&=&\Set{\emptyset \mapsto 1}, \\
\cch{}{A}{g}&=&\Set{\emptyset \mapsto 1, A \mapsto \frac{4}{3}}, \\
\cch{}{B}{g}&=&\Set{\emptyset \mapsto 1, B \mapsto 2}, \\
\cch{}{C}{g}&=&\Set{\emptyset \mapsto 1, C \mapsto \frac{4}{3}}, \\
\cch{}{AB}{g}&=&\Set{\emptyset \mapsto 1, A \mapsto 2, B \mapsto 3, AB \mapsto 4}, \\
\cch{}{AC}{g}&=&\Set{\emptyset \mapsto 1, A,C \mapsto \frac{3}{2}, AC \mapsto 2}, \\
\cch{}{BC}{g}&=&\Set{\emptyset,C \mapsto 1, B \mapsto \frac{3}{2}, BC \mapsto 2} \mbox{ and} \\
\cch{}{ABC}{g}&=&\Set{\emptyset,C \mapsto 1, A,B,AC \mapsto 2, BC \mapsto 3, AB,ABC \mapsto 4}.
\end{eqnarray*}
\exend \end{example}

The concise change profiles can be interpreted as affine axis-parallel
subspaces of $\RN^{\Abs{\Frequent{\sigma,\DB}}}$ (i.e., affine
hyperplanes in $\RN^{\Abs{\Frequent{\sigma,\DB}}}$) that are indexed
\begin{itemize}
\item
by itemsets $Y$ such that $X \cap Y=\emptyset$ and $X \cup Y \in
\Frequent{\sigma,\DB}$ in the specializing case, and 
\item
by itemsets $Y$
such that $Y \subseteq X$ in the generalizing case. 
\end{itemize}
The concise change profiles for a frequent itemset collection
$\Frequent{\sigma,\DB}$ can be computed efficiently by
Algorithm~\ref{a:Change-Profiles}.

\begin{algorithm}
\caption{Generation of concise change profiles. \label{a:Change-Profiles}}
\begin{algorithmic}[1]
\Input{The collection $\Frequent{\sigma,\DB}$ of $\sigma$-frequent
itemsets in a transaction database $\DB$ and their frequencies.}
\Output{The concise specializing change profiles and the concise
generalizing change profiles of $\Frequent{\sigma,\DB}$.}
\Function{Change-Profiles}{$\Frequent{\sigma,\DB},\freq$}
\ForAll{$X \in \Frequent{\sigma,\DB}$}
 \ForAll{$Y \subseteq X$}
  \State $\cch{Y}{X \setminus Y}{s} \leftarrow \fr{X,\DB}/\fr{X\setminus Y,\DB}$
  \State $\cch{Y}{X}{g} \leftarrow \fr{X\setminus Y,\DB}/\fr{X,\DB}$
 \EndFor
\EndFor
\State \textbf{return} $\Tuple{\cch{}{}{s},\cch{}{}{g}}$
\EndFunction
\end{algorithmic}
\end{algorithm}

As shown in Example~\ref{ex:redccp}, the neighborhoods of even the
concise change profiles can be too large.

\begin{example}[redundancy in concise change profiles] \label{ex:redccp}
Let $X$ be an itemset in the collection $\Frequent{\sigma,\DB}$. Then
$\Abs{\cch{}{\emptyset}{s}} \geq 2^{\Abs{X}}$ and
$\Abs{\cch{}{X}{g}}\geq 2^{\Abs{X}}$ in $\Frequent{\sigma,\DB}$.
\exend \end{example}

Thus, the following definitions of association rules, we define
\emph{simple specializing change profiles} and \emph{simple
generalizing change profiles}:

\begin{definition}[simple change profiles]
A simple specializing (generalizing) change profile $\sch{}{X}{s}$
($\sch{}{X}{g}$) is restriction of $\cch{}{X}{s}$ ($\cch{}{X}{g}$) to
singleton itemsets $Y$.
\end{definition}

\begin{example}[simple change profiles] \label{ex:sch}
Let us consider the collection $\Frequent{\sigma,\DB}$ of the
$\sigma$-frequent itemsets in $\DB$ with $\sigma=1/4$ where $\DB$ is
as shown in Table~\ref{t:ex:ch:db}, and the $\sigma$-frequent itemsets
and their frequencies as shown in Table~\ref{t:ex:ch:frequent}.  The
simple specializing change profiles of $\Frequent{\sigma,\DB}$ are
\begin{eqnarray*}
\sch{}{\emptyset}{s}&=&\Set{A,C \mapsto \frac{3}{4}, B \mapsto \frac{1}{2}}, \\
\sch{}{A}{s}&=&\Set{C \mapsto \frac{2}{3}, B \mapsto \frac{1}{3}}, \\
\sch{}{B}{s}&=&\Set{C \mapsto 1, A \mapsto \frac{1}{2}}, \\
\sch{}{C}{s}&=&\Set{A,B \mapsto \frac{2}{3}}, \\
\sch{}{AB}{s}&=&\Set{C \mapsto 1}, \\
\sch{}{AC}{s}&=&\Set{B \mapsto \frac{1}{2}}, \\
\sch{}{BC}{s}&=&\Set{A \mapsto \frac{1}{2}} \mbox{ and} \\
\sch{}{ABC}{s}&=&\Set{}
\end{eqnarray*}
and the simple generalizing change profiles of $\Frequent{\sigma,\DB}$
are
\begin{eqnarray*}
\sch{}{\emptyset}{g}&=&\Set{}, \\
\sch{}{A}{g}&=&\Set{A \mapsto \frac{4}{3}}, \\
\sch{}{B}{g}&=&\Set{B \mapsto 2}, \\
\sch{}{C}{g}&=&\Set{C \mapsto \frac{4}{3}}, \\
\sch{}{AB}{g}&=&\Set{A \mapsto 2, B \mapsto 3}, \\
\sch{}{AC}{g}&=&\Set{A,C \mapsto \frac{3}{2}}, \\
\sch{}{BC}{g}&=&\Set{C \mapsto 1, B \mapsto \frac{3}{2}} \mbox{ and} \\
\sch{}{ABC}{g}&=&\Set{C \mapsto 1, A,B \mapsto 2}.
\end{eqnarray*}
\exend \end{example}

The number of bits needed for representing a simple change profile is
at most $\Abs{\Items}\log\Abs{\DB}$: Each change profile can be
described as a length-$\Abs{\Items}$ vector of changes as the number
of singleton subsets of the set $\Items$ of items is
$\Abs{\Items}$. Each change can be described using at most $\log
\Abs{\DB}$ bits since there are at most as many different possible
changes from a given itemset to any other itemset as there are are
transactions in $\DB$.  This upper bound can sometimes be quite loose
as shown by Example~\ref{ex:loosesgcp}.

\begin{example}[loose upper bounds for simple generalizing change profiles] \label{ex:loosesgcp}
Let the set $\Items$ of items be large. Then the above upper bound is
often very loose: The number of itemsets in the collection
$\Frequent{\sigma,\DB}$ of $\sigma$-frequent itemsets in $\DB$ is
exponential in the cardinality of the largest itemset in
$\Frequent{\sigma,\DB}$. Thus, the largest itemset $X$ in
$\Frequent{\sigma,\DB}$ has to be moderately small in order to be able
to represent the collection $\Frequent{\sigma,\DB}$ in a reasonable
space.  Thus, in this case, the upper bound for binary description of
a simple generalizing change profile of an itemset $X \in
\Frequent{\sigma,\DB}$ should rather be $\Abs{X}\Paren{\log
\Abs{\Items}} \Paren{\log \Abs{\DB}}$.  \exend \end{example}

\section{Clustering the Change Profiles}

In order to be able to find groups of similar change profiles, it
would be useful to be able to somehow measure the (similarity or)
dissimilarity between change profiles $\ch{}{X}{}$ and $\ch{}{Y}{}$.

The dissimilarity between the change profiles $\ch{}{X}{}$ and
$\ch{}{Y}{}$ can be defined to be their distance in their common
domain $\Dom{\ch{}{X}{}} \cap \Dom{\ch{}{Y}{}}$ with respect to some
distance function $\distance$.  A complementary approach would be to
focus on the differences in the structure of the pattern collection,
e.g., to measure the difference between two change profiles by
computing the symmetric difference of their domains. This kind of
dissimilarity function concentrates solely on the structure of the
pattern collection and thus neglects the frequencies.  A sophisticated
dissimilarity should probably consist of both points of view.

We shall focus on the first one. The only requirements we have for a
distance function are given by Definition~\ref{d:distances}.

\begin{definition}[a distance function] \label{d:distances}
A function $\distance$ is a distance function if
\begin{displaymath}
\dist{\ch{}{X}{},\ch{}{Y}{}}=0 \iff \ch{Z}{X}{}=\ch{Z}{Y}{}
\end{displaymath}
holds for all $Z \in \Dom{\ch{}{X}{}} \cap \Dom{\ch{}{Y}{}}$.
\end{definition}

There are several ways to define what is a good clustering and each
approach has its own strengths and
weaknesses~\cite{a:estivill-castro02,i:kleinberg02}.  A simple way to
group the change profiles based on a dissimilarity function defined in
their (pairwise) common domains is to allow two change profiles
$\ch{}{X}{}$ and $\ch{}{Y}{}$ to be in the same group only if
$\dist{\ch{}{X}{},\ch{}{Y}{}}=0$. Thus, the problem can be formulated
as follows.

\begin{problem}[change profile packing] \label{p:profilepacking}
Given a collection $\Ch$ of change profiles and a dissimilarity
function $\distance$, find a partition of $\Ch$ into groups
$\Ch_1,\ldots,\Ch_k$ with the smallest possible $k$ such that
$\dist{\ch{}{X}{},\ch{}{Y}{}}=0$ holds for all $\ch{}{X}{},\ch{}{Y}{}
\in \Ch_i$ with $1 \leq i \leq k$.
\end{problem}

Unfortunately, the problem seems to be very difficult. Namely, it can
be shown to be at least as difficult as the minimum graph coloring
problem:
\begin{problem}[minimum graph coloring~\cite{b:ausiello99}] \label{p:mincoloring}
Given a graph $G=\Tuple{V,E}$, find a labeling $\lbl{} : V \to
\mathbb{N}$ of the vertices with smallest number $\Abs{\lbl{V}}$ of
different labels such that if $u,v \in V$ are adjacent then
$\lbl{u}\neq\lbl{v}$.
\end{problem}

\begin{theorem} \label{t:packing}
The change profile packing problem is at least as hard as the minimum
graph coloring problem.
\end{theorem}
\begin{proof}
Let $G=\Tuple{V,E}$ be an instance of the minimum graph coloring
problem where $V=\Set{v_1,\ldots,v_n}$ is the set of vertices and
$E=\Set{e_1,\ldots,e_m}$ is the set of edges.

We reduce the minimum graph coloring problem
(Problem~\ref{p:mincoloring}) to the change profile packing problem
(Problem~\ref{p:profilepacking}) by first constructing an instance
$\Tuple{\sigma,\DB}$ of the frequent itemset mining problem and then
showing that the collection $\Ch$ of specializing change profiles
computed from the collection $\Frequent{\sigma,\DB}$ of the
$\sigma$-frequent itemsets in $\DB$ and their frequencies can be
partitioned into $k+2$ subcollections $\Ch_1,\ldots,\Ch_{k+2}$ if and
only if the graph $G$ is $k$-colorable.  To simplify the description,
we shall consider, without loss of generality, simple change profiles
instead of change profiles in general.

The set $\Items$ of items consists of elements in $V \cup E$. For each
vertex $v_i \in V$ there are $3n$ transactions with transaction
identifiers $\Tuple{i,1},\ldots,\Tuple{i,3n}$. Thus, in total there
are $3n^2$ transactions in $\DB$.

Each transaction $\Tuple{\Tuple{i,j},X}$ contains the vertex $v_i \in
V$.  Transactions $\Tuple{\Tuple{i,3(j-1)+1},X}$ and
$\Tuple{\Tuple{i,3(j-1)+2},X}$ contain an edge $\Set{v_i,v_j} \in E$
if and only if $i<j$. The transaction $\Tuple{\Tuple{i,3j},X}$
contains an edge $\Set{v_i,v_j} \in E$ if $i>j$.

Let the minimum frequency threshold $\sigma$ be $1/\Paren{3n^2}$. Then
the collection $\Frequent{\sigma,\DB}$ consists of the empty itemset
$\emptyset$ the singleton itemsets
$\Set{v_1},\ldots,\Set{v_n},\Set{e_1},\ldots,\Set{e_m}$ and 2-itemsets
$\Set{v_i,e}$ where $v_i \in e \in E$.  Thus, the cardinality of
$\Frequent{\sigma,\DB}$ is polynomial in the number of vertices of
$G$. The simple change profiles of $\Frequent{\sigma,\DB}$ are the
following ones:
\begin{displaymath}
\begin{array}{r c l l}
\sch{x}{\emptyset}{s}&=&
\left\{
\begin{array}{l}
1/n \\
1/n^2
\end{array}
\right.
\quad &
\begin{array}{l}
\mbox{if } x \in V \\
\mbox{if } x \in E
\end{array}
\\
\sch{v}{e}{s} &=&
1 \quad &
\begin{array}{l}
\mbox{if } v \in e
\end{array}
\\
\sch{\Set{v_i,v_j}}{v_i}{s} &= &
\left\{
\begin{array}{l}
2/\Paren{3n} \\
1/\Paren{3n}
\end{array}
\right.
\quad &
\begin{array}{l} 
\mbox{if } i<j, \Set{v_i,v_j} \notin E \\
\mbox{if } i>j, \Set{v_i,v_j} \notin E
\end{array}
\end{array}
\end{displaymath}

Clearly,
\begin{itemize}
\item
$\dist{\sch{}{\emptyset}{s},\sch{}{v}{s}}> 0$ for
$\sch{}{\emptyset}{s}$ and all $\sch{}{v}{s}$ where $v \in V$,
\item
$\dist{\sch{}{\emptyset}{s},\sch{}{e}{s}}> 0$ for
$\sch{}{\emptyset}{s}$ and all $\sch{}{e}{s}$ where $e \in E$, and
\item
$\dist{\sch{}{v}{s},\sch{}{e}{s}}> 0$ for all $\sch{}{v}{s}$ and
$\sch{}{e}{s}$ where $v \in V$ and $e \in E$.
\end{itemize}

On one hand, no two of $\ch{}{\emptyset}{s}$, $\ch{}{v}{s}$ and
$\sch{}{e}{s}$ can be in the same group for any $v \in V, e \in E$.
On the other hand, all $\sch{}{e}{s}$ can be packed into one set
$\Ch_{k+1}$ and $\sch{}{\emptyset}{s}$ always needs its own set
$\Ch_{k+1}$.

Hence, it is sufficient to show that the simple specializing change
profiles $\sch{}{v}{s}$ can be partitioned into $k$ sets
$\Ch_1,\ldots,\Ch_k$ without any error if and only if the graph $G$ is
$k$-colorable.  No two simple specializing change profiles
$\sch{}{v_i}{s}$ and $\sch{}{v_j}{s}$ with $\Set{v_i,v_j} \in E$ can
be in the same group since
$\sch{\Set{v_i,v_j}}{v_i}{s}\neq\sch{\Set{v_i,v_j}}{v_j}{s}$. If
$\Set{v_i,v_j} \not \in E$ then $\Dom{\sch{}{v_i}{s}} \cap
\Dom{\sch{}{v_j}{s}} =\emptyset$, i.e., $\sch{}{v_i}{s}$ and
$\sch{}{v_j}{s}$ can be in the same group.

As the minimum graph coloring problem can be mapped to the change
profile packing for specializing change profiles in polynomial time,
the latter is at least as hard as the minimum graph coloring problem.
\end{proof}

The minimum graph coloring problem is hard to approximate within
$\Abs{V}^{1-\epsilon}$ for any $\epsilon>0$ unless
\NP=\ZPP~\cite{a:feige98:graphcoloring}. (Recall that the complexity
class \ZPP\ consists of the decision problems that have randomized
algorithms that always make the right decision and run in expected
polynomial time~\cite{b:papadimitriou95}.)  Assuming that the graph is
connected we get from the above mapping from graphs to change profiles
the following rough upper bound
\begin{displaymath}
\Abs{\Ch}=1+\Abs{V}+\Abs{E}\leq 1 + \Abs{V} + \Choose{\Abs{V}}{2} =
\Oh{\Abs{V}^2}.
\end{displaymath}
Therefore, the change profile packing problem is hard to approximate
within $\Om{\Abs{\Ch}^{\Paren{1/2}-\epsilon}}$ for any
$\epsilon>0$ unless \NP=\ZPP.

Although the inapproximability results seem to be devastating, there
are efficient heuristics, such as the first-fit and the best-fit
heuristics~\cite{a:coffman02}, that might be able to find sufficiently
good partitions efficiently. However, the usefulness of such
heuristics depends on the actual transaction databases inducing the
collections of frequent itemsets.

The requirement that two change profiles $\ch{}{X}{}$ and $\ch{}{Y}{}$
can be in the same group $\Ch_i$ only if
$\dist{\ch{}{X}{},\ch{}{Y}{}}=0$ might be too strict. This restriction
can be relaxed also by discretizing the frequencies of the frequent
itemsets or the changes in the change profiles. (Recall that in
Section~\ref{s:discretization} we have seen that discretizations
minimizing several different loss functions can be found efficiently.)

Instead of minimizing the number of clusters, one could minimize the
error for a fixed number of clusters. This kind of clustering is
called a \emph{$k$-clustering}. The problem of finding good
$k$-clusterings is well-studied and good approximation algorithms are
known if the dissimilarity function is a
metric~\cite{i:dasgupta02,i:delavega03,i:feder88}. The problem of
finding the $k$-clustering of change profiles that minimizes the sum
of intracluster distances can be defined as follows:

\begin{problem}[minimum sum of distances $k$-clustering of change profiles] \label{p:k-clustering}
Given a collection $\Ch$ of change profiles, a distance function
$\distance : \Ch \times \Ch \to \RN$ and a positive integer $k$, find
the partition of $\Ch$ into $k$ groups $\Ch_1,\ldots,\Ch_k$ such that
\begin{displaymath}
\sum_{i=1}^k \sum_{\ch{}{X}{},\ch{}{Y}{} \in \Ch_i} \dist{\ch{}{X}{},\ch{}{Y}{}}
\end{displaymath}
is minimized.
\end{problem}

Unfortunately, it turn out that a dissimilarity function that is
defined to consist of the dissimilarities between the change profiles
in their common domains cannot be a metric since it cannot satisfy
even the triangle inequality in general:

\begin{proposition}\label{p:nonmetric}
A function $\distance$ that measures the distance between the change
profiles $\ch{}{X}{}$ and $\ch{}{Y}{}$ in their common domain
$\Dom{\ch{}{X}{}} \cap \Dom{\ch{}{Y}{}}$ is not a metric.
\end{proposition}
\begin{proof}
Let $\ch{}{X}{}$, $\ch{}{Y}{}$ and $\ch{}{Z}{}$ be three change
profiles such that
\begin{displaymath}
\Dom{\ch{}{X}{}} \cap
\Dom{\ch{}{Y}{}}=\emptyset=\Dom{\ch{}{Y}{}} \cap \Dom{\ch{}{Z}{}}
\end{displaymath}
but $\dist{\ch{}{X}{},\ch{}{Z}{}}>0$ (and thus $\Dom{\ch{}{X}{}} \cap
\Dom{\ch{}{Z}{}}$). The distance between these change profiles do not
satisfy triangle inequality since
\begin{displaymath}
\dist{\ch{}{X}{},\ch{}{Z}{}}>0=\dist{\ch{}{X}{},\ch{}{Y}{}}+\dist{\ch{}{Y}{},\ch{}{Z}{}}.
\end{displaymath}
Thus, such a distance cannot be not a metric.
\end{proof}

It turns out that the minimum $k$-clustering of specializing change
profiles is even worse than the change profile packing problem in the
sense of approximability as combining Theorem~\ref{t:packing} and
Proposition~\ref{p:nonmetric} we get:

\begin{theorem}
The minimum sum of distances $k$-clustering
(Problem~\ref{p:k-clustering}) of specializing change profiles cannot
be approximated within any ratio.
\end{theorem}
\begin{proof}
If we could approximate $k$-clustering of specializing change
profiles, then we could, by Theorem~\ref{t:packing} and
Proposition~\ref{p:nonmetric}, solve the minimum graph coloring
problem exactly. Namely, if a graph is $k$-colorable, then the
corresponding change profiles have $k$-clustering with the sum of
intracluster distances being zero. Thus, an approximation algorithm
with any approximation guarantees would find a solution with error
zero if and only if the corresponding graph is $k$-colorable.
\end{proof}

A major goal in the clustering of the change profiles is to further
understand the relationships between the frequent itemsets (and
collection of interesting patterns in general). As the nature of
pattern discovery is exploratory, defining a maximum number of
clusters or a maximum dissimilarity threshold might be difficult and
unnecessary. Fixing these parameters in advance can be avoided by
searching for a hierarchical clustering, instead~\cite{b:hastie01}.

A hierarchical clustering of $\Ch$ is a recursive partition of the
elements to $1,2,\ldots,\Abs{\Ch}$ clusters. It is most fortunate for
the exploratory data analysis point of view that in the case of
hierarchical clustering, the clusterings of all cardinalities can be
visualized in the same time by a tree (often called a
\emph{dendrogram}).

There are two main types of hierarchical clustering: agglomerative and
divisive (see also Example~\ref{ex:hierdisc}). The first begins with
$\Abs{\Ch}$ singleton clusters and recursively merges them and the
latter recursively partitions the set $\Ch$. Both are optimal in
certain sense: each agglomerative (divisive) hierarchical clustering
of $\Ch$ into $k$ groups is optimal with respect to the clustering
into $k+1$ groups ($k-1$ groups) determined by the same agglomerative
(divisive) hierarchical clustering.

\begin{example}[a hierarchical clustering of change profiles]
Let us consider subsets of the simple change profiles of
Example~\ref{ex:sch}. As the distance function between the change
profiles, we use the sum of absolute distances in the common
domain. (For brevity, we write the simple change profiles as
$3$-tuples. The positions denote the changes with respect to $A$, $B$
and $C$, respectively, $*$ denoting undefined value.)

First, let us consider the simple specializing change profiles
\begin{eqnarray*}
\sch{}{A}{s} &=& \Tuple{*,\frac{1}{3},\frac{2}{3}}, \\
\sch{}{B}{s} &=& \Tuple{\frac{1}{2},*,1} \quad \mbox{and} \\
\sch{}{C}{s} &=& \Tuple{\frac{2}{3},\frac{2}{3},*}.
\end{eqnarray*}
The sums of the absolute differences in their common domains are
\begin{eqnarray*}
\dist{\sch{}{A}{s},\sch{}{B}{s}} &=& \Abs{\sch{C}{A}{s}-\sch{C}{B}{s}}=\Abs{\frac{2}{3}-1}=\frac{1}{3}, \\
\dist{\sch{}{A}{s},\sch{}{C}{s}} &=& \Abs{\sch{B}{A}{s}-\sch{B}{C}{s}}=\Abs{\frac{1}{3}-\frac{2}{3}}=\frac{1}{3} \quad \mbox{and} \\
\dist{\sch{}{B}{s},\sch{}{C}{s}} &=& \Abs{\sch{A}{B}{s}-\sch{A}{C}{s}}=\Abs{\frac{1}{2}-\frac{2}{3}}=\frac{1}{6}.
\end{eqnarray*}
Agglomerative and divisive hierarchical clusterings suggest both that
the clustering into two groups is $\Set{\sch{}{A}{s}}$ and
$\Set{\sch{}{B}{s},\sch{}{C}{s}}$ with the sums of distances $0$ and
$1/6$, respectively.

Second, let us consider the simple generalizing change profiles
\begin{eqnarray*}
\sch{}{AB}{g} &=& \Tuple{2,3,*},\\
\sch{}{AC}{g} &=& \Tuple{\frac{3}{2},*,\frac{3}{2}} \quad \mbox{and} \\
\sch{}{BC}{g} &=& \Tuple{*,\frac{3}{2},1}.
\end{eqnarray*}
The sums of the absolute differences in their common domains are
\begin{eqnarray*}
\dist{\sch{}{AB}{g},\sch{}{AC}{g}} &=& \Abs{\sch{A}{AB}{g}-\sch{A}{AC}{s}}=\Abs{2-\frac{3}{2}}=\frac{1}{2}, \\
\dist{\sch{}{AB}{g},\sch{}{BC}{g}} &=& \Abs{\sch{B}{AB}{s}-\sch{B}{BC}{s}}=\Abs{3-\frac{3}{2}}=\frac{3}{2} \quad \mbox{and} \\
\dist{\sch{}{AC}{g},\sch{}{BC}{g}} &=& \Abs{\sch{C}{AC}{s}-\sch{C}{BC}{s}}=\Abs{\frac{3}{2}-1}=\frac{1}{2}.
\end{eqnarray*}
This time there are two equally good clusterings to two groups: the
only requirement is that $\sch{}{AB}{g}$ and $\sch{}{BC}{}$ are in
different clusters. The sums of the distances for the singleton
cluster and the cluster of two change profiles are $0$ and $1/2$,
respectively.

The dendrogram visualizations of the hierarchical clusterings are
shown in Figure~\ref{f:hierclust}.
\begin{figure}[h!] \centering 
\includegraphics[width=0.4\textwidth]{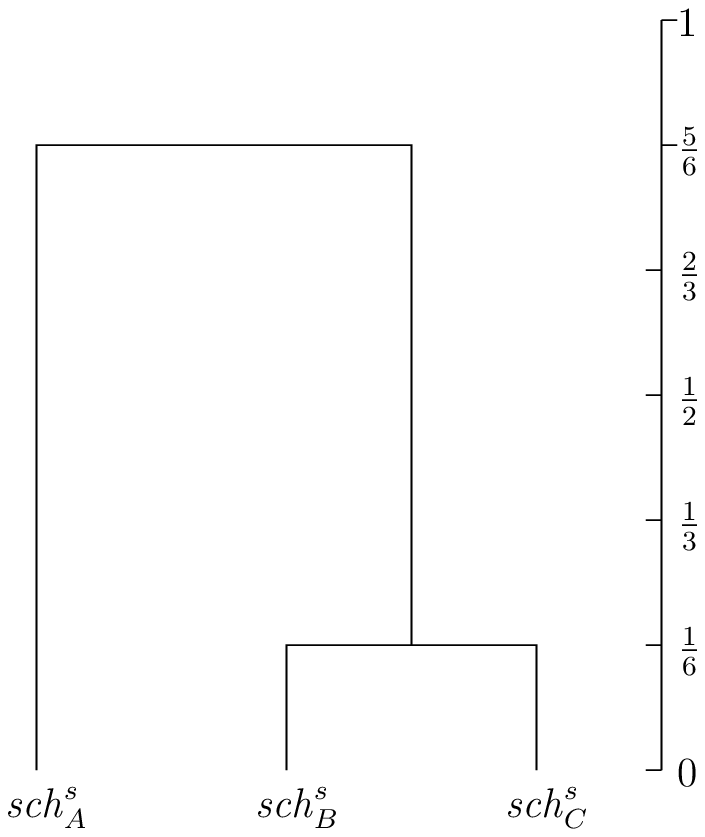} \hfill
\includegraphics[width=0.4\textwidth]{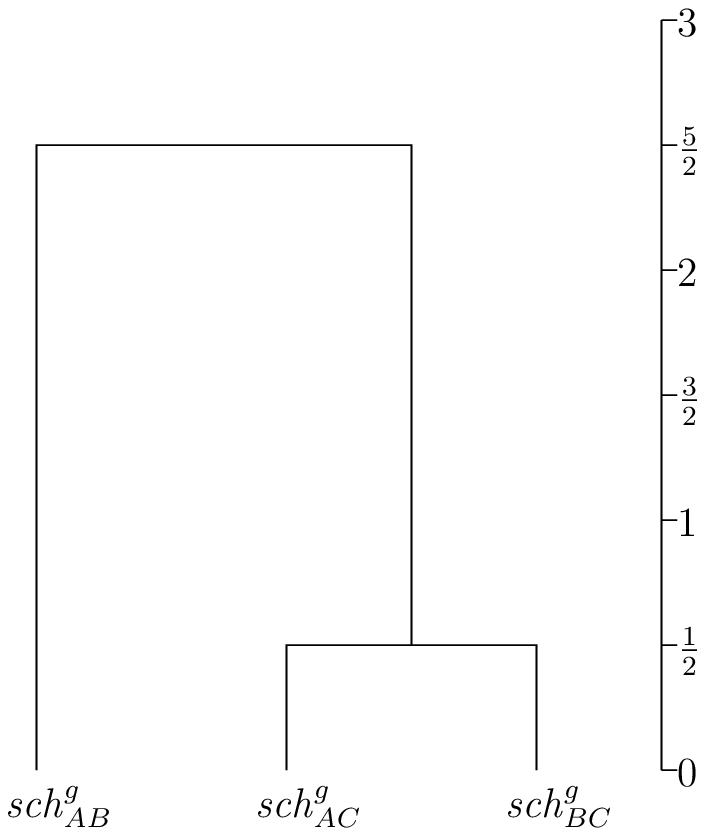}
\caption{The dendrogram of the hierarchical clusterings of the simple
specializing change profiles $\mathit{sch}^s_{A}$,
$\mathit{sch}^s_{B}$ and $\mathit{sch}^s_{C}$, and the simple
generalizing change profiles $\mathit{sch}^g_{AB}$,
$\mathit{sch}^g_{AC}$ and $\mathit{sch}^g_{BC}$, respectively. The
$y$-axis corresponds to the sum of absolute
errors.\label{f:hierclust}}
\end{figure}
\exend \end{example}

The divisive strategy seems to be more suitable for clustering the
change profiles since the dissimilarity functions we consider are
defined to be distances between the change profiles in their
(pairwise) common domains: The agglomerative clustering first puts
more or less arbitrarily the change profiles with disjoint domains
into the clusters. The choices made in the first few merges can cause
major differences in the clusterings into smaller number of clusters,
although the groups of change profiles with disjoint domains are
probably quite unimportant for determining the complete hierarchical
clustering. Contrary to the agglomerative clustering, the divisive
clustering concentrates first on the nonzero distances and thus the
change profiles with disjoint domains do not bias the whole
hierarchical clustering.

\begin{realexample}[hierachical clustering of the simple specializing change profiles of the $34$ most frequent courses in the course completion database] \label{rex:hierprofs}
To illustrate the hierarchical clustering of change profiles, let us
consider the simple specializing change profiles of the $34$ most
frequent items (i.e., the courses shown in Table~\ref{t:freqitems}) in
the collection consisting of all $1$- and $2$-subsets of the $34$ most
frequent items in the course completion database (see
Subsection~\ref{ss:ds}).

The agglomerative clustering of the simple change profiles using the
average distances between the courses as the merging criterion (i.e.,
the average linkage hierarchical clustering) is shown in
Figure~\ref{f:rex:hierprofs}.

\begin{figure}[p]
\includegraphics[height=\textwidth,angle=270]{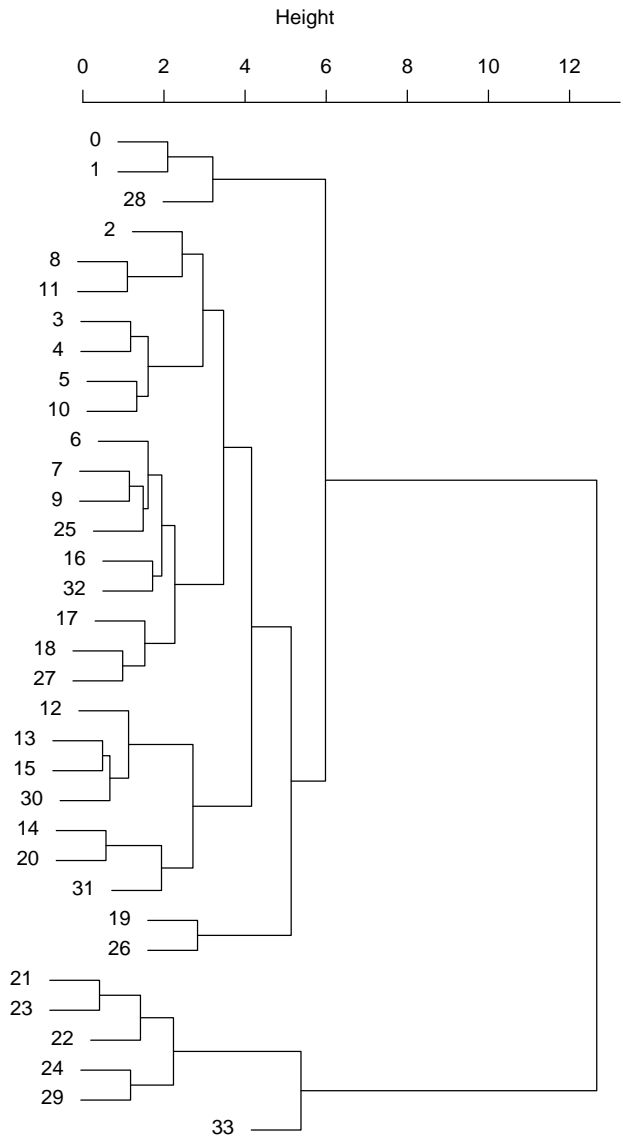}
\caption{The hierarchical clustering of the $34$ most frequent items
based on their simple specializing change
profiles.\label{f:rex:hierprofs}}
\end{figure}

The clustering of the specializing change profiles captures many
important dependencies between the courses. For example, the courses
$8$ (English Oral Test) and $11$ (Oral and Written Skills in Swedish)
are close to each other. Also, the courses $16$ (Approbatur in
Mathematics I) and $32$ (Approbatur in Mathematics II) are in the same
branch although their ranking with respect to their frequencies differ
quite much. Furthermore, the courses $18$ (Discrete Mathematics I) and
$27$ (Logic I) are close to each other as their content overlap
considerably and they form two thirds of an alternative for the
courses $16$ and $32$ to obtain Approbatur in Mathematics.

The courses $14$ (Scientific Writing) and $20$ (Maturity Test in
Finnish) are naturally close to each other since it is very customary
to take the maturity test in the end of the Scientific Writing
course. Also the course $31$ (Software Engineering Project) is close
to the course $14$. The explanation for this is that both courses have
almost the same prerequisites and both are needed for the Bachelor of
Science degree with Computer Science as the major subject.

The courses $5$ (Information Systems) and $10$ (Programming in Pascal)
are deprecated and they replaced in the current curriculum by the
courses $21$ (Introduction to Application Design), $23$ (Introduction
to Databases), $24$ (Introduction to Programming) and $22$
(Programming in Java). Similarly, the course $12$ (Information Systems
Project) has been replaced by the course $33$ (Database Application
Project). The courses close to the course $12$, namely the courses
$13$ (Concurrent Systems), $15$ (Databases Systems I) and $30$ (Data
Communications), are also deprecated versions although there are
courses with the same names in the current curriculum.

\begin{figure}
\includegraphics[height=\textwidth,angle=270]{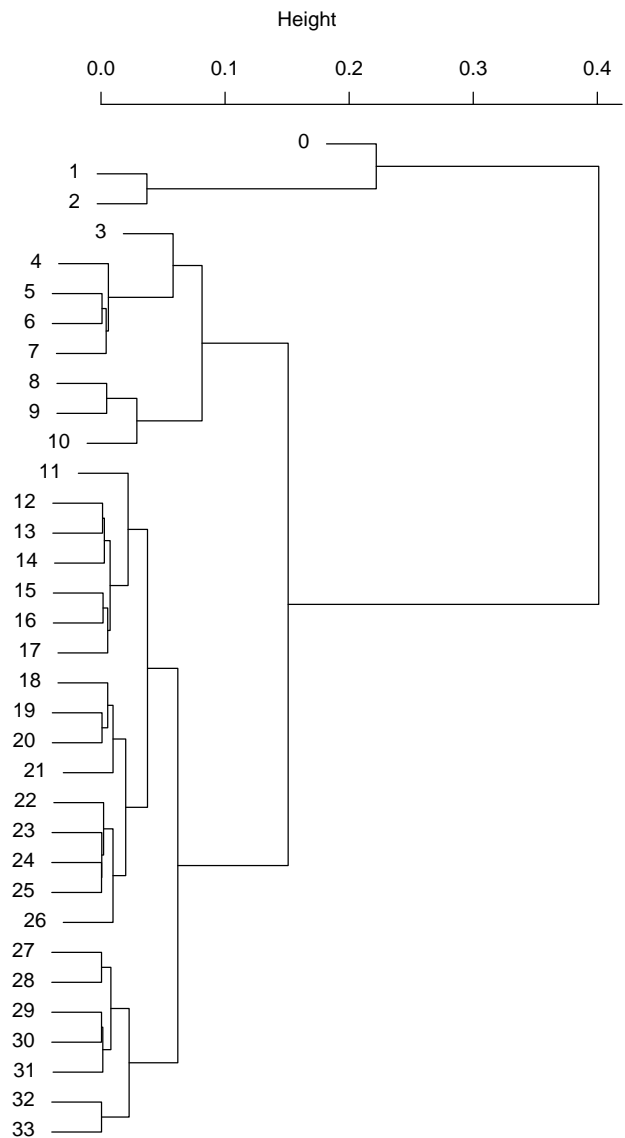}
\caption{The hierarchical clustering of the $34$ most frequent items
based on the absolute differences between their frequencies.\label{f:rex:hierfreqs}}
\end{figure}

As the simplest comparison, the clustering of items based on the
absolute differences between their frequencies is shown in
Figure~\ref{f:rex:hierfreqs}. However, the clustering based on
frequencies does not capture much of the relationships between the
courses. This is not very surprising since the frequencies of the
courses contain quite little information about the courses.

\begin{figure}
\includegraphics[height=\textwidth,angle=270]{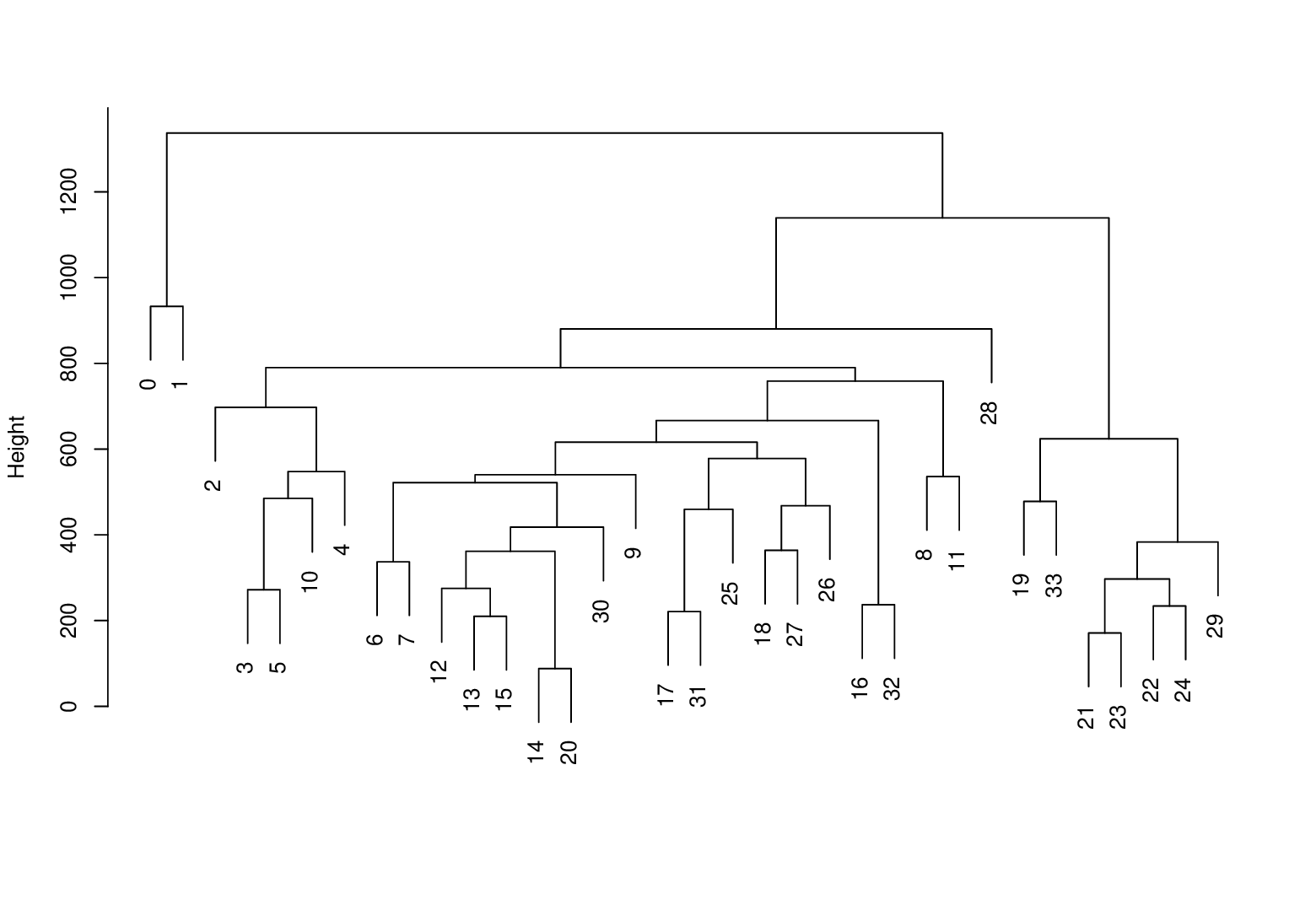}
\caption{The hierarchical clustering of the $34$ most frequent items
based on their Hamming distances.\label{f:rex:hiercorrs}}
\end{figure}

A more realistic comparison would be the average linkage hierarchical
clustering based on the Hamming distances between the items.  The
Hamming distance between the two items in a transaction database is
the number transactions in the database containing one of the items
but not both of them, i.e., the Hamming distance between items $A$ and
$B$ in a transaction database $\DB$ is
\begin{eqnarray*}
\distance_H(A,B,\DB)&=&\Abs{\cover{A,\DB} \setminus \cover{B,\DB}}+ \\
&&\Abs{\cover{B,\DB} \setminus \cover{A,\DB}}.
\end{eqnarray*}
Such a clustering is shown in Figure~\ref{f:rex:hiercorrs}.

The results obtained using Hamming distance are quite similar to the
results obtained using the change profiles. There are slight
differences, however. For example, the courses $0$, $1$ and $28$ that
are close to each other in Figure~\ref{f:rex:hierprofs}, are quite far
from each other in in Figure~\ref{f:rex:hiercorrs}. The courses $17$
and $31$ are close to each other in Figure~\ref{f:rex:hiercorrs},
whereas the course 31 is in the same cluster with the courses $14$ and
$20$ in Figure~\ref{f:rex:hierprofs}.

The courses $18$, $26$ and $27$ form a cluster in
Figure~\ref{f:rex:hiercorrs} forming an alternative Approbatur in
Mathematics but in Figure~\ref{f:rex:hierprofs} the course $26$ is
together with the course $19$ which is mathematically demanding for
many students. (In Figure~\ref{f:rex:hiercorrs} the course $19$ is in
the same cluster with the course $33$.)

In general, the hierarchical clustering with respect to Hamming
distances seems to capture courses forming entities (for example,
pairs of courses that earlier formed one course), whereas the
hierarchical clustering of change profiles seems to be related more
closely to the essence of the courses in a broader way. This is in
line with the fact that the Hamming distances between the items
compare the co-occurrences of the items directly, whereas the distances
between the change profiles measure the similarity of the behavior of
the items with respect to their whole neighborhoods except each
other. Note that the change profiles do not depend on the actual
frequencies of the items but the Hamming distances are strongly
affected by the frequencies.

\begin{figure}
\includegraphics[height=\textwidth,angle=270]{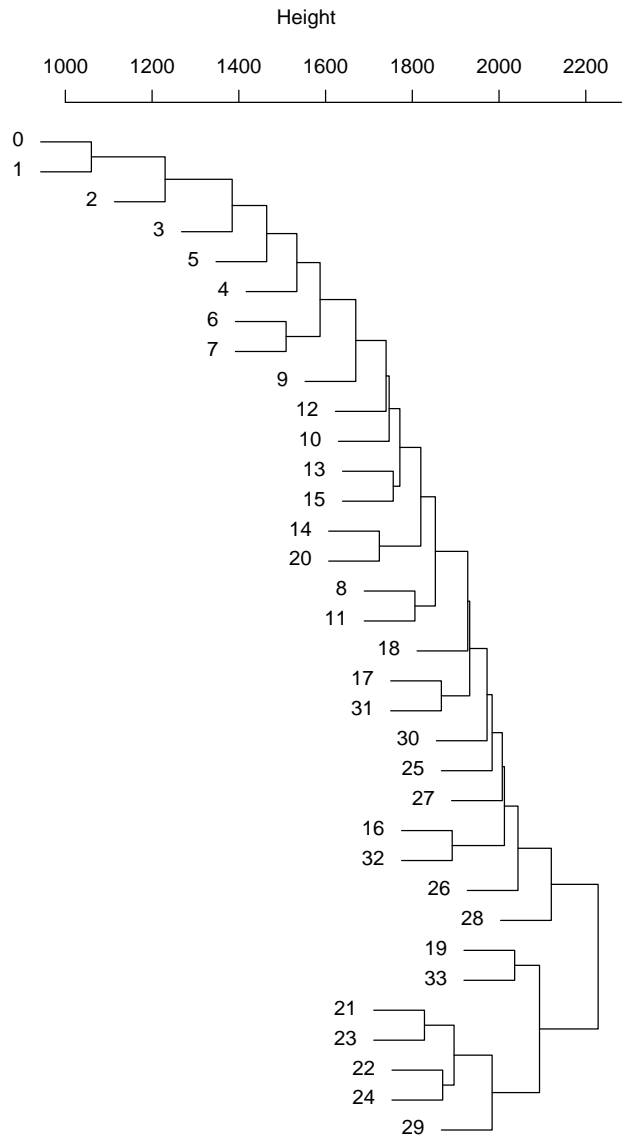}
\caption{The hierarchical clustering of the $34$ most frequent items
based on the their scalar products.\label{f:rex:hierdots}}
\end{figure}

\begin{figure}
\includegraphics[height=\textwidth,angle=270]{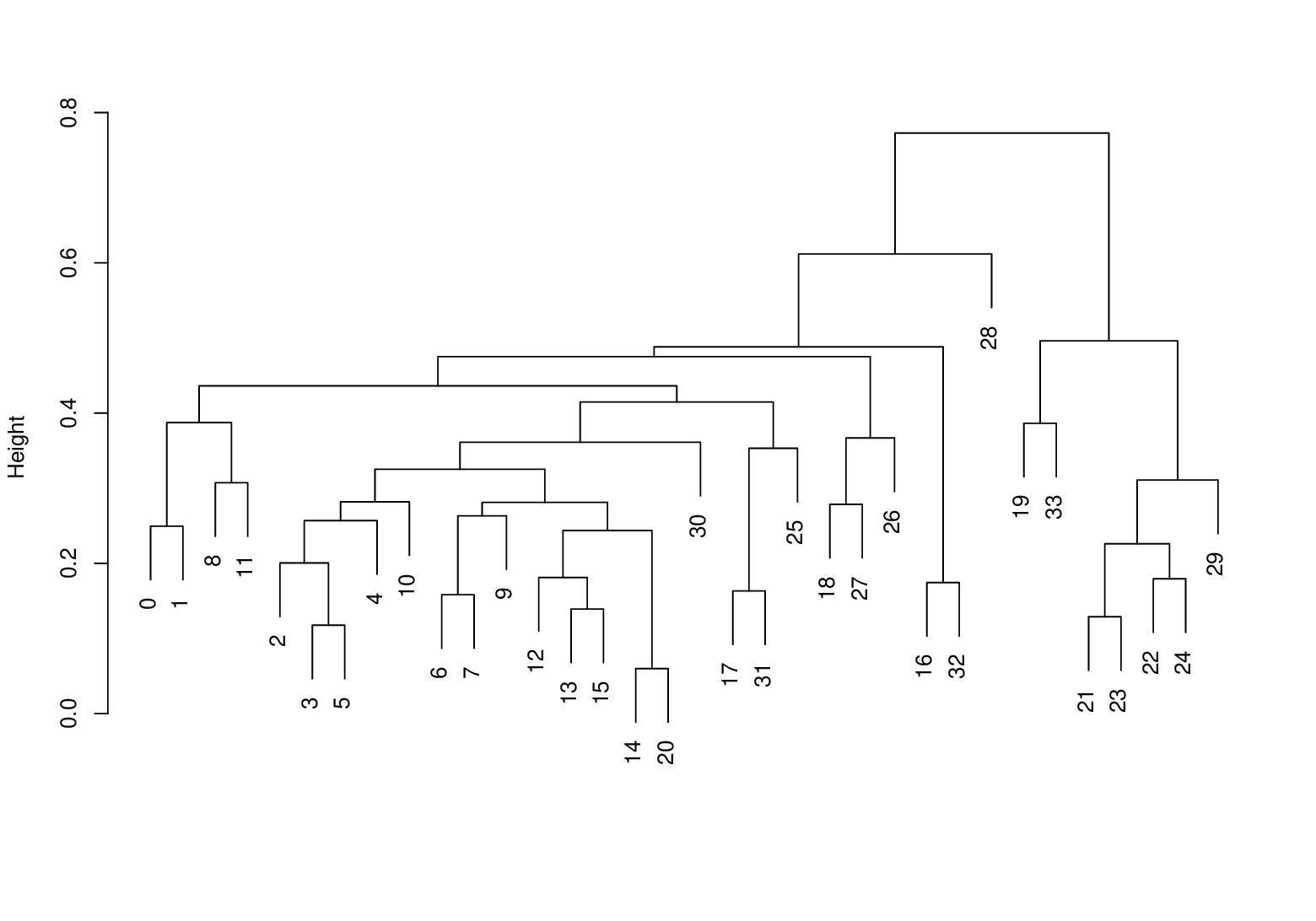}
\caption{The hierarchical clustering of the $34$ most frequent items
based on their cosine distances.\label{f:rex:hiercoss}}
\end{figure}

The Hamming distance is symmetric with respect to whether the item is
contained in the transaction. As the transaction databases often
correspond to sparse binary matrices, this assumption about the
symmetry of presence and absence is not always justified. The
similarity between two items could be measured by the number of
transactions containing them both instead of counting the number of
transactions containing either both or neither of them. This is also
equal to the scalar product between the binary vectors representing
the covers of the items. To transform similarity to dissimilarity, we
subtract the similarity value from the cardinality of the
database. Thus, the dissimilarity is
\begin{displaymath}
\Abs{\DB}-\Abs{\cover{A,\DB} \cap \cover{B,\DB}}=\supp{\emptyset,\DB}-\supp{AB,\DB}.
\end{displaymath}
The hierarchical clustering for this dissimilarity is shown in
Figure~\ref{f:rex:hierdots}. The results are unfortunately similar to
the clustering based on the frequencies (Figure~\ref{f:rex:hierfreqs})
although also some related courses, such as the courses $16$ and $32$,
are close to each other in the dendrogram regardless of their
dissimilar frequencies.

The change profiles used in the clustering in
Figure~\ref{f:rex:hierprofs} can be computed from the frequencies of
the $34$ most frequent items (Figure~\ref{f:rex:hierfreqs}) and the
frequencies of the $2$-itemsets formed from the $34$ most frequent
itemsets (Figure~\ref{f:rex:hierdots}). Thus, in this particular case,
the specializing simple change profiles can be considered as
normalizations of the frequencies of the $2$-itemsets by the
frequencies of the items. Another approach to normalize the
frequencies of the $2$-itemsets by the frequencies of the items is as
follows. The supports of the $2$-itemsets can be considered the scalar
products between the the items. By normalizing the scalar product by
the euclidean lengths of the vectors corresponding to the covers of
the items, we get the cosine of the angle between the vector. The
cosine of the angle between two (non-zero) binary vectors is always in
the interval $\IntC{0,1}$. Thus, the cosine distance between two
items $A$ and $B$ in a transaction database $\DB$ is
\begin{eqnarray*}
\distance_{\cos}(A,B,\DB)
&=&1-\frac{\Abs{\cover{A,\DB} \cap \cover{B,\DB}}}{\sqrt{\Abs{\cover{A,\DB}}}\sqrt{\Abs{\cover{B,\DB}}}} \\
&=&1-\frac{\supp{AB,\DB}}{\sqrt{\supp{A,\DB}}\sqrt{\supp{B,\DB}}} \\
&=&1-\frac{\supp{AB,\DB}/\supp{\emptyset,\DB}}{\sqrt{\supp{A,\DB}}\sqrt{\supp{B,\DB}}/\supp{\emptyset,\DB}} \\
&=&1-\frac{\fr{AB,\DB}}{\sqrt{\fr{A,\DB}\fr{B,\DB}}}.
\end{eqnarray*}
The hierarchical clustering of the $34$ most frequent items based
cosine distances is shown in Figure~\ref{f:rex:hiercoss}. The
clustering is very close to the one shown in
Figure~\ref{f:rex:hiercorrs}. The main difference between these two
clusterings is that in the clustering shown in
Figure~\ref{f:rex:hiercorrs} the courses $0$ and $1$ are very
different to everything (including each other), whereas the clustering
shown in Figure~\ref{f:rex:hiercoss} grasps the similarity between the
courses $0$, $1$, $8$ and $11$.
\exend \end{realexample}

\section{Estimating Frequencies from Change Profiles \label{s:estch}}

The change profiles can be used as a basis for condensed
representations of frequent itemsets. Furthermore, several known
condensed representations can be adapted to change profiles. One
interesting approach to condense the change profiles (and thus the
underlying pattern collections, too) is to choose a small set of
representative change profiles (using, e.g., hierarchical clustering)
and replace the original change profiles by the chosen
representatives. Then the frequencies of the frequent itemsets can be
estimated from the approximate change profiles.

Representing the frequencies of the frequent itemsets by approximate
change profiles can be seen as a condensed representation of the
collection of frequent itemsets as the approximate change profiles can
(potentially) fit into smaller space than the exact change profiles or
even the frequent itemsets. Also, the condensed representations can be
applied to further condense the approximate change profiles.

In addition to the fact that the frequencies can be estimated from the
approximate change profiles, the change profiles themselves can
benefit from the frequency estimation. Namely, the quality of the
approximate change profiles can be assessed by evaluating how well the
frequencies can be approximated from them.

For the rest of the section we consider only the case where no change
profile is missing but the changes are not exact. The methods
described in this section can be generalized to handle missing change
profiles and missing changes.

Given the approximations of the change profiles for the collection
$\Frequent{\sigma,\DB}$ of the $\sigma$-frequent itemsets in $\DB$, it
is possible to estimate the frequencies of the itemsets in
$\Frequent{\sigma,\DB}$ from the approximate change profiles. The
estimation can be done in many ways and the quality of each
estimation method depends on how the approximations of the change
profiles are obtained.

Next we describe an approach based on the estimates given by different
paths (in the graph determined by the changes of the change profiles)
from the empty itemset $\emptyset$ to the itemset $X$ whose frequency
is under estimation. Especially, we concentrate on computing the
average frequencies given by the paths from $\emptyset$ to $X$. The
methods are described using simple specializing change profiles, but
their generalization to other kinds of change profiles is
straightforward.

Without loss of generality, let $X=\Set{1,\ldots,k}$. In principle, we
could compute the frequency estimate $\fr{X}$ of the itemset $X$ in $\Frequent{\sigma,\DB}$, the average of the
frequencies suggested by paths from $\emptyset$ to $X$.
Let $\Pi_k$ be the collection of all permutations of $\Set{1,\ldots,k}$. Then the frequency estimate can be written as
\begin{equation} \label{eq:avgpath}
\fr{X}=\frac{1}{\Abs{X}!}\sum_{\pi \in \Pi_k} \sch{\pi(1)}{\emptyset}{s} \prod_{i=2}^k
\sch{\pi(i)}{\pi(i-1)}{s}.
\end{equation}

The main practical difficulty of this formula is the number of paths:
The number of paths from $\emptyset$ to $X$ is equal to the number of
permutations of items in $X$, i.e., the number of paths from
$\emptyset$ to $X$ is $\Abs{X}!$. This can be superpolynomial in
$\Abs{\Frequent{\sigma,\DB}}$.

\begin{example}[the number of paths given by simple change profiles is superpolynomial]
Let $\Frequent{\sigma,\DB}$ consist of an itemset $X$ and all of its
subitemsets.  Then
\begin{displaymath}
\Abs{\Frequent{\sigma,\DB}}=2^{\Abs{X}}
\end{displaymath}
and
\begin{displaymath}
\Abs{X}!=\sqrt{2\pi \Abs{X}}\Paren{\frac{\Abs{X}}{e}}^{\Abs{X}} \Paren{1+\Th{\Abs{X}^{-1}}}.
\end{displaymath}

Hence,
\begin{displaymath}
\frac{\Abs{X}!}{\Abs{\Frequent{\sigma,\DB}}}=\sqrt{2\pi \Abs{X}}\Paren{\frac{\Abs{X}}{2e}}^{\Abs{X}} \Paren{1+\Th{\Abs{X}^{-1}}}
\end{displaymath}
which is clearly exponential in $\Abs{X}$.
\exend \end{example}

The frequency estimate $\fr{X}$ of $X$ as the average over all paths
from $\emptyset$ to $X$ can be computed much faster by observing that
the frequency of $X$ is the average of the frequencies of the itemsets
$X \setminus \Set{A}, A \in X$, scaled by the changes $\sch{\Set{A}}{X
\setminus \Set{A}}{s}$, i.e.,
\begin{displaymath}
\fr{X}=\frac{1}{\Abs{X}}\sum_{Y \subset X,
\Abs{Y}=\Abs{X}-1}\fr{Y}\sch{X \setminus Y}{Y}{s}.
\end{displaymath}

This observation readily gives a dynamic programming solution described
as Algorithm~\ref{a:DP-from-schs}.

\begin{algorithm}
\caption{A dynamic programing solution for frequency estimation from (inexact) change profiles. \label{a:DP-from-schs}}
\begin{algorithmic}[1]
\Input{An itemset $X$ and the simple specializing change profiles (at least) for $X$ and all
of its subsets.}
\Output{The frequency estimate $\fr{X}$ of $X$ as described by
Equation~\ref{eq:avgpath}.}
\Function{DP-from-schs}{$X,\mathit{sch}_s$}
\State $\fr{\emptyset} \leftarrow 1$
\For{$i=1,\ldots,\Abs{X}$}
 \ForAll{$Y\subseteq X, \Abs{Y}=i$}
  \State $\fr{Y}=0$
  \ForAll{$Z \subset Y, \Abs{Z}=\Abs{Y}-1$}
   \State $\fr{Y} \leftarrow \fr{Y}+\fr{Z}\sch{Y\setminus Z}{Z}{s}$
  \EndFor
  \State $\fr{Y} \leftarrow \fr{Y}/\Abs{Y}$
 \EndFor
\EndFor
\EndFunction
\end{algorithmic}
\end{algorithm}

As the frequency estimate has to be computed also for all subsets of
$X$ and the frequency estimate of $Y$ can be computed from the
frequency estimates of the subsets of $Y$ in time $\Oh{\Abs{Y}}$ the
time complexity of Algorithm~\ref{a:DP-from-schs} is
\begin{displaymath}
\Oh{\Abs{X}2^{\Abs{X}}}=\Oh{\Abs{\Frequent{\sigma,\DB}}\log
\Abs{\Frequent{\sigma,\DB}}}.
\end{displaymath}

Even this can be too much for a restive data analyst. The estimation
can be further speeded up by sampling uniformly from the paths from
$\emptyset$ to $X$ as described by Algorithm~\ref{a:Sample-from-schs}.

\begin{algorithm}
\caption{A randomized algorithm for frequency estimation from (inexact) change profiles. \label{a:Sample-from-schs}}
\begin{algorithmic}[1]
\Input{An itemset $X$, the simple specializing change profiles (at least) for $X$ and all
of its subsets, and a positive integer $k$.}
\Output{An estimate of the frequency estimate $\fr{X}$ of $X$ as
described by Equation~\ref{eq:avgpath}.}
\Function{Sample-from-schs}{$X,\mathit{sch}_s,k$}
\State $\fr{\emptyset} \leftarrow 1$
\State $\fr{X} \leftarrow 0$
\For{$j=1,\ldots,k$}
 \State $Y \leftarrow \emptyset$
 \For{$i=1,\ldots,\Abs{X}-1$}
  \State $A \leftarrow $ \Call{Random-Element}{$X \setminus Y$}
  \State $\fr{Y \cup \Set{A}} \leftarrow \fr{Y}\sch{\Set{A}}{Y}{s}$
  \State $Y \leftarrow Y \cup \Set{A}$
 \EndFor
 \State $\fr{X}\leftarrow \fr{X}+\fr{Y}\sch{X \setminus Y}{Y}{s}$
\EndFor
\State $\fr{X} \leftarrow \fr{X}/k$
\EndFunction
\end{algorithmic}
\end{algorithm}

The time complexity of Algorithm~\ref{a:Sample-from-schs} is
$\Oh{k\Abs{X}}$ where $k$ is the number of randomly chosen paths in
the estimate. Note that the algorithm can be easily modified to be an
any-time algorithm. This would sometimes be useful in interactive data
mining and for resource bounded data mining in general.

Algorithm~\ref{a:DP-from-schs} and Algorithm~\ref{a:Sample-from-schs}
can be adapted to other kinds of estimates, too. Especially, if upper
and lower bounds for the changes $\sch{A}{Y}{s}$ are given for all $Y
\subseteq X$ such that $A \in X \setminus Y$, then it is possible to
compute the upper and lower bounds for the frequency of $X$ for all
itemsets $X$ reachable from $\emptyset$ by changes of the change
profiles.  Namely, the frequency of the itemset $X$ is at most the
minimum of the upper bound estimates and at least the maximum of the
lower bound estimates determined by the change paths from $\emptyset$
to $X$.

\section{Condensation by Change Profiles}

The usefulness of approximate change profiles, the stability of the
frequency estimation algorithms proposed in Section~\ref{s:estch} and
the accuracy of the path sampling estimates were evaluated by
estimating frequencies from noisified simple specializing change
profiles in the transaction databases Internet Usage and IPUMS Census
(see Subsection~\ref{ss:ds}).

In order to study how the estimation methods (i.e.,
Algorithm~\ref{a:DP-from-schs} and Algorithm~\ref{a:Sample-from-schs})
tolerate different kinds of noise, the simple specializing change
profiles were noisified in three different ways:
\begin{itemize}
\item
randomly perturbing the changes of the change profiles by $\pm
\epsilon$,
\item
adding uniform noise from the interval $\IntC{-\epsilon,\epsilon}$ to
the changes of the change profiles, and
\item
adding Gaussian noise with zero mean and standard deviation $\epsilon$ to
the changes of the change profiles.
\end{itemize}

The changes of the noisified change profiles were truncated to the
interval $\IntC{0,1}$ since, by the definition of specializing change
profiles (Definition~\ref{d:chs}), the changes in the specializing
change profiles must be in the interval $\IntC{0,1}$.

We tested the dependency of the approximation on the number of sample
paths by evaluating the absolute difference between the correct and
the estimated frequencies for the dynamic programming solution
corresponding to the average frequency estimate over all paths, and
the sample solution corresponding to the average frequency estimate
over the sampled paths. The experiments were repeated with different
number of randomly chosen paths, minimum frequency thresholds and
noise levels $\epsilon$. 

\begin{figure}[p] \centering
\includegraphics[width=\columnwidth]{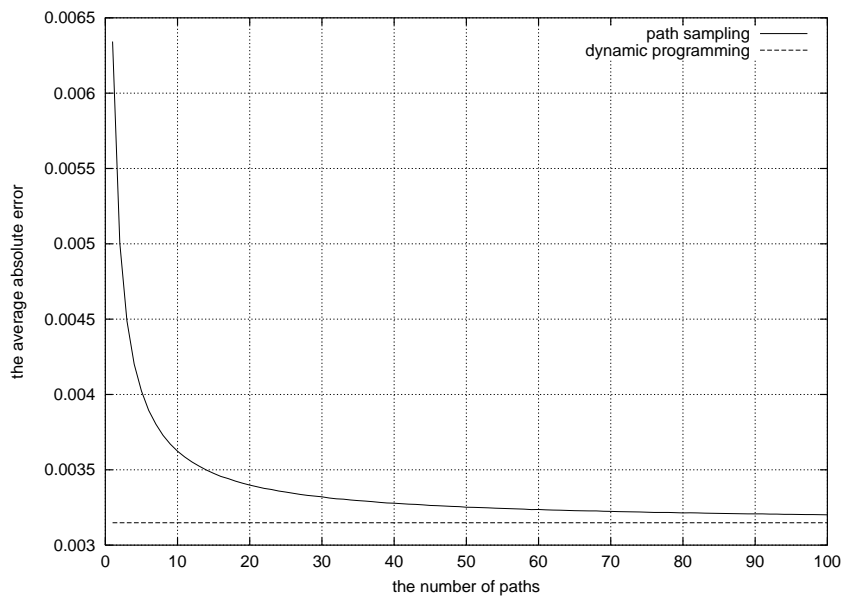}
\caption{Internet Usage data, Gaussian noise. \label{f:profiles:internet:g}}
\end{figure}

\begin{figure}[p] \centering
\includegraphics[width=\columnwidth]{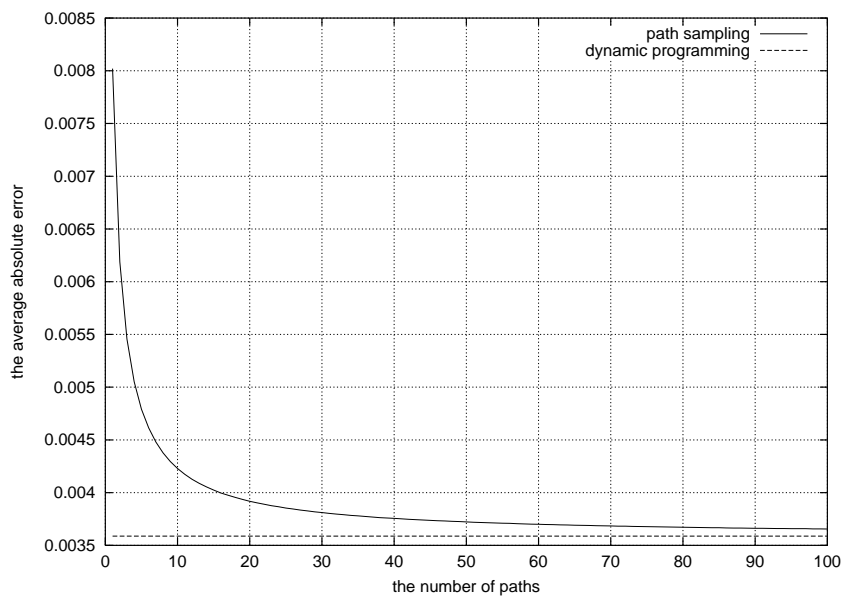}
\caption{IPUMS Census data, Gaussian noise. \label{f:profiles:ipums:g}}
\end{figure}

\begin{figure}[p] \centering
\includegraphics[width=\columnwidth]{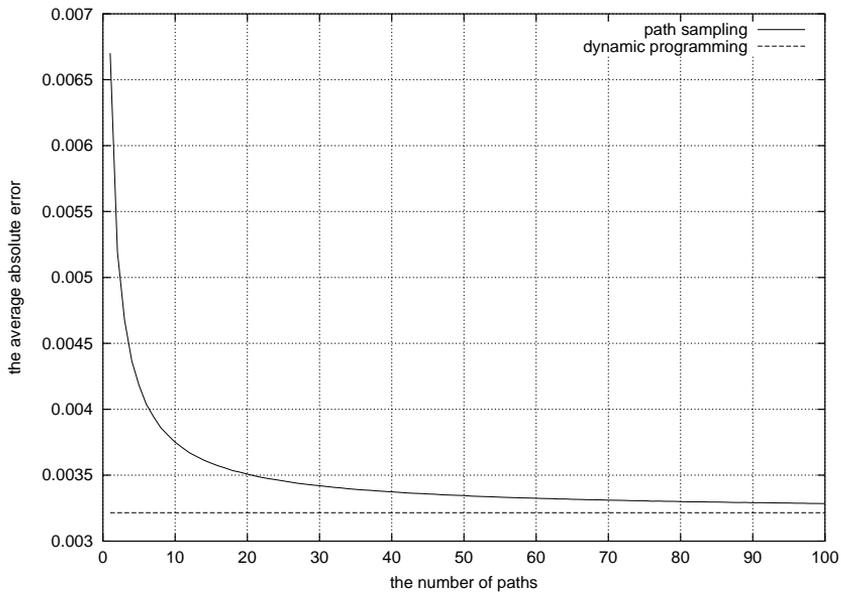}
\caption{Internet Usage data, perturbation. \label{f:profiles:internet:p}}
\end{figure}

\begin{figure}[p] \centering
\includegraphics[width=\columnwidth]{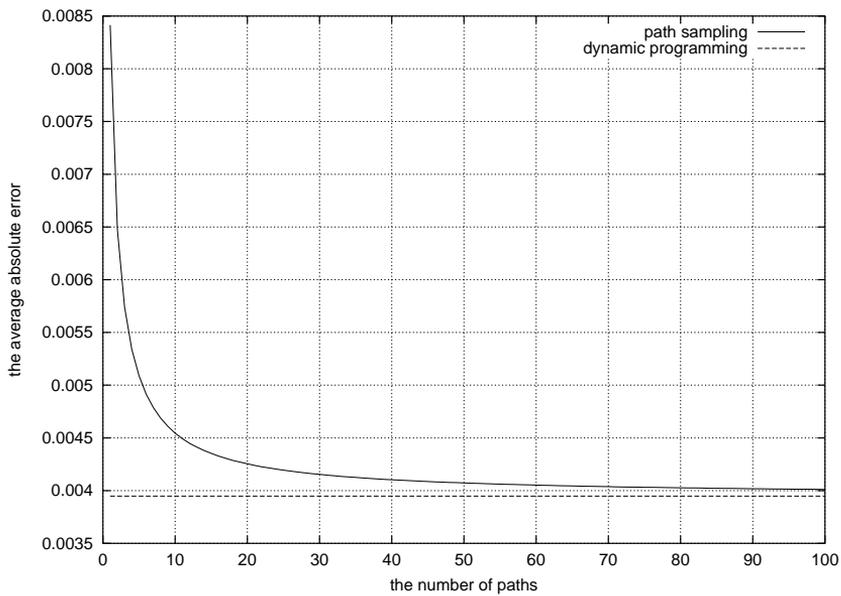}
\caption{IPUMS Census data, perturbation. \label{f:profiles:ipums:p}}
\end{figure}

\begin{figure}[p] \centering
\includegraphics[width=\columnwidth]{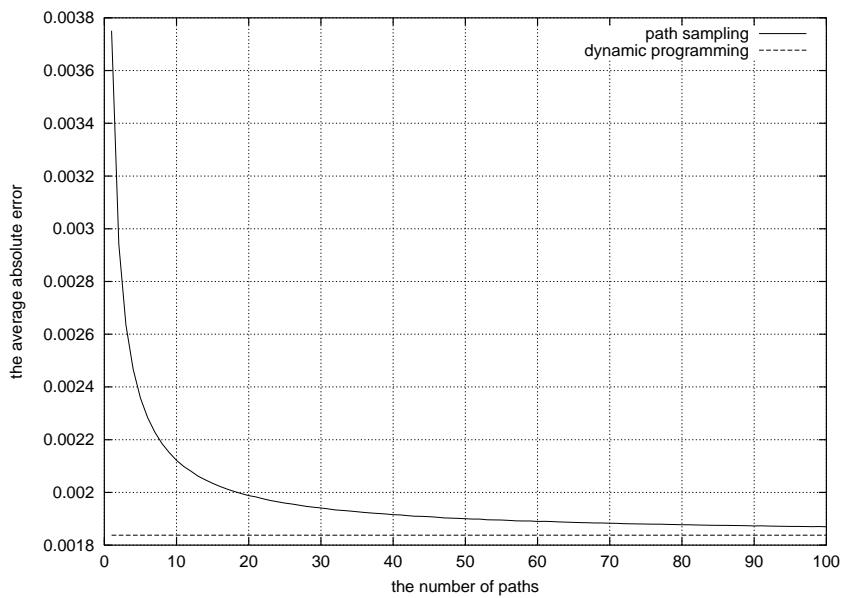}
\caption{Internet Usage data, uniform noise.\label{f:profiles:internet:u}}
\end{figure}

\begin{figure}[p] \centering
\includegraphics[width=\columnwidth]{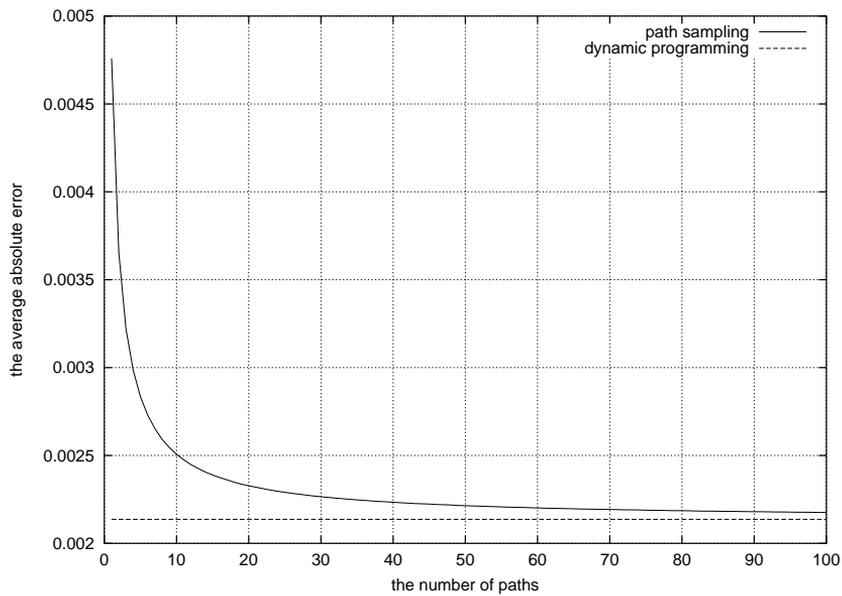}
\caption{IPUMS Census data, uniform noise. \label{f:profiles:ipums:u}}
\end{figure}

The results for Internet Usage data with minimum frequency threshold
$0.20$ are shown in Figures~\ref{f:profiles:internet:g},
\ref{f:profiles:internet:p} and \ref{f:profiles:internet:u}, and for
IPUMS Census data with minimum frequency threshold $0.30$ are shown in
Figures~\ref{f:profiles:ipums:g}, \ref{f:profiles:ipums:p} and
\ref{f:profiles:ipums:u}, with noise level $\epsilon=0.01$. The each
of the curves are averages of $1000$ random experiments. The results
were similar with the other minimum frequency thresholds, too.

The results show that already a quite small number of random paths
suffices to give frequency approximations closed to the dynamic
programming solution. Furthermore, the average absolute errors
achieved by dynamic programming were relatively small, especially as
the errors in the changes cumulate multiplicatively as the frequencies
are estimated as the paths.

%% file: ipd.tex

\chapter{Inverse Pattern Discovery \label{c:ipd}}


The problem of discovering interesting patterns from data has been
studied very actively in data mining. (See, e.g.,
Chapter~\ref{c:discovery} for more details.) For such a well-studied
problem, it is natural to study also the inverse version of the
problem, i.e., the problem of the inverse pattern discovery. That is,
to study the problem of finding a database compatible with a given
collection of interesting patterns.

In addition of being an important class of tasks for data mining as a
scientific discipline, inverse pattern discovery might also have some
practical relevancy.

First, the existence of databases compatible with the given collection
of patterns is usually highly desirable since the collection of
interesting patterns is often assumed to be a summary of some
database. Deciding whether there exists a database compatible with the
given collection of interesting patterns (and their quality values)
can be considered as a very harsh quality control.  An efficient
method for answering to that question could have also some practical
implications to pattern discovery since several instances are not
willing to share their data but might sell some patterns claiming they
are interesting in their database. (If interaction with the pattern
provider would be allowed, then also, e.g., zero knowledge proofs
deciding whether they are interesting or not could be
considered~\cite{t:goldreich02}.)

Second, if the number of compatible databases could be counted, then
the pattern provider could evaluate how well the pattern user could
detect the correct database from the patterns: without any background
information, a randomly chosen database compatible with the patterns
would be the original one with probability $1/k$ where $k$ is the
number of databases compatible with the patterns. If there is more
background information, however, then the probability of finding the
original database can sometimes made higher but still the number of
compatible databases is likely to tell about the difficulty of finding
the original database based on the patterns.  Furthermore, the number
of compatible databases can be used as a measure of how well the
pattern collection characterizes the database.

Third, the pattern collection can be considered as a collection of
queries that should be answered correctly. Thus, the patterns can be
used to optimize the database with respect to, e.g., query efficiency
or space consumption. The optimization task could be expressed as
follows: given the pattern collection, find the smallest database that
gives the correct quality values for the patterns.

In this chapter, the computational complexity of inverse pattern
discovery is studied in the special case of frequent
itemsets. Deciding whether there is a database compatible with the
frequent itemsets and their frequencies is shown to be \NP-complete
although some special cases of the problem can be solved in polynomial
time. Furthermore, finding the smallest compatible transaction
database for an itemset collection consisting only of two disjoint
maximal itemsets and all their subitemsets is shown to be \NP-hard.

Obviously inverse frequent itemset mining is just one example of
inverse pattern discovery.  For example, let us assume that the data
is a $d$-dimensional matrix (i.e., a $d$-dimensional data
cube~\cite{i:gray96}) and the summary of the data consists of the sums
over each coordinate (i.e., all $d-1$-dimensional sub-cubes).  The
computational complexity of the inverse pattern discovery for such
data and patterns, i.e., the computational complexity of the problem
of reconstructing a multidimensional table compatible with the sums,
has been studied in the field of discrete tomography~\cite{p:dt99}:
the problem is solvable in polynomial time if the matrix is
two-dimensional binary matrix~\cite{a:kuba89} and \NP-hard
otherwise~\cite{a:chrobak01,a:gritzmann00,a:irving94}. In this
chapter, however, we shall focus on inverting frequent itemset mining.

This chapter is based on the article ``On Inverse Frequent Set
Mining''~\cite{i:mielikainen03:ifsm}. Some similar results were
independently shown by Toon Calders~\cite{i:calders04}. Recently, a
heuristic method for the inverse frequent itemset mining problem has
been proposed~\cite{i:wu05}.

\section{Inverting Frequent Itemset Mining}
The problem of deducing a transaction database compatible with a given
collection frequent itemsets are their supports can be formulated as
follows.

\begin{problem}[inverse frequent itemset mining] \label{p:ifim}
Given a downward closed collection $\Fr$ of itemsets and the support
$\supp{X}$ for each itemset $X \in \Fr$, find a transaction database
$\DB$ compatible with the supports of the collection, i.e., a
transaction database $\DB$ such that $\supp{X,\DB}=\supp{X,\Fr}$ for
all $X \in \Fr$.
\end{problem}

\begin{example}[inverse frequent itemset mining]
The collection $\Fr=\Set{\emptyset,A,B,C,AB,BC}$ with supports
\begin{eqnarray*}
\supp{\emptyset,\Fr}&=&6, \\ 
\supp{A,\Fr}&=&4, \\
\supp{B,\Fr}&=&4, \\
\supp{C,\Fr}&=&4, \\
\supp{AB,\Fr}&=&3 \quad \mbox{ and} \\
\supp{BC,\Fr}&=&3
\end{eqnarray*}
restrict the collection of transaction databases compatible with these
constraints. For example, the following constraints can be deduced from the support constraints:
\begin{itemize}
\item
The support of the empty itemset tells that the number of transactions
in any compatible databases is six.
\item
There are exactly one transaction with $A$ and without $B$
($\supp{A,\Fr}-\supp{AB,\Fr}=1$), one with $B$ and without $C$
($\supp{B,\Fr}-\supp{BC,\Fr}=1$), and vice versa
($\supp{C,\Fr}-\supp{BC,\Fr}=1$).
\item
The support of $AC$ is at least two since there are at most two
transactions that contain $B$ but not $A$ and $C$.
\end{itemize}

One transaction database compatible with these constraints is
\begin{displaymath}
\DB=\Set{\Tuple{1,ABC},\Tuple{2,ABC},\Tuple{3,AB},\Tuple{4,BC},\Tuple{5,A},\Tuple{6,C}}
\end{displaymath}
since the supports 
\begin{eqnarray*}
\supp{\emptyset,\DB}&=&6, \\
\supp{A,\DB}&=&4, \\
\supp{B,\DB}&=&4, \\
\supp{C,\DB}&=&4, \\
\supp{AB,\DB}&=&3, \\
\supp{AC,\DB}&=&2, \\
\supp{BC,\DB}&=&3 \quad \mbox{ and} \\
\supp{BC,\DB}&=&3
\end{eqnarray*}
determined by $\DB$ agree with the given supports.

There are also other databases compatible with the constraints. For
example,
\begin{displaymath}
\DB=\Set{\Tuple{1,ABC},\Tuple{2,ABC},\Tuple{3,ABC},\Tuple{4,A},\Tuple{5,B},\Tuple{6,C}}
\end{displaymath}
is one such database. \exend
\end{example}

The reason why we use supports instead of frequencies in the inverse
frequent itemset mining problem (Problem~\ref{p:ifim}) is that
supports are slightly more informative. On one hand, the frequencies
of the frequent itemsets $X \in \Frequent{\sigma,\DB}$ can be computed
from their supports since
$\fr{X,\DB}=\supp{X,\DB}/\supp{\emptyset,\DB}$. On the other hand, the
supports cannot be computed from the frequencies: the number of
transactions in the database, i.e., $\supp{\emptyset,\DB}$ is not
revealed by the frequencies of the itemsets.

\section{Frequent Itemsets and Projections}

To determine the complexity of the inverse frequent itemset mining
problem (Problem~\ref{p:ifim}), let us consider an intermediate
representation between the frequent itemsets and the transaction
database.

\begin{definition}[projections of transaction databases]
The \emph{projection} of the transaction database $\DB$ onto itemset
$X$ is a restriction
\begin{displaymath}
\pr{X,\DB}=\Set{ \Tuple{i,X \cap Y} : \Tuple{i,Y} \in \DB}
\end{displaymath}
of the database $\DB$. The collection of projections $\pr{X,\DB}$ onto
itemsets $X \in \Fr$ is denoted by
\begin{displaymath}
\pr{\Fr,\DB}=\Set{\pr{X,\DB} : X \in \Fr}.
\end{displaymath}

Two projections $\pr{X,\DB}$ and $\pr{X,\DB'}$ are considered to be
equivalent if and only if $\Abs{\DB}=\Abs{\DB'}$ and there is a
bijective mapping $\eqmap{}$ from $\tid{\DB}$ to $\tid{\DB'}$ such
that for each $\Tuple{i,Y} \in \DB$ there is $\Tuple{\eqmap{i},Y'} \in
\DB'$ with $X \cap Y=X \cap Y'$. (That is, the mapping $\eqmap{}$ is a
permutation since we can assume that
$\tid{\DB}=\tid{\DB'}=\Set{1,\ldots,\Abs{\DB}}$; see
Definition~\ref{d:tdb}.)
\end{definition}

The projections of transaction databases have many desirable
similarities to itemsets.  For example, neglecting the transaction
identifiers, the projections of the database $\DB$ onto maximal
$\sigma$-frequent itemsets contain the same information than the
$\sigma$-frequent itemsets and their supports.
\begin{theorem} \label{t:projs}
The frequent itemsets in $\Frequent{\sigma,\DB}$ and their supports in
$\DB$ can be computed from the projections
$\pr{\Maximal{\sigma,\DB},\DB}$ and the projections equivalent to
$\pr{\Maximal{\sigma,\DB},\DB}$ can be computed from the frequent
itemsets in $\Frequent{\sigma,\DB}$ and their supports in $\DB$.
\end{theorem}
\begin{proof}
For each $X \in \Frequent{\sigma,\DB}$ and each $Y \supseteq X$ we have 
\begin{eqnarray*}
\supp{X,\DB}&=&\Abs{\Set{\Tuple{i,Z} \in \DB : X \subseteq Z}} \\
&=&
\Abs{\Tuple{i,Y \cap Z} \in \DB : X \subseteq \Paren{Y \cap Z}} \\
&=&
\supp{X,\pr{Y,\DB}}.
\end{eqnarray*}

By definition, each $\sigma$-frequent itemset $X \in
\Frequent{\sigma,\DB}$ is contained in some maximal $\sigma$-frequent
itemset $Y \in \Maximal{\sigma,\DB}$. Furthermore, no
$\sigma$-infrequent itemset is contained in any of the maximal
$\sigma$-frequent itemsets in $\DB$. Thus, the collection
$\Frequent{\sigma,\DB}$ of the $\sigma$-frequent itemsets and their
supports in $\DB$ can be computed from the collection
$\pr{\Maximal{\sigma,\DB},\DB}$ of projections of the transaction
database $\DB$ onto the maximal $\sigma$-frequent itemsets in $\DB$.

The projections equivalent to $\pr{\Maximal{\sigma,\DB},\DB}$ can be
computed from the collection $\Frequent{\sigma,\DB}$ of the
$\sigma$-frequent itemsets and their supports in $\DB$ by
Algorithm~\ref{a:To-Projections}. The running time of the algorithm is
polynomial in $\Abs{\Frequent{\sigma,\DB}}$, $\Abs{\Items}$ and
$\Abs{\DB}=\supp{\emptyset,\DB}$.

The running time can be further improved if the transaction database
$\DB'$ has a primitive for inserting $k$ transactions consisting of an
itemset $X$ into $\DB'$ in time polynomial in $\Abs{\DB'}$ and in
$\Abs{X}$ but not in $k$ at all. Namely, then the running time of
Algorithm~\ref{a:To-Projections} can be expressed as a polynomial of
$\Abs{\Frequent{\sigma,\DB}}$ and $\Abs{\Items}$, i.e., not depending
on the actual number of transactions in the transaction database
$\DB$.

The efficient insertion of $k$ transactions with the itemset $X$ into
$\DB$ can be implemented, e.g., by ``run-length encoding'' the
database, i.e., by describing the transactions
$\Tuple{i,X},\ldots,\Tuple{i+k-1,X}$ by the triple $\Tuple{i,k,X}$. Then the
insertion of $k$ transactions with the itemset $X$ to $\DB'$ can be
implemented by inserting the tuple $\Tuple{\Abs{\DB}+1,k,X}$ to $\DB'$.
\end{proof}

\begin{algorithm}
\caption{An algorithm to compute projections equivalent to
$\pr{\Maximal{\sigma,\DB},\DB}$ from $\Frequent{\sigma,\DB}$ and their
supports. \label{a:To-Projections}}
\begin{algorithmic}[1]
\Input{The collection $\Frequent{\sigma,\DB}$ $\sigma$-frequent
itemsets in a transaction database $\DB$ and their supports.}
\Output{The projections $\pr{\Maximal{\sigma,\DB},\Frequent{\sigma,\DB}}$ equivalent to
projections $\pr{\Maximal{\sigma,\DB},\DB}$.}
\Function{To-Projections}{$\Frequent{\sigma,\DB},\support$}
\State $\Maximal{\sigma,\DB}=\Set{X \in \Frequent{\sigma,\DB} : Y \supset X \Rightarrow Y \notin \Frequent{\sigma,\DB}}$
\ForAll{$ X \in \Maximal{\sigma,\DB}$}
 \State $\DB' \leftarrow \emptyset$
 \State $\Mx \leftarrow \Set{X}$
 \State $\Fr \leftarrow \Set{ Y \in \Frequent{\sigma,\DB} : Y \subseteq X}$
 \ForAll{$Y \in \Fr$}
  \State $\supp{Y,\Fr} \leftarrow \supp{Y,\DB}$
 \EndFor
 \While{$\Mx\neq \emptyset$}
  \State $\Fr \leftarrow \Fr \setminus \Mx$
  \ForAll{$Y \in \Mx$}
   \State $\DB' \leftarrow \DB' \cup \Set{\Tuple{\Abs{\DB'}+1,Y}, \ldots, \Tuple{\Abs{\DB'}+\supp{Y,\Fr},Y}}$
   \ForAll{$Z \in \Fr, Z \subset Y$}
    \State $\supp{Z,\Fr} \leftarrow \supp{Z,\Fr}-\supp{Y,\Fr}$
    \If{$\supp{Z,\Fr}=0$}
     \State $\Fr \leftarrow \Fr \setminus \Set{Z}$
    \EndIf
   \EndFor
  \EndFor
  \State $\Mx \leftarrow \Set{Y \in \Fr : Y \subset Z \Rightarrow Z \notin \Fr}$
 \EndWhile
 \State $\pr{X,\Frequent{\sigma,\DB}} \leftarrow \pr{X,\DB'}$
\EndFor
\State \Return $\pr{\Maximal{\sigma,\DB},\Frequent{\sigma,\DB}}$
\EndFunction
\end{algorithmic}
\end{algorithm}

Theorem~\ref{t:projs} also implies at if
$\Frequent{\sigma,\DB}=2^{\Items}$, then the whole transaction
database $\DB$ (although without the correct transaction identifiers)
can be reconstructed from the collection $\Frequent{\sigma,\DB}$ and
their supports in $\DB$ in time polynomial in
$\Abs{\Frequent{\sigma,\DB}}$ and $\Abs{\DB}$ since the collection
$\Maximal{\sigma,\DB}$ consists only of the itemset
$\Items$. Furthermore, the supports of frequent itemsets can determine
(implicitly) also supports of some infrequent
itemsets~\cite{i:calders04:bounds,i:calders02}.

Let us denote the projections determined by the downward closed
itemset collection $\Fr$ by $\pr{\Mx,\Fr}$. The number of different
itemsets in the transactions of $\pr{\Mx,\Fr}$ can be considerably
smaller than the number itemsets in $\Fr$. Thus, each projection
$\pr{X,\DB}$ of $\DB$ onto $X \in \Maximal{\sigma,\DB}$ represented as
a list of tuples $\Tuple{\cnt{Y,\DB},Y}, Y \subseteq X$, can be used
as a condensed representation of the collection
$\Frequent{\sigma,\DB}$ and their supports. Such projections provide
sometimes very small representations compared to
$\Frequent{\sigma,\DB}$~\cite{i:mielikainen03:faosi}.

As projections constructed from the collection $\Frequent{\sigma,\DB}$
of the $\sigma$-frequent itemsets and their supports in $\DB$ are (at
least seemingly) closer to the original transaction database than the
collection $\Frequent{\sigma,\DB}$ and their supports, the projections
could be useful to make the inverse frequent itemset mining problem
more comprehensible by an equivalent formulation of the problem.

\begin{problem}[database reconstruction from projections] \label{p:reco}
Given a collection $\pr{\Mx,\Fr}$ of projections onto maximal
itemsets, find a transaction database $\DB$ such that
$\pr{\Mx,\Fr}=\pr{\Mx,\DB}$.
\end{problem}

There are, however, collections of projections that cannot be realized
as downward closed itemset collections.  We should be able to ensure
in time polynomial in the sum of the cardinalities of transactions in
the projections that the collection of projections can be realized as
a downward closed itemset collection with some supports. Fortunately,
there are simple conditions that are necessary and sufficient to
ensure that there is a downward closed itemset collection compatible
with a given collection of projections.

\begin{theorem}
The projections $\pr{X_1,\Fr_1},\ldots,\pr{X_m,\Fr_m}$ have the
compatible collection $\Fr$ of itemsets, i.e., a collection $\Fr$ such
that 
\begin{displaymath}
\supp{Y,\pr{X_i,\Fr_i}}=\supp{Y,\Fr}
\end{displaymath}
for all $Y \subseteq X_i, 1\leq i \leq m$, if and only if
\begin{displaymath}
\pr{X_i \cap X_j, \Fr_i}=\pr{X_i \cap X_j, \Fr_j}
\end{displaymath}
for all $1\leq i,j\leq m$.
\end{theorem}
\begin{proof}
If there is a downward closed itemset collection $\Fr$ such that
\begin{displaymath}
\supp{Y,\pr{X_i,\Fr_i}}=\supp{Y,\Fr}
\end{displaymath}
for all $Y \subseteq X_i, 1 \leq i \leq m$, then
\begin{displaymath}
\pr{X_i \cap X_j,\Fr_i}=\pr{X_i \cap X_j,\Fr_j}
\end{displaymath}
for all $1 \leq i,j\leq m$. Otherwise $\pr{X_i \cap X_j,\Fr_i}$ and
$\pr{X_i \cap X_j,\Fr_j}$ would determine different supports for some
itemset $Y \subseteq X_i \cap X_j$ where $1\leq i,j\leq m$.

If 
\begin{displaymath}
\pr{X_i \cap X_j,\Fr_i}=\pr{X_i \cap X_j,\Fr_j}
\end{displaymath}
for all $1 \leq
i,j\leq m$, then 
\begin{displaymath}
\supp{Y,\pr{X_i \cap X_j,\Fr_i}}=\supp{Y,\pr{X_i \cap X_j,\Fr_j}}
\end{displaymath}
for all itemsets $Y \subseteq X_i \cap X_j$ where $1 \leq i,j \leq m$.
\end{proof}

The number of transactions in the transaction database $\DB$ can be
exponential in the number of frequent itemsets (and thus also in the
sum of the cardinalities of the frequent itemsets).

\begin{example}[a transaction database being exponentially larger than the frequent itemset collection]
Let the itemset collection consist of just one itemset $\emptyset$
with support exponential in $\Abs{\Items}$.  Then the number of
transactions in $\DB$ is exponential in $\Abs{\Items}$.  \exend
\end{example}

This fact does not have to be considered as a drawback since most of
the results shown in this chapter are hardness results. Furthermore,
it is reasonable to assume that if one is trying to reconstruct a
transaction database then the number of transaction in the database is
not considered to be unfeasibly large.

\section{The Computational Complexity of the Problem}

In this section we show that Problem~\ref{p:reco} is difficult in
general but some of its special cases can be solved in polynomial time
and even in logarithmic space.  Our first hardness result shows that
Problem~\ref{p:reco} is \NP-hard in general. The hardness is shown by
a reduction from the graph $3$-colorability problem:

\begin{problem}[graph $3$-colorability~\cite{b:garey79}] \label{p:3-colorability}
Given a graph $G=\Tuple{V,E}$, decide whether there is a good
$3$-coloring, i.e., a labeling $\lbl{} : V \to \Set{r,g,b}$ such that
$\lbl{u}\neq\lbl{v}$ for all $\Set{u,v} \in E$.
\end{problem}

\begin{theorem} \label{t:reco:NP}
The problem of deciding whether there is a transaction database $\DB$
compatible with the projections $\pr{\Mx,\Fr}$ (i.e., the decision
version of Problem~\ref{p:reco}) is \NP-complete even when the
compatible transaction databases consist of only six transactions.
\end{theorem}
\begin{proof}
The problem is clearly in \NP\ since it can be verified in time
polynomial in the sizes of $\pr{\Mx,\Fr}$ and $\DB$ whether a certain
transaction database $\DB$ is compatible with projections
$\pr{\Mx,\Fr}$ simply by computing the projections $\pr{\Mx,\DB}$.

We show the \NP-hardness of Problem~\ref{p:reco} by a reduction from
an instance $G=\Tuple{V,E}$ of the graph $3$-colorability problem
(Problem~\ref{p:3-colorability}) to projections $\pr{\Mx,\Fr}$ in
$\pr{\Mx,\Fr}$ are compatible with the projections
$\pr{\Maximal{\sigma,\DB},\DB}$ of some transaction database $\DB$ if
and only if $G$ is $3$-colorable.

Let the set $\Items$ of items be $\Set{r_v,g_v,b_v : v \in V}$.  The
projections are constructed as follows. For each edge $\Set{u,v} \in
E$ we define a projection
\begin{eqnarray*}
\pr{\Set{r_u,g_u,b_u,r_v,g_v,b_v},\Fr}&=&
\left\{\Tuple{1,\Set{r_u,g_v}},\Tuple{2,\Set{r_u,b_v}},\right. \\
&&\left. \Tuple{3,\Set{g_u,r_v}},\Tuple{4,\Set{g_u,b_v}}, \right. \\
&&\left. \Tuple{5,\Set{b_u,r_v}},\Tuple{6,\Set{b_u,g_v}}\right\}.
\end{eqnarray*}

If the graph $G=\Tuple{V,E}$ is not $3$-colorable then there is no
transaction database $\DB$ compatible with the projections: for every
$3$-coloring of $G$, there is an edge $\Set{u,v} \in E$ with
$\lbl{u}=\lbl{v}$ but none of the pairs $\Set{r_u,r_v}$,
$\Set{g_u,g_v}$, and $\Set{b_u,b_v}$ appear in the projection
$\pr{\Set{r_u,g_u,b_u,r_v,g_v,b_v},\Fr}$. Thus there is not even a
partial solution of one transaction compatible the projections.

If the graph $G$ is $3$-colorable then there is a transaction database
$\DB$ that is compatible with the projections: the six transactions in
the database $\DB$ are the six permutations of a $3$-coloring $\lbl{}$
such that $\lbl{u}\neq\lbl{v}$ for all $\Set{u,v} \in E$.
\end{proof}

As mentioned in the beginning of the chapter, it would be desirable to
be able to estimate how many compatible databases there exist. The
proof of Theorem~\ref{t:reco:NP} can also be adapted to give the
hardness result for the counting version of
Problem~\ref{p:reco}. (See~\cite{b:papadimitriou95} for more details
on counting complexity.)
\begin{theorem}
The problem of counting the number of transaction databases $\DB$
compatible with the projections $\pr{\Mx,\Fr}$ is $\#P$-complete.
\end{theorem}
\begin{proof}
The problem is in $\#P$ since its decision version is in \NP. Using
the reduction described in the proof of Theorem~\ref{t:reco:NP}, the
number of good $3$-colorings could be counted: the number of good
$3$-colorings is $1/6!=1/720$ times the number of transaction
databases compatible with the projections corresponding to the given
graph $G$. As counting the number of good $3$-colorings is
$\#P$-hard~\cite{b:garey79}, so is counting the number of compatible
databases.
\end{proof}

Although the database reconstruction problem is \NP-complete in
general, there are some special cases that can be solved in polynomial
time. In one of the most simplest such cases the instance consists of
only two projections (with arbitrary number of items).

\begin{theorem} \label{t:ifim2}
It can be decided in polynomial time whether there is a transaction
database $\DB$ that is compatible with given projections
$\pr{X_1,\Fr_1}$ and $\pr{X_2,\Fr_2}$. Furthermore, the number of
compatible transaction databases $\DB$ can be computed in polynomial
time.
\end{theorem}
\begin{proof}
By definition, the projection $\pr{X_1,\Fr_1}$ is compatible with a
transaction database $\DB$ if and only if $\pr{X_1,\Fr_1}=\pr{X_1,\DB}$
and the projection $\pr{X_2,\Fr_2}$ is compatible with $\DB$ if and only
if $\pr{X_2,\Fr_2}=\pr{X_1,\DB}$. The database $\DB$ compatible with both
projections if and only if 
\begin{displaymath}
\pr{X_1 \cap X_2,\Fr_1}=\pr{X_1 \cap
X_2,\DB}=\pr{X_1 \cap X_2,\Fr_2}, 
\end{displaymath}
\begin{displaymath}
\pr{X_1 \setminus X_2,\Fr_1}=\pr{X_1
\setminus X_2,\DB}
\end{displaymath}
and
\begin{displaymath}
\pr{X_2 \setminus X_1,\Fr_2}=\pr{X_2 \setminus
X_1,\DB}.
\end{displaymath}

A transaction database $\DB$ compatible with the two projections
$\pr{X_1,\Fr_1}$ and $\pr{X_2,\Fr_2}$ can be found by sorting the
transactions in the projections $\pr{X_1,\Fr_1}$ and $\pr{X_2,\Fr_2}$
with respect to the itemsets in $\pr{X_1 \cap X_2,\Fr_1}$ and $\pr{X_1
\cap X_2,\Fr_2}$, respectively. This can be implemented to run in time
$\Oh{\Abs{X_1 \cap X_2}\Abs{\DB}}$~\cite{b:knuth:3}.  This method for
constructing the compatible database is shown as
Algorithm~\ref{a:From-Two-To-One}.  The running time of the algorithm
is linear in the size of the input, i.e., in the sum of the
cardinalities of the transactions in the projections.

\begin{algorithm}
\caption{An algorithm for constructing a transaction database $\DB$
compatible with projections $\pr{X_1,\Fr_1}$ and $\pr{X_2,\Fr_2}$.\label{a:From-Two-To-One}}
\begin{algorithmic}[1]
\Input{Projections $\pr{X_1,\Fr_1}$ and $\pr{X_2,\Fr_2}$.}  
\Output{A transaction database $\DB$ compatible with $\pr{X_1,\Fr_1}$
and $\pr{X_2,\Fr_2}$, or $\emptyset$ if such a database does not
exist.}
\Function{From-Two-To-One}{$\pr{X_1,\Fr_1},\pr{X_2,\Fr_2}$}
\State $\Pat_1 \leftarrow \emptyset$
\ForAll{$\Tuple{i,Y} \in \pr{X_1,\Fr_1}$}
 \State $X \leftarrow Y \cap X_2$
 \State $\Pat_1 \leftarrow \Pat_1 \cup \Set{X}$ 
 \State $\SubPat^1_X \leftarrow \SubPat^1_X \cup \Set{\Tuple{i,Y}}$
\EndFor
\State $\Pat_2 \leftarrow \emptyset$
\ForAll{$\Tuple{j,Z} \in \pr{X_2,\Fr_2}$}
 \State $X \leftarrow Z \cap X_1$
 \State $\Pat_2 \leftarrow \Pat_2 \cup \Set{X}$
 \State $\SubPat^2_Y \leftarrow \SubPat^2_X \cup \Set{\Tuple{j,Z}}$
\EndFor
\If{$\Pat_1\neq \Pat_2$}
\State \Return $\emptyset$
\EndIf
\State $\DB \leftarrow \emptyset$
\ForAll{$X \in \Pat_1$}
 \If{$\Abs{\SubPat^1_X}\neq \Abs{\SubPat^2_X}$}
  \State \Return $\emptyset$
 \EndIf
 \While{$\SubPat^1_X \neq \emptyset$}
  \State Choose $\Tuple{i,Y} \in \SubPat^1_X$ and $\Tuple{j,Z} \in \SubPat^2_X$ arbitrarily.
  \State $\DB \leftarrow \DB \cup \Tuple{\Abs{\DB}+1,Y \cup Z}$
  \State $\SubPat^1_X \leftarrow \SubPat^1_X \setminus \Set{\Tuple{i,Y}}$
  \State $\SubPat^2_X \leftarrow \SubPat^2_X \setminus \Set{\Tuple{j,Z}}$
 \EndWhile
\EndFor
\State \Return $\DB$
\EndFunction
\end{algorithmic}
\end{algorithm}

The number of transaction databases compatible with the projections
$\pr{X_1,\DB}$ and $\pr{X_2,\DB}$ of a given transaction database $\DB$
can be computed from the counts $\cnt{X,\pr{X_1 \cap X_2,\DB}}$,
$\cnt{Y_1,\pr{X_1,\DB}}$ and $\cnt{Y_2,\pr{X_2, \DB}}$ for all $X$,
$Y_1$ and $Y_2$ such that $X =Y_1 \cap X_2=Y_2 \cap X_1$, $Y_1
\subseteq X_1$, $Y_2 \subseteq X_2$, $\cnt{Y_1,\pr{X_1,\DB}}>0$ and
$\cnt{Y_2,\pr{X_2,\DB}}>0$.

The collection 
\begin{displaymath}
\SubPat=\Set{X \subseteq X_1 \cap X_2 :
\cnt{X,\pr{X_1 \cap X_2,\DB}}>0}
\end{displaymath}
partitions the transactions in $\pr{X_1,\DB}$ and $\pr{X_2,\DB}$ into
equivalence classes of transactions with the same projections to $X_1
\cap X_2$.  The partition can be further refined by the collections
\begin{displaymath}
\SubPat^1_X=\Set{Y_1 \subseteq X_1 : Y_1 \cap X_2=X,
\cnt{Y_1,\pr{X_1,\DB}}>0}
\end{displaymath}
 and
\begin{displaymath}
\SubPat^2_X=\Set{Y_2 \subseteq X_2
: Y_2 \cap X_2=X, \cnt{Y_2,\pr{X_2,\DB}}>0}.
\end{displaymath}

Using these collections, the number of compatible databases can be
computed as follows.  The transaction identifiers can be partitioned
to classes $X \in \SubPat$ in 
\begin{displaymath}
c=\frac{\Abs{\DB}!}{\prod_{X \in \SubPat}
\cnt{X,\pr{X_1 \cap X_2,\DB}}!}
\end{displaymath}
ways.  In each class $X \in \SubPat$,
the transaction identifiers can be further partitioned into classes $Y
\in \SubPat^1_X$ in 
\begin{displaymath}
a_{X}=\frac{\cnt{X,\pr{X_1 \cap X_2,\DB}}!}{\prod_{Y_1 \in
\SubPat^1_X} \cnt{Y_1,\pr{X_1,\DB}}!}
\end{displaymath}
ways. Now we have counted the number of different projections
$\pr{X_1,\DB}$.  The number of different databases that can be
obtained by merging the transactions in $\pr{X_2,\DB}$ to the
transactions of $\pr{X_1,\DB}$ using the transaction identifiers of
$\pr{X_1,\DB}$ is
\begin{displaymath}
b_{X}=\frac{\cnt{X,\pr{X_1 \cap
X_2,\DB}}!}{\prod_{Y_2 \in \SubPat^2_X} \cnt{Y_2,\pr{X_2,\DB}}!}.
\end{displaymath}
Thus, the total number of transaction databases compatible with
$\pr{X_1,\DB}$ and $\pr{X_2,\DB}$ is $c\prod_{X \in \SubPat} a_Xb_X$.
\end{proof}

The practical relevancy of this positive result
(Theorem~\ref{t:ifim2}) depends on how much the domains $X_1$ and
$X_2$ overlap. If $\Abs{X_1 \cap X_2}$ is very small but $\Abs{X_1
\cup X_2}$ is large then there is a great danger that there are
several compatible transaction databases. Fortunately, in the case of
two projections we are able to efficiently count the number of
compatible databases and thus to evaluate the usefulness of the found
database.

In the simplest case of the database reconstruction problem all
projections $\pr{X_1,\Fr_1},\ldots,\pr{X_m,\Fr_m}$ are disjoint since
in that case any database with projections
$\pr{X_1,\Fr_1},\ldots,\pr{X_m,\Fr_m}$ is compatible one.
Unfortunately this also means that the number compatible databases is
very large. Thus, one should probably require something more than mere
compatibility.

One natural restriction, applying the Occam's razor, is to search for
the compatible database with the smallest number of transactions with
different itemsets. This kind of database is (in some sense) the
simplest hypothesis based on the downward closed itemset collection.
This can be beneficial for both analyzing the data and actioning using
the database. 

Unfortunately, it can be shown that finding the transaction database
with the smallest number of different transactions is \NP-hard for
already two disjoint projections. We show the \NP-hardness by a
reduction from $3$-partition problem:

\begin{problem}[$3$-partition~\cite{b:garey79}] \label{p:3-partition}
Given a set $A$ of $3l$ elements, a bound $B \in \NN$, and a size
$\sz{a}\in \NN$ for each $a \in A$ such that $B/4 < \sz{a} <B/2$ and
such that $\sum_{a \in A}\sz{a}=lB$, decide whether or not $A$ can be
partitioned into $l$ disjoint sets $A_1,\ldots,A_l$ such that for each
$\sum_{a \in A_i}=B$ for all $1 \leq i \leq l$.
\end{problem}

\begin{theorem}
It is \NP-hard to find a transaction database consisting of the
smallest number of different transactions and being compatible with
the projections $\pr{X_1,\Fr}$ and $\pr{X_2,\Fr}$ such that $X_1 \cap
X_2=\emptyset$.
\end{theorem}
\begin{proof}
We show the \NP-hardness of the problem by reduction from the
$3$-partition problem (Problem~\ref{p:3-partition}).

As Problem~\ref{p:3-partition} is known to be strongly \NP-complete,
we can assume that the sizes $\sz{a}$ of all elements $a \in A$ are
bounded above by polynomial in $l$.

The instance $\Tuple{A,B,\sz{}}$ of $3$-partition can be encoded as
two projections as follows. Without loss of generality, let the
elements of $A$ be $1,\ldots,3l$. Then
\begin{displaymath}
X_1=\Set{1,\ldots,\Ceil{3\log l}}
\end{displaymath}
and
\begin{displaymath}
X_2=\Set{\Ceil{\log 3l}+1,\ldots,\Ceil{\log 3l}+\Ceil{\log l}}. 
\end{displaymath}

Again, let us denote the binary coding of $x \in \NN$ as a set
consisting the positions of ones in the binary code by $\bin{x}$.
Then projection $\pr{X_1,\Fr}$ consists of $\sz{a}$ transactions
consisting of the itemset $\bin{a} \subseteq X_1, a \in A$. Projection
$\pr{X_2,\Fr}$ consists of $B$ transactions consisting of the itemset
$\bin{b}+\Ceil{\log 3l}\subseteq X_2,b \in \Set{1,\ldots,l}$.

Clearly there is a $3$-partition for $\Tuple{A,B,\sz{}}$ if and only
if there is a database $\DB$ with $3l$ different transactions that is
compatible with projections $\pr{X_1,\Fr}$ and $\pr{X_2,\Fr}$.
\end{proof}

Finally, let us note that if the number of items is fixed, then a
compatible transaction database can be found in time polynomial in the
number of transactions in the projections: Finding a transaction
database compatible with the projections can be formulated as a linear
integer programming task where the variables are the possible
different itemsets in the transactions.  The number of possible
different itemsets is $2^{\Abs{\Items}}$. The linear integer
programming tasks with a fixed number of variables can be solved in
time polynomial in the size of the linear
equations~\cite{a:lenstra83}.

%% file: conclusions.tex

\chapter{Conclusions \label{c:conclusions}}

Pattern discovery is an important subfield of data mining that
attempts to discover interesting (or high-quality) patterns from
data. There are several efficient techniques to discover such patterns
with respect to different interestingness measures. Merely discovering
the patterns efficiently is rarely the ultimate goal, but the patterns
are discovered for some purpose. One important use of patterns is to
summarize data, since the pattern collections together with the
quality values of the patterns can be considered a summaries of the
data.

In this dissertation we have studied how the pattern collections could
be summarized. Our approach has been five-fold.

First, we studied how to cast views to pattern collections by
simplifying the quality values of the patterns. In particular, we gave
efficient algorithms for optimally discretizing the quality
values. Furthermore, we described how the discretizations can be used
in conjunction with pruning of redundant patterns to simplify the
pattern collections.

Second, continuing with the theme of simplifying pattern collections,
we considered the trade-offs between the understandability and the
accuracy of the pattern collections and their quality values. As a
solution that supports exploratory data analysis, we proposed the
pattern orderings. A pattern ordering of a pattern collection lists
the patterns in such an order that each pattern improves our estimate
about the whole pattern collection as much as possible (with respect
to given loss function and estimation method). Furthermore, we showed
that under certain reasonable assumptions each length-$k$ prefix of
the pattern ordering provides a $k$-subcollection of patterns that is
almost as good description of the whole pattern collection as the best
$k$-subcollection. We illustrated the applicability of pattern
orderings in approximating pattern collections and data.

Third, we examined how the structural properties (especially partial
orders) of the pattern collections can be exploited to obtain
clusterings of the patterns and more concise descriptions of the
pattern collections.  The same techniques can be used to simplify also
transaction databases.

Fourth, we proposed a generalization of association rules: change
profiles. A change profile of a pattern describes how the quality
value of the pattern has to be changed to obtain the quality values of
neighboring patterns. The change profiles can be used to compare
patterns with each other: patterns can be considered similar, if their
change profiles are similar. We studied the computational complexity
of clustering patterns based on their change profiles. The problem
turned out to be quite difficult if some approximation quality
requirements are given. This does not rule out the use of heuristic
clustering methods or hierarchical clustering.  We illustrated the
hierarchical clusterings of change profiles using real data.  In
addition to clustering change profiles, we considered frequency
estimation from approximate change profiles that could be used as
building blocks of condensed representations of pattern collections.
We provided efficient algorithms for the frequency estimation from the
change profiles and evaluated empirically the noise tolerance of the
methods.

Fifth, we studied the problem of inverse pattern discovery, i.e., the
problem of constructing data sets that could have induced the given
patterns and their quality values. More specifically, we studied the
computational complexity of inverse frequent itemset mining.  We
showed that the problem of finding a transaction database compatible
with a given collection of frequent itemsets and their supports is
\NP-hard in general, but some of its special cases are solvable in
polynomial time.

Although the problems studied in this dissertation are different, they
have also many similarities.  Frequency simplifications, pattern
orderings, pattern chains and change profiles are all techniques for
summarizing pattern collections.  Frequency simplifications and
pattern orderings provide primarily approximations of the pattern
collections, whereas pattern chains and change profiles describe the
pattern collection by slightly more complex patterns obtained by
combining the patterns of the underlying pattern collection.

There are also many other ways to group the techniques.  For example,
the following similarities and dissimilarities can be observed:
\begin{itemize}
\item
Pattern orderings, pattern chains and change profiles make use of the
relationships between the patterns directly, whereas frequency
simplifications do not depend on the actual patterns.
\item
Frequency simplifications, pattern orderings and change profiles can
be used to obtain an approximate description of the pattern collection,
whereas pattern chains provide an exact description.
\item
Frequency simplifications, pattern orderings and pattern chains
describe the quality values of the patterns, whereas change profiles
describe the changes in the quality values.
\item
Frequency simplifications, pattern chains and change profiles can be
used to cluster the patterns, whereas the interpretation of pattern
orderings as clusterings is not so straightforward.
\end{itemize}
Also inverse pattern discovery has similarities with the other
problems, as all the problems are related to the problem of evaluating
the quality of the pattern collection.  Furthermore, all problems are
closely related to the two high-level themes of the dissertation,
namely post-processing and condensed representations of pattern
collections.

As future work, exploring the possibilities and limitations of
condensed representations of pattern collections is likely to be
continued. One especially interesting question is how the pattern
collections should actually be represented. Some suggestions are
provided
in~\cite{i:mielikainen04:ssi,i:mielikainen04:automata,i:mielikainen04:implicit}.
Also, measuring the complexity of the data and its relationships to
condensed representations seems to be an important and promising
research topic.

As data mining is inherently exploratory process involving often huge
data sets, a proper data management infrastructure seems to be
necessary.  A promising model for that, and for data mining as whole,
is offered by inductive databases~\cite{p:dsdma04}.  There are many
interesting questions related to inductive databases.  For example, it
is not completely clear what inductive databases are or what they
should be~\cite{i:mielikainen04:idb}.

Recently also the privacy issues of data mining have been recognized
to be of high importance~\cite{a:pinkas02,a:verykios04:survey}. There
are two very important topics in privacy preserving data
mining. First, sometimes no one has access to the whole data but still
the data owners are interested in mining the data. There has been
already many proposals for secure computation of many data mining
results, for example frequent
itemsets~\cite{i:evfimievski02,i:freedman04,i:goethals04,i:vaidya02}.
Second, in addition to computing the data mining results securely, it
is often very important that the data mining results themselves are
secure, i.e., that they do not leak any sensitive information about
the
data~\cite{a:farkas02,i:mielikainen04:anonymity,i:oliveira04,a:saygin01,a:verykios04}.

Another important problem related to inductive databases is finding
the underlying general principles of pattern
discovery~\cite{a:mannila97:levelwise}. There are many pattern
discovery algorithms, but it is still largely open what are the
essential differences between the methods and how to choose the
technique for some particular pattern discovery task. Some preliminary
evaluation of the techniques in the case of frequent itemset mining
has recently been done~\cite{p:fimi03}, but the issues of a general
theory of pattern discovery are still largely open.


%% file: asarja.tex

%
\def\loppu{$^{\rm\scriptsize\dag}$}
\def\aloitus{\topsep0pt\partopsep0pt\itemsep0pt\parsep\parskip
\rightmargin0pt\listparindent0pt\itemindent0pt
\leftmargin14mm\labelsep1mm\labelwidth13mm
\def\makelabel##1{##1\hfill}}
\makeatletter
\def\ps@headings{\let\@mkboth\@gobbletwo
 \def\@oddfoot{} \def\@evenfoot{}
 \def\@oddhead{\hbox{}\hfil\hbox to 0pt{\bf\enspace\scriptsize\thepage\hss}}
 \def\@evenhead{{\bf\scriptsize\thepage\enspace}\hfil}
}
\def\ps@fpage{\let\@mkboth\@gobbletwo
 \def\@oddhead{} \def\@evenhead{}
 \let\@evenfoot\@oddfoot
}
\makeatother
\pagestyle{empty} 
%
\textwidth 125mm
\textheight 195mm
\footskip 30bp 
\parindent0pt \parskip 5pt plus 2pt minus 1pt
%
%
\pagebreak
\thispagestyle{fpage}

\scriptsize
\tabcolsep0pt
\begin{tabular*}{\textwidth}{l@{\extracolsep{\fill}}l}
TIETOJENK\"ASITTELYTIETEEN LAITOS & DEPARTMENT OF COMPUTER SCIENCE\\
PL 68 (Gustaf Hällströmin katu 2 b) & P.O. Box 68 (Gustaf Hällströmin katu 2 b)\\
00014 Helsingin yliopisto & FIN-00014 University of Helsinki, {\sc Finland}
\\
\strut & \\
JULKAISUSARJA {\bf A} & SERIES OF PUBLICATIONS {\bf A}
\end{tabular*}
\vskip6bp
\underline{Reports may be ordered from}: Kumpula Science Library,
P.O. Box 64, FIN-00014 University
of Helsinki, {\sc Finland}.

\begin{list}{}{\aloitus}

\item[A-1996-1] R.~Kaivola: Equivalences, preorders and compositional
verification for linear time temporal logic and concurrent systems.
185~pp. (Ph.D.\ thesis).
\item[A-1996-2] T.~Elomaa: Tools and techniques for decision
tree learning. 140~pp. (Ph.D.\ thesis).
\item[A-1996-3] J.~Tarhio \& M.~Tienari~(eds.): Computer Science
at the University of Helsinki 1996. 89~pp.
\item[A-1996-4] H.~Ahonen: Generating grammars for structured
documents using grammatical inference methods. 107~pp.
(Ph.D.\ thesis).
\item[A-1996-5] H.~Toivonen: Discovery of frequent patterns
in large data collections. 116~pp. (Ph.D.\ thesis).
\item[A-1997-1] H. Tirri: Plausible prediction by Bayesian
inference. 158 pp. (Ph.D.\ thesis).
\item[A-1997-2] G. Lind\'en: Structured document transformations.
122~pp. (Ph.D.\ thesis).
\item[A-1997-3] M. Nyk\"anen: Querying string databases with modal
logic. 150~pp. (Ph.D.\ thesis).
\item[A-1997-4] E. Sutinen, J. Tarhio, S.-P. Lahtinen, A.-P. Tuovinen,
E. Rautama \& V. Meisalo: Eliot -- an algorithm animation 
environment. 49~pp.
\item[A-1998-1] G. Lind\'en \& M. Tienari (eds.): Computer Science 
at the University of Helsinki 1998. 112~pp. 
\item[A-1998-2] L. Kutvonen: Trading services in open
distributed environments. 231 + 6 pp. (Ph.D.\ thesis). 
\item[A-1998-3] E. Sutinen: Approximate pattern matching with 
the q-gram family. 116 pp. (Ph.D.\ thesis). 
\item[A-1999-1] M. Klemettinen: A knowledge discovery methodology
for telecommunication network alarm databases. 137 pp. (Ph.D.\ thesis). 
\item[A-1999-2] J. Puustj\"arvi: Transactional workflows. 104 pp.
(Ph.D.\ thesis).
\item[A-1999-3] G. Lind\'en \& E. Ukkonen (eds.): Department of 
Computer Science: annual report 1998. 55~pp. 
\item[A-1999-4] J. K\"arkk\"ainen: Repetition-based text indexes.
106 pp. (Ph.D.\ thesis).
\item[A-2000-1] P. Moen: Attribute, event sequence, and event type
similarity notions for data mining. 190+9 pp. (Ph.D. thesis).
\item[A-2000-2] B. Heikkinen: Generalization of document structures and
document assembly. 179 pp. (Ph.D. thesis). 
\item[A-2000-3] P. K\"ahkipuro: Performance modeling framework for CORBA
based distributed systems. 151+15 pp. (Ph.D. thesis).
\item[A-2000-4] K. Lemstr\"om: String matching techniques for music
retrieval. 56+56 pp. (Ph.D.Thesis).
\item[A-2000-5] T. Karvi: Partially defined Lotos specifications and
their refinement relations. 157 pp. (Ph.D.Thesis).
\item[A-2001-1] J.~Rousu: Efficient range partitioning 
in classification learning. 68+74 pp. (Ph.D.\ thesis)
\item[A-2001-2] M.~Salmenkivi: Computational methods for
intensity models. 145 pp. (Ph.D.\ thesis)
\item[A-2001-3] K.~Fredriksson: Rotation invariant template
matching. 138 pp. (Ph.D.\ thesis)
\item[A-2002-1] A.-P.~Tuovinen: Object-oriented engineering of
visual languages. 185 pp. (Ph.D.\ thesis)
\item[A-2002-2] V.~Ollikainen: Simulation techniques for disease
gene localization in isolated populations. 149+5 pp. (Ph.D.\ thesis)
\item[A-2002-3] J.~Vilo: Discovery from biosequences. 149 pp. (Ph.D.\ thesis)
\item[A-2003-1] J.~Lindstr\"om: Optimistic concurrency control methods
for real-time database systems. 111 pp. (Ph.D.\ thesis)
\item[A-2003-2] H.~Helin: Supporting nomadic agent-based applications
in the FIPA agent architecture. 200+17 pp. (Ph.D.\ thesis)
\item[A-2003-3] S.~Campadello: Middleware infrastructure for distributed
mobile applications. 164 pp. (Ph.D.\ thesis)
\item[A-2003-4] J.~Taina: Design and analysis of a distributed
database architecture for IN/GSM data. 130 pp. (Ph.D.\ thesis)
\item[A-2003-5] J.~Kurhila: Considering individual differences in 
computer-supported special and elementary education. 135 pp. (Ph.D.\ thesis)
\item[A-2003-6] V.~M\"akinen: Parameterized approximate string matching and 
local-similarity-based point-pattern matching. 144 pp. (Ph.D.\ thesis)
\item[A-2003-7] M.~Luukkainen: A process algebraic reduction strategy
     for automata theoretic
       verification of untimed and timed concurrent systems. 
       141 pp. (Ph.D.\ thesis)
\item[A-2003-8] J.~Manner: Provision of quality of service
     in IP-based mobile access networks. 
       191 pp. (Ph.D.\ thesis)
\item[A-2004-1] M.~Koivisto: Sum-product algorithms for the analysis of genetic risks.
     155 pp. (Ph.D.\ thesis)
\item[A-2004-2] A.~Gurtov: Efficient data transport in wireless overlay networks.
     141 pp. (Ph.D.\ thesis)
\item[A-2004-3] K.~Vasko: Computational methods and models for paleoecology.
     176 pp. (Ph.D.\ thesis)
\item[A-2004-4] P.~Sevon: Algorithms for Association-Based Gene Mapping.
     101 pp. (Ph.D.\ thesis)
\item[A-2004-5] J.~Viljamaa: Applying Formal Concept Analysis to Extract Framework Reuse Interface Specifications from Source Code.
     206 pp. (Ph.D.\ thesis)
\item[A-2004-6] J.~Ravantti: Computational Methods for Reconstructing Macromolecular Complexes from Cryo-Electron Microscopy Images.
     100 pp. (Ph.D.\ thesis)
\item[A-2004-7] M.~K\"a\"ari\"ainen: Learning Small Trees and Graphs that Generalize.
     45+49 pp. (Ph.D.\ thesis)
\item[A-2004-8] T.~Kivioja: Computational Tools for a Novel Transcriptional Profiling Method.
     98 pp. (Ph.D.\ thesis)
\item[A-2004-9] H.~Tamm: On Minimality and Size Reduction of One-Tape and Multitape Finite Automata.
     80 pp. (Ph.D.\ thesis)
\end{list}